\documentclass[aps,pra,reprint,superscriptaddress]{revtex4-2}

\usepackage[english]{babel}

\usepackage{amsfonts,amsmath,amssymb,amsthm}
\usepackage{graphicx}
\usepackage[dvipsnames]{xcolor}
\usepackage{array}
\usepackage{enumitem}
\usepackage{mathpazo}
\usepackage{setspace}
\usepackage{bm}
\usepackage{amssymb}
\usepackage{amsthm}
\usepackage{mathtools}
\usepackage{multirow}
\usepackage{cases}

\usepackage[colorlinks=true,linkcolor=blue,citecolor=red,linktocpage=true,breaklinks=true]{hyperref}
\usepackage{mathptmx}
\usepackage[framemethod=default]{mdframed}
\usepackage{tcolorbox}
\usepackage{tikz} 
\usepackage{url}
\usepackage{varwidth}

\usepackage{pgfplots}

\usepackage{youngtab}
\usepackage{subfigure}
\usepackage{verbatim}
\usepackage[capitalise]{cleveref}
\usepackage{float}
\usepackage{qcircuit}

\usepackage{tabularx}
\usepackage{listings}

\RequirePackage[framemethod=default]{mdframed}

\newcounter{dummy-example}
\newcounter{dummy-exercise} 
\newcounter{dummy-knowledge}

\newtheoremstyle{maincolornumbox}
{0pt}
{0pt}
{\normalfont}
{}
{\small
	\sffamily\color{maincolor}}
{\;}
{0.25em}
{\sffamily\bfseries\color{maincolor}\thmname{#1} \thmnumber{#2}\thmnote{\sffamily\mdseries\color{black}{ ---}\nobreakspace #3.}\newline\noindent} 

\theoremstyle{maincolornumbox}
\newtheorem*{ProofT*}{Proof}
\newtheorem{KnowledgeT}[dummy-knowledge]{Knowledge}
\newtheorem{exampleT}[dummy-example]{Example}
\newtheorem{exerciseT}[dummy-exercise]{Exercise}

\newmdenv[skipabove=7pt,
skipbelow=7pt,
rightline=false,
leftline=true,
topline=false,
bottomline=false,
linecolor=maincolor,
backgroundcolor=maincolor!10,
innerleftmargin=5pt,
innerrightmargin=5pt,
innertopmargin=5pt,
leftmargin=0cm,
rightmargin=0cm,
linewidth=3pt,
innerbottommargin=5pt]{dBox}

\newmdenv[skipabove=7pt,
skipbelow=7pt,
rightline=false,
leftline=true,
topline=false,
bottomline=false,
backgroundcolor=lightgrey,
linecolor=maincolor,
innerleftmargin=5pt,
innerrightmargin=5pt,
innertopmargin=5pt,
innerbottommargin=5pt,
leftmargin=0cm,
rightmargin=0cm,
linewidth=3pt]{eBox}

\definecolor{aqua1}{HTML}{4FC1E9}
\definecolor{aqua2}{HTML}{3BAFDA}
\definecolor{mint1}{HTML}{48CFAD}
\definecolor{mint2}{HTML}{37BC9B}
\definecolor{bittersweet1}{HTML}{FC6E51}
\definecolor{bittersweet2}{HTML}{E9573F}
\definecolor{lightgrey1}{HTML}{F5F7FA}
\definecolor{lightgrey2}{HTML}{E6E9ED}
\definecolor{grapefruit1}{HTML}{ED5565}
\definecolor{grapefruit2}{HTML}{DA4453}
\definecolor{sunflower1}{HTML}{FFCE54}
\definecolor{sunflower2}{HTML}{F6BB42}
\definecolor{bluejeans1}{HTML}{5D9CEC}
\definecolor{bluejeans2}{HTML}{4A89DC}
\definecolor{lavander1}{HTML}{AC92EC}
\definecolor{lavander2}{HTML}{967ADC}
\definecolor{pinkrose1}{HTML}{EC87C0}
\definecolor{pinkrose2}{HTML}{D770AD}
\definecolor{grass1}{HTML}{A0D468}
\definecolor{grass2}{HTML}{8CC152}
\definecolor{IoPpurple}{HTML}{A89CCE}
\definecolor{IoPdarkpurple}{HTML}{7E6FB0}

\definecolor{bsblue}{HTML}{0d6efd}
\definecolor{bsindigo}{HTML}{6610f2}
\definecolor{bspurple}{HTML}{6f42c1}
\definecolor{bspink}{HTML}{d63384}
\definecolor{bsred}{HTML}{dc3545}
\definecolor{bsorange}{HTML}{fd7e14}
\definecolor{bsyellow}{HTML}{ffc107}
\definecolor{bsgreen}{HTML}{28a745}
\definecolor{bsteal}{HTML}{20c997}
\definecolor{bscyan}{HTML}{17a2b8}

\colorlet{maincolor}{bspurple}
\colorlet{altcolor}{bsorange}
\colorlet{lightgrey}{black!5}

\colorlet{varcolour}{bsorange}
\colorlet{dualcolour}{bsred}
\colorlet{datacolour}{bsblue}
\colorlet{linkcolour}{bspurple}

\newcommand{\equalcontribution}{These authors contributed equally to this work.}

\begin{document}
\title{From Computing to Quantum Mechanics: Accessible and Hands-On Quantum Computing Education for High School Students}

\author{Qihong Sun}
\thanks{\equalcontribution}

\author{Shuangxiang Zhou}
\thanks{\equalcontribution}

\author{Ronghang Chen}
\thanks{\equalcontribution}

\affiliation{College of Physics and Electronic Engineering, Center for Computational Sciences,  Sichuan Normal University, Chengdu 610068, China}

\author{Guanru Feng}

\affiliation{Shenzhen SpinQ Technology Co., Ltd., Shenzhen, China}

\author{Shi-Yao Hou}
\email[Corresponding author:]{hshiyao@sicnu.edu.cn}
\affiliation{College of Physics and Electronic Engineering, Center for Computational Sciences,  Sichuan Normal University, Chengdu 610068, China}
\affiliation{Department of Physics, The Hong Kong University of Science and Technology, Clear Water Bay, Kowloon, Hong Kong, China}

\author{Bei Zeng}
\affiliation{Department of Physics, The Hong Kong University of Science and Technology, Clear Water Bay, Kowloon, Hong Kong, China}

\begin{abstract}

This paper outlines an alternative approach to teaching quantum computing at the high school level, tailored for students with limited prior knowledge in advanced mathematics and physics. This approach diverges from traditional methods by building upon foundational concepts in classical computing before gradually introducing quantum mechanics, thereby simplifying the entry into this complex field. The course was initially implemented in a program for gifted high school students under the Hong Kong Education Bureau and received encouraging feedback, indicating its potential effectiveness for a broader student audience. A key element of this approach is the practical application through portable NMR quantum computers, which provides students with hands-on experience. The paper describes the structure of the course, including the organization of the lectures, the integration of the hardware of the  portable nuclear magnetic resonance (NMR) quantum computers, the Gemini/Triangulum series, and detailed lecture notes in an appendix. The initial success in the specialized program and ongoing discussions to expand the course to regular high schools in Hong Kong and Shenzhen suggest the viability of this approach for wider educational application. By focusing on accessibility and student engagement, this approach presents a valuable perspective on introducing quantum computing concepts at the high school level, aiming to enhance student understanding and interest in the field.

\end{abstract}

\maketitle


\section{Introduction}

Quantum computing stands at the forefront of the next technological revolution, offering unprecedented computational power and potential for a variety of applications \cite{22_aru,23_al,24_pre}. However, its complexity and reliance on advanced concepts in mathematics and physics have traditionally kept it confined to higher education, specifically at the graduate level. Recent trends, however, indicate a shift towards introducing quantum computing concepts at the high school level, recognizing the importance of early exposure to these future-oriented technologies \cite{25_Schuld}.

Initiatives such as MIT's ``Qubit by Qubit" program have been pioneers in bringing quantum computing education to high school students through summer camps and year-long courses \cite{mit_qubit}. Similarly, hands-on workshops designed for grades 9-12 offer innovative and engaging activities to teach basic quantum computing concepts \cite{qc_workshops}. These efforts reflect a broader movement in K-12 education to incorporate quantum computing instruction, which was previously reserved for advanced graduate programs \cite{k12_quantum}. The urgency of this shift is underscored by the growing consensus among educators on the necessity of teaching quantum information science at the high school level to maintain technological competitiveness \cite{quantum_highschool_urgency}. Collaborative efforts between educational institutions and industry, like the partnership between Stanford Quantum Computing Association and qBraid, further exemplify this trend by developing introductory quantum computing curricula for high school students \cite{stanford_qbraid}.

In this context, our paper introduces an innovative approach to teaching quantum computing at the high school level. Designed to cater to students without a strong background in advanced mathematics or physics, this method focuses on making quantum computing accessible and engaging. The course spans eight weeks and is divided into four parts: Introduction to Quantum Computing, Matrices for Quantum Computing, Quantum Circuit Model, and How to Make a Quantum Computer. Each part builds upon the previous, gradually introducing students to more complex concepts in a simplified and comprehensible manner.

The pedagogical approach of this course is strategically designed to bridge the gap between conventional computing and the cutting-edge field of quantum computing. This method is particularly beneficial for high school students who may not possess extensive backgrounds in advanced mathematics or physics. The course commences by laying a solid foundation in the fundamentals of computing, ensuring that students understand the basic principles of binary numbers, Boolean logic \cite{31_sve,32_sheldon,33_john}, and computer architecture \cite{34_sta}. This grounding in classical computing concepts is crucial as it sets the stage for a smoother transition into the more abstract and less intuitive quantum computing concepts.

As the course progresses, students are gradually introduced to quantum-specific topics. This gradual introduction is pivotal in avoiding the overwhelming impact that immediate exposure to complex topics like linear algebra and quantum mechanics \cite{41_gri,42_r,43_sak,44_squ} can often have. By contextualizing quantum computing within the broader landscape of general computing and then incrementally introducing quantum concepts, the course maintains student engagement and curiosity. This methodology is crucial for sustaining student interest and ensuring a comprehensive understanding of the subject matter.

A distinctive feature of this course is the incorporation of hands-on experience with quantum computing hardware. This practical aspect is realized through portable NMR quantum computers, the Gemini/Triangulum series. The inclusion of such devices provides students with a tangible connection to the theoretical concepts discussed in the course. It allows them to observe and experiment with real quantum computing processes, thereby reinforcing their learning and enhancing their comprehension of quantum mechanics \cite{43_sak} in a practical setting.
The hands-on learning aspect of the course plays a crucial role in clarifying quantum computing concepts \cite{4_sx}, making them more accessible and concrete for students.

This paper's organization mirrors the planning and strategy behind the quantum computing course tailored for high school students. Sec. II. provides an in-depth introduction to the organization of the lectures, outlining how each component of the course builds upon the previous to offer a gradual and comprehensive understanding of quantum computing. This section details the pedagogical strategies employed to transition from fundamental computing concepts to the more advanced topics of quantum mechanics, ensuring that these complex ideas are presented in an accessible and engaging manner \cite{45_bru,46_Reiser}.

Following this, in Sec. III., the paper introduces the hardware of the portable quantum computer, a key element of the course that offers students hands-on experience with real quantum computing. The inclusion of portable NMR quantum computers is a pivotal aspect of the course, allowing students to directly engage with the quantum computing processes they learn about in theory. This practical experience is crucial in solidifying students' understanding of quantum computing principles and bridging the gap between theoretical knowledge and real-world applications.

The paper is complemented by supplementary materials, which include detailed lecture notes. These notes are an essential resource for both educators and students, offering a complete guide to the course content. They are carefully designed to match the course's structured approach, providing clear, step-by-step instructions to navigate the intricate field of quantum computing.

\section{Organization of Lectures}

By adopting a teaching methodology that strategically introduces concepts of quantum mechanics after establishing a strong foundation in quantum computing, this paper describes a course that reverses the traditional approach to instruction \cite{48_pun,49_ste}. Quantum computing leverages quantum principles, where the postulates of quantum mechanics—state, time evolution, and measurement—play a crucial role. Unlike traditional courses that begin with the abstract postulates of quantum mechanics, our course starts with tangible quantum computing concepts, making it more accessible to students with limited computer science background.

The correspondence between the postulates of quantum mechanics and the concepts in quantum computing is illustrated in Table \ref{tab:cor}, which provides a straightforward mapping to facilitate understanding:

\begin{table*}
	\centering
    \setlength{\tabcolsep}{5mm}
    \renewcommand\arraystretch{1.5}
	\begin{tabularx}{\textwidth}{|X|X|}
        \hline
		Postulates in quantum mechanics & Concepts in quantum computing \\\hline
The state of a quantum mechanical system is completely specified by the wavefunction. & The state of a qubit can be described using a two-dimensional complex vector, typically represented as $|\psi\rangle= \alpha|0\rangle + \beta |1\rangle$, where $|0\rangle$ and $|1\rangle$ are computational basis states, and $\alpha$ and $\beta$ are complex probability amplitudes. Thus, the wavefunction postulate corresponds to the qubit concept in quantum computation.\\\hline
The evolution of a quantum system (not during measurement) is given by a unitary operator or transformation. & In quantum computing, the evolution of qubits is controlled by quantum logic gates, which correspond to unitary transformations in the Hilbert space. Therefore, the evolution postulate corresponds to gates in quantum computing.\\\hline
When we measure a quantum system, the system collapses to one of its possible states, and the probability of each state occurring is given by the square of the modulus of the wavefunction. & In quantum computing, when we measure a qubit, we obtain a result of either $|0\rangle$ or $|1\rangle$. Thus, the measurement postulate corresponds to the readout process in quantum computing.\\\hline
	\end{tabularx}
	\caption{\label{tab:cor}Correspondence between several postulates in quantum mechanics and fundamental concepts in quantum computing.}
\end{table*}

In our course, concepts such as qubits, gates, and measurement are introduced in a simplified manner, enabling students to grasp the fundamental principles of quantum computing without delving into the more complex mathematics typically associated with quantum mechanics. This approach not only makes the field of quantum computing more approachable and comprehensible for younger students but also aims to ignite a passion for quantum science and technology.

After laying the groundwork with quantum computing, we seamlessly transition into the basics of quantum mechanics. This progression allows students to see how the concepts they've learned in quantum computing, such as qubits, gates, and measurement, are underpinned by quantum mechanical principles. By concluding with a module on building a quantum computer, we connect these principles to their practical applications, helping students not only to understand the mechanics behind quantum computing but also to gain a comprehensive insight into quantum mechanics itself. This full-circle approach, moving from quantum computing to quantum mechanics, provides students with a complete understanding of the subject, linking theoretical concepts with practical uses.

The course is structured to span eight weeks, with each week comprising approximately $3$ hours of instruction. Designed for students with a foundational understanding of algebra but without prior knowledge in linear algebra, analytic geometry, or quantum mechanics, this curriculum stands out from traditional quantum computing courses which often assume advanced knowledge.

The curriculum is divided into four main sections, each spanning two weeks:

\begin{enumerate}
    \item[A.] Part I: Introduction to Quantum Computing (Week 1 and Week 2)
    \item[B.] Part II: Matrices for Quantum Computing (Week 3 and Week 4)
    \item[C.] Part III: Quantum Circuit Model (Week 5 and Week 6)
    \item[D.] Part IV: How to Build a Quantum Computer (Week 7 and Week 8)
\end{enumerate}

Each section is designed to progressively build upon the previous one, ensuring a smooth transition from basic computing concepts to the advanced theories of quantum mechanics. The paper will detail the rationale behind the structure of these sections, providing a clear and comprehensive pathway for student learning. Detailed course notes for each section are included in the appendix to support this educational journey.

\subsection{Part I: Introduction to Quantum Computing}

The first week of this section starts with an exploration of the history of computers and computing. Students are introduced to the fundamentals of binary numbers, Boolean logic, computer architecture, and circuits \cite{2_sx}. This foundation lays the groundwork for understanding classical computing, explaining how computation is executed using basic logical gates and the construction of computers from these gates. Additionally, the concept of Landauer's principle is introduced \cite{3_sx}, illustrating the energy costs associated with irreversible computing and leading to a discussion on the potential of reversible computing, exemplified by quantum gates like CNOT and Toffoli gates \cite{4_sx}. This initial week is structured to be accessible, requiring no prior knowledge of linear algebra or quantum mechanics.

In the second week, the focus shifts to providing an overview of fundamental concepts in quantum computing. This includes discussing the historical development of quantum computing, featuring insights from renowned physicists such as Richard Feynman and David Deutsch \cite{5_sx,6_sx}. Their perspectives highlight the significance and complexity of quantum mechanics in the context of computing, helping to motivate the study of this field. The core content of this week revolves around the ``Quantum Circuit Model," which delves into quantum circuits, their key components like the Quantum Processing Unit (QPU), and quantum reversible circuits. The introduction to quantum gates encompasses single qubit gates (NOT, Hadamard, and Phase gates), two-qubit gates (Controlled-NOT, Controlled-Z, Controlled-U), and three-qubit gates (Toffoli and Fredkin gates) \cite{4_sx}. Given the students' lack of prior knowledge in complex numbers, all gates introduced operate in real space. Moreover, complex concepts such as qubits, quantum gates, and quantum algorithms are simplified for better comprehension. A qubit is presented as a quantum analogue of a classical bit capable of existing in states 0, 1, or both simultaneously (superposition), with coefficients assumed to be $\pm 1$. The section avoids delving into the normalization of state vector coefficients and entanglement at this stage. One of the key concepts introduced is quantum measurement, simplified to focus on the probability of obtaining certain states rather than the measured physical operators. With this approach, students can calculate the output of simple quantum circuits and predict probabilities during quantum measurements by counting binary numbers. This foundational knowledge enables students to understand concepts like the DiVincenzo Criteria and Deutsch's Algorithm, making basic quantum algorithms accessible and comprehensible without the need for complex quantum mechanics and linear algebra, thus maintaining and potentially increasing student interest.

\subsection{Part II: Matrices for Quantum Computing}

The first week of this section introduces the concept of matrix representation of quantum gates \cite{4_sx}. An essential aspect of understanding quantum computing involves grasping the concepts of matrices and vectors. To make this introduction suitable for high school students, the course begins with a simple explanation of how bits and qubits can be represented as vectors. Here, 0 and 1 are represented as vertical two-dimensional vectors, with the quantum states $|0\rangle$ and $|1\rangle$ depicted in a similar manner, but also incorporating the concept of superposition. The introduction to gates starts with the classical NOT gate as an example, leading to a detailed explanation of matrix arithmetic. This foundation enables the exploration of vector representations for two classical bits and matrix representations for two-bit classical gates, with a focus on the NAND gate. The reversible nature of the NAND gate naturally leads to the introduction of the quantum CNOT gate, followed by a generalization to the Controlled-Z gate. By introducing single-qubit and two-qubit classical gates, and their corresponding matrix representations, students are guided towards understanding the tensor product \cite{9_sx}. The rules for computing the tensor product are explained, enabling the extension of these concepts to multiple quantum bits. Through this approach, students gain a fundamental understanding of how gates and states are represented using matrices and vectors in the realm of real numbers, laying the groundwork for more advanced quantum computing topics.

In the second week, the course builds upon the foundational knowledge from previous weeks to introduce matrix representation for general quantum circuits. This week's lecture is dedicated to Grover's algorithm, a pivotal quantum search algorithm known for its ability to find a marked item in an unsorted database more efficiently than any classical algorithm \cite{s1}. The lecture breaks down the algorithm into understandable segments, starting with phase separation and progressing to inversion about the average. These steps are crucial for comprehending how quantum states can be manipulated to solve problems more efficiently compared to classical computing methods. The introduction of Grover's algorithm at this stage is strategic, as it consolidates the students' understanding of matrix operations in quantum computing, while simultaneously demonstrating the practical applications of these concepts in solving complex problems.

\subsection{Part III: Quantum Circuit Model}

This part of the course delves into more advanced topics: the quantum circuit model and the universal quantum gates, using the quantum Fourier transform as an illustrative example.

The first week's lecture introduces the concept of universal quantum gates. Up until this point, the course has primarily dealt with gates in the realm of real numbers. To expand upon this, the lecture begins with an introduction to complex numbers, which is essential for understanding more sophisticated quantum computing concepts \cite{s11preskill2023quantum}. This includes covering the fundamentals of complex numbers, operational rules, and Euler's formula. Building on this foundation, the discussion then progresses to the Bloch Sphere, a critical tool for visualizing the state of a single qubit. This comprehensive introduction sets the stage for understanding single qubit universal quantum gates, a cornerstone concept in quantum computing.

In the second week, the focus shifts to unitary matrices and the introduction of two-qubit quantum gates. This lecture aims to consolidate the students' understanding of unitary transformations, a key concept in quantum mechanics, and how they apply to quantum computing. With a solid grasp of single qubit gates, students are now ready to explore how these gates can be combined with multi-qubit gates to construct more complex quantum circuits. This week culminates with an introduction to the quantum Fourier transform, a transformation widely used in various quantum algorithms \cite{s2}. To demonstrate its application, the lecture covers the period finding algorithm and the Shor's algorithm as optional reading materials, both of which highlight the power and potential of quantum computing in solving problems that are intractable for classical computers \cite{s3}. The inclusion of these algorithms not only deepens the students' understanding of quantum computing principles but also provides a glimpse into the practical applications and future possibilities of the field.

\subsection{Part IV: How to Build a Quantum Computer}

This final part of the course introduces students to quantum physics, particularly focusing on the underlying mechanisms of a quantum computer. This section builds on the previously introduced concepts, offering a unique perspective on quantum computing.

The first week's lecture departs from the traditional approach of starting with wave dynamics, often used in undergraduate textbooks. Such an approach can sometimes lead students to become overly focused on solving complex differential equations, losing sight of the fundamental principles of quantum mechanics. With a foundation already established in basic gates, quantum states, and measurements, we introduce the fundamental postulates of quantum mechanics. Given the students' familiarity with multi-qubit quantum systems, extending to the mathematical description of many-body systems and quantum entanglement is a logical next step. This discussion naturally leads to an understanding of the quantum no-cloning theorem, a key theorem in quantum information science, which forms the core of the first week's lecture \cite{s4}.

In the second week, the focus shifts to the practical aspects of quantum computing, specifically the construction of a basic NMR quantum computer \cite{s5}. Recognizing that concepts such as Hamiltonians and Hamiltonian mechanics might be beyond the scope of high school students, the lecture utilizes unitary transformations as a fundamental postulate to derive the Schrödinger equation. This approach simplifies the introduction of the Hamiltonian as a Hermitian operator, representing the system's energy. This level of explanation is tailored to be comprehensible for high school students, while more ambitious learners can explore further into matrix and operator functions. This theoretical groundwork provides a complete framework for understanding the energy levels in NMR quantum computing \cite{s5}.

Building upon this theoretical foundation, the course seamlessly transitions into a practical exploration of quantum computing hardware, using a portable quantum computer as a prime example. This portable quantum computer serves not just as a tool for demonstration but as a real-world application of the concepts covered in the lectures. By examining the portable quantum computer, students gain insights into the actual construction and functioning of quantum computers \cite{s5}. This part of the course is designed to bridge the theoretical principles with their tangible implementation, allowing students to see firsthand how the abstract concepts they have learned materialize in a functioning quantum computing device. The hands-on experience with the portable quantum computer enhances the learning process, enabling students to interact with and better understand the intricacies of quantum computing technology, turning theoretical knowledge into practical understanding.

\section{The Portable Quantum Computer}

As we transition from the theoretical aspects of quantum computing to its practical applications, the Gemini/Triangulum series, the two/three-qubit portable quantum computers, play a pivotal role in our course. They are not merely a tool for demonstration but an integral part of the curriculum, used extensively to showcase quantum algorithms and hardware usage \cite{s7}. In this section, we delve into the specifics of the portable quantum computers, providing a detailed overview of their capabilities and how they are employed in the course to bring abstract quantum concepts to life.

The following subsections will provide a comprehensive overview of the Gemini/Triangulum series, covering their overall system and user interface:

\begin{itemize}
    \item \textbf{System}: This subsection will explore the physical components and architecture of the Gemini/Triangulum series, explaining how they contribute to their functionality as a quantum computer. It will cover the essential hardware elements that enable quantum computing, providing students with a clear understanding of the machine's operation at a physical level.

    \item \textbf{User Interface}: This subsection will focus on the user interface of the Gemini/Triangulum series, explaining how users can interact with the quantum computer to run experiments and observe results. 
    
\end{itemize}

Through exploring the Gemini/Triangulum series, students will acquire a comprehensive understanding of the nature and operation of quantum computers, and how theoretical concepts in quantum computing are implemented in actual hardware and software solutions.

\subsection{System}

\begin{figure}[!htb]
	\includegraphics[scale=0.3]{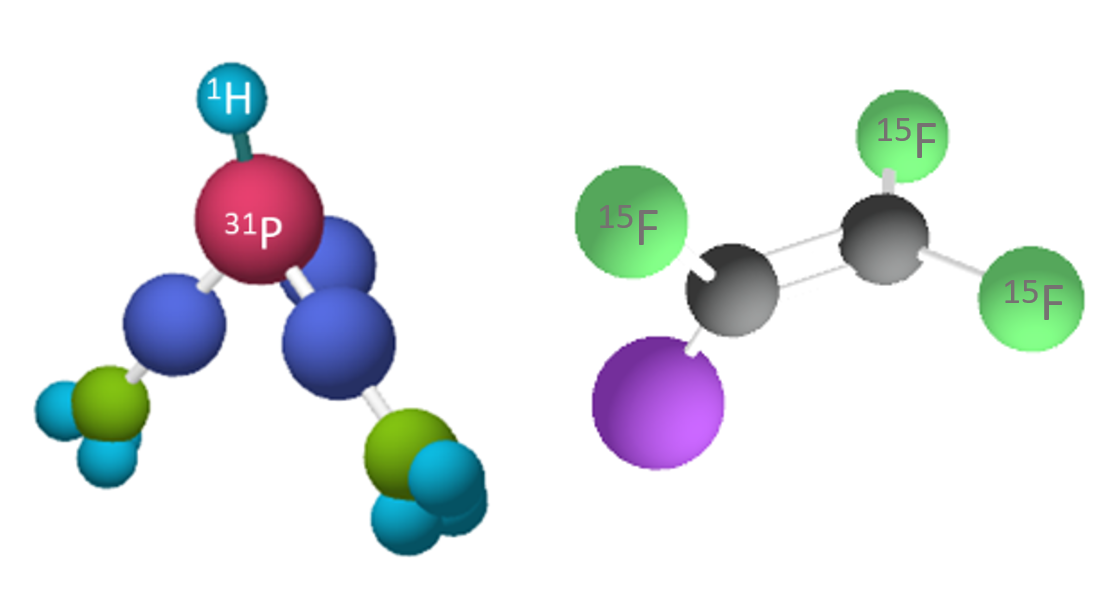}
	\centering
	\caption{\label{molecule}The molecule structures of samples used in Gemini/Triangulum series. In Gemini Mini/Mini Pro and Gemini, the $^{31}$P and $^{1}$H nuclear spins in Dimethylphosphite ((CH$_3$O)$_2$PH) labeled in the figure are used as the two qubits. In Triangulum Mini and Triangulum, the three $^{15}$F nuclear spins in Iodotrifluoroethylene (C$_2$F$_3$I) are used as the three qubits.}
\end{figure}

\begin{figure*}
	\includegraphics[scale=0.54]{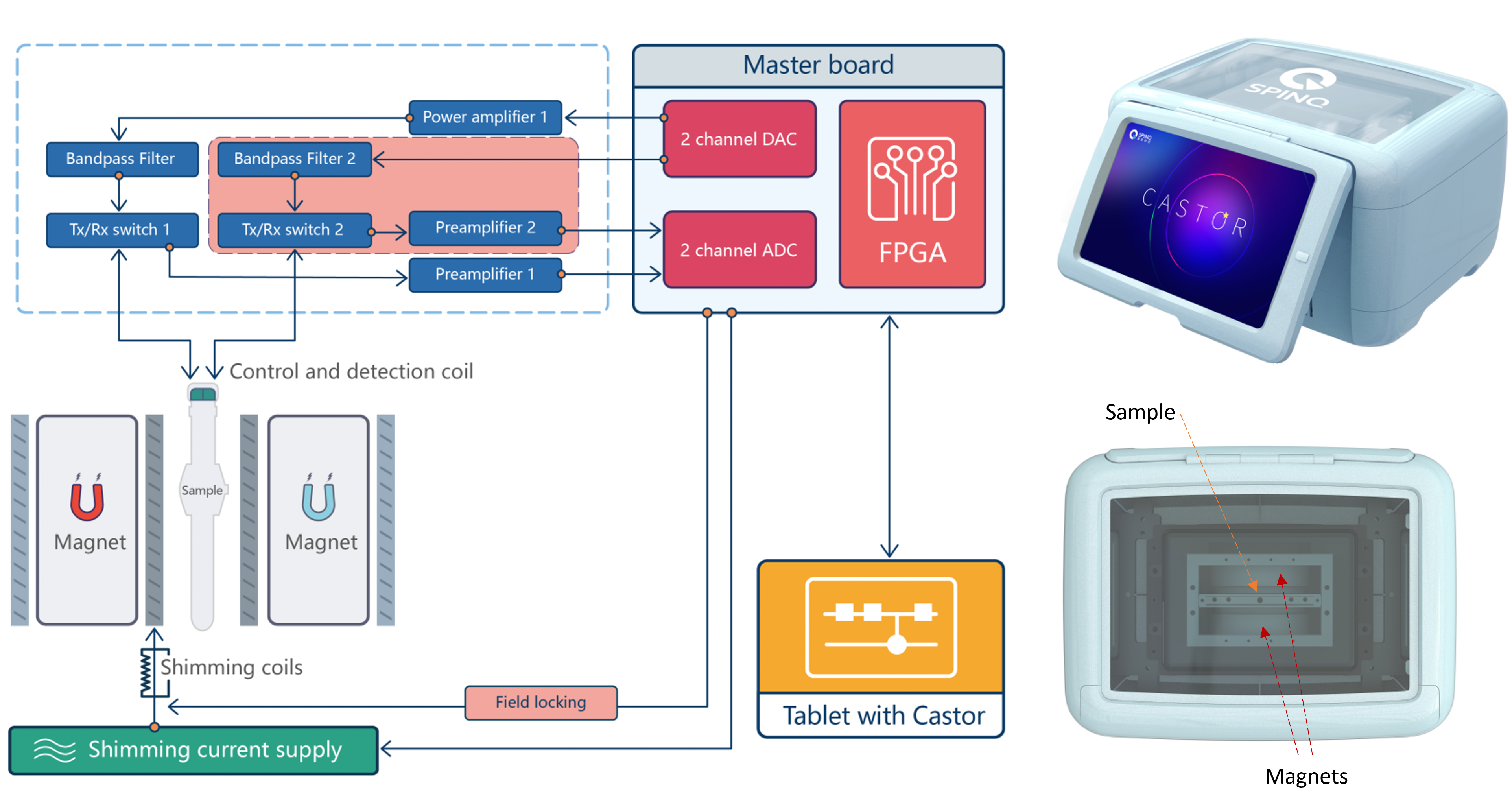}
	\centering
	\caption{\label{structure} Left: The schematics of Gemini Mini/Mini Pro and Triangulum Mini. Right: The side view and top view of the device. The device include a tablet which provides the user interface Castor, a master board, an RF transmission system (enclosed by the dashed box), a pair of parallel plate magnets, a sample tube, probe coils and shimming coils, etc. The RF transmission system provides RF pulse and signal paths for both qubit control and field locking. Depending on different frequencies in different models, the number of RF channels might be different. There is a removable cover on top of the device. Students can observe the inner structure after removing the cover. }
\end{figure*}

\begin{figure*}
	\includegraphics[scale=0.7]{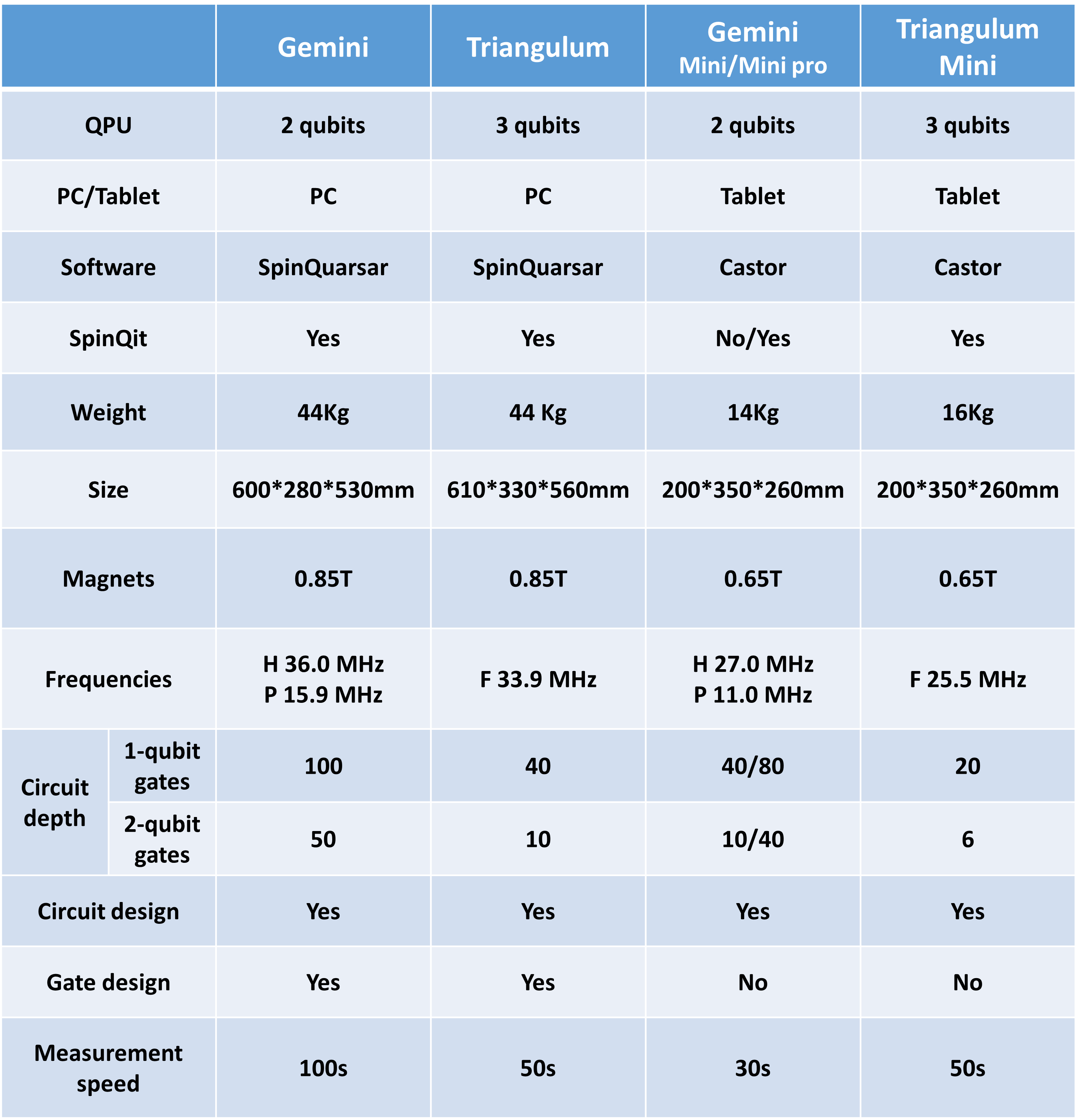}
	\centering
	\caption{\label{specs} Specifications of Gemini and Triangulum portable products, including Gemini, Triangulum, Gemini Mini/Mini Pro and Triangulum Mini. The Gemini and Triangulum specifications displayed here represent the latest updates, which may vary slightly from the initially launched version as in Refs. \cite{s5,s7}. The magnetic fields of Gemini and Triangulum are higher than those of their mini versions. Additionally, they require a PC equipped with SpinQuarsar for operation. These features allow for more precise quantum control, manifested as larger circuit depths, and low-level control such as quantum gate design. }
\end{figure*}

Gemini/Trangulum series are based on Nuclear magnetic resonance (NMR). A nuclear spin which has a spin number $1/2$ has two energy levels in a static magnetic field, and thus can be used as a qubit {s8jones2000geometric,s9jones2001quantum}. The $^{31}$P and $^{1}$H nuclear spins in Dimethylphosphite ((CH$_3$O)$_2$PH) which are connected directly (see Figure \ref{molecule}) are used as the two qubits in Gemini series \cite{s5}. The three $^{15}$F nuclear spins in the  Iodotrifluoroethylene (C$_2$F$_3$I) molecule (see Figure \ref{molecule}) are used as the three qubits in Triangulum series \cite{s7}. The Dimethylphosphite or Iodotrifluoroethylene molecules are in the liquid form and contained in a sample tube. The sample tube is placed in a pair of parallel plate NdFeB magnets (see Figure \ref{structure}). The nuclear spins in the magnetic field have their own frequencies (see Figure \ref{specs}). Therefore, nuclear spin qubits can be controlled using microwave pulses on resonance with them. Quantum gates are realized by microwave pulses together with the help of the natural couplings mediated by bonds between different nuclear spins. The measurement of a quantum state is realized by detecting the microwave signals irradiated by the nuclear spin qubits \cite{s5}.

The Gemini/Triangulum series include Gemini Mini/Mini Pro, Gemini, Triangulum Mini and Triangulum. Gemini and Triangulum, launched earlier, were introduced in details in \cite{s5,s7} . Hence here we only show the structure of the Mini ones in Figure \ref{structure}. As show in Figure \ref{structure}, apart from the sample tube and the magnets, the overall structure of the device includes a tablet, a master board, a radio frequency (RF) transmission system, and the shimming and locking systems. As mentioned above, the sample tube and the magnets provide the qubit system. The master board incorporates an FPGA, a digital-analog converter (DAC) and an analog-digital converter (ADC). The master board realizes the function of control pulse generation, i.e., the generation of quantum control pulses. The generated pulses go through the RF transmission path which include filters and amplifiers, and finally reach the probe coil near the qubit system to realize quantum control gates. The signal irradiated by the qubit system can be collected by the probe coil and goes through the RF transmission path and also gets filtered and amplified. Then the detected signal reaches the master board. It get digitized and processed by the master board and returned to users in the form of entries of quantum state density matrices. The master board also realizes the functions of magnetic field shimming and field locking with the help of the shimming coils \cite{10_sx}, which ensures a homogeneous and stable magnetic field  and thus stable and homogeneous qubit frequencies for better quantum control \cite{11_sx}.

Figure \ref{specs} lists specifications of Gemini and Triangulum series. Compared with Gemini and Triangulum, the most significant difference the Mini ones have is the small size and weight. They are also more cost-effective. They are more suitable to be used in high schools for K12 education because of their more portable and cost-effective features. There are also compromises. The magnetic field is weaker and thus the qubit frequencies are smaller in Mini ones \cite{17_sx,18_sx,19_du}.  Because of multiple factors related to a lower magnetic field, the Mini ones have  slightly worse control performance and a shallower circuit depth \cite{20_sx,21_russ}. In spite of this, they are still very good tools for demonstrating the basic ideas of quantum gates and quantum algorithms. Most of the experiments in this course are realized on Gemini Mini and Triangulum Mini. However, for experiments related to low-level control, Gemini and Triangulum are needed. For example, in the experiment of Week 8,  Gemini is applied to demonstrate how a CNOT gate is composed by hardware-level pulse sequences. In view of this, compared to their mini counterparts,  Gemini and Triangulum are more suitable to scientific research where quantum gate design is usually important.

\subsection{User Interface}
\begin{figure*}
	\includegraphics[scale=0.9]{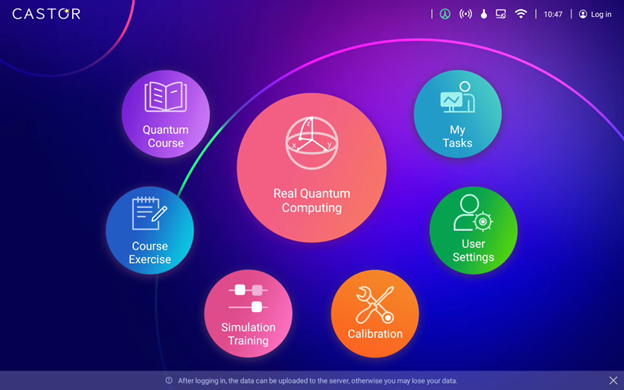}
	\centering
	\caption{\label{castor} Castor Homepage.}
\end{figure*}

\begin{figure*}
	\includegraphics[scale=0.9]{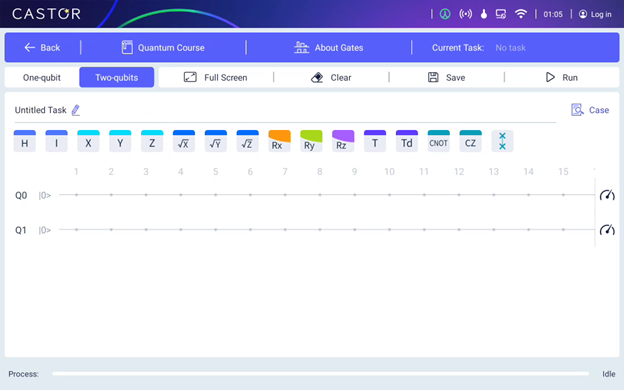}
	\centering
	\caption{\label{circuit} The circuit construction interface in  ``Real Quantum Computing" of Castor.}
\end{figure*}

\begin{figure*}
	\includegraphics[scale=0.7]{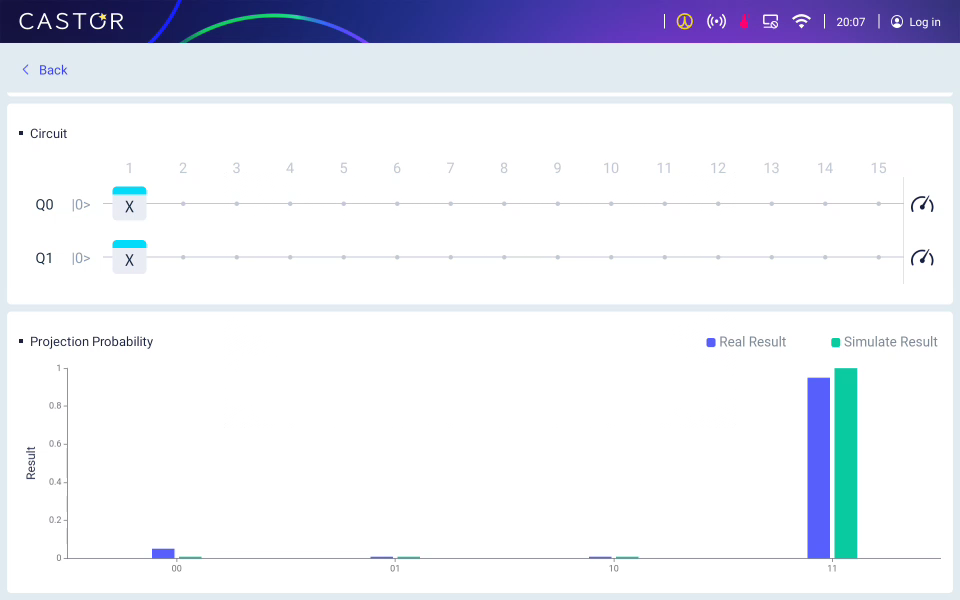}
	\centering
	\caption{\label{results} The results of a task is in the form of projection probabilities of each basis states. \bf{Please change this to English}}
\end{figure*}

As introduced in Refs. \cite{s5,s7}, Gemini and Triangulum integrate a PC with a user-friendly software SpinQuasar. Here we introduce the user interface of Gemini Mini/Mini Pro and Triangulum Mini. They integrate a pad that runs a new operating system named Castor. This provides a more compact and convenient user interface for managing quantum computing tasks. Figure \ref{castor} shows the Homepage of Castor. 
The two/three-qubit QPU can be accessed through the module ``Real Quantum Computing" and an eight-qubit simulator can be accessed through ``Simulating Training". In both the two modules, users edit and construct a quantum circuit using a drag-and-drop manner (see Figure \ref{circuit}). There are multiple built-in quantum algorithms as well that users can select from the ``Case" button (see Figure \ref{circuit}) and run directly, such as the Grover's algorithm and the Deutsch's algorithm. All experimental or simulation task information can be managed in ``My Tasks" module. For example, the quantum circuit and results for each task that users have run can be revisited in ``My Tasks". The results for each task contain the diagonal elements of the final quantum state density matrices of the task, i.e. the probabilities for each basis states (Figure \ref{results}). In ``Calibration" module, users can calibrate the NMR signals and quantum control parameters by themselves, or they can also run a standard automatic calibration instead. 

In Castor users can also access a quantum course that was developed based on the curriculum introduced in this paper, in the module ``Quantum Course". There is a course exercise module as well, ``Course Exercise". The ``User Setting" enables the configuration of various parameters for the user account, such as WIFI setting and remote control setting.

In addition  to the graphic user interface Castor, Gemini Mini Pro and Triangulum Mini provide access via SpinQit (https://spinquanta.com/products-solutions/spinqKit). SpinQit is a quantum software development kit (SDK) created by SpinQ (https://spinquanta.com/products-solutions/spinqKit). SpinQit supports three types of programming syntaxes, native SpinQit Python syntax, OpenQASM 2.0 syntax, and IBM Qiskit Python syntax. It also offers a rich interface for quantum algorithms, and facilitates terminal connections with different types of back-ends, such as quantum computers, quantum simulators, and quantum computing cloud platforms.

The Gemini and Triangulum series serve as a tangible example for students, bridging the gap between theory and practice. Their use throughout the course allows students to directly observe and understand the application of quantum algorithms and the intricacies of quantum computing hardware. This hands-on experience is invaluable in enhancing students' comprehension and appreciation of the field.

As a real quantum computer is used in experiments, students could easily observe that the results of the experiments differ from the theoretical predictions, as shown in Figure \ref{results}. It can be explained to the students that errors can occur in quantum computing, and to correct these errors, quantum error correction was invented. Quantum error correction \cite{gaitan2008quantum,gottesman2010introduction,djordjevic2012quantum,devitt2013quantum,lidar2013quantum,campbell2017roads,terhal2015quantum} can rectify errors that occur during computations, making the construction of quantum computers possible. While the knowledge of quantum error correction is beyond the scope of this course, informing students about errors and correction can help them develop a better understanding of the subject matter.

\section{Discussion}

The initial offering of this course under the Hong Kong Education Bureau's program for gifted students in 2022 provided significant feedback on its impact and attractiveness. Catering to 40 gifted high school students, the program received favorable reviews, demonstrating substantial engagement and enthusiasm for quantum computing. Such feedback underlines the course's potential for wider adoption in high school curricula.

A pivotal element in the course's success is its pedagogical journey from quantum computing to quantum mechanics, requiring no extensive background in mathematics or physics. By initiating with fundamental computing principles and progressively unveiling quantum concepts, the curriculum effectively maintains student interest and enhances comprehension. This gradual progression culminates in the ``How to Build a Quantum Computer" module, where theoretical learning transitions into practical application. Here, students not only grasp the underpinnings of quantum mechanics but also engage in constructing and operating a quantum computer. The incorporation of portable quantum computing hardware allows students to directly apply what they have learned, bridging the gap between abstract theory and tangible experience. This hands-on approach not only solidifies their understanding of quantum computing concepts but also empowers them with the skills to assemble and use quantum computers, thereby offering a comprehensive and immersive learning experience that embodies the course's innovative teaching methodology.

Given the encouraging response from the gifted program, there is a significant opportunity to include this curriculum in standard high school programs. Efforts are underway to introduce the course in schools in Hong Kong and Shenzhen, reflecting a growing recognition of the importance of early exposure to advanced scientific concepts in general education. This expansion aligns with global educational trends and aims to prepare students for emerging technological challenges and opportunities.

Integrating this course into the standard high school curriculum could significantly widen student exposure to quantum computing. Making these advanced concepts accessible to a broader audience helps create a more diverse and well-prepared cohort of future learners and professionals, especially as quantum computing's relevance grows across various sectors.

Moreover, the course's structure and content could serve as a model for other educational institutions aiming to incorporate quantum computing into their syllabi. It shows how complex scientific topics can be tailored to younger learners, aligning the subject's intricacies with the students' educational level.

In conclusion, the favourable reception of the quantum computing course in the gifted program establishes a solid foundation for its adoption in a broader high school context. While the course is designed to deepen high school students' understanding of quantum computing, its widespread integration could markedly enhance the educational framework in this emerging field. As the course is adopted in more schools and regions, it presents an excellent opportunity to augment the quantum computing learning experience for students globally.

\section{Acknowledgement}

S-Y. Hou is supported by National Natural Science Foundation of China under Grant No. 12105195. B.Z. Acknowledges the Gifted Education Fund by the Hong Kong Education Bureau, Programme No.: 2022-07.


\appendix


%

\bibliography{sample}

\end{document}


\maketitle

\tableofcontents

\newpage

\chapter{Week 1: A Brief History of Computers}
\section{What is computing}

The term ``computing" often brings to mind the image of computers, machines that we typically associate with performing computational tasks \cite{sun1piccinini2007computing}. However, computing encompasses far more than just the operations carried out by electronic devices. In essence, computing is a fundamental aspect of numerous processes and activities in the world. Our brain, for instance, can be considered one of the earliest computing tools, constantly engaged in computational processes. Everyday activities like arithmetic calculations, strategizing in chess, or decision-making in games like mah-jongg are all instances where our brain naturally engages in abstraction and computation.

When we consider the computational ability of a computer, we are referring to its capacity for mathematical induction and transformation. This ability involves transforming abstract and complex mathematical expressions or numbers into forms that are comprehensible to us, using various mathematical methods. In this sense, computing is not just about the hardware (computers) but also about the fundamental process of transforming and understanding information, a process that is central to both artificial machines and the human mind.

Computers are renowned for their capability to perform a wide array of tasks at speeds far surpassing human capabilities. This efficiency and speed in processing complex calculations and managing vast amounts of data is one of the primary reasons for their ubiquitous presence in modern life.

Historically, humans have always sought tools to aid in computation. From the abacus, an ancient tool for arithmetic, to the concepts of binary logic, and eventually to the Analytical Engine conceptualized by Charles Babbage, the journey of computational tools has been long and transformative.

\begin{figure}[H]
    \centering
    \subfigure[Intelligent identification]{
        \includegraphics[scale=0.2]{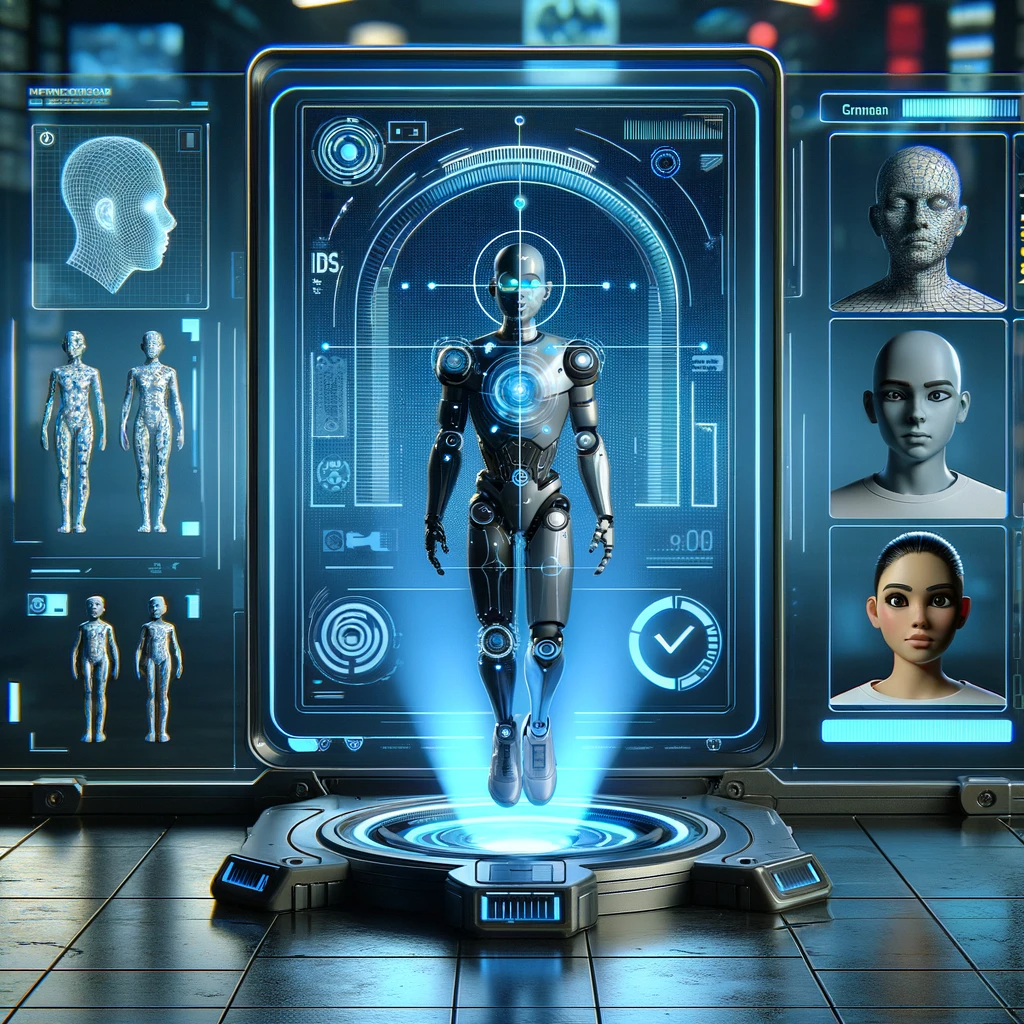}
        \label{1_1}
    }
    \subfigure[Mobile phone]{
        \includegraphics[scale=0.2]{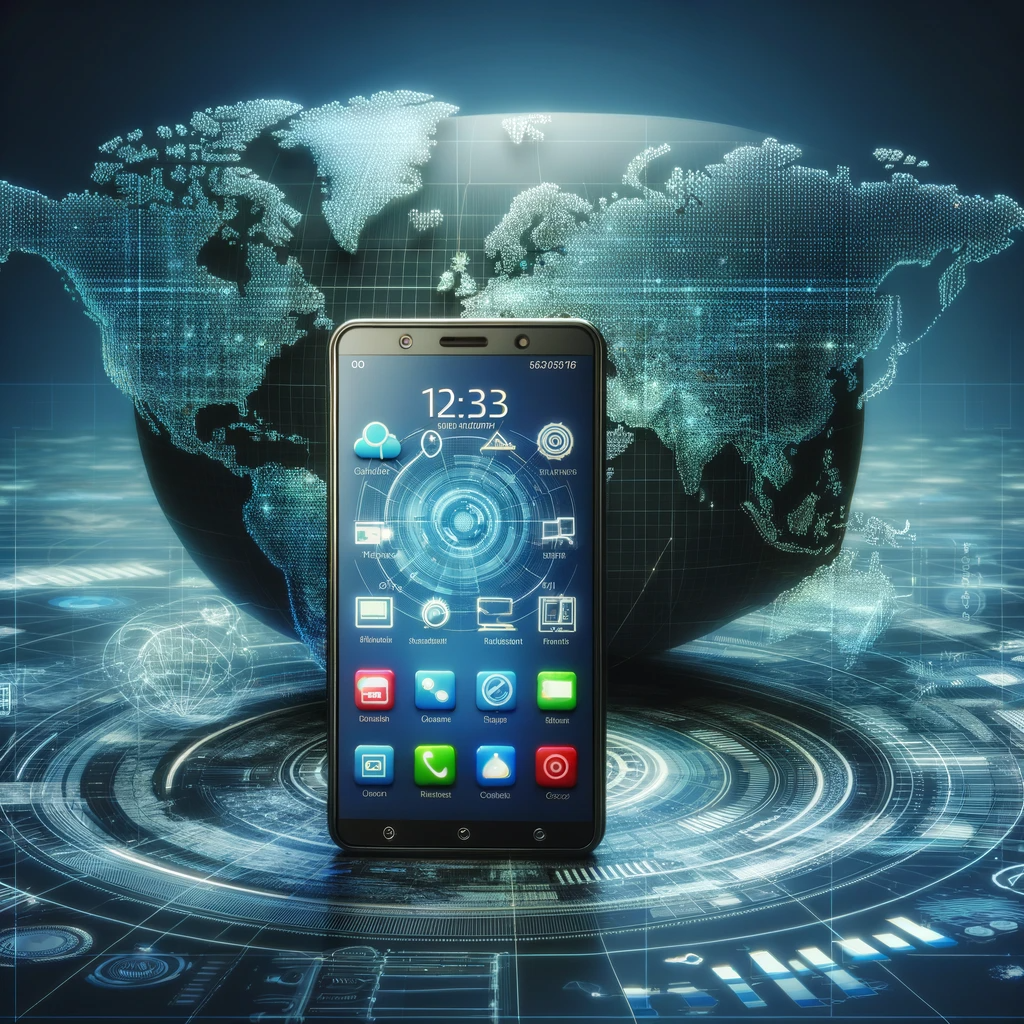}
        \label{1_2}
    }
    \caption{(a) An illustration of a futuristic intelligent identification system, showcasing advanced technology with a sleek, high-tech design. This system is depicted scanning and identifying cartoon-style human figures, demonstrating the process of data analysis and recognition in a sophisticated yet approachable manner.(b)A depiction of global connectivity, represented by a stylized world map in the background, with a modern smartphone in the foreground showcasing advanced computing features. This image symbolizes the widespread use of smartphones across the globe, highlighting their capabilities in data processing, communication, and multimedia, and emphasizing their role in a globally connected world.}
    \label{fig:main}
\end{figure}

Looking ahead, it becomes increasingly clear that both individuals and societies, in the context of countries, science, or everyday life, will continue to rely heavily on data and computing. Data has emerged as a cornerstone of competitiveness in numerous fields, making the ability to process and understand this data a critical skill. This is evident in technologies such as intelligent identification systems, as illustrated in Figure~\ref{1_1}, and the ubiquitous use of smartphones, as shown in Figure~\ref{1_2}. These are prime examples of data processing capabilities in action. The demand for computing power is a constant in an ever-evolving technological landscape, indicating that there will always be a need for more advanced and efficient computational tools.

\section{Binary numbers and Boolean logic}~{}

The foundation of modern computing lies in binary arithmetic. Therefore, it is essential to understand binary numbers~\cite{sun40combet1965computation}.

A binary number is a number expressed in the base-2 numeral system, which uses only two digits, 0 and 1. All numbers can be represented in binary form.

There are two primary methods for converting between binary and decimal systems. The first method is converting from decimal to binary. This is achieved using the ``division by 2" method, which involves dividing the decimal number by 2 to get a quotient and a remainder. This process is repeated, using the quotient for subsequent divisions, until the quotient becomes 0. The binary number is then formed by arranging the remainders in reverse order, starting from the last remainder obtained (the least significant bit) to the first (the most significant bit).

The second method is converting from binary to decimal. In this method, each digit of the binary number is multiplied by the corresponding power of 2, starting from the rightmost digit. The sum of these products gives the decimal equivalent of the binary number.

Table~\ref{1_t1} illustrates the conversion of binary numbers from 0 to 8, providing a clear reference for understanding this fundamental concept in computing.

\begin{table}[H]
	\centering
	\setlength{\tabcolsep}{1mm}
	\begin{tabular}{|c|c|c|c|c|c|c|c|c|c|} 
		\hline
		Decimal number & 0 & 1 & 2 & 3 & 4 & 5 & 6 & 7 & 8 \\
		\hline
		Binary number & 0 & 1 & 10 & 11 & 100 & 101 & 110 & 111 & 1000 \\
		\hline
	\end{tabular}
	\caption{\label{1_t1} The convert of binary and decimal from 0 to 8}
\end{table}

Next, we explore the conversion of decimal fractions to binary. This process involves two main parts: converting the integer part and converting the fractional part of a decimal number to binary.

\textbf{Converting the Integer Part of a Decimal to Binary:}
The conversion of the integer part is relatively straightforward and can be done using the division-by-2 method. The steps include:

\begin{enumerate}
    \item Divide the integer part of the decimal number by 2 and write down the remainder. This remainder is the least significant bit of the binary representation.
    \item Continue dividing the quotient by 2 and write down the remainder as the next higher bit in the binary number.
    \item Repeat the process until the quotient becomes 0.
\end{enumerate}

\textbf{Converting the Fractional Part of a Decimal to Binary:}
The conversion of the fractional part employs the multiplication-by-2 method. The steps are:

\begin{enumerate}
    \item Multiply the fractional part of the decimal number by 2. The integer part of the result is the most significant bit of the binary fraction.
    \item Use the fractional part of the result for the next multiplication by 2 and record the integer part as the next binary digit.
    \item Continue this process until the fractional part becomes 0 or reaches a repeating pattern.
\end{enumerate}

\textbf{Converting Decimal Fractions to Binary:}
To convert a decimal fraction to binary, convert the numerator and denominator into binary separately. Then, represent the binary fraction with the binary numerator and denominator separated by the fraction line.

Table~\ref{1_t2} shows the binary conversion and approximation of fractions, including the approximation of infinite recurring decimals using geometric series. In the series, ``a" represents the first term, and ``r" is the common ratio. The sum of an infinite geometric series is given by:

\begin{equation}
  \sum_{k=0}^{\infty}ar^{k} = \frac{a}{1-r} 
\end{equation}

This mathematical principle helps in understanding the approximation of recurring decimals in binary representation.

\begin{table}[H]
	\centering
	\resizebox{1.0\columnwidth}{!}{
		\begin{tabular}{|c|c|c|c|} 
			\hline
			Fraction & Decimal & Binary & Fraction Approximation \\
			\hline
			$1/1$ & $1$ or $0.999...$ & $1$ or $0.111...$ & $1/2+1/4+1/8$ \\
			\hline
			$1/2$ & $0.5$ or $0.4999...$ & $0.11$ or $0.0111...$ & $1/4+1/8+1/16$ \\
			\hline
			$1/3$ & $0.333$ or $0.010101$ & $1$ or $0.111...$ & $1/4+1/16+1/64$ \\
			\hline
	\end{tabular}}
	\caption{\label{1_t2} Binary conversion and approximation of fractions}
\end{table}

\begin{exercise}	
	conversion Between Decimal and Binary:
	
	According to the example of convent decimal and binary, fill in the correct result of binary conversion in the blank in table \ref{1_t3}
	
	\begin{table}[H]
		\centering
		\setlength{\tabcolsep}{5mm}
		\begin{tabular}{|c|c|c|c|} 
			\hline
			Decimal & Binary & Decimal & Binary \\
			\hline
			0 & 0 & 8 & 1000 \\
			\hline
			1 & 1 & 9 &      \\
			\hline
			2 & 10 & 10 &    \\
			\hline
			3 & 11 & 11 &    \\
			\hline
			4 & 100 & 12 &   \\
			\hline
			5 & 101 & 13 &   \\
			\hline
			6 & 110 & 16 &   \\
			\hline
			7 & 111 & 15 &   \\
			\hline
			16 &    & 99 &    \\
			\hline
			100 &&&    \\
			\hline          
		\end{tabular}
		\caption{\label{1_t3} Exercise 1 conversion Between Decimal and Binary}
	\end{table}
\end{exercise}

Now let's learn binary arithmetic, including addition, subtraction, multiplication, and division:

\textbf{Binary Addition:}
The rules for binary addition are as follows:
\begin{align*}
    0 + 0 &\to 0 \\
    0 + 1 &\to 1 \\
    1 + 0 &\to 1 \\
    1 + 1 &\to 0 \text{ (carry 1)}
\end{align*}

\begin{example}
Addition of two binary numbers:
\begin{tabular}{cccccccc} 
    &   & 0 & 1 & 1 & 0 & 1 &\\
    & + & 1 & 0 & 1 & 1 & 1 &\\
    \hline
    =  & 1 & 0 & 0 & 1 & 0 & 0 & = 36 \\
\end{tabular}
\end{example}

\textbf{Binary Subtraction:}
The rules for binary subtraction are as follows:
\begin{align*}
    0 - 0 &\to 0 \\
    0 - 1 &\to 1 \text{ (borrow 1)} \\
    1 - 0 &\to 1 \\
    1 - 1 &\to 0
\end{align*}

\begin{example}
Subtraction of two binary numbers:
\begin{tabular}{ccccccccc} 
    &   & 1 & 1 & 0 & 1 & 1 & 1 & 0 \\
    & - &   &   & 1 & 0 & 1 & 1 & 1 \\
    \hline
    &=  & 1 & 0 & 1 & 0 & 1 & 1 & 1 \\
\end{tabular}
\end{example}

\textbf{Binary Multiplication:}
\begin{example}
Multiplication of two binary numbers:
\begin{tabular}{ccccccccc} 
    &  &   &   &   & 1 & 0 & 1 & 1  \\
    &  &   &   & x & 1 & 0 & 1 & 0  \\
    \hline
    &  &   &   &   & 0 & 0 & 0 & 0 \\   
    & +&   &   & 1 & 0 & 1 & 1 &   \\
    & +&   & 0 & 0 & 0 & 0 &   &   \\
    & +& 1 & 0 & 1 & 1 &   &   &   \\
    \hline
    & =& 1 & 1 & 0 & 1 & 1 & 1 & 0 \\
\end{tabular}
\end{example}

\textbf{Binary Division:}
\begin{example}
Division of two binary numbers:
$\begin{array}{lrl} 
    & 101 \\ 
    101 & \overline{)11011} \\ 
    & \underline{-101} \\ 
    & 111 \\
    & \underline{-101}\\ 
    & 010
\end{array}$
\end{example}

\begin{exercise}
Four operations in binary.
Please calculate and fill in the correct result in the blank of the following vertical operations:

(1) Binary Addition
\begin{tabular}{ccccccccc} 
    &  & 1 & 1 & 0 & 1 & 1 & 1 & 0 \\
    &  &   &   & + & 1 & 1 & 0 & 0 \\
    \hline
    &  &   &   &   &   &   &   &   \\
\end{tabular}

(2) Binary Subtraction
\begin{tabular}{ccccccccc} 
    &  & 1 & 1 & 0 & 1 & 1 & 1 & 0 \\
    &  &   &   & - & 1 & 1 & 0 & 0 \\
    \hline
    &  &   &   &   &   &   &   &   \\
\end{tabular}

(3) Binary Multiplication
\begin{tabular}{cccccc}
    & & 1 & 1 & 0 & 1 \\
    & &   & 1 & 0 & 1 \\
    \hline
    & &   &   &   &   \\
\end{tabular}

(4) Binary Division
$\begin{array}{lrl} 
    1101 & \overline{)1101110} \\ 
\end{array}$
\end{exercise}


Boolean logic is a fundamental concept in computer science and mathematics, involving set algebra and various basic operators to symbolize and algebraize logical processes. Set algebra, a branch of mathematics, focuses on operations involving sets. The primary operations defined in set algebra include:

\textbf{Union}: The union of two sets \(A\) and \(B\), denoted by \(A \cup B\), is the set of elements that are in \(A\), in \(B\), or in both \(A\) and \(B\). For example, if \(A = \{1, 2\}\) and \(B = \{2, 3\}\), then \(A \cup B = \{1, 2, 3\}\).

\textbf{Intersection}: The intersection of two sets \(A\) and \(B\), denoted by \(A \cap B\), is the set of elements that are in both \(A\) and \(B\). For instance, if \(A = \{1, 2\}\) and \(B = \{2, 3\}\), then \(A \cap B = \{2\}\).

\textbf{Complement}: The complement of a set \(A\), denoted by \(A'\), is the set of elements not in \(A\) but present in the universal set \(U\) (the set containing all possible elements). For example, if \(A = \{1, 2\}\) and \(U = \{1, 2, 3, 4\}\), then \(A' = \{3, 4\}\).

A Venn Diagram~\cite{sun42heberle2015interactivenn} is a visual tool used to represent sets and their relationships. It employs overlapping circles or other shapes to illustrate the logical relationships between two or more sets. Each circle represents a set, and the areas where the circles overlap indicate the intersection of these sets.

\begin{figure}[H]
	\includegraphics[scale=0.45]{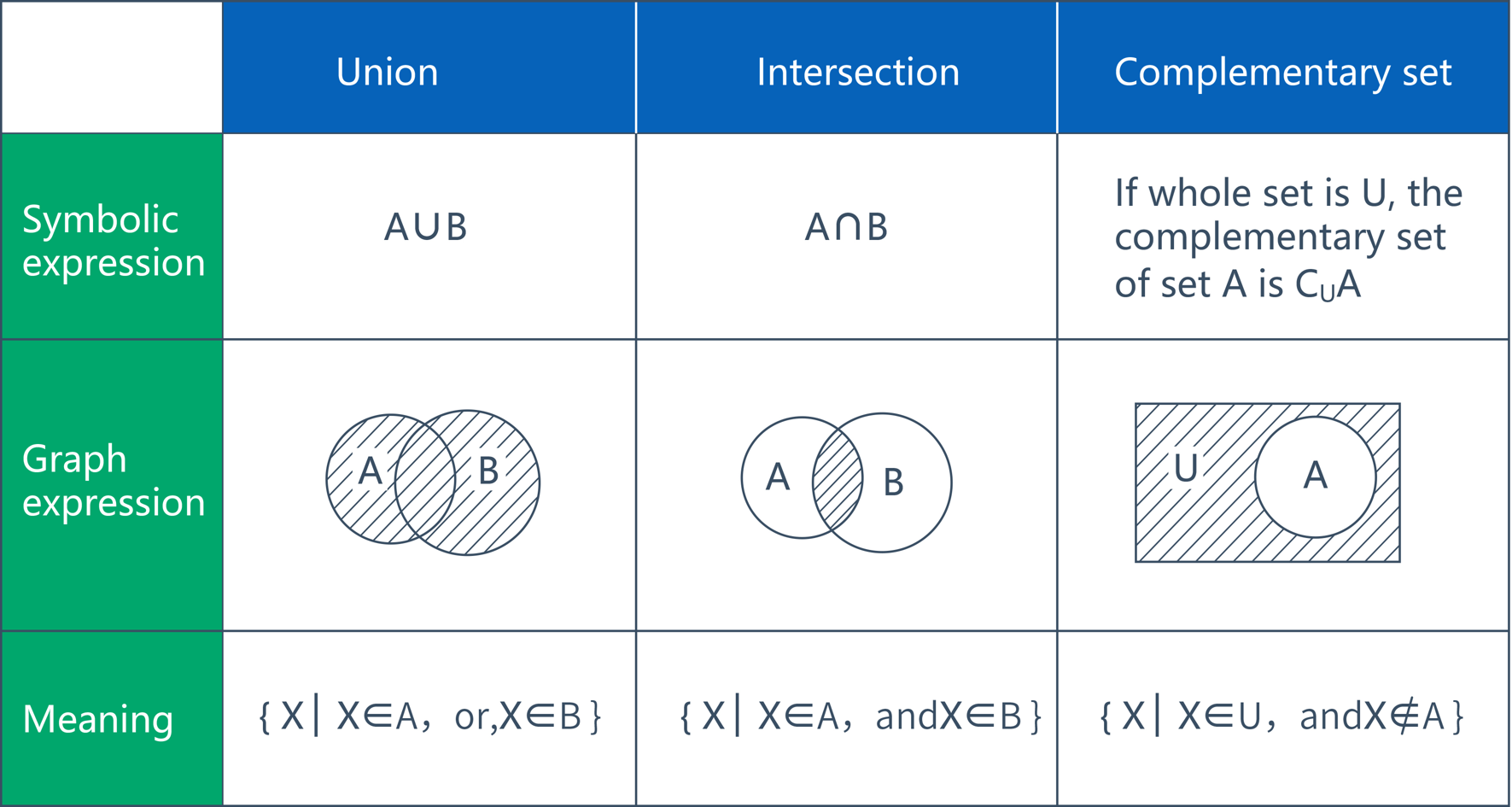}
	\centering
	\caption{\label{1_20} Set algebra in this figure we show the union, intersection and complement}
\end{figure}

The aforementioned operations, along with their corresponding Venn diagrams, are illustrated in Figure~\ref{1_20}, demonstrating the principles of set algebra in Boolean logic.

Boolean algebra~\cite{sun41halmos2009introduction}, also known as logical algebra, involves variables that represent the states of ``True" and ``False”, typically denoted as ``1" for True and ``0" for False. This form of algebra is foundational in computer science and digital circuit design. There are three basic operational logics in Boolean algebra:

\textbf{AND}: Symbolized as \(\land\). For \(x \land y\), the output is 1 (True) if and only if both \(x = 1\) and \(y = 1\); otherwise, \(x \land y = 0\) (False).

\textbf{OR}: Symbolized as \(\vee\). For \(x \vee y\), the output is 0 (False) if and only if both \(x = 0\) and \(y = 0\); in all other cases, \(x \vee y = 1\) (True).

\textbf{NOT}: Symbolized as \(\neg\). For \(\neg x\), if \(x = 1\), then \(\neg x = 0\) (False), and if \(x = 0\), then \(\neg x = 1\) (True).

The truth tables for the ``AND", ``OR" and ``NOT" operations are shown in Table \ref{1_t4}:

\begin{table}[H]
	\centering
	\setlength{\tabcolsep}{5mm}
	\begin{tabular}{|c|c|c|c|} 
		\hline
		\(x\) & \(y\) & \(x \land y\) & \(x \vee y\) \\
		\hline
		0 & 0 & 0 & 0 \\
		\hline
		0 & 1 & 0 & 1 \\
		\hline
		1 & 0 & 0 & 1 \\
		\hline
		1 & 1 & 1 & 1 \\
		\hline
		\(x\) & \(\neg x\) &  &   \\
		\hline
		0 & 1 &  &   \\
		\hline
		1 & 0 &  &   \\
		\hline           
	\end{tabular}
	\caption{\label{1_t4} The truth table of the three basic operational logics: ``AND", ``OR", ``NOT"}
\end{table}

The logical operations ``AND", ``OR" and ``NOT" in Boolean algebra can be effectively illustrated using Venn diagrams. Venn diagrams provide a visual representation of these operations by showing the relationship between different sets.

In a Venn diagram:
\begin{itemize}
    \item The ``AND" operation (\(\land\)) is represented by the intersection of sets, highlighting the common elements.
    \item The ``OR" operation (\(\vee\)) is depicted by the union of sets, encompassing all elements that belong to either set.
    \item The ``NOT" operation (\(\neg\)) is visualized by showing the complement of a set, which includes all elements not in the set within a given universal context.
\end{itemize}

\begin{figure}[H]
	\includegraphics[scale=0.35]{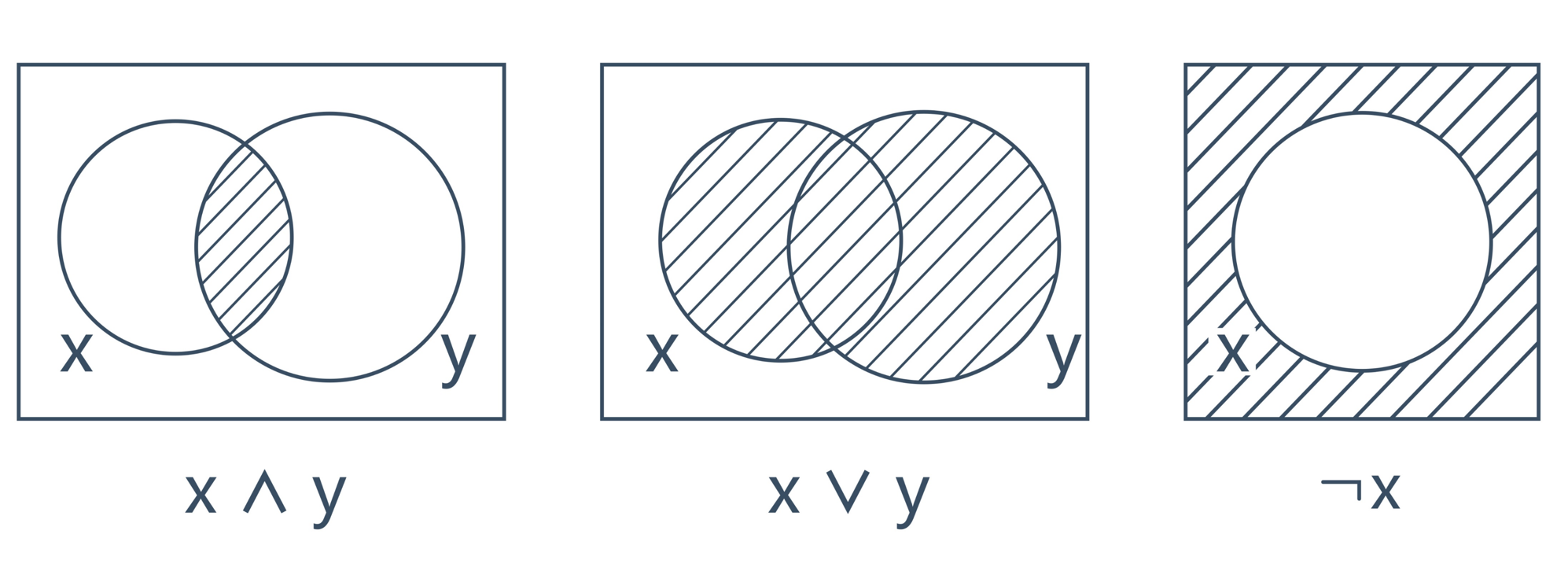}
	\centering
	\caption{\label{1_9}Venn Graph of ``AND", ``OR", ``NOT"}
\end{figure}

These representations help in understanding how Boolean operations combine and manipulate sets, reflecting their applications in logical processes and digital circuits. The Venn diagrams corresponding to the ``AND", ``OR" and ``NOT" operations are presented in Figure~\ref{1_9}.

Boolean operations can be used to construct various other logical operations. For example:

\begin{align}
    x \to y &= \neg x \vee y\\
    x \oplus y &= (x \vee y) \land (\neg x \vee \neg y) = (x \land \neg y) \vee (\neg x \land y)\\
    x \equiv y &= (x \land y) \vee (\neg x \land \neg y)
\end{align}

The truth tables for these operations are shown in Table~\ref{1_t41}:

\begin{table}[H]
	\centering
	\setlength{\tabcolsep}{5mm}
	\begin{tabular}{|c|c|c|c|c|} 
		\hline
		$x$ & $y$ & $x \to y$ & $x \oplus y$ & $x \equiv y$ \\
		\hline
		0 & 0 & 1 & 0 & 1 \\
		\hline
		0 & 1 & 1 & 1 & 0 \\
		\hline
		1 & 0 & 0 & 1 & 0 \\
		\hline
		1 & 1 & 1 & 0 & 1 \\
		\hline       
	\end{tabular}
	\caption{\label{1_t41} The truth table for ``AND", ``OR", and ``NOT"}
\end{table}

De Morgan's theorem is another crucial concept in Boolean algebra:

\begin{align}
    x \land y &= \neg(\neg x \vee \neg y)\\
    x \vee y &= \neg(\neg x \land \neg y)
\end{align}

\begin{exercise}
	
	Truth table of De Morgan's Law:
	
	Using the theorem formula, please calculate and fill in the correct result in the blank of the following truth tables in Tables~\ref{1_t5} and ~\ref{1_t6}:

	Formula one: \(x \land y = \neg(\neg x \vee \neg y)\)

	\begin{table}[H]
		\centering
		\setlength{\tabcolsep}{2mm}
		\begin{tabular}{|c|c|c|c|c|c|c|} 
			\hline
			$x$ & $y$ & $x \land y$ & $\neg x$ & $\neg y$ & $\neg x \vee \neg y$ & $\neg(\neg x \vee \neg y)$\\
			\hline
			0 & 0 & 0 & 1 & 1 & 1 & 0\\
			\hline
			0 & 1 & 0 & 1 & 0 & 1 & 0\\
			\hline
			1 & 0 & 0 & 0 & 1 & 1 & 0\\
			\hline
			1 & 1 & 1 & 0 & 0 & 0 & 1\\
			\hline          
		\end{tabular}
		\caption{\label{1_t5} The truth table for De Morgan's Law (Formula one)}
	\end{table}

	Formula two: \(x \vee y = \neg(\neg x \land \neg y)\)

	\begin{table}[H]
		\centering
		\setlength{\tabcolsep}{2mm}
		\begin{tabular}{|c|c|c|c|c|c|c|} 
			\hline
			$x$ & $y$ & $x \vee y$ & $\neg x$ & $\neg y$ & $\neg x \land \neg y$ & $\neg(\neg x \land \neg y)$\\
			\hline
			0 & 0 & 0 & 1 & 1 & 1 & 0\\
			\hline
			0 & 1 & 1 & 1 & 0 & 0 & 1\\
			\hline
			1 & 0 & 1 & 0 & 1 & 0 & 1\\
			\hline
			1 & 1 & 1 & 0 & 0 & 0 & 1\\
			\hline          
		\end{tabular}
		\caption{\label{1_t6} The truth table for De Morgan's Law (Formula two)}
	\end{table}
\end{exercise}

A Boolean function~\cite{sun43wegener1987complexity} is defined as a logical calculation that takes Boolean inputs to produce a Boolean output. Essentially, a Boolean function can only return True (1) or False (0). For a Boolean function with $k$ inputs, it is defined as:
\begin{equation}
    f:\{0, 1\}^{k} \to \{0, 1\}
\end{equation}

In cases where a Boolean function has multiple outputs, $k$ represents the number of inputs and $m$ the number of outputs:
\begin{equation}
    f:\{0, 1\}^{k} \to \{0, 1\}^{m}
\end{equation}

\begin{example}
For a Boolean function $f$ with $k=2$ (input of two numbers) and $m=1$:

If $f$ represents the ``OR" operation, then:
\begin{align*}
    f(00) &= 0, \\
    f(01) &= 1, \\
    f(10) &= 1, \\
    f(11) &= 1
\end{align*}

For a Boolean function $f$ with $k=3$ (input of three numbers) and $m=3$:

$f$ can be defined as:
\begin{align*}
    f(000) &= 000, \\
    f(011) &= 011, \\
    f(110) &= 111, \\
    f(111) &= 110
\end{align*}
\end{example}

The basic unit of computer operation is the logic circuit~\cite{sun44nelson1995digital}, which typically contains ``AND", ``OR" and ``NOT" gates. These gates function as Boolean operations.

Consider the ``Parity" Boolean function, represented by the following truth table (Table~\ref{1_t7}):
For a variable $x$, it can be written as:
\begin{equation}
    x = x_{2}x_{1}x_{0} , \text{ where } x_{2}, x_{1}, x_{0} \in \{ 0, 1 \}
\end{equation}

\begin{table}[H]
	\centering
	\setlength{\tabcolsep}{2mm}
	\begin{tabular}{|c|c|c|c|c|c|c|c|c|} 
		\hline
		x & 000 & 001 & 010 & 011 & 100 & 101 & 110 & 111 \\
		\hline
		$f$ & 0   & 1   & 1   & 0   & 1   & 0   & 0   & 1   \\
		\hline         
	\end{tabular}
	\caption{\label{1_t7} Truth table of the ``Parity" Boolean function}
\end{table}

The Boolean function $f$ in terms of Boolean logic can be expressed as:
\begin{equation}
    f = (\neg x_{2} \land \neg x_{1} \land x_{0}) \vee (\neg x_{2} \land x_{1} \land \neg x_{0}) \vee (x_{2} \land \neg x_{1} \land \neg x_{0}) \vee (x_{2} \land x_{1} \land x_{0})
\end{equation}

Alternatively, it can be simplified as:
\begin{equation}
    f = x_{2} \oplus x_{1} \oplus x_{0}
\end{equation}

\section{Computer architecture and logic circuits}~{}

\begin{figure}[H]
	\includegraphics[scale=0.5]{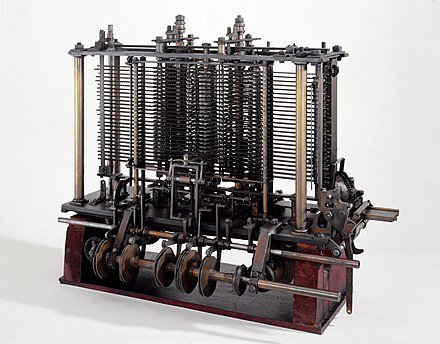}
	\centering
	\caption{\label{1_3}Charles Babbage invented an analysis engine based on Binary, the figure is from \url{https://commons.wikimedia.org/wiki/File:Babbages_Analytical_Engine,_1834-1871._(9660574685).jpg}.}
\end{figure}

Human computing tools have a history spanning thousands of years ~\cite{sun2nardi1996context}. From primitive tools like fingers to ancient calculators such as the abacus, and later to mechanical computers, the evolution of these tools has been remarkable. Early tools like the abacus were based on the Decimal system, while modern computers operate on Binary logic. For instance, Figure~\ref{1_3} illustrates a tool based on Binary.

Decimal: Historically, the Decimal system has been pivotal in mathematical computing. Although the human brain is a powerful computing tool, we have always needed auxiliary tools to enhance and accelerate the computing process.

Binary: A significant advancement was the invention of binary logic around 1702, which laid the technical groundwork for the development of modern computers.

Before modern computer was invented, the word ``computer" is already exist, it represented a person who perform computing. Until 1946, when the first general digital computer was invented, the word ``Computer” refer to the machine which perform computing tasks.


Before the invention of modern computers, the term ``computer” referred to a person who performed calculations. This definition changed in 1946 with the invention of the first general digital computer, after which ``computer" came to denote the machine itself.

The Turing Machine is a pivotal mathematical model of computation, defining an abstract machine that manipulates symbols on a strip of tape according to a table of rules~\cite{sun3de2018turing}. As a general representation of a CPU, the Turing Machine controls all data manipulation in a computer, using sequential memory for data storage. Despite its simplicity, a Turing Machine can simulate the logic of any computer algorithm.

With an understanding of computation models, the next question is: how were computers built?

\begin{figure}[H]
	\includegraphics[scale=0.1]{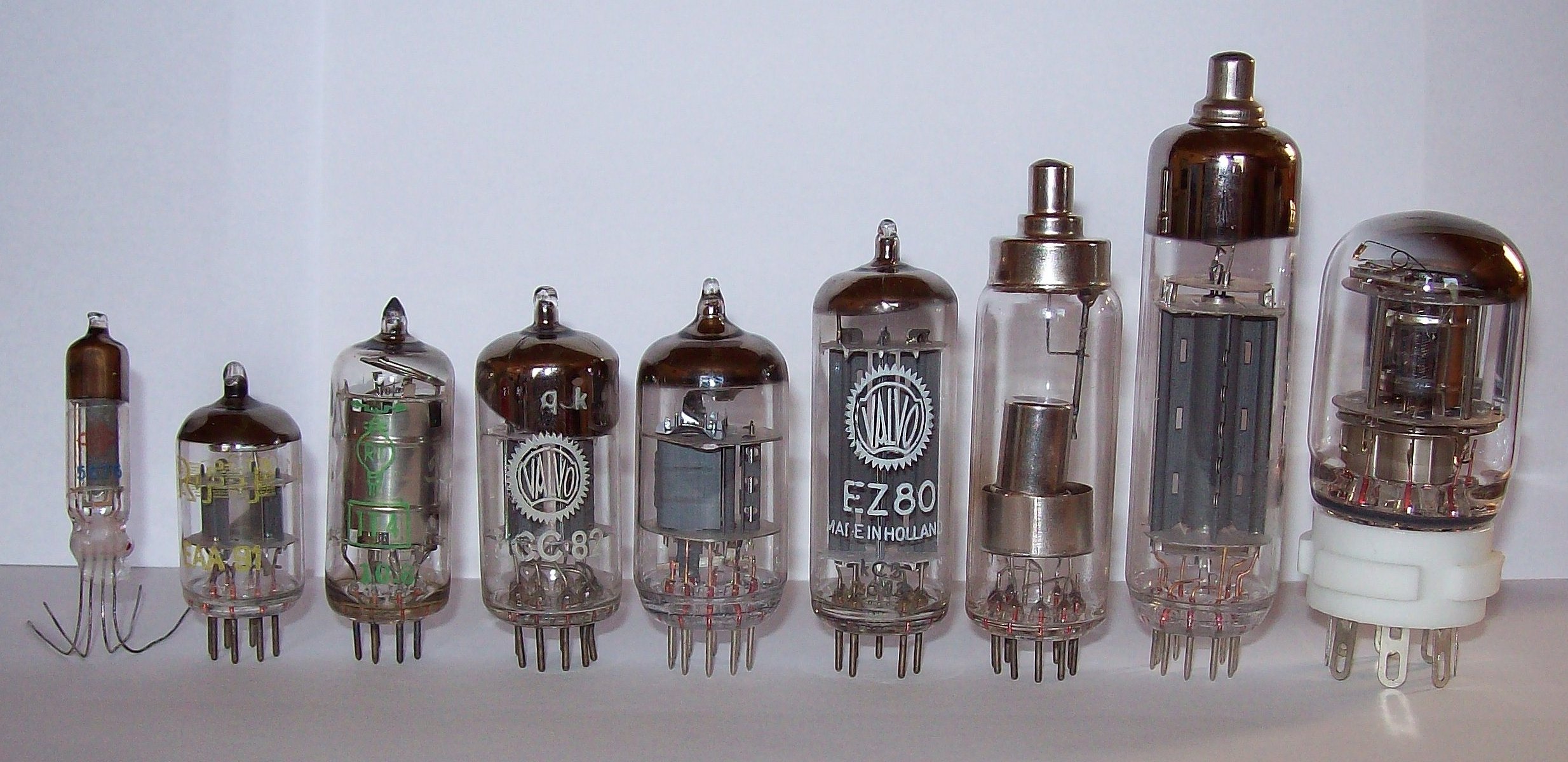}
	\centering
	\caption{\label{1_5} A vacuum tube, electron tube. The figure is from 
    \url{https://commons.wikimedia.org/wiki/File:Elektronenroehren-auswahl.jpg}.}
\end{figure}

From 1937 to 1960, the digital computer was developed, heralding the birth of computer science. The earliest electronic computers were built using vacuum tubes, with high and low levels representing binary digits (1 and 0, respectively). These tubes facilitated the implementation of binary operation rules. Figure~\ref{1_5} shows various vacuum tubes.

\begin{figure}[H]
	\includegraphics[scale=0.25]{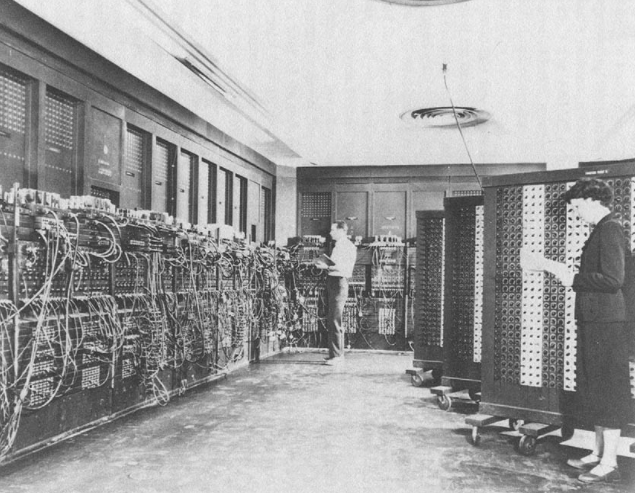}
	\centering
	\caption{\label{1_4}The world's first electronic computer was based on Binary called ENIAC. The figure is from  \url{https://commons.wikimedia.org/wiki/File:Glen_Beck_and_Betty_Snyder_program_the_ENIAC_in_building_328_at_the_Ballistic_Research_Laboratory.jpg}.}
\end{figure}

In 1946, the world's first electronic computer based on binary logic, the ENIAC, was introduced. Figure~\ref{1_4} shows the ENIAC.

\begin{figure}[H]
	\includegraphics[scale=0.1]{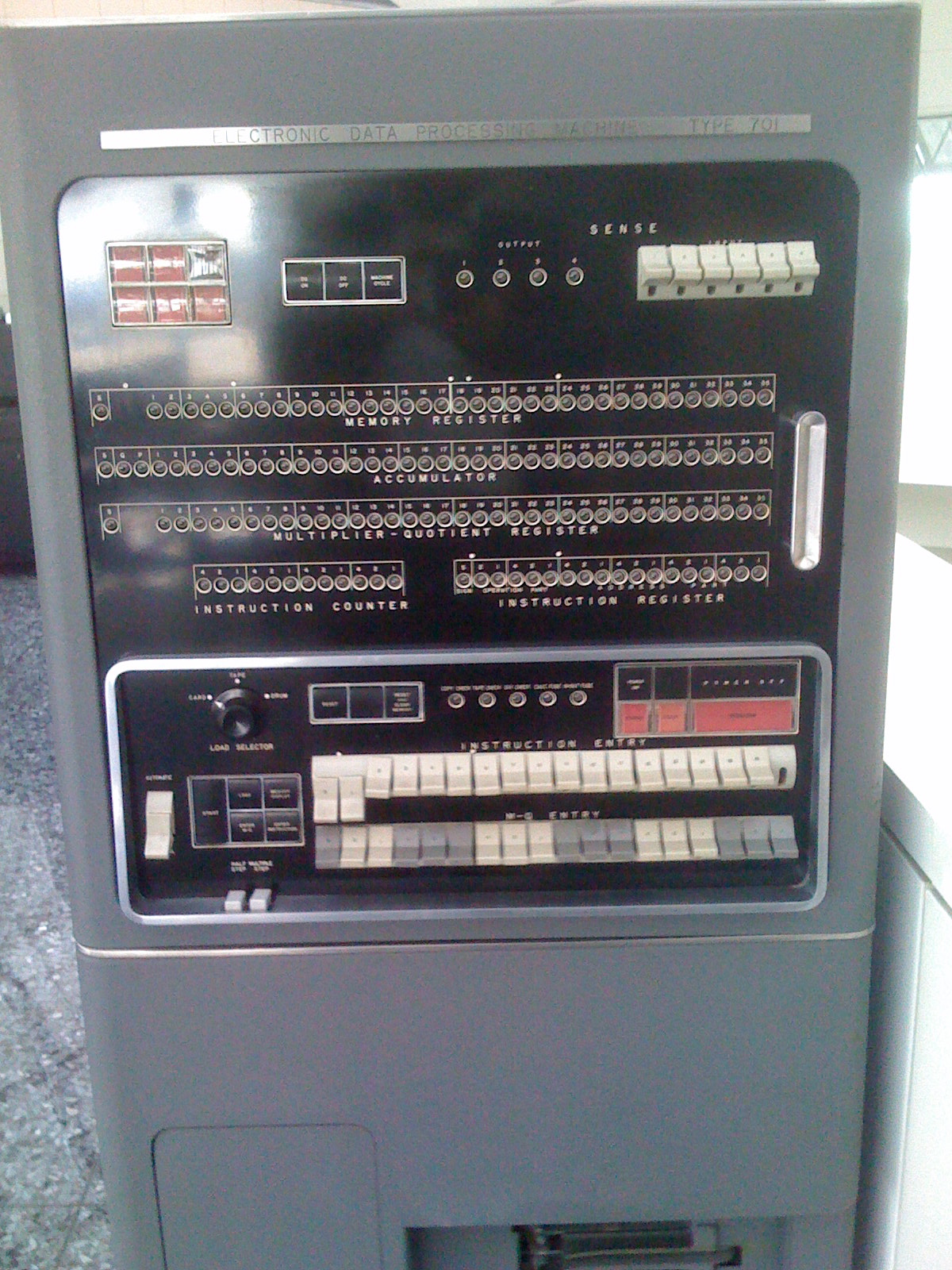}
	\centering
	\caption{\label{1_6}The first commercial electronic computer was made by IBM :IBM 701 (1952 year). The figure is from \url{https://commons.wikimedia.org/wiki/File:IBM_701console.jpg}.}
\end{figure}

IBM constructed the first commercial computer, the IBM 701, in 1952. Figure~\ref{1_6} showcases this pioneering machine.

Modern computers, products of the third industrial revolution, include PCs, mobile phones, laptops, and other devices typically built using semiconductor technology. These computers are based on the "Von Neumann Architecture".

\begin{figure}[H]
	\includegraphics[scale=0.5]{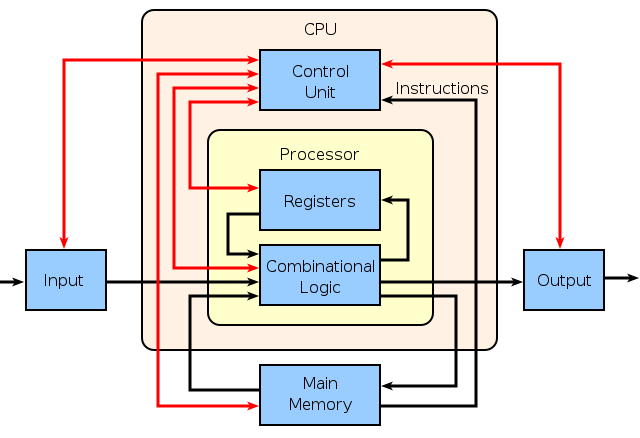}
	\centering
	\caption{\label{1_7}Von Neumann Architecture. The figure is from  
    \url{https://commons.wikimedia.org/wiki/File:ABasicComputer.svg}.}
\end{figure}

Von Neumann Architecture: Despite significant changes in the manufacturing process, size, and shape of electronic computers, their fundamental architecture has remained consistent. This architecture encompasses input, output, control unit, computing unit, and storage unit. Figure~\ref{1_7} illustrates the Von Neumann Architecture.

The basic unit of computer operation is the logic circuit. The basic logic circuit contains ``AND", ``OR", ``NOT" 3 gates. ``AND", ``OR", ``NOT" are also the basic operations of Boolean algebra~\cite{sun4whitesitt2012boolean,sun41halmos2009introduction}.

\begin{figure}[H]
	\includegraphics[scale=0.3]{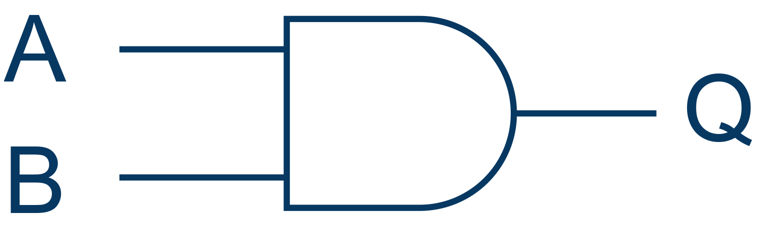}
	\centering
	\caption{\label{1_10}The ``AND" gate, $Q=A\land B$}
\end{figure}

\begin{figure}[H]
	\includegraphics[scale=0.3]{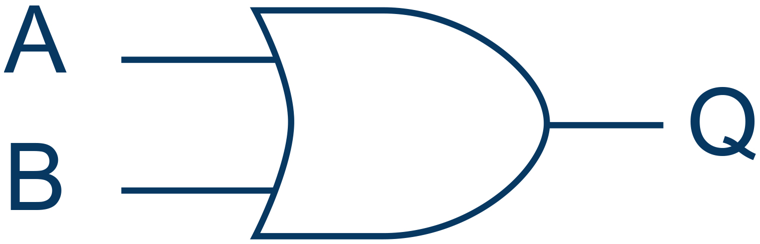}
	\centering
	\caption{\label{1_11}The ``OR" gate, $Q=A\vee B$}
\end{figure}

\begin{figure}[H]
	\includegraphics[scale=0.3]{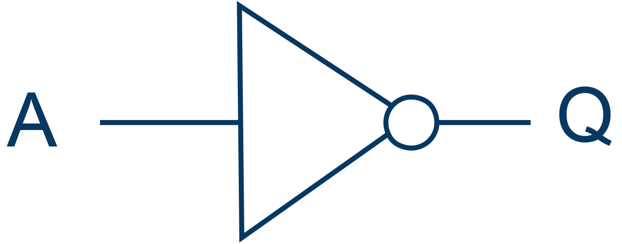}
	\centering
	\caption{\label{1_12}The ``NOT" gate, $Q=\neg A$}
\end{figure}

In the ``AND" gate, the output $Q=A\land B$. In the ``OR" gate, the output $Q=A\vee B$. In the ``NOT" gate, the output $Q=\neg A$. The ``AND" gate, ``OR" gate, ``NOR" gate shown in Figure~\ref{1_10}, Figure~\ref{1_11} and Figure \ref{1_12}.

The truth table of ``AND" gate, ``OR" gate, ``NOT" gate in table~\ref{1_t8}:

\begin{table}[H]
	\centering
	\setlength{\tabcolsep}{5mm}
	\begin{tabular}{|c|c|c|c|} 
		\hline
		x & y & $x\land y$ & $x\vee y$ \\
		\hline
		0 & 0 & 0          & 0         \\
		\hline    
		1 & 0 & 0          & 1         \\
		\hline
		0 & 1 & 0          & 1         \\
		\hline
		1 & 1 & 1          & 1         \\
		\hline
		\multicolumn{2}{|c|}{x} & \multicolumn{2}{|c|}{$\neg x$} \\
		\hline
		\multicolumn{2}{|c|}{0} & \multicolumn{2}{|c|}{1} \\
		\hline
		\multicolumn{2}{|c|}{1} & \multicolumn{2}{|c|}{0} \\  
		\hline
	\end{tabular}
	\caption{\label{1_t8} The truth table of ``AND" gate, ``OR" gate, ``NOT" gate}
\end{table}	

\begin{figure}[H]
	\includegraphics[scale=0.25]{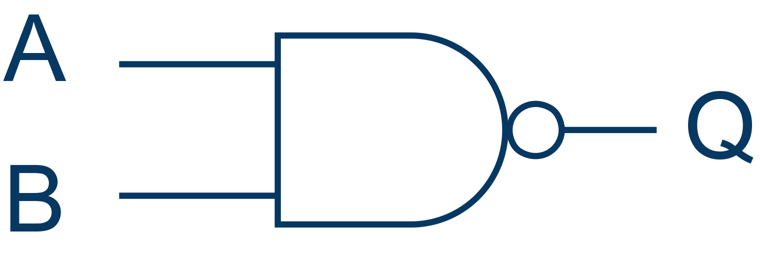}
	\centering
	\caption{\label{1_13} ``NAND" gate $Q= \neg(A\land B)$}
\end{figure}

\begin{figure}[H]
	\includegraphics[scale=0.25]{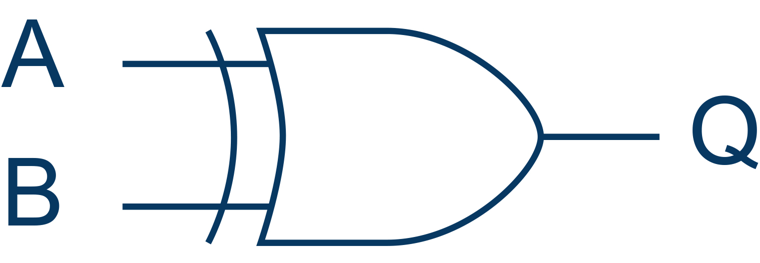}
	\centering
	\caption{\label{1_14} ``XOR" gate $Q=(A\land\neg B)\vee(\neg A\land B)=A\oplus B$
	}
\end{figure}

\begin{figure}[H]
	\includegraphics[scale=0.25]{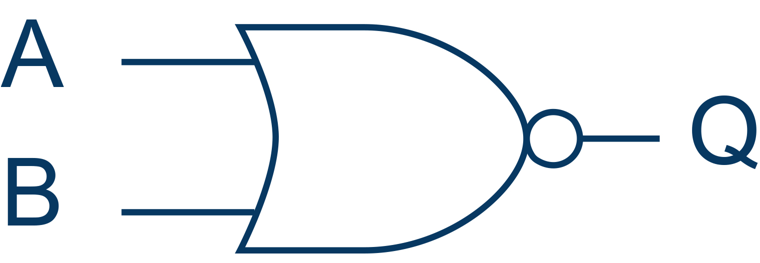}
	\centering
	\caption{\label{1_15} ``NOR"gate $Q=\neg(A\vee B)$}
\end{figure}

\begin{figure}[H]
	\includegraphics[scale=0.25]{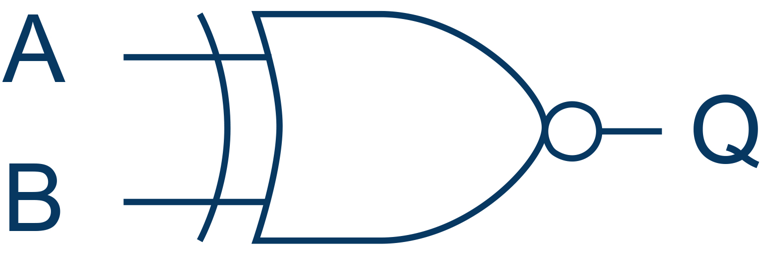}
	\centering
	\caption{\label{1_16} ``XNOR"gate $Q=\neg(A\oplus B)$}
\end{figure}

Other operations ``NAND", ``NOR", ``XOR", ``XNOR" also is common logic circuits. In the ``NAND" gate, the output $Q= \neg(A\land B)$. In the ``XOR" gate, the output $Q=(A\land\neg B)\vee(\neg A\land B)=A\oplus B$. In the ``NOR" gate, the output $Q=\neg(A\vee B)$. In the ``XNOR" gate the output $Q=\neg(A\oplus B)$. The gate ``NAND", ``NOR", ``XOR", ``XNOR" are show in Figure~\ref{1_13}, Figure~\ref{1_14}, Figure~\ref{1_15} and Figure~\ref{1_16}.

\begin{exercise}

	Truth table of Logic Gate
	
	According to the logic gate formula provided, please calculate and fill the correct result in the following truth table the truth table are in table~\ref{1_t9} and table~\ref{1_t10}.
	
	1. ``XOR" gate $Q=(A\land\neg B)\vee(\neg A\land B)=A\oplus B$
	
	\begin{table}[H]
		\centering
		\setlength{\tabcolsep}{5mm}
		\begin{tabular}{|c|c|c|c|c|} 
			\hline
			A & B & $A\land\neg B$ & $\neg A\land B$ & $A\oplus B$\\
			\hline
			0 & 0 &           &         & \\
			\hline    
			0 & 1 &           &         & \\
			\hline
			1 & 0 &           &         & \\
			\hline
			1 & 1 &           &         & \\
			\hline   
		\end{tabular}
		\caption{\label{1_t9} ``XOR" gate $Q=(A\land\neg B)\vee(\neg A\land B)=A\oplus B$}
	\end{table}	
	
	2. ``XNOR"gate $Q=\neg(A\oplus B)$
	
	\begin{table}[H]
		\centering
		\setlength{\tabcolsep}{5mm}
		\begin{tabular}{|c|c|c|c|} 
			\hline
			A & B & $A\oplus B$ & $\neg(A\oplus B)$\\
			\hline
			0 & 0 &           &        \\
			\hline    
			0 & 1 &           &        \\
			\hline
			1 & 0 &           &        \\
			\hline
			1 & 1 &           &        \\
			\hline   
		\end{tabular}
		\caption{\label{1_t10} ``XOR" gate $Q=(A\land\neg B)\vee(\neg A\land B)=A\oplus B$}
	\end{table}	
\end{exercise}

Now we look at some examples of logic circuits.

\textbf{Half adder:} 

\begin{figure}[H]
	\includegraphics[scale=0.4]{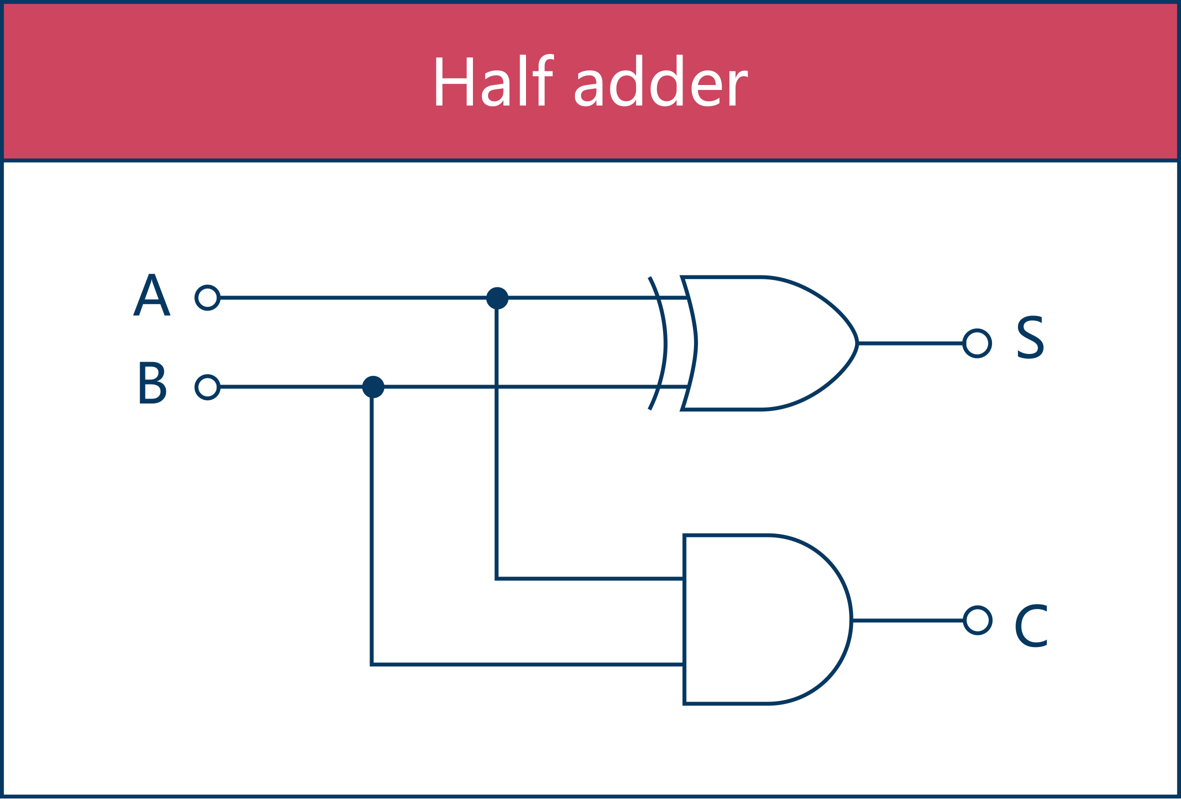}
	\centering
	\caption{\label{3_2}The half adder only calculates the sum of the base bits, and does not consider the carry from the lower bits, so it is called a half adder.}
\end{figure}

All computing algorithms can be implemented using logic circuits. The basic units in arithmetic circuits are half adders and full adders. A half adder has two inputs and two outputs, the inputs can be identified as A, B, and the outputs are usually identified as Sum (Sum) and Carry (Carry). The input is S after the XOR operation AND C after the and operation. The half adder has two binary inputs, which add the input values and output the result to Sum and Carry. Although the half adder can produce carry values, the half adder itself cannot process carry values. The Figure~\ref{3_2} show the half adder.

\begin{exercise}
The truth table of half adder in table~\ref{3_t1}. Based on the logic circuit shown in Figure~\ref{3_2}, calculate and fill in the correct result in the blanks of the truth table.
\begin{table}[H]
	\centering
	\setlength{\tabcolsep}{3mm}
	\begin{tabular}{|c|c|c|c|} 
		\hline
		A & B & Sum of Half add "S" & Carry Number "C"\\
		\hline
		0 & 0 &           &        \\
		\hline    
		1 & 0 &           &        \\
		\hline
		0 & 1 &           &        \\
		\hline
		1 & 1 &           &        \\
		\hline   
	\end{tabular}
	\caption{\label{3_t1} The truth table of half adder}
\end{table}	
\end{exercise}

\textbf{Full adder:}

\begin{figure}[H]
	\includegraphics[scale=0.4]{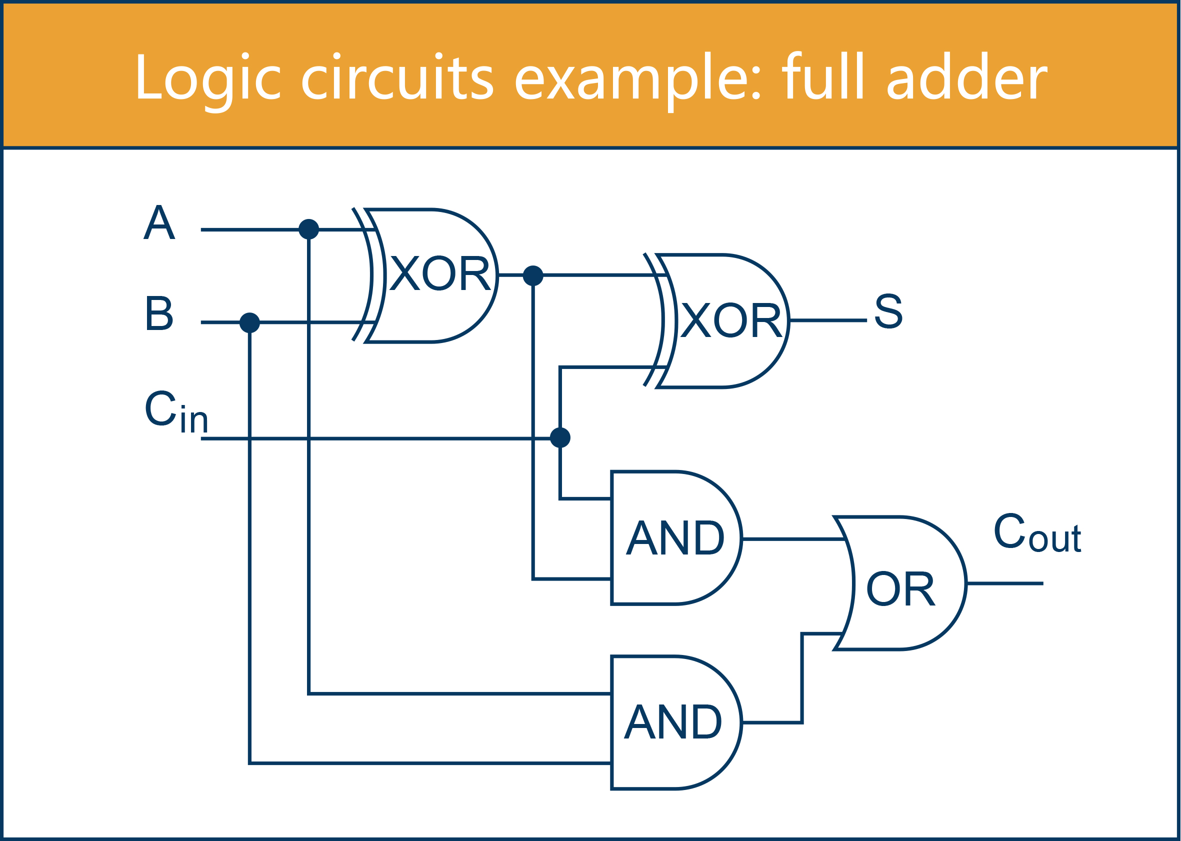}
	\centering
	\caption{\label{3_3} Logic circuits example : Full adder }
\end{figure}

The English name of the full adder is full-adder, which is a combination line that uses a gate circuit to add two binary numbers and find the sum, called a full adder. The one-bit full adder can handle the low carry and output the standard add carry. Multiple one-bit full adders can be cascaded to obtain a multi-bit full adder. All computing algorithms can be implemented using logic circuits. The basic units in arithmetic circuits are half adders and full adders. There is an example of full adder in Figure~\ref{3_3} 

\begin{exercise}

	Truth table Full adder
	
	Based on the logic circuit shown in Figure~\ref{3_3}, calculate and fill in the correct result in the blanks of the truth table. The truth table of full adder in table~\ref{3_t2}
	
	\begin{table}[H]
		\centering
		\setlength{\tabcolsep}{3mm}
		\begin{tabular}{|c|c|c|c|c|} 
			\hline
			A & B & low-order carry & Sum S & Carry number \\
			\hline
			0 & 0 & 0 &&\\
			\hline    
			0 & 0 & 1 &&\\
			\hline
			0 & 1 & 0 &&\\
			\hline
			0 & 1 & 1 &&\\
			\hline  
			1 & 0 & 0 &&\\
			\hline
			1 & 0 & 1 &&\\
			\hline
			1 & 1 & 0 &&\\
			\hline
			1 & 1 & 1 &&\\
			\hline              
		\end{tabular}
		\caption{\label{3_t2} The truth table of half adder}
	\end{table}	
\end{exercise}

\begin{figure}[H]
	\includegraphics[scale=0.6]{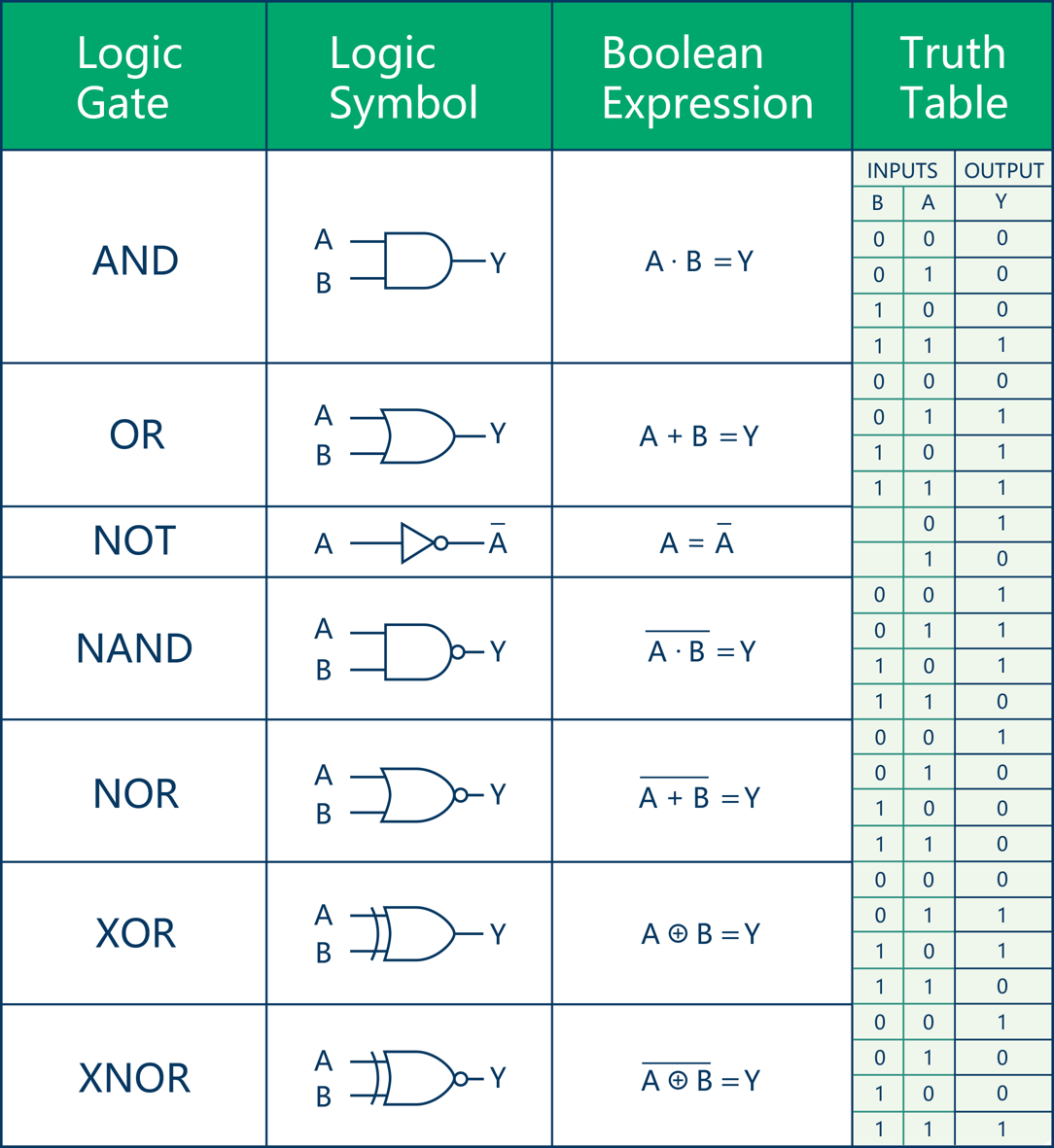}
	\centering
	\caption{\label{1_21}The summary of Logical circuit truth table are as follows}
\end{figure}

The summary of Logical circuit truth table shown in Figure~\ref{1_21} :

Logical operator ``And": $A\land B$ can be write as: $A\cdot B$. ``OR": $A\vee B$ can be write as: $A+B$. ``NOT": $\neg A$ can be write as: $\bar{A}$.

``NAND" gates alone (or alternatively ``NOR" gates alone) can be used to reproduce the functions of all the other logic gates.

So NAND is ``universal” to compute any Boolean function. ``NAND" or ``NOR" can form all logic gates independently (Charles Sanders Peirce,during 1880–1881).
Therefore, the ``NAND"  gate is a complete logic gate.

\begin{figure}[H]
	\includegraphics[scale=0.3]{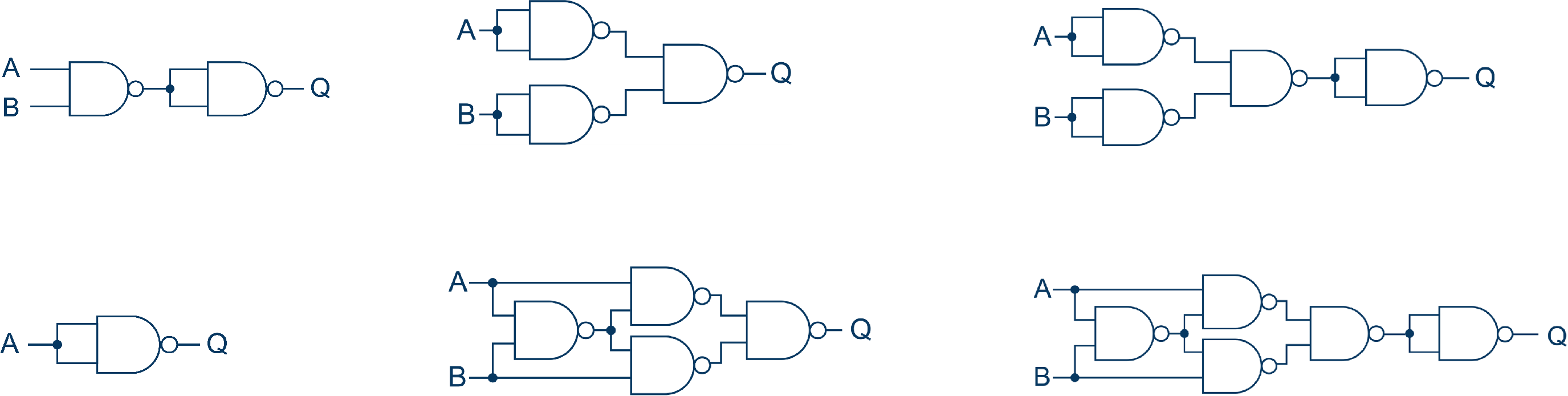}
	\centering
	\caption{\label{1_17} Use the ``NAND" gate to build the six gate(``AND" gate, ``OR" gate, ``NOR" gate, ``NOT" gate, ``XOR" gate, ``XNOR" gate). The order is left to right, top to bottom.}
\end{figure}

In the Figure~\ref{1_17} the six gate(``AND" gate, ``OR" gate, ``NOT" gate, ``NAND", ``XOR", ``XNOR") 
are use ``NAND" gate to build.

\section{Computers with integrated circuits}~{}


The earliest electronic computers utilized vacuum tubes in their logic circuits. However, with the invention of the transistor in 1947, these circuits began incorporating transistors in the 1950s. A significant milestone was the invention of the integrated circuit in 1959, which revolutionized logic circuit manufacturing, with transistors being the core component. Integrated circuit technology has advanced dramatically, allowing for an increasing number of transistors to be integrated per unit area, thereby significantly enhancing chip computing power. In 1965, Gordon Moore forecasted that the number of transistors on a chip would double approximately every 18 months, a prediction known as Moore's Law~\cite{sun45schaller1997moore}. This trend continued until around 2020, when the diminishing scale of semiconductor processes approached the quantum limit (1 nanometer), causing Moore's Law to reach its threshold.

From 1945 to 2020, computing power witnessed an astonishing increase of four thousand billion times. However, this exponential growth faces inevitable limits as transistors near atomic sizes and quantum effects at microscopic levels pose new challenges. 



While Moore's Law~\cite{sun45schaller1997moore} has primarily focused on hardware developments to enhance computing power, reaching its limit could shift the focus to software improvements, such as cloud computing. This approach involves dividing a task into numerous parts and executing them simultaneously, vastly improving efficiency. For example, a task that would take 1000 hours for one person can be completed in an hour by 1000 people, as seen in Bitcoin mining pools. However, the energy consumption of traditional electronic computers is increasing, leading to projections that by 2040, global computer electricity consumption might equal the current global power generation.

The evolution of software that has improved computing power includes:

1950s: Machine language and programming languages like Fortran.

1950s: Finite state machines and formal languages.
1960s: Computational complexity.

1970s: Computational time and circuit complexity; VLSI models; algorithms and data structures.

1980s and 1990s: Parallel computing; distributed computing; cryptography.

2000s and beyond: Cloud computing and machine learning.

\begin{figure}[H]
	\includegraphics[scale=0.12]{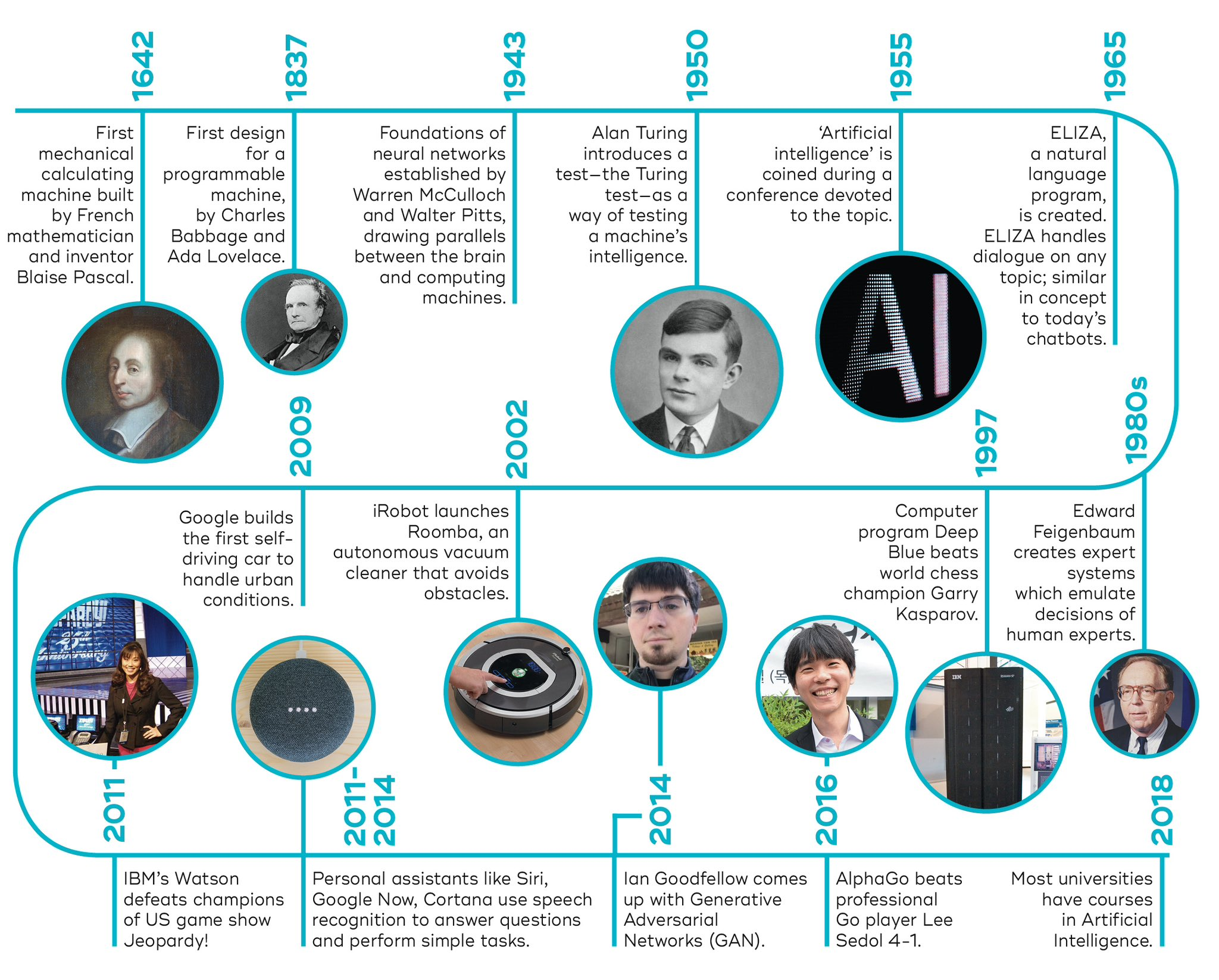}
	\centering
	\caption{\label{1.1gai} The brief history of neural networks. The figure is from \url{https://qbi.uq.edu.au/files/40697/The-Brain-Intelligent-Machines-AI-timeline.jpg}.}
\end{figure}

\begin{figure}[H]
	\includegraphics[scale=0.2]{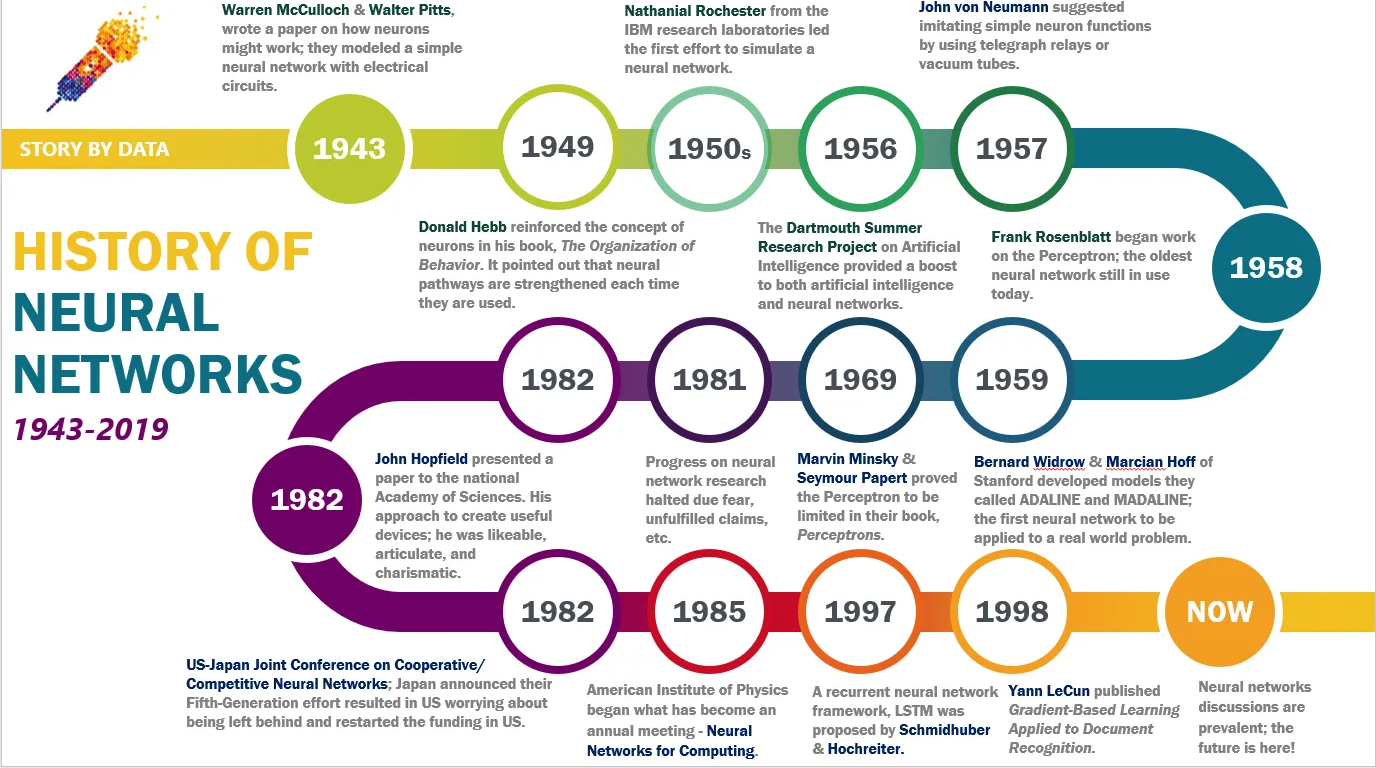}
	\centering
	\caption{\label{1.2gai1} The brief history of neural networks. The figure is from \url{https://medium.com/analytics-vidhya/brief-history-of-neural-networks-44c2bf72eec}.}
\end{figure}

\begin{figure}[H]
	\includegraphics[scale=0.35]{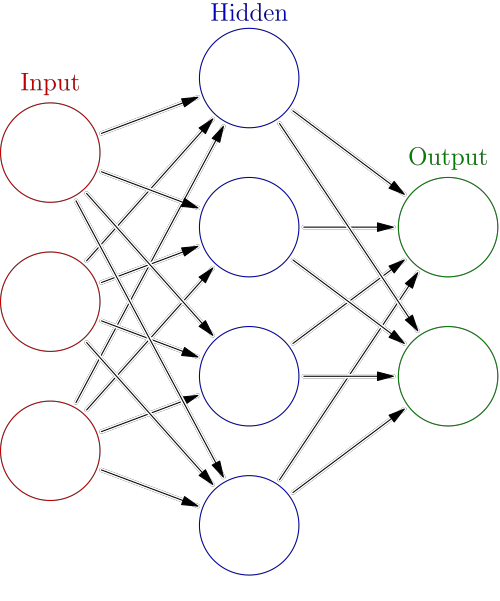}
	\centering
	\caption{\label{1.3gai} Artificial neural network with layer coloring. The figure is from \url{https://commons.wikimedia.org/wiki/File:Colored_neural_network.svg}.}
\end{figure}

Artificial Neural Networks (ANNs)~\cite{sun46dongare2012introduction} have been a focal point in artificial intelligence research since the 1980s. The development and structure of ANNs are depicted in Figures~\ref{1.1gai}, ~\ref{1.2gai1}, and ~\ref{1.3gai}, illustrating the progression and architecture of these networks.

A neural network has three layer structure. The artificial neural networks structure is shown in the Figure~\ref{1.3gai}.

\section{Reversible computation}~{}

Reversible computing~\cite{sun47toffoli1980reversible} is a model of computation where the computational process to some extent is time-reversible. In a model that uses deterministic transitions from one state of the abstract machine to another, a necessary condition for reversibility is that the mapping from states to their successors must be one-to-one. Reversible computing is a form of unconventional computing. Due to the unitarity of quantum mechanics, quantum circuits are usually reversible.

For a computational process to be physically reversible, it must also be logically reversible.

Landauer’s principle~\cite{sun7maroney2009generalizing} is the rigorously valid observation that the oblivious erasure of \( n \) bits of known information must always incur a cost of \( nkT\) in thermodynamic entropy, where \( k \) is the Boltzmann constant and \( T \) is the temperature. A discrete, deterministic computational process is logically reversible if its transition function, mapping old computational states to new ones, is a one-to-one function. This means the output logical states uniquely determine the input logical states of the computational operation.

\begin{figure}[H]
	\includegraphics[scale=0.25]{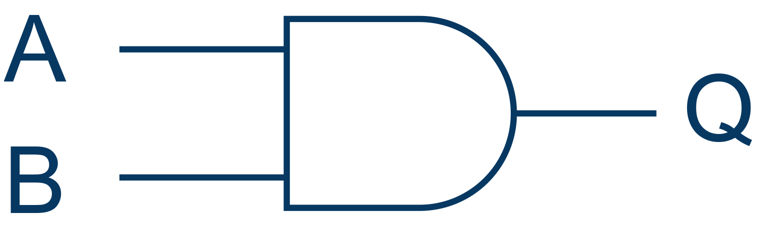}
	\centering
	\caption{\label{3_7} TThe ``AND" gate }
\end{figure}

\begin{figure}[H]
	\includegraphics[scale=0.25]{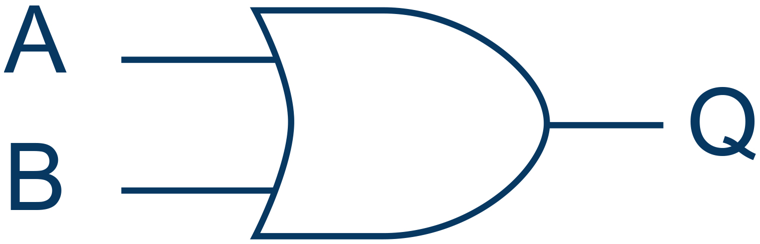}
	\centering
	\caption{\label{3_8}The ``OR" gate}
\end{figure}

According to the principles of informatics, modern computing processes based on ``AND" and ``OR" gates must consume energy. Both ``AND" and ``OR" gates, having two inputs and one output, necessitate the erasure of one bit of information during operation. This process consumes energy, resulting in an increase in entropy. Figures \ref{3_7} and \ref{3_8} show the ``AND" and ``OR" gates, respectively.

If a reversible computing process can be constructed, then, from the informatics perspective, the computing process can potentially consume no energy. For a computational process to be physically realizable as reversible, it must also be logically reversible (Randall, 1962, IBM).

\begin{figure}[H]
	\includegraphics[scale=0.6]{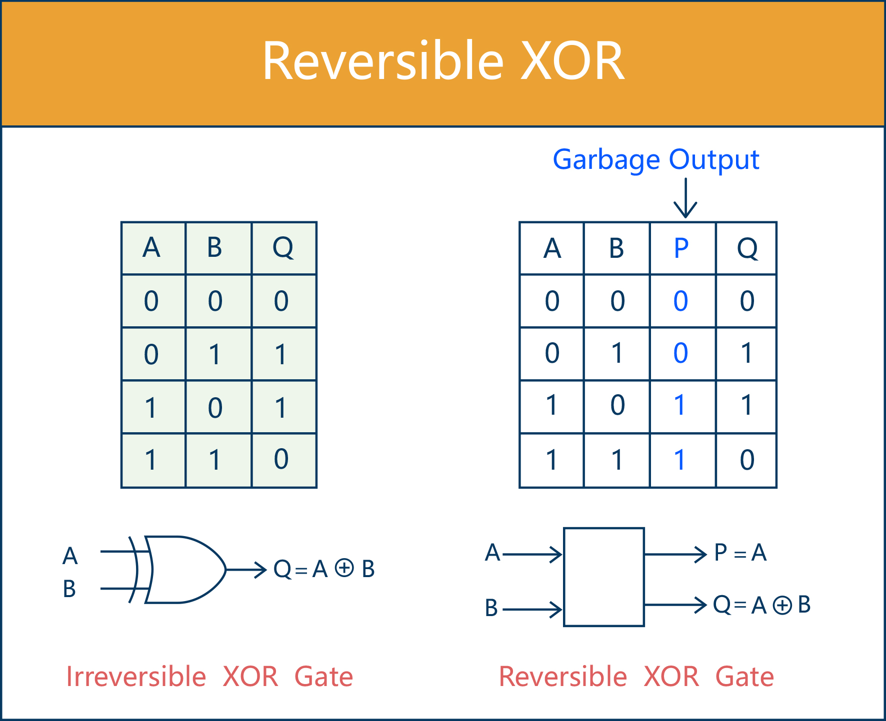}
	\centering
	\caption{\label{3_9} Example:reversible XOR gate}
\end{figure}

\begin{figure}[H]
	\includegraphics[scale=0.25]{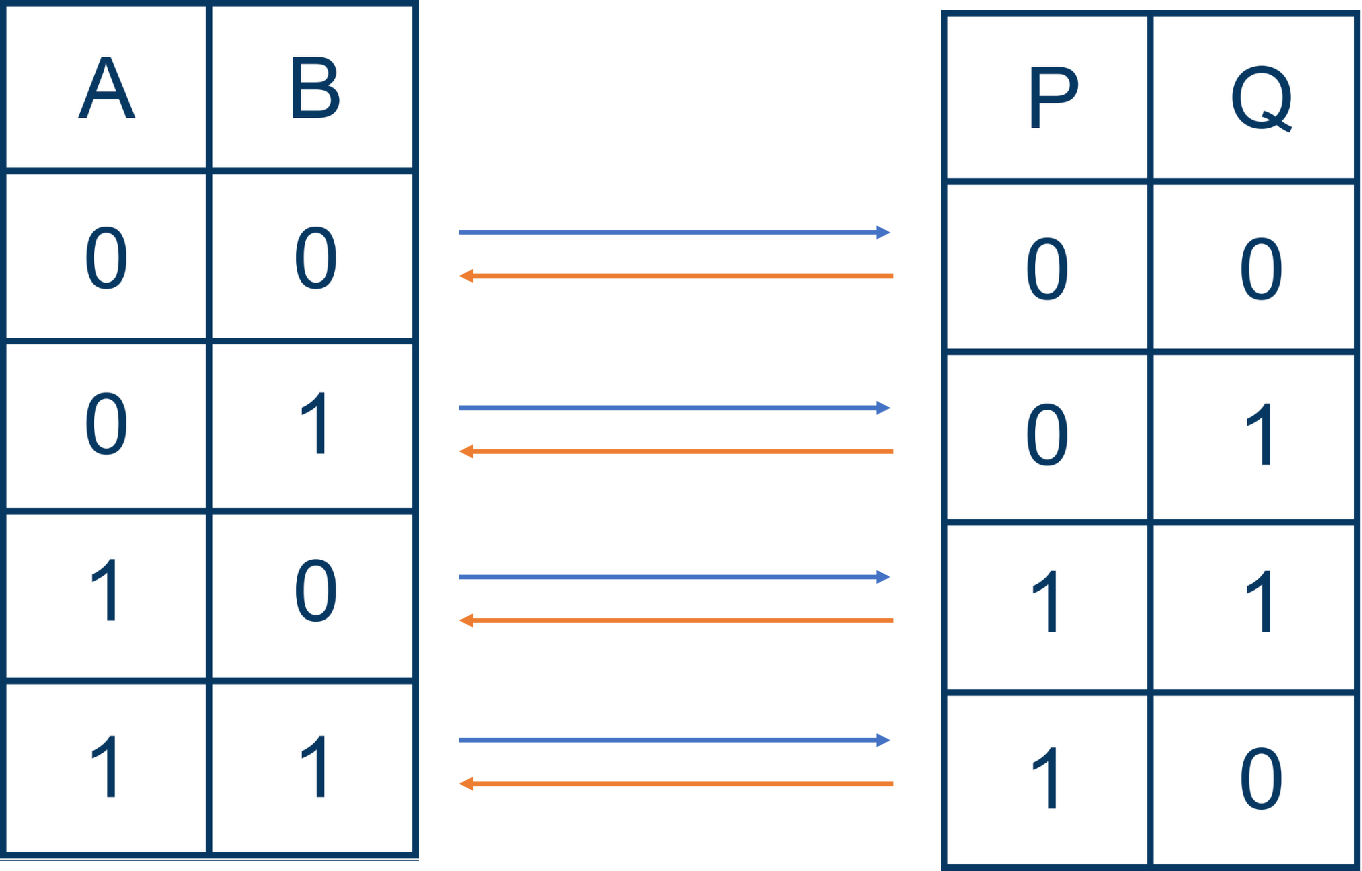}
	\centering
	\caption{\label{3_10} The relationship of input and output. }
\end{figure}

An example of reversible computing is illustrated in Figure~\ref{3_9}, which shows the reversible XOR gate. Figure~\ref{3_10} demonstrates the relationship between the inputs (\( A \) and \( B \)) and the outputs (\( P \) and \( Q \)). If we take the outputs \( P \) and \( Q \) as new inputs and pass them through the reversible XOR gate again, we find that the new outputs are \( P' = A \) and \( Q' = B \), thereby demonstrating the reversibility of the process.

Now we consider some examples of reversible computing gates.

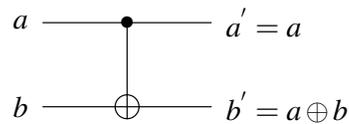
\begin{figure}[H]
    \[ \Qcircuit @!C @!R{
        &\lstick{a} &\ctrl{1}&\rstick{a^{'}=a}\qw\\
        &\lstick{b} &\targ&\rstick{b^{'}=a\oplus b}\qw}\]
    \caption{\label{1_18}The reversible computation gate CNOT gate}
\end{figure}

\textbf{Control-NOT(XOR) gate}:

Now lei's see the example of reversible: The Control-NOT(XOR) gate. This gate is a universal reversible logic gate it's contains 2 input qubits and 2 output qubits, the first input qubit are called control qubit, and the second qubit is called the controlled qubit. When and only control qubits are both 1, the controlled qubit performs a ``NO" operation, otherwise, the state of the controlled qubit remains unchanged. The Control-NOT(XOR) gate are shown in Figure~\ref{1_18}.

\textbf{Toffoli gate:}

\begin{figure}[H]
	\includegraphics[scale=0.45]{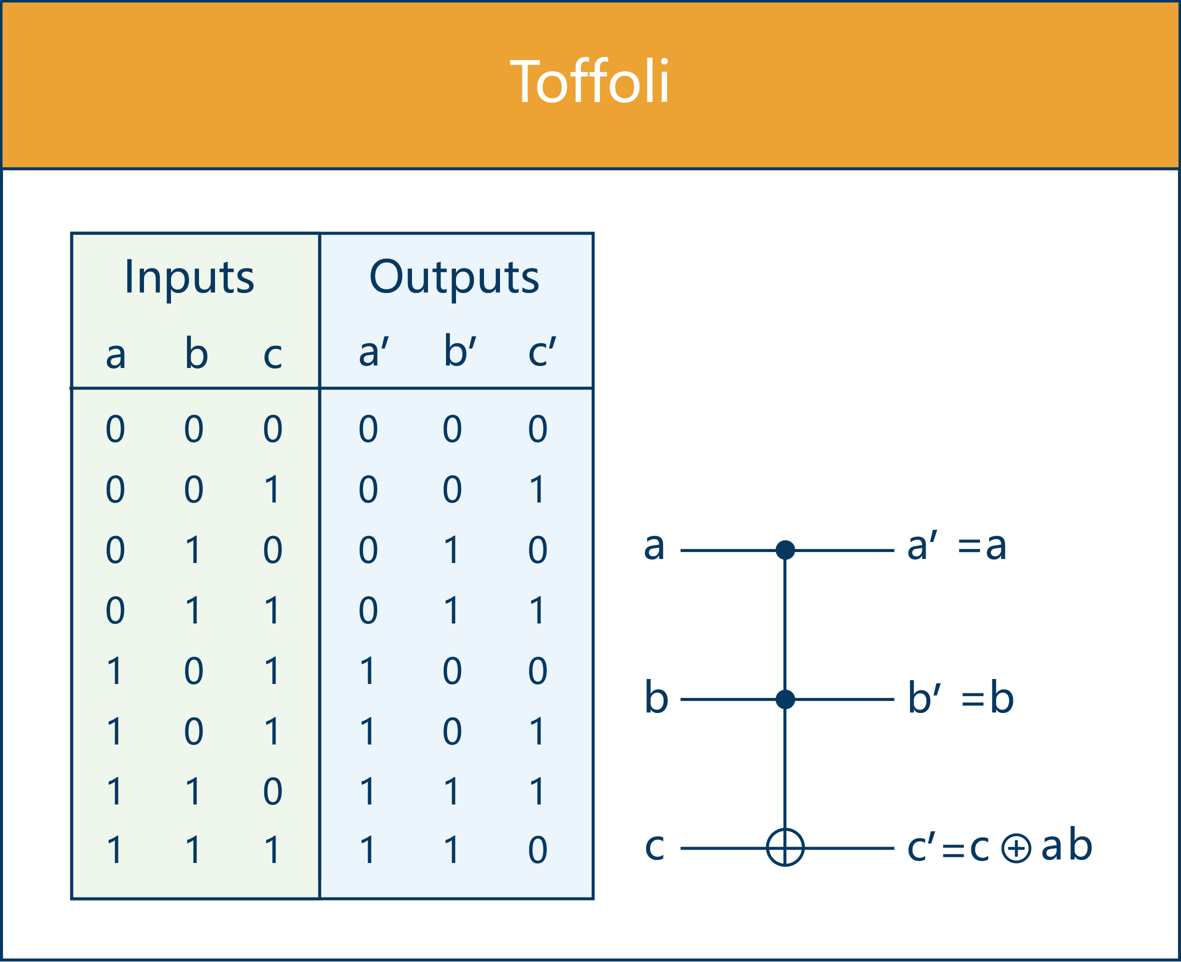}
	\centering
	\caption{\label{3_11} Toffoli gate}
\end{figure}

The Toffoli gate is a universal reversible logic gate it's contains 3 input qubits and 3 output qubits, the first two input qubits are called control qubits, and the third qubit is called the controlled qubit. When and only two control qubits are both 1, the controlled qubit performs a ``NO" operation, otherwise, the state of the controlled qubit remains unchanged. The Toffoli gate is a complete general logic gate. The Figure~\ref{3_11} show a Toffoli gate.

\begin{figure}[H]
	\includegraphics[scale=0.45]{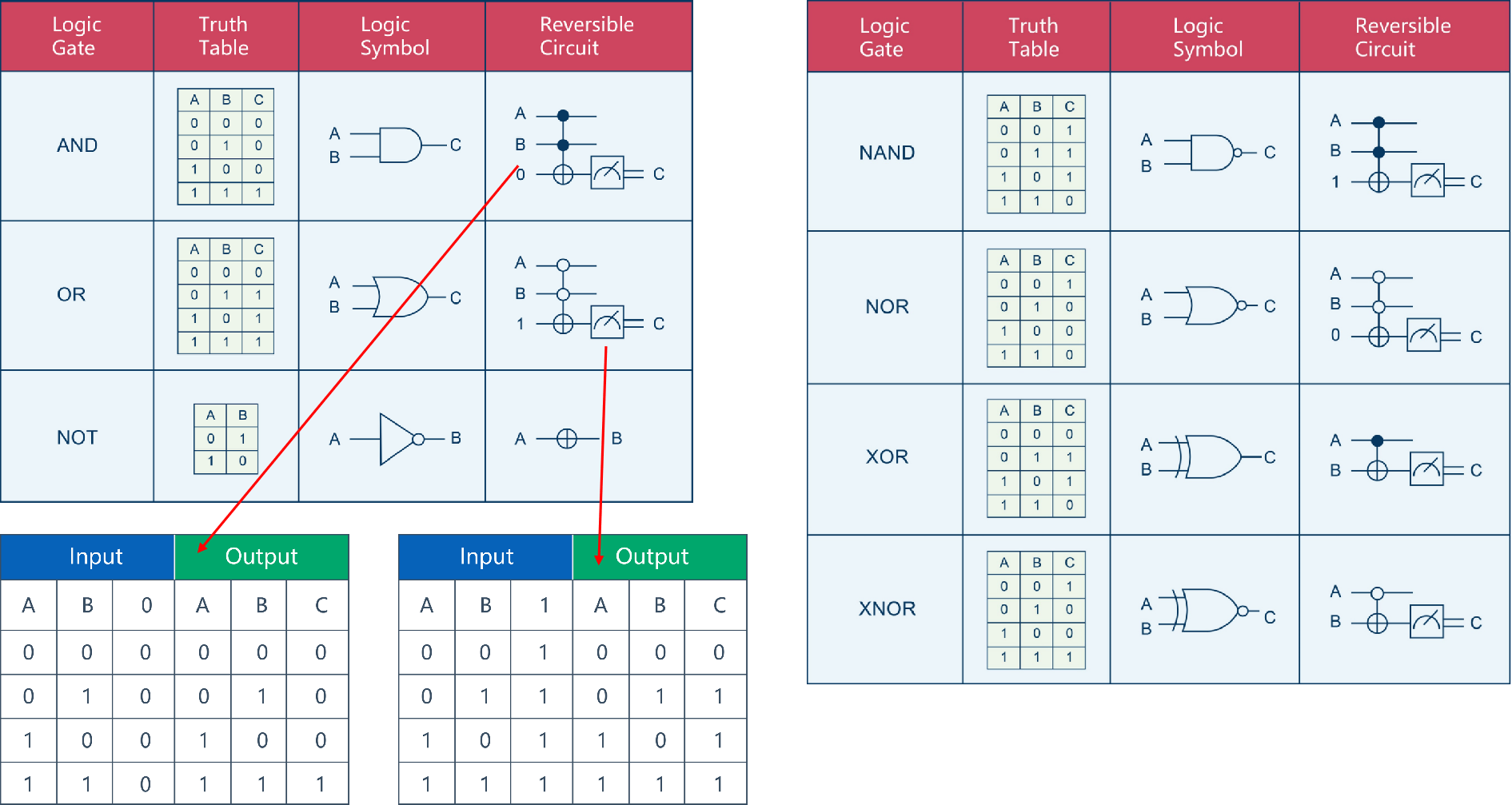}
	\centering
	\caption{\label{3_26} Any classical logic gate can be converted into a reversible logic gate}
\end{figure}

Reversible Computation: Any classical logic gate can be converted into a reversible logic gate. The Figure~\ref{3_26} shown that any classical logic gate can be converted into a reversible logic gate. For example in the reversible AND gate if we let the output A, B and C as a new input pass the reversible AND gate then the new output and input are the same. reversible OR gate is the same as reversible AND gate.

\begin{figure}[H]
	\includegraphics[scale=0.35]{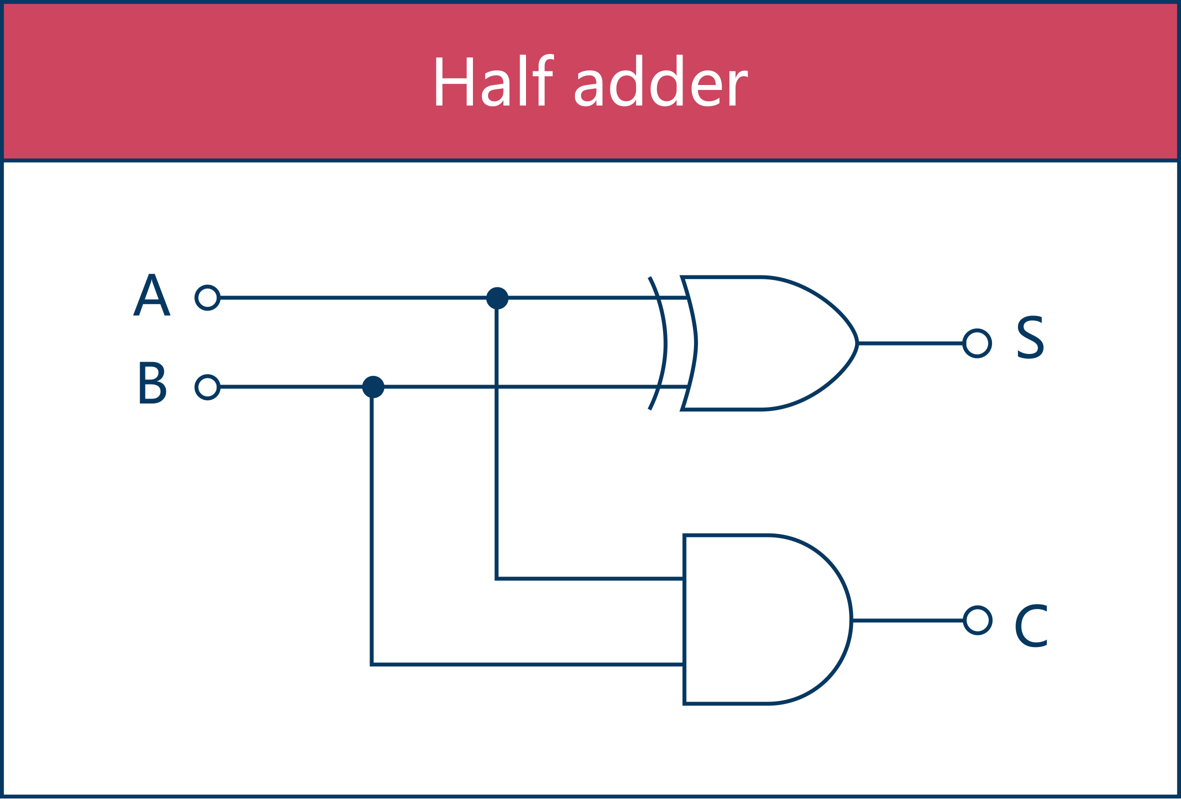}
	\centering
	\caption{\label{3_27} Classical half adder}
\end{figure}

Implementing a half adder with reversible computational logic gates. The classical half adder in Figure~\ref{3_27}.

\begin{figure}[H]
	\includegraphics[scale=0.35]{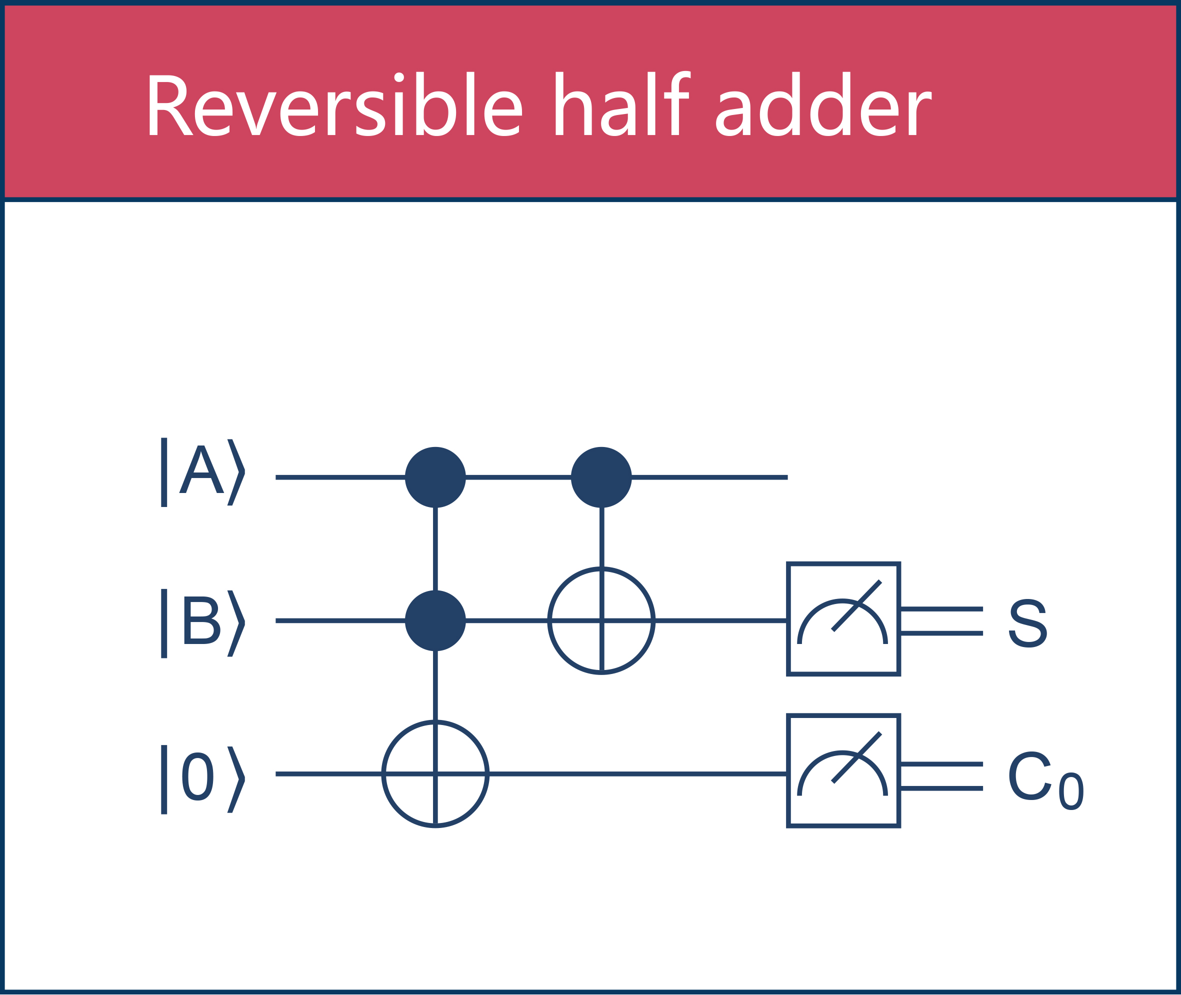}
	\centering
	\caption{\label{3_28} Half adder implemented with reversible logic gates}
\end{figure}

The half adder implemented with reversible logic gates shown in Figure~\ref{3_28}.






\newpage

\chapter{Week 2: A Brief Introduction to Quantum Computing}

\section{Reversible computing and quantum computing}~{}

The early history of quantum computing~\cite{sun8hidary2021brief} is a fascinating tale of theoretical exploration and groundbreaking ideas, particularly centered around the concepts of reversible computing and the pioneering work of figures like Richard Feynman and David Deutsch.

Reversible computing~\cite{sun47toffoli1980reversible}, a cornerstone concept in quantum computing, emerged from Rolf Landauer's principle~\cite{sun7maroney2009generalizing} formulated in 1961. Landauer, an IBM researcher, postulated that information erasure is inherently irreversible and that computation could, in theory, be energy-efficient if it were reversible. This principle set the stage for Charles Bennett, also of IBM, to demonstrate in the 1970s that reversible computing is not just a theoretical possibility but can be practically implemented. Bennett's work showed that any computation could be made reversible, thereby reducing the energy requirements of computational processes.

While reversible computing~\cite{sun47toffoli1980reversible,sun10frank2005introduction} was developing, Richard Feynman, a renowned physicist, brought a new perspective in the early 1980s. At the First Conference on the Physics of Computation at MIT in 1981, Feynman proposed the idea of a quantum computer. He argued that classical computers were inadequate for simulating quantum phenomena. Feynman's insight was that a computer based on quantum mechanical principles would be more adept at such tasks, highlighting the need for quantum computers to accurately simulate the natural world.

In 1985, David Deutsch, a physicist at the University of Oxford, built upon Feynman's ideas. Deutsch proposed the theoretical framework for a universal quantum computer, suggesting that such machines, operating on the principles of quantum mechanics, would possess remarkable capabilities beyond the reach of classical Turing machines. His work was instrumental in establishing the field of quantum complexity theory, exploring the potential of quantum computers to solve problems intractably complex for classical computers.

Paul Benioff's work in the early 1980s also played a crucial role. He proposed a quantum mechanical model of the Turing machine, bridging the gap between reversible computing and quantum computing. Benioff's model showed that a computing machine could operate in a quantum mechanical fashion, utilizing the principles of superposition and entanglement inherent to quantum computing.

The early history of quantum computing is thus marked by a series of theoretical developments, from Landauer's and Bennett's work on reversible computing to Feynman's and Deutsch's foundational contributions to the concept and theory of quantum computing. These advancements laid the groundwork for the subsequent development and realization of quantum computing technologies, marking the beginning of a new era in computational capability and understanding.

\section{Quantum gates}~{}

To understand how a qubit transitions from a state of \(\ket{0}\) or \(\ket{1}\) to a state of superposition, we use quantum gates. The Hadamard gate, denoted as \(H\), is particularly significant in this process. The Hadamard gate transforms a qubit's state from a definite state (\(\ket{0}\) or \(\ket{1}\)) to a superposition of states \cite{sun11shepherd2006role}.

When the Hadamard gate is applied to a qubit in the state \(\ket{0}\), it results in the qubit transitioning to a state which can be described as \(\ket{0} + \ket{1}\). Similarly, applying the Hadamard gate to a qubit in the state \(\ket{1}\) transforms it to \(\ket{0} - \ket{1}\). For simplicity, in this explanation (and throughout Week 2), we are omitting the normalization factor of \(\sqrt{2}\) that is typically included in these expressions.

The operation can be represented mathematically as:
\[ H\ket{0} \to \ket{0} + \ket{1}, \]
\[ H\ket{1} \to \ket{0} - \ket{1}. \]

The action of the Hadamard gate can be visualized using quantum circuit diagrams. In these diagrams, the Hadamard gate is represented by the symbol \(H\), applied to the initial qubit state, leading to the resultant superposition state. 

\begin{figure}[H]
    \centering
    \[
    \Qcircuit @C=1em @R=.7em {
        & \lstick{\ket{0}} & \gate{H} & \rstick{\ket{0} + \ket{1}} \qw    }
    \]
    \[
    \Qcircuit @C=1em @R=.7em {
        & \lstick{\ket{1}} & \gate{H} & \rstick{\ket{0} - \ket{1}} \qw    }
    \]
    \caption{\label{4_3} Sample diagram of the \(H\) gate operation.}
\end{figure}

From Figure~\ref{4_3}, we understand the action of the Hadamard gate \(H\) on the basis states \(\ket{0}\) and \(\ket{1}\):

\begin{align}
    H\ket{0} = \ket{0} + \ket{1}, \\
    H\ket{1} = \ket{0} - \ket{1}.
\end{align}

Now, let's verify the effect of applying the Hadamard gate twice on these basis states:

\begin{align}
    H(H\ket{0}) = H(\ket{0} + \ket{1}) = \ket{0}, \\
    H(H\ket{1}) = H(\ket{0} - \ket{1}) = \ket{1}.
\end{align}

This demonstrates the interesting property of the Hadamard gate where applying it twice returns the qubit to its original state.

Using two Hadamard gates, we can generate a superposition of two qubits, representing four states simultaneously. Figure~\ref{4_4} shows how passing two qubits through \(H\) gates results in a system with four states. In contrast, Figure~\ref{4_5} illustrates that when only the first qubit passes through an \(H\) gate, the system represents two states.


\begin{figure}[H]
    \centering
    \[ \Qcircuit @C=1em @R=.7em {
        & \lstick{\mid0\rangle} & \gate{H} & \qw \\
        &&&\rstick{\mid00\rangle+\mid01\rangle+\mid10\rangle+\mid11\rangle}\\
        & \lstick{\mid0\rangle} & \gate{H} & \qw & }\]
    \caption{\label{4_4} Two qubit pass by $H$ gate there will be 4 states}
\end{figure}
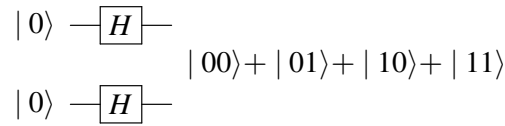
    

\begin{figure}[H]
    \centering
    \[ \Qcircuit @C=1em @R=.7em {
        & \lstick{\mid0\rangle} & \gate{H} & \qw \\
        &&&\rstick{\mid00\rangle+\mid10\rangle}\\
        & \lstick{\mid0\rangle} & \qw & \qw & }\]
    \caption{\label{4_5} Only first qubit pass by $H$ gate there will be 2 states}
\end{figure}
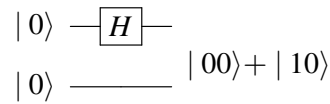


In Figure~\ref{4_5} we ignore the other two states $\mid01\rangle$ and $\mid11\rangle$, which probabilities are 0.




Using \(n\) Hadamard gates (\(H\)), we can construct a superposition of \(n\) qubits. This superposition state can represent \(2^{n}\) different states simultaneously. These states correspond to all the possible combinations of the \(n\) qubits, ranging from \(\ket{0}\) to \(\ket{2^{n} - 1}\), where each qubit can be either in the state \(\ket{0}\) or \(\ket{1}\). The states are often labeled using decimal numbers for simplicity. 

\begin{figure}[H]
    \centering
    \[ \Qcircuit @C=1em @R=.7em {
        &\lstick{\mid0\rangle} &\qw &\gate{H} &\qw &\qw \\
        &\lstick{\mid0\rangle} &\qw &\gate{H} &\qw &\qw \\
        &&&\dstick{\cdots}&&\\
        &&&&&\rstick{\mid0\cdots0\rangle+\mid0\cdots1\rangle+\cdots+\mid1\cdots1\rangle}
        \\
        &\lstick{\mid0\rangle} &\qw &\gate{H} &\qw &\qw\\
        & &\push{\downarrow}& & & \\
        &\lstick{{\mid0\rangle}^n} &{/}\qw &\gate{H} &\qw&\qw &\rstick{\sum_{x}\mid x\rangle}}\]
    \caption{\label{4_6} Use $n$ H gates to build the superposition of $n$ qubits}
\end{figure}
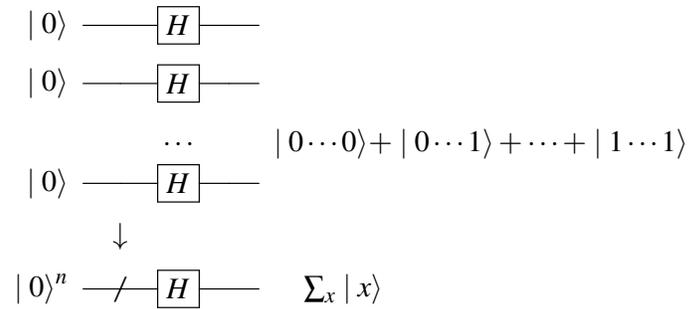

For instance, with \(n\) qubits in superposition, the states they represent can be enumerated as \(\ket{0}\), \(\ket{1}\), \(\ket{2}\), ..., \(\ket{2^{n}-1}\). As illustrated in Figure~\ref{4_6}, a quantum system with multiple qubits in superposition can be represented in two ways: the more explicit left representation showing each qubit's contribution, and the simplified right representation using a single ket notation to denote the entire superposed state.

\section{Bell states}~{}

Bell states~\cite{sun12shukla2014protocols} are a fundamental concept in quantum information theory, describing the four maximally entangled states of a two-qubit system. In a system that is in one of the Bell states, measuring the state of one qubit immediately determines the state of the other qubit .

The Bell states are defined as follows (omitting the normalization factor of \(\sqrt{2}\) for simplicity):

\begin{equation}
    \ket{\beta_{00}} = \ket{00} + \ket{11}
\end{equation}
\begin{equation}
    \ket{\beta_{10}} = \ket{00} - \ket{11}
\end{equation} 
\begin{equation}
    \ket{\beta_{01}} = \ket{01} + \ket{10}
\end{equation}
\begin{equation}
    \ket{\beta_{11}} = \ket{01} - \ket{10}
\end{equation}

\begin{figure}[H]
    \centering
    \[ \Qcircuit @C=1em @R=.7em {
         &&&&&\push{\mid0\rangle+\mid1\rangle}&&&&&&\\
         & \lstick{\mid0\rangle} & \gate{H} & \qw & \qw & \qw &\qw & \qw & \ctrl{2} &\qw &\qw &\qw \\
         &&&&&\push{\mid0\rangle}&&&&&&\\
         & \lstick{\mid0\rangle} & \qw & \qw & \qw & \qw & \qw & \qw & \targ & \qw &\qw &\qw \\
         &\lstick{\mid00\rangle}& & & &\push{\mid00\rangle+\mid10\rangle} & & & & & &\push{\mid00\rangle+\mid11\rangle}}\]
    \caption{\label{4_9} use $H$ gate and CNOT gate to build Bell states: $\mid\beta_{00}\rangle={\mid00\rangle+\mid11\rangle}$}
\end{figure}
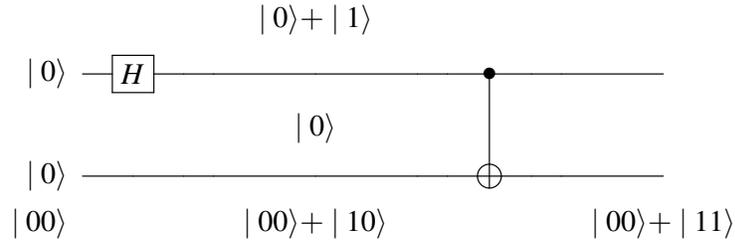

To generate Bell states, we can use a combination of Hadamard (H) and Controlled-NOT (CNOT) gates. For instance, to create the Bell state \(\ket{\beta_{00}} = \ket{00} + \ket{11}\), a specific sequence of H and CNOT gates is applied. This process is illustrated in Figure~\ref{4_9}, where we demonstrate how to use the H gate and CNOT gate to construct the Bell state \(\ket{\beta_{00}}\).

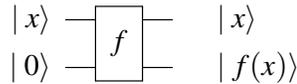
\begin{figure}[H]
    \centering
        \[ \Qcircuit @C=1em @R=.7em{
            &\lstick{\mid x\rangle} & \multigate{1}{f} &\qw &\rstick{\mid x \rangle} \\
            &\lstick{\mid 0\rangle} & \ghost{f}        &\qw &\rstick{\mid f(x)\rangle}}\]
    \caption{\label{5.17gai}A quantum circuit with function $f$.}
\end{figure}

In quantum circuits, the Hadamard ($H$) gate is often used to create superposition states~\cite{sun11shepherd2006role}. However, the introduction of an arbitrary controlled operation, denoted as \( f \), adds another layer of complexity and utility to the circuit. Figure~\ref{5.17gai} presents an example of such a configuration.

In Figure~\ref{5.17gai}, we introduce a function \( f \), treated as a black box operation, within a two-qubit quantum circuit. This setup allows us to observe the effects of applying \( f \) in conjunction with other quantum gates. After the operation of the circuit and upon measurement, the state of the first qubit remains as \(\ket{x}\), indicating that it is unaffected by \( f \). However, the state of the second qubit evolves to \( f(x) \), demonstrating the application of the function \( f \) to the input state. This configuration exemplifies how quantum circuits can implement complex operations and how the state of one qubit can be manipulated based on the state of another qubit, a fundamental aspect of quantum computation.

Now, let's explore the effect of adding the \( H \) gate to our quantum circuit.

\begin{figure}[H]
    \centering
\[ \Qcircuit @C=1em @R=.7em{
   &\lstick{{\mid0\rangle}^n}&\qw {/^n}\qw &\gate{H^n} &\multigate{2}{f}&\qw\\
   &&&&&\rstick{\sum_{x}\mid x,f(x)\rangle}\\
   &\lstick{{\mid0\rangle}^n}&\qw {/^n}\qw &\qw        &\ghost{f} &\qw}\]
   \caption{\label{5.18gai}Add the quantum dot path after the H gate.}
\end{figure}
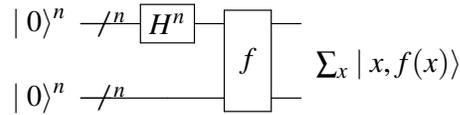

In Figure~\ref{5.18gai}, which builds upon the setup in Figure~\ref{5.17gai}, we initialize all the qubits to the \(\ket{0}\) state. Then, we apply the \( H \) gate to the first qubit. This addition of the \( H \) gate is crucial for generating superposition. After this operation and subsequent measurement, the resulting state of the system can be represented as a superimposed state \(\sum_{x} \ket{x, f(x)}\). This notation implies that for each possible value of \( x \) (generated by the \( H \) gate's action on the first qubit), the circuit outputs a corresponding \( f(x) \) on the second qubit. Hence, the entire quantum system is in a superposition of all possible inputs \( x \) and their corresponding outputs \( f(x) \) from the function.

In quantum circuits, the \( H \) gate, also known as the Hadamard gate, continues to play a pivotal role in building superposition states. The Quantum Processing Unit (QPU) leverages this capability of the \( H \) gate to perform calculations utilizing quantum superposition. This feature is integral to the power of quantum computing, allowing the QPU to process and represent multiple states simultaneously through superposition, which is a fundamental aspect of quantum mechanics.

\section{Quantum circuit model}~{}

\textbf{Single Qubit Gates}~{}

In quantum computing, qubits 
are the fundamental units of information. Unlike classical bits which are either \( \ket{0} \) or \( \ket{1} \), qubits can exist in a superposition of these states, represented as~\cite{4_sx}:

\begin{equation}
    \ket{\phi} = a\ket{0} + b\ket{1}
\end{equation}

Understanding this quantum superposition principle is key to grasping the behavior of qubits. Now, what about quantum ``gates"? Quantum gates are reversible operations that transform the states of qubits. They are analogous to logic gates in classical computing but operate under the principles of quantum mechanics. Here are some examples of single qubit gates~\cite{4_sx}:

The ``NOT" gate, denoted by $X$, flips the state of a qubit:
\begin{equation}
	\begin{cases}
		X\ket{0} = \ket{1}, \\
		X\ket{1} = \ket{0}.
	\end{cases}
\end{equation}

The ``Hadamard" gate, denoted by $H$, creates superpositions:
\begin{equation}
	\begin{cases}
		H\ket{0} = \ket{0} + \ket{1}, \\
		H\ket{1} = \ket{0} - \ket{1}.
	\end{cases}
\end{equation}

The ``Phase" gate, denoted by $Z$, applies a phase shift:
\begin{equation}
	\begin{cases}
	    Z\ket{0} = \ket{0}, \\
		Z\ket{1} = -\ket{1}.
	\end{cases}
\end{equation}

Interestingly, quantum gates~\cite{4_sx,sun13divincenzo1998quantum} can be combined to build other gates. For instance, the Phase gate can be constructed using the Hadamard gate and the NOT gate, and vice versa. Figure~\ref{5.1gai} illustrates this concept:

\begin{figure}[H]
    \centering
    \[ \Qcircuit @C=1em @R=.7em {
        & \gate{H} & \gate{Z} & \gate{H} & \qw & \push{=} &  & \gate{X} & \qw \\
        & \gate{H} & \gate{X} & \gate{H} & \qw & \push{=} &  & \gate{Z} & \qw 
    }\]
    \caption{\label{5.1gai} Construction of Phase gate and NOT gate using combinations of quantum gates.}
\end{figure}


\textbf{Two-Qubit Gates}~{}

Let's explore some examples of two-qubit gates in quantum computing. 

\begin{figure}[H]
	\centering
	\includegraphics[scale=0.7]{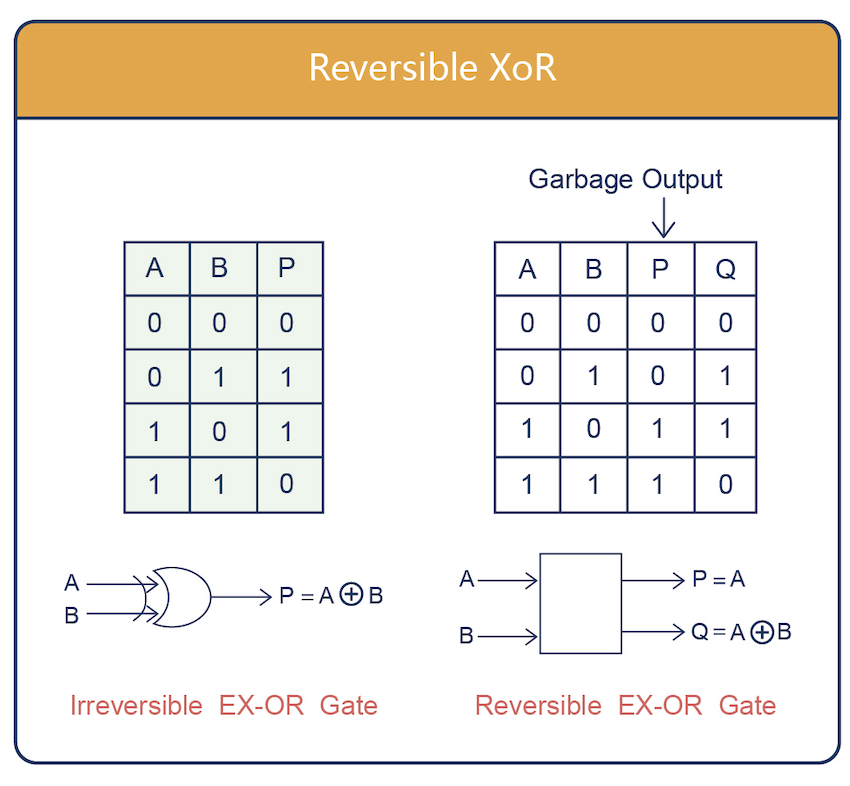}
	\caption{\label{5.2gai} Reversible XOR}
\end{figure}

Figure~\ref{5.2gai} illustrates the Reversible XOR, also an example of Reversible Computation.

The Reversible XOR, also known as the ``Controlled-NOT (CNOT)", is a fundamental two-qubit gate. It has two input qubits: a control qubit and a target qubit. As shown on the left of Figure~\ref{5.3gai}, the top \(\ket{x}\) is the control qubit and the bottom \(\ket{y}\) is the target qubit.

Figure~\ref{5.3gai} presents two examples of the CNOT gate.

CNOT with the first qubit as the control:
\begin{equation}
    \ket{x} \otimes \ket{y} \to \ket{x} \otimes \ket{y \oplus x}
\end{equation}
Similarly, a CNOT gate with the second qubit as the control:
\begin{equation}
   \ket{x} \otimes \ket{y} \to \ket{x \oplus y} \otimes \ket{y}
\end{equation}

\begin{figure}[H]
    \centering
    \[\Qcircuit @!C @!R{
        &\lstick{\ket{x}} &\ctrl{1}&\rstick{\ket{x}}\qw \\
        &\lstick{\ket{y}} &\targ&\rstick{\ket{y\oplus x}}\qw
    }\]
    \[\Qcircuit @!C @!R{
        &\lstick{\ket{x}} &\targ&\rstick{\ket{x\oplus y}}\qw\\
        &\lstick{\ket{y}} &\ctrl{-1}&\rstick{\ket{y}}\qw
    }\]
    \caption{\label{5.3gai} CNOT gate }
\end{figure}

Controlled-Z (CZ) transforms input states as follows:

\begin{align}
    \ket{00} &\to \ket{00}\\
    \ket{01} &\to \ket{01}\\
    \ket{10} &\to \ket{10}\\
    \ket{11} &\to -\ket{11}.
\end{align}

Since the CZ operation is symmetric between the two qubits, there's no need to specify control and target qubits. Figure~\ref{5.4gai} shows the circuit for the CZ gate.

\begin{figure}[H]
    \centering
    \[\Qcircuit @!C @!R {
        \lstick{\ket{x}} &\ctrl{1}&\rstick{\ket{x}}\qw \\
        \lstick{\ket{y}} &\control\qw&\rstick{(-1)^{xy}\ket{y}}\qw
    }\]
    \caption{\label{5.4gai} CZ operation}
\end{figure}

Let \( U \) be any single-qubit unitary operation. Then, the controlled-\( U \) operation is a two-qubit operation. Figure~\ref{5.5gai} illustrates the controlled-\( U \) gate.

\begin{figure}[H]
    \centering
    \[\Qcircuit @!C @!R {
        &\lstick{\ket{x}} &\ctrl{1}&\rstick{\ket{x}}\qw \\
        &\lstick{\ket{y}} &\gate{U}&\rstick{U^{x}\ket{y}}\qw
    }\]
    \caption{\label{5.5gai} Controlled-\(U\)}
\end{figure}

The CNOT is effectively a controlled-\(X\). Figure~\ref{5.6gai} depicts this operation.

\begin{figure}[H]
    \centering
    \[\Qcircuit @!C @!R {
        &\lstick{\ket{x}} &\ctrl{1}&\rstick{\ket{x}}\qw \\
        &\lstick{\ket{y}} &\gate{X}&\rstick{X^{x}\ket{y}}\qw
    }\]
    \caption{\label{5.6gai} Controlled-\(X\)}
\end{figure}

Lastly, Figure~\ref{5.7gai} presents the controlled-\(Z\) gate.

\begin{figure}[H]
    \centering
    \[\Qcircuit @!C @!R {
        &\lstick{\ket{x}} &\ctrl{1}&\rstick{\ket{x}}\qw \\
        &\lstick{\ket{y}} &\gate{Z}&\rstick{Z^{x}\ket{y}}\qw
    }\]
    \caption{\label{5.7gai} Controlled-\(Z\)}
\end{figure}

\textbf{Multi-qubit gate(three-qubit gate)}~{}

\begin{figure}[H]
    \centering
        \[\Qcircuit @!C @!R {
        &\lstick{\mid x\rangle} &\ctrl{2}&\rstick{\mid x\rangle}\qw \\
        &\lstick{\mid y\rangle} &\ctrl{1}&\rstick{\mid y\rangle}\qw \\
        &\lstick{\mid z\rangle} &\targ   &\rstick{\mid z\oplus(x\land y)\rangle}\qw}\]
    \caption{\label{5.8gai} A Toffoli gate}
\end{figure}
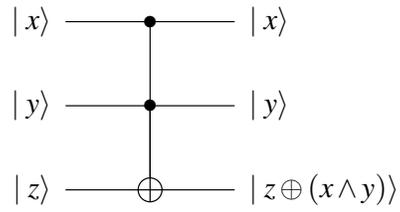

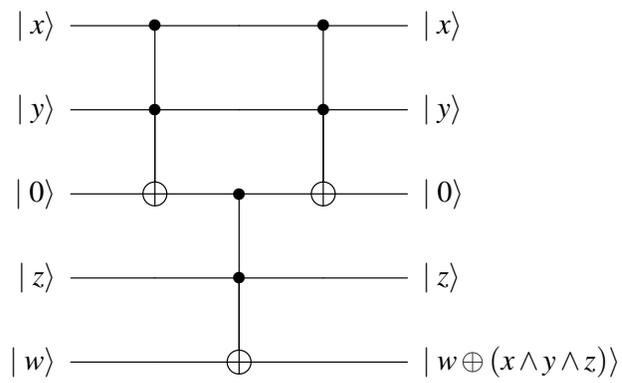
\begin{figure}[H]
    \centering
        \[\Qcircuit @!C @!R {
        &\lstick{\mid x\rangle} &\ctrl{2} &\qw &\ctrl{2}&\rstick{\mid x\rangle}\qw \\
        &\lstick{\mid y\rangle} &\ctrl{1} &\qw &\ctrl{1}&\rstick{\mid y\rangle}\qw \\
        &\lstick{\mid 0\rangle} &\targ    &\ctrl{2} &\targ&\rstick{\mid 0\rangle}\qw\\
        &\lstick{\mid z\rangle} &\qw      &\ctrl{1} &\qw  &\rstick{\mid z\rangle}\qw\\
        &\lstick{\mid w\rangle} &\qw      &\targ    &\qw  &\rstick{\mid w\oplus(x\land y\land z)\rangle}\qw}\]
    \caption{\label{5.9gai} A complex circuit}
\end{figure}

Now, let's expand from a two-qubit gate to a three-qubit gate. For example the Toffoli gate~\cite{sun15shende2008cnot} is an example of three qubit like Figure~\ref{5.8gai} and Figure~\ref{5.9gai}

Toffoli gate: when the first two qubits (control qubits) are both 1, flips the third qubit (target qubit). The Figure~\ref{5.8gai} shown that.

A complex circuit consisting of a toffoli gate is shown in Figure~\ref{5.9gai}.

\begin{figure}[H]
    \centering
        \[\Qcircuit @!C @!R {
        &\lstick{\mid x\rangle} &\ctrl{2}&\rstick{\mid x\rangle}\qw \\
        &\lstick{\mid y\rangle} &\ctrl{1}&\rstick{\mid y\rangle}\qw \\
        &\lstick{\mid 0\rangle} &\targ   &\rstick{\mid x\land y\rangle}\qw}\]

         \[\Qcircuit @!C @!R {
        &\lstick{\mid 1\rangle} &\ctrl{2}&\rstick{\mid 1\rangle}\qw \\
        &\lstick{\mid 1\rangle} &\ctrl{1}&\rstick{\mid 1\rangle}\qw \\
        &\lstick{\mid z\rangle} &\targ   &\rstick{\mid\neg z\rangle}\qw}\]

        \[\Qcircuit @!C @!R {
        &\lstick{\mid 1\rangle} &\ctrl{2}&\rstick{\mid 1\rangle}\qw \\
        &\lstick{\mid y\rangle} &\ctrl{1}&\rstick{\mid y\rangle}\qw \\
        &\lstick{\mid 0\rangle} &\targ   &\rstick{\mid y\rangle}\qw}\]
    \caption{\label{5.10gai} Universal gate}
\end{figure}
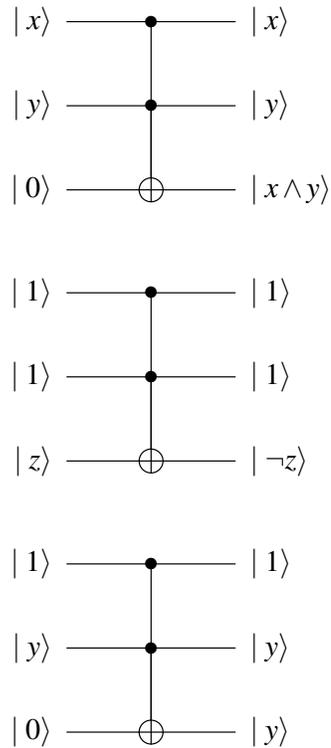

Toffoli gate is universal for example in the Figure~\ref{5.10gai}.

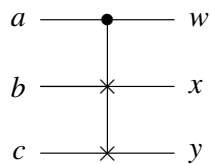
\begin{figure}[H]
    \centering
    \[\Qcircuit @!C @!R{
    &\lstick{a}&\ctrl{1}&\rstick{w}\qw\\
    &\lstick{b}&\qswap\qw&\rstick{x}\qw\\
    &\lstick{c}&\qswap\qwx&\rstick{y}\qw}\]
    \caption{\label{1_19} The circuit of Fredkin gate}
\end{figure}

The Fredkin gate~\cite{sun16patel2016quantum} also is a kind of three-qubit gate. The circuit of Fredkin gate shown in Figure~\ref{1_19}.

In Fredkin gate~\cite{sun16patel2016quantum} when $a=0$, the output $w=a, x=b, y=c$, when $a=1$, $b,c$ perform the exchange operation the output $w=a, x=c, y=b$. In other word, when input $a=1$, $b,c$ exchange as output $x=c,y=b$, when $a=0$, $b=x,c=y$ as output.

\textbf{Measurements}~{}

For $\mid\psi\rangle = \sum\alpha_{x}\mid x\rangle$ , measure in the basis of \{$\mid x\rangle$\} returns $\mid x\rangle$ with probability $p_{x} = \mid\alpha_{x}\mid^{2}$. The Figure~\ref{5.11gai} shown this.

\begin{figure}[H]
    \centering
        \[\Qcircuit @C=1em @R=.7em {
        &\lstick{\mid\phi\rangle}&=&\push{\sum{\alpha_{x}\mid x\rangle}} &{/}\qw &\meter&=&\rstick{\mid x\rangle}}\]
    \caption{\label{5.11gai} Measurements $\mid\psi\rangle = \sum\alpha_{x}\mid x\rangle$}
\end{figure}


For $\mid\psi\rangle = \sum\alpha_{x}\mid x\rangle\mid\phi_{x}\rangle$ , measure the first register in the basis of $\{\mid x\rangle\}$ returns $\mid x\rangle_{n}\mid\psi x\rangle$ with probability $p_{x} = \mid\alpha_{x}\mid^{2}$ Figure\ref{5.12gai} shown that.

\begin{figure}[H]
    \centering
        \[\Qcircuit @C=1em @R=.7em {
        &&&&{/}\qw&\meter&\cw&\rstick{\mid x\rangle}\\
        &\push{\mid\psi\rangle=\sum\alpha_{x}\mid x
        \rangle\mid\phi\rangle}&&\\
        &&&&{/}\qw&\qw&\qw&\rstick{\mid\phi_{x}\rangle}}\]
    \caption{\label{5.12gai} Measurements $\mid\psi\rangle = \sum\alpha_{x}\mid x\rangle\mid\phi_{x}\rangle$}
\end{figure}


\begin{example}

Find the output state $\mid\phi\rangle$ of the following circuit:

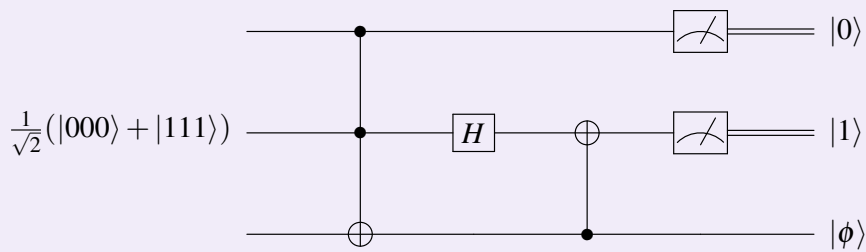
\begin{figure}[H]
    \centering
	\[\Qcircuit @!C @!R {
        &\lstick{} &\ctrl{1} &\qw &\qw &\meter &\rstick{|0\rangle}\cw\\
        &\lstick{\frac{1}{\sqrt{2}}(|000\rangle+|111\rangle)} &\ctrl{1} &\gate{H} &\targ &\meter &\rstick{|1\rangle}\cw\\
        &\lstick{} &\targ &\qw &\ctrl{-1} &\qw &\rstick{|\phi\rangle}\qw
        }\]
    \caption{\label{5.13gai1} The example of quantum circuit}
\end{figure}


The circuit Figure~\ref{5.13gai1} operation process is shown below:
\begin{align}
 \frac{1}{\sqrt{2}}(|000\rangle + |111\rangle&\stackrel{\text{Toffoli}}{\longrightarrow}\frac{1}{\sqrt{2}}(|000\rangle + |110\rangle)\nonumber\\
&\stackrel{\text{H}_2}{\longrightarrow}\frac{1}{\sqrt{2}}[|0\rangle\frac{1}{\sqrt{2}}(|0\rangle+|1\rangle)+|1\rangle\frac{1}{\sqrt{2}}(|0\rangle-|1\rangle)|0\rangle]\nonumber\\ &\qquad  =\frac{1}{2}(|000\rangle + |010\rangle + |100\rangle - |110\rangle)\nonumber\\ 
&\stackrel{\text{CNOT}_{32}}{\longrightarrow}\frac{1}{2}(|000\rangle + |010\rangle + |100\rangle - |110\rangle) \to |\phi\rangle=|0\rangle.   
\end{align}
\end{example}

\textbf{DiVincenzo Criteria}~{}

David DiVincenzo, a prominent figure in quantum computing, proposed a set of requirements essential for the realization of a quantum computer. Known as the DiVincenzo criteria~\cite{sun17perez2011quantum}, these guidelines outline the foundational aspects necessary for the construction and operation of a quantum computer. The five specific criteria are:

\begin{enumerate}
    \item A scalable physical system with well-characterized qubits, allowing for the construction of larger quantum systems while maintaining control over individual qubits.
    
    \item The ability to initialize the state of the qubits to a simple fiducial state, which is crucial for setting a known starting point for quantum computations.

    \item Long relative decoherence times, significantly longer than the gate-operation time, to ensure that quantum states maintain coherence throughout the computation process.

    \item A universal set of quantum gates, enabling the performance of any quantum computation through a combination of these fundamental operations.

    \item Qubit-specific measurement capability, allowing for the extraction of information from individual qubits without disturbing the entire quantum system.
\end{enumerate}

Meeting these criteria is essential for any physical system to be considered a viable quantum computer, as they address the key challenges in harnessing and manipulating quantum mechanics for computational purposes.

\section{Deutsch's algorithm}~{}

Deutsch's algorithm~\cite{sun18vianna2018classical}, proposed by British scientist David Deutsch in 1985, holds a significant place in the field of quantum computing. It is celebrated for demonstrating the potential of quantum computing to outperform classical computing in specific scenarios. Deutsch's algorithm is often considered one of the most fundamental quantum algorithms, known for its ability to determine properties of functions at a quantum speedup.

The algorithm utilizes a combination of quantum gates, namely the X, H (Hadamard), and CNOT gates, to solve a specific type of problem. This problem involves determining the nature of a given function, showcasing how quantum algorithms can achieve tasks more efficiently compared to their classical counterparts.

Deutsch's problem can be formulated as follows:

Consider a binary function \( f \) defined as:

\begin{equation}
    f: \{0, 1\} \to \{0, 1\}
\end{equation}

\begin{figure}[H]
	\includegraphics[scale=0.5]{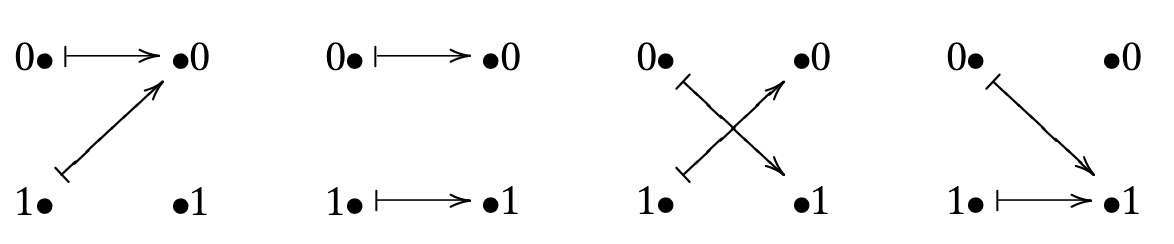}
	\centering
	\caption{\label{5.14gai} The kind of case of the function $f: \{0, 1\} \to \{0, 1\}$ }
\end{figure}

The function \( f \) has four possible configurations, each representing different scenarios. These scenarios are illustrated in Figure~\ref{5.14gai} and include cases where the function outputs are both equal to $0$, invariant under input change, flipped based on the input, or both equal to $1$.

Based on the input and output results, the four possible configurations of the function \( f \) can be divided into two categories:

\begin{enumerate}
    \item The function \( f \) is \textit{balanced} if \( f(0) \neq f(1) \). This category includes the ``invariant" and ``flipped" cases, where the output changes with the input.
    
    \item The function \( f \) is \textit{constant} if \( f(0) = f(1) \). This category encompasses the cases where the output is ``both equal to 0" or ``both equal to 1", indicating the output remains the same regardless of the input.
\end{enumerate}

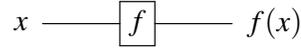
\begin{figure}[H]
    \centering
    \[\Qcircuit @!C @!R{
        &\lstick{x}&\gate{f}&\rstick{f(x)}\qw
    }\]
    \caption{\label{5.15gai} Function \( f: \{0, 1\} \to \{0, 1\} \) as a black box.}
\end{figure}

Now, consider the function \( f: \{0, 1\} \to \{0, 1\} \) provided as a black box. In this scenario, one can evaluate the function for a given input but cannot inspect the internal mechanism defining \( f \). The challenge is to determine whether \( f \) is balanced or constant using the least number of evaluations. Figure~\ref{5.15gai} illustrates the function \( f \) as a black box.

\begin{figure}[H]
    \centering
        \[\Qcircuit @!C @!R{
        &\lstick{x}&\gate{f}&\rstick{f(x)}\qw}\]
        \[\Qcircuit @!C @!R{
        &\lstick{\mid x\rangle}&\multigate{1}{U_{f}}&\rstick{x}\qw \\
        &\lstick{\mid y\rangle}&\ghost{U_{f}}&\rstick{\mid y\oplus f(x)\rangle}\qw}\]    
    \caption{\label{5.16gai}The reversible function convert the initial state to $\mid x,y\oplus f(x)\rangle$.}
\end{figure}
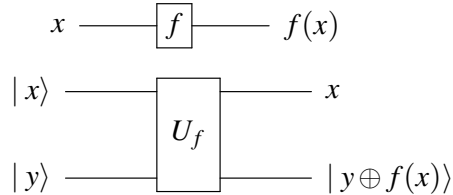

If the function \( f \) is reversible, we can represent its operation in a quantum circuit as depicted in Figure~\ref{5.16gai}. In the context of a quantum computer, consider an initial state \(\ket{x, y}\). The goal is to transform this state using \( f \) in a way that the second part of the state, \( y \), becomes \( y \oplus f(x) \). This operation is a quantum representation of the function \( f \).

To achieve this, we define a unitary operation \( U_{f} \) that maps the initial state \(\ket{x, y}\) to the transformed state \(\ket{x, y \oplus f(x)}\). The mapping is expressed as:

\begin{equation}
    U_{f} : \ket{x, y} \to \ket{x, y \oplus f(x)}
\end{equation}

This quantum operation \( U_{f} \) effectively encodes the classical function \( f \) into a quantum circuit, allowing for the superposition and interference properties of quantum mechanics to be harnessed in evaluating the nature of \( f \).

Next, let's consider the quantum circuit depicted in Figure~\ref{5.5gai1}. In this setup, we add an \( H \) gate to the first qubit and then apply the operation \( U_{f} \). As a result, the state of the system after these operations, ignoring the factor of \(\sqrt{2}\) for simplicity, will be:

\begin{equation}
    \ket{\phi_{2}} = \ket{0, f(0)} + \ket{1, f(1)}
\end{equation}

This equation represents the output state of the circuit, where the first qubit is in a superposition of \(\ket{0}\) and \(\ket{1}\), and the second qubit is transformed based on the function \( f \) applied to the corresponding state of the first qubit.

\begin{figure}[H]
    \centering
        \[\Qcircuit @!C @!R{
        &\lstick{\mid 0\rangle}&\qw&\gate{H}&\qw&\multigate{1}{U_{f}}&\qw&\qw\\
        &\lstick{\mid 0\rangle}&\qw&\qw     &\qw&\ghost{U_{f}}       &\qw&\meter\\
        &                      &\dstick{\mid\phi_{0}\rangle}& &\dstick{\mid\phi_{1}\rangle}& &\dstick{\mid\phi_{2}\rangle}&&
        }\]
    \caption{\label{5.5gai1}The circuit.}
\end{figure}
    

The process of the Deutsch's algorithm in the quantum circuit can be described step-by-step as follows, ignoring the normalization factor of \(\sqrt{2}\):

\begin{equation}
    \ket{\phi_{0}} = \ket{0, 0}
\end{equation}

This is the initial state of the two qubits in the system.

\begin{equation}
    \ket{\phi_{1}} = (\ket{0} + \ket{1})\ket{0}
\end{equation}

Here, the first qubit is put into a superposition of \(\ket{0}\) and \(\ket{1}\) using the \( H \) gate, while the second qubit remains in the state \(\ket{0}\).

\begin{equation}
    \ket{\phi_{2}} = \ket{0, f(0)} + \ket{1, f(1)}
\end{equation}

In this step, the \( U_{f} \) operation is applied, resulting in the transformation of the second qubit based on the function \( f \) evaluated at the state of the first qubit.

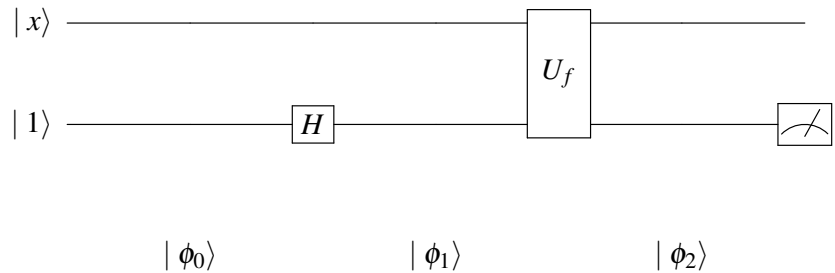
\begin{figure}[H]
    \centering
        \[\Qcircuit @!C @!R{
        &\lstick{\mid x\rangle}&\qw&\qw&\qw&\multigate{1}{U_{f}}&\qw&\qw\\
        &\lstick{\mid 1\rangle}&\qw&\gate{H}     &\qw&\ghost{U_{f}}       &\qw&\meter\\
        &                      &\dstick{\mid\phi_{0}\rangle}& &\dstick{\mid\phi_{1}\rangle}& &\dstick{\mid\phi_{2}\rangle}&&
        }\]
    \caption{\label{5.6gai1}The circuit of $\mid\phi_{0}\rangle=\mid x,1\rangle$.}
\end{figure}

Now, consider the first qubit \(\ket{x}\) and the second qubit \(\ket{1}\). Apply the \( H \) gate to the second qubit, followed by the \( U_{f} \) operation, and then measure the second qubit to obtain the state \(\ket{\phi}\). The circuit for this process is depicted in Figure~\ref{5.6gai1}.

The process can be further described as follows:

\begin{equation}
    \ket{\phi_{0}} = \ket{x, 1}
\end{equation}

This represents the initial state of the system where the first qubit is \(\ket{x}\) and the second qubit is \(\ket{1}\).

\begin{equation}
    \ket{\phi_{1}} = \ket{x} (\ket{0} - \ket{1}) = \ket{x, 0} - \ket{x, 1}
\end{equation}

After applying the \( H \) gate to the second qubit, the second qubit is in a superposition of \(\ket{0}\) and \(\ket{1}\) states.

\begin{align}
        \ket{\phi_{2}} &= \ket{x, 0 \oplus f(x)} - \ket{x, 1 \oplus f(x)} \nonumber \\
        &= \ket{x, f(x)} - \ket{x, \bar{f(x)}} \nonumber \\
        &= \ket{x} (\ket{f(x)} - \ket{\bar{f(x)}}).
\end{align}

Here, the \( U_{f} \) operation is applied, affecting the second qubit based on the value of \( f(x) \).

\begin{equation}
    \ket{\phi_{2}} = 
    \begin{cases}
        \ket{x} (\ket{0} - \ket{1}), & \text{if } f(x) = 0, \\
        \ket{x} (\ket{1} - \ket{0}), & \text{if } f(x) = 1.
    \end{cases}
\end{equation}

This equation illustrates the state of the system after applying \( U_{f} \), depending on the function \( f \).

\begin{equation}
    \ket{\phi_{2}} = (-1)^{f(x)} \ket{x} (\ket{0} - \ket{1})
\end{equation}

This final form shows the quantum state right before measurement, indicating how the function \( f \) influences the state.

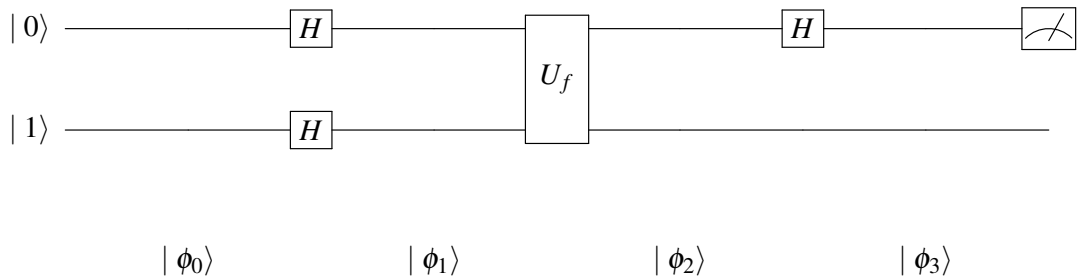
\begin{figure}[H]
    \centering
        \[\Qcircuit @!C @!R{
        &\lstick{\mid 0\rangle}&\qw&\gate{H}&\qw&\multigate{1}{U_{f}}&\qw&\gate{H}&\qw&\meter\\
        &\lstick{\mid 1\rangle}&\qw&\gate{H}&\qw&\ghost{U_{f}}&\qw&\qw&\qw&\qw\\
        &                      &\dstick{\mid\phi_{0}\rangle}& &\dstick{\mid\phi_{1}\rangle}& &\dstick{\mid\phi_{2}\rangle}&&\dstick{\mid\phi_{3}\rangle}
        }\]
    \caption{\label{5.7gai1}The quantum circuit implementation of Deutsch's algorithm.}
\end{figure}

By making slight modifications to the circuit shown in Figure~\ref{5.6gai1}, we can derive a quantum circuit that implements Deutsch's algorithm. The quantum circuit implementation of Deutsch's algorithm is depicted in Figure~\ref{5.7gai1}.

In the quantum circuit implementation of Deutsch's algorithm, the process unfolds as follows:

\begin{equation}
    \ket{\phi_{0}} = \ket{0, 1}
\end{equation}

This is the initial state with the first qubit in \(\ket{0}\) and the second qubit in \(\ket{1}\).

\begin{equation}
    \ket{\phi_{1}} = (\ket{0} + \ket{1})(\ket{0} - \ket{1}) = \ket{0, 0} - \ket{0, 1} + \ket{1, 0} - \ket{1, 1}
\end{equation}

After applying the \( H \) gate to both qubits, we get the first qubit in a superposition of \(\ket{0}\) and \(\ket{1}\), and the second qubit in a superposition of \(\ket{0}\) and \(\ket{1}\), but with a negative sign on \(\ket{1}\).
\begin{align}
        \ket{\phi_{2}} &= \ket{0, f(0)} - \ket{0, \bar{f(0)}} + \ket{1, f(1)} - \ket{1, \bar{f(1)}} \nonumber \\
        &= (-1)^{f(0)}\ket{0}(\ket{0} - \ket{1}) + (-1)^{f(1)}\ket{1}(\ket{0} - \ket{1}) \nonumber \\
        &= [(-1)^{f(0)}\ket{0} + (-1)^{f(1)}\ket{1}](\ket{0} - \ket{1}).
\end{align}

After applying the \( U_{f} \) operation, the state reflects the outcome of the function \( f \) applied to both possible inputs.

\begin{equation}
	\ket{\phi_{2}} = \begin{cases}
		(\pm1)(\ket{0} + \ket{1})(\ket{0} - \ket{1}), & \text{when } f(0) = f(1), \\
		(\pm1)(\ket{0} - \ket{1})(\ket{0} - \ket{1}), & \text{when } f(0) \neq f(1).
	\end{cases}
\end{equation}

This state indicates whether \( f \) is balanced or constant.

\begin{equation}
	\ket{\phi_{3}} = \begin{cases}
		(\pm1)\ket{0}(\ket{0} - \ket{1}), & \text{when } f(0) = f(1), \\
		(\pm1)\ket{1}(\ket{0} - \ket{1}), & \text{when } f(0) \neq f(1).
	\end{cases}
\end{equation}

By measuring the first qubit, we can determine the relationship between \( f(0) \) and \( f(1) \). This quantum circuit can determine if \( f(x) \) is balanced or constant by evaluating \( f(x) \) only once, showcasing an advantage over classical methods.

\textbf{Conclusion:}
The Deutsch algorithm only needs to calculate \( f(x) \) once to determine whether \( f(0) = f(1) \).

The key to the Deutsch algorithm's efficiency is the ability to set qubits to a superposition state \(\ket{0} + \ket{1}\) in quantum computing. Thus, \( f(\ket{0}) \) and \( f(\ket{1}) \) can be calculated simultaneously.

While the Deutsch algorithm can determine whether \( f(0) = f(1) \), it does not output the specific values of \( f(0) \) and \( f(1) \). To know these specific values, we must calculate \( f(x) \) twice, aligning with classical computation.

The Deutsch algorithm is seminal in demonstrating the potential superiority of quantum computing. Its extension, the Deutsch-Jozsa algorithm, further explores these concepts. Further study of these algorithms can offer deeper insights into quantum computation.

\textbf{Brief Summary:}

The Deutsch algorithm marks a pivotal point in demonstrating the advantages of quantum computing. It can perform two classical computations simultaneously in a single operation. The algorithm underscores the significance of quantum superposition in computational processes.

\section{Experiment}~{}
Now let's do some experiment of the quantum computer. The experiments in this section are all basic experiments, aiming to give students an introductory understanding of quantum computing.(The experiments are all completed on real quantum computers. Due to the influence of noise, there will be a difference between real data and analog data.)

\textbf{The experiment of reversible logic}~{}

Reversible logic~\cite{sun19peres1985reversible} plays a crucial role in quantum computing due to its fundamental alignment with the principles of quantum mechanics. In this context, let's delve into an experiment demonstrating reversible logic.

Consider two inputs, A and B. These inputs are fed into a quantum circuit, resulting in two outputs, P and Q. To test the reversibility of the logic, we then use P and Q as new inputs and pass them back through the circuit. The outputs obtained from this second pass are denoted as \( P' \) and \( Q' \). 

\begin{figure}[H]
    \centering
    \[\Qcircuit@!C @!R{
    &\lstick{A}&\gate{X}&\ctrl{1}&\rstick{P}\qw&\lstick{P}&\ctrl{1}&\gate{X}&\rstick{P^{'}}\qw\\
    &\lstick{B}&\qw     &\gate{X}&\rstick{Q}\qw&\lstick{Q}&\gate{X}&\qw   &\rstick{Q^{'}}\qw}\]
    \caption{\label{e_10}If $A=P^{'}, B=Q^{'}$ we finish the reversible logic experiment.}
\end{figure}
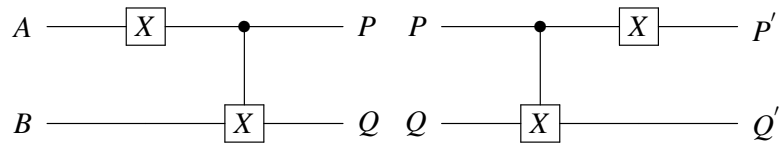

The experiment is considered successful in demonstrating reversible logic if the outputs \( P' \) and \( Q' \) are identical to the original inputs A and B, respectively. This means that \( A = P' \) and \( B = Q' \), indicating that the process is indeed reversible. Figure~\ref{e_10} illustrates this reversible logic process.

Now let's see an exmaple of reversible logic(XOR also called CNOT gate).

\begin{figure}[H]
    \centering
    \[\Qcircuit @C=1em @R=.7em{
    &\lstick{A}&\push{\to}&\multigate{1}{U}&\qw&\push{\to}&\rstick{P=A}\\
    &\lstick{B}&\push{\to}&\ghost{U}&\qw&\push{\to}&\rstick{Q=A\oplus B}}\]
    \caption{\label{e_9} The example of XOR gate.}
\end{figure}
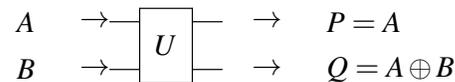

The XOR gate as Figure~\ref{e_9}. The gate of U is XOR(CNOT) gate.

\begin{figure}[H]
    \centering
    \includegraphics[scale=0.55]{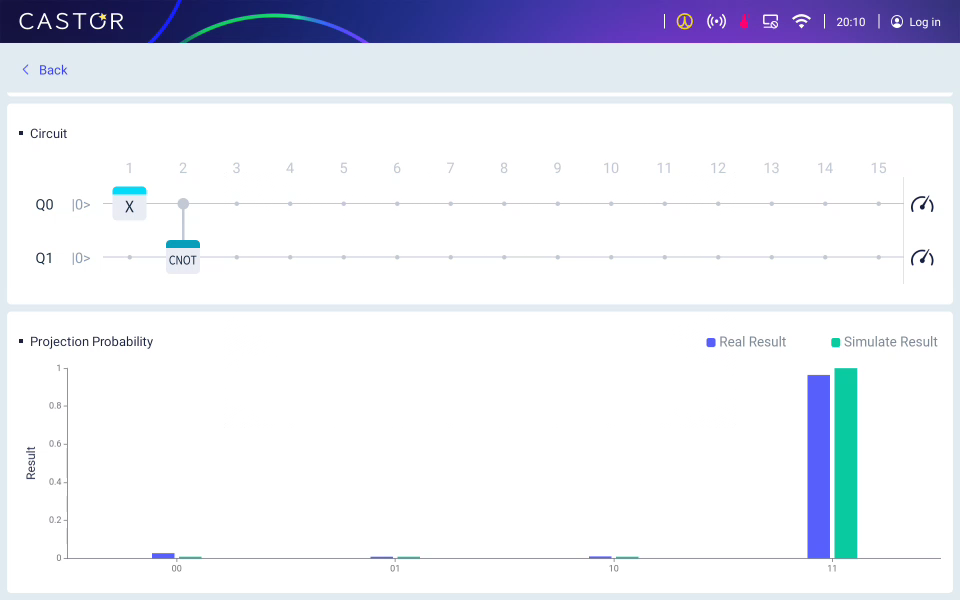}
    \caption{\label{e_11} Build with reversible logic ircuit}
\end{figure}

Thus, it is feasible to construct and implement reversible logic circuits on an actual quantum computer. As depicted in Figure~\ref{e_11}, we have designed and executed a reversible logic circuit on a real quantum computing platform. This practical execution allows us to observe and verify the principles of reversible logic in a real-world quantum computing environment, thereby concretely demonstrating the concept. The successful completion of this experiment on a quantum computer not only validates the theoretical aspects of reversible logic but also showcases the practical capabilities and potential of quantum computing technology in processing complex computations.


\textbf{The experiment of Single qubit gates ($X$, $H$ and $Z$)}~{}

In quantum computing the gates $X$, $H$ and $Z$ are three important gates. 

\begin{figure}[H]
    \centering
    \[\Qcircuit @!C @!R {
    &\lstick{\mid0\rangle}&\gate{H}&\qw&\rstick{\mid0\rangle+\mid1\rangle}\qw}\]
    \[\Qcircuit @!C @!R {
    &\lstick{\mid0\rangle}&\gate{X}&\qw&\rstick{\mid1\rangle}\qw}\]
    \[\Qcircuit @!C @!R {
    &\lstick{\mid1\rangle}&\gate{Z}&\qw&\rstick{-\mid1\rangle}\qw}\]
    \caption{\label{e_1}The circuit of gate H, Z and X}
\end{figure}
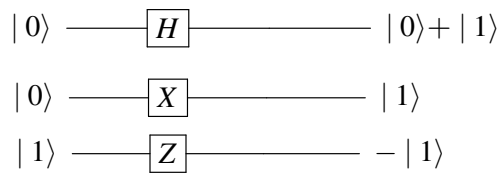

First let's build a quantum circuit of gates $X$, $H$ and $Z$. We shown the circuit in Figure~\ref{e_1}. We set $\mid0\rangle$ as input.

\begin{figure}[H]
    \centering
    \includegraphics[scale=0.55]{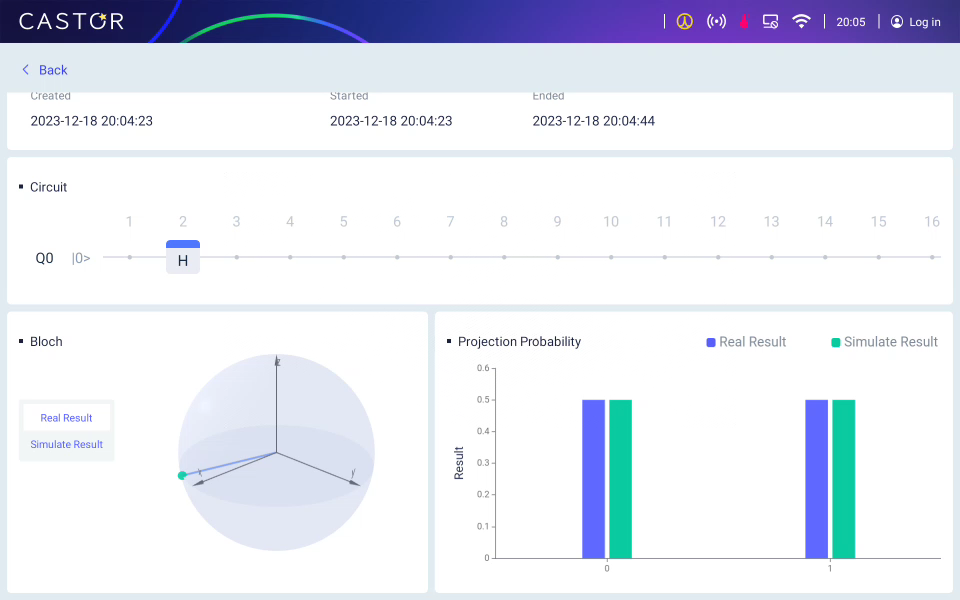}
    \caption{\label{e_2}Build with $H$ gate circuit in quantum circuit.}
\end{figure}

\begin{figure}[H]
    \centering
    \includegraphics[scale=0.55]{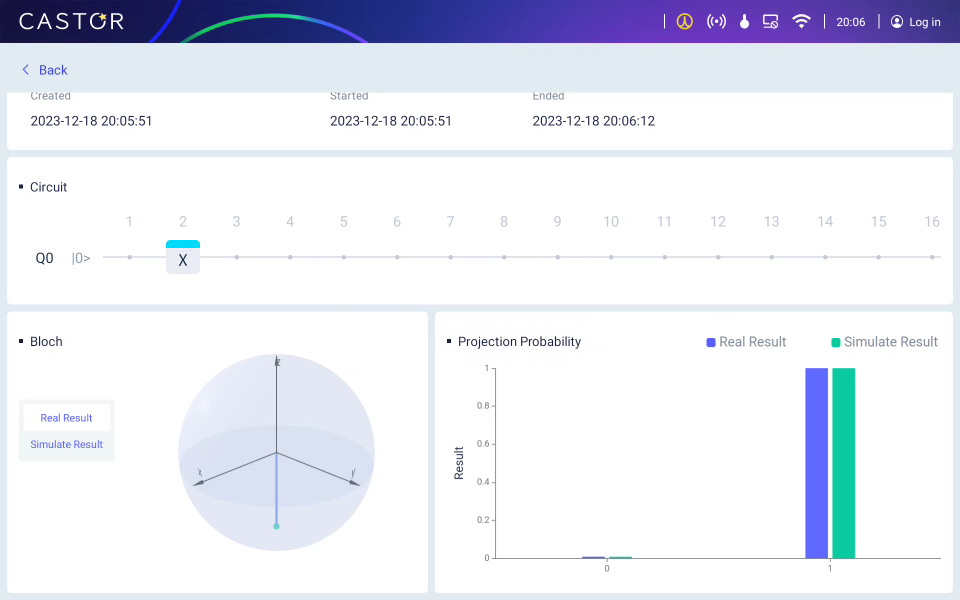}
    \caption{\label{e_3}Build with $X$ gate circuit in quantum circuit.}
\end{figure}

Second we build the $H$ and $X$ circuit on a real quantum computer in Figure~\ref{e_2} we build $H$ gate, in Figure~\ref{e_3} we build $X$ gate.

Now, let's consider the question: How can we construct a circuit incorporating the \( Z \) gate on an actual quantum computer? To provide a hint, remember that we can convert the state \(\ket{0}\) to \(\ket{1}\) using the \( X \) gate. This conversion is key to demonstrating the effect of the \( Z \) gate, as its impact is more pronounced when applied to the \(\ket{1}\) state.

\begin{figure}[H]
    \centering
    \includegraphics[scale=0.55]{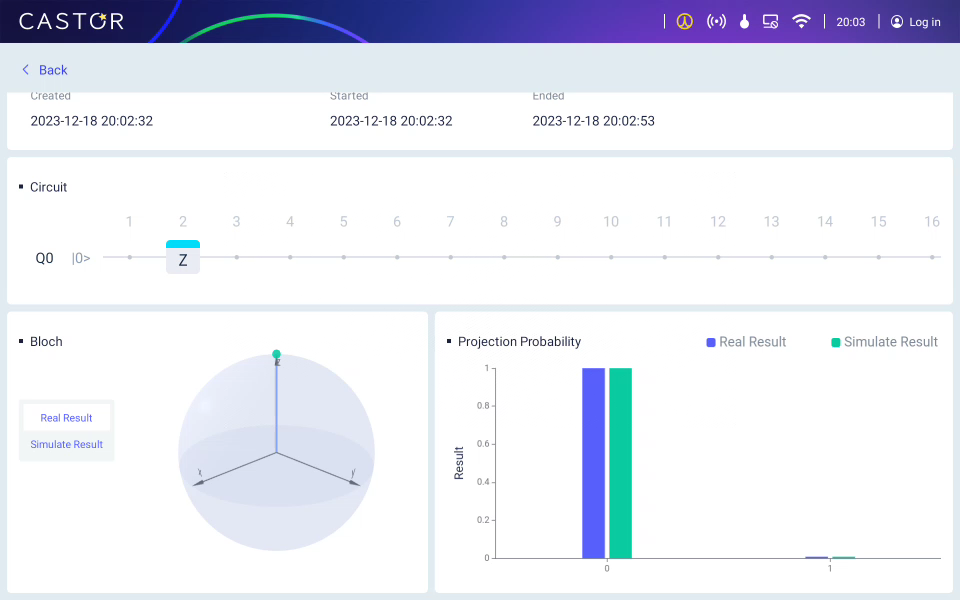}
    \caption{\label{e_4}Build with Z gate circuit in quantum circuit.}
\end{figure}

Figure~\ref{e_4} illustrates the process of building and implementing a circuit with the \( Z \) gate on a real quantum computer. This example helps to understand how the \( Z \) gate operates within a quantum circuit and how its effects can be observed and analyzed in an actual quantum computing setup.

Finally we make a measurement. In this way we have completed the gates of $H$, $X$ and $Z$ experiments.

\textbf{The experiment of Two-qubit gates (CNOT)}~{}

The Controlled-NOT (CNOT) gate 
is a fundamental element in quantum computing, playing a pivotal role in the manipulation and entanglement of qubits.

To explore the functionalities and implications of the CNOT gate, we will conduct an experiment contrasting it with the classical XOR gate. The first step in this process is to construct a quantum circuit that incorporates the CNOT gate.

\begin{figure}[H]
    \centering
    \[\Qcircuit @!C @!R{
    &\lstick{\mid x\rangle}&\ctrl{1}&\rstick{\mid x\rangle}\qw\\
    &\lstick{\mid y\rangle}&\targ&\rstick{\mid x\oplus y\rangle}\qw}\]
    \caption{\label{e_5}The CNOT gate circuit}
\end{figure}
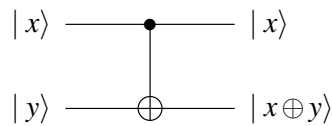

Figure~\ref{e_5} illustrates the designed circuit that integrates the CNOT gate. 

\begin{figure}[H]
    \centering
    \includegraphics[scale=0.55]{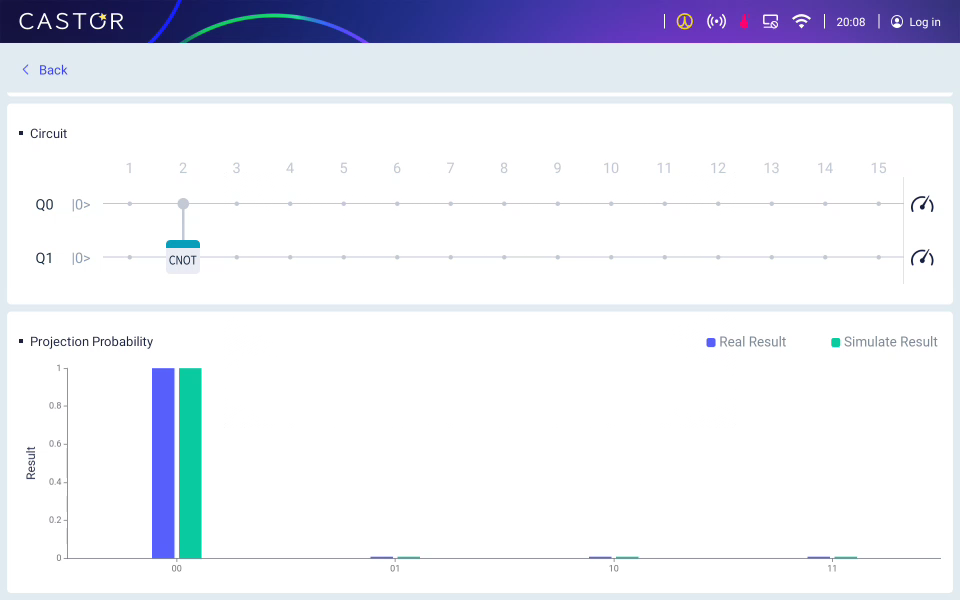}
    \caption{\label{e_6} The CNOT gate circuit on a real quantum computer}
\end{figure}

Second we build with CNOT gate circuit on a real quantum computer shown in Figure~\ref{e_6}.

Finally we make a measurement. In this way we have completed the CNOT gate experiment.

\textbf{The Experiment of Bell states}~{}

\begin{figure}[H]
	\includegraphics[scale=0.55]{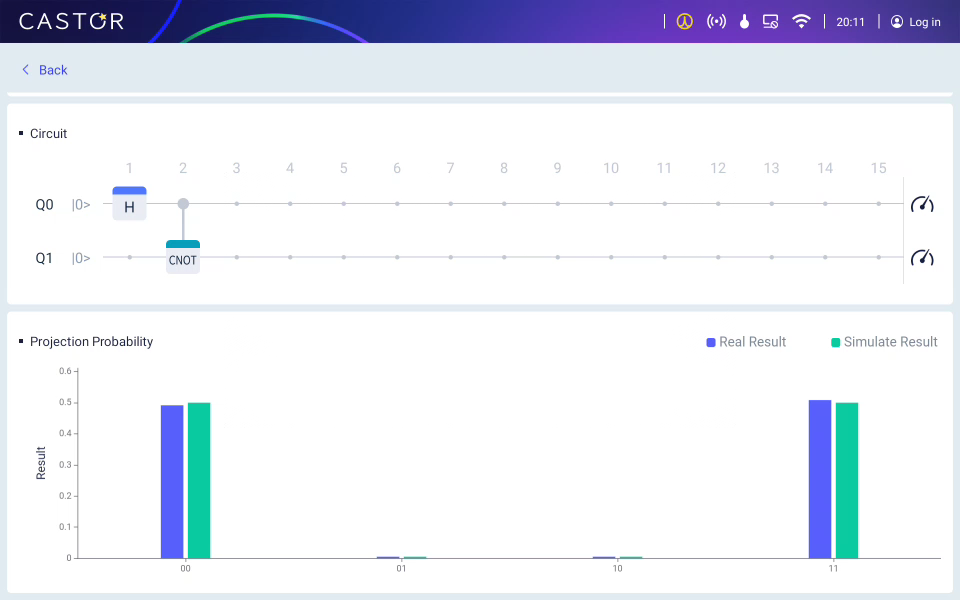}
	\centering
	\caption{\label{4_5.2} The SpinQ experiments for Bell states.}
\end{figure}

The Figure\ref{4_5.2} shown the SpinQ experiments for Bell states~\cite{sun12shukla2014protocols}. With SpinQ we can achieve the Bell states

\textbf{The Experiment of Deutsch's Algorithm}~{}

Deutsch's Algorithm~\cite{sun21de2005implementing} is a pivotal algorithm in the field of quantum computing, known for demonstrating the potential of quantum algorithms to outperform classical ones in specific scenarios.

To explore the practical implementation and effectiveness of Deutsch's Algorithm, we will conduct an experiment by constructing and executing its quantum circuit. This experiment aims to demonstrate the algorithm's ability to determine properties of functions more efficiently than classical computing methods.

\begin{figure}[H]
    \centering
        \[\Qcircuit @!C @!R{
        &\lstick{\mid 0\rangle}&\qw&\gate{H}&\qw&\multigate{1}{U_{f}}&\qw&\gate{H}&\qw&\meter\\
        &\lstick{\mid 1\rangle}&\qw&\gate{H}&\qw&\ghost{U_{f}}&\qw&\qw&\qw&\qw\\
        &                      &\push{\mid\phi_{0}\rangle}& &\push{\mid\phi_{1}\rangle}& &\push{\mid\phi_{2}\rangle}&&\push{\mid\phi_{3}\rangle}
        }\]
    \caption{\label{e_7}The quantum circuit implementation of Deutsch's algorithm.}
\end{figure}
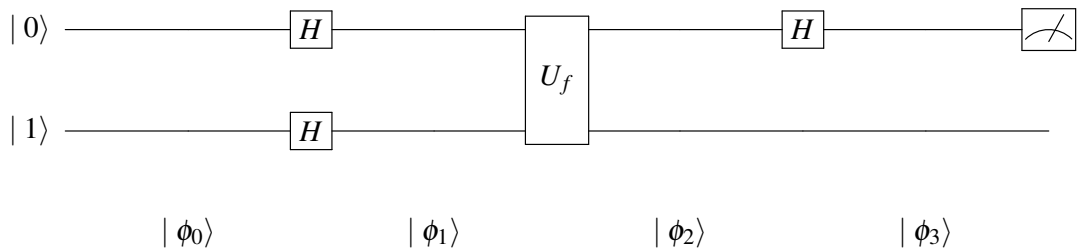

The first step in this experiment is to build the quantum circuit that represents Deutsch's Algorithm. The design of this circuit, which is crucial for the experiment, is depicted in Figure~\ref{e_7}. This circuit serves as the foundation for demonstrating the algorithm's functionality and showcases how quantum superposition and interference are harnessed to achieve computational advantages in quantum computing.


Next, we proceed to implement this circuit on an actual quantum computer. This step involves interacting with the quantum computer's interface. Specifically, we select the "Case" option on the right side of the interface, choose the appropriate algorithm category, and then select "Deutsch's Algorithm" from the list. Once selected, we initiate the simulation to observe the algorithm in action.

\begin{figure}[H]
    \centering
    \includegraphics[scale=0.5]{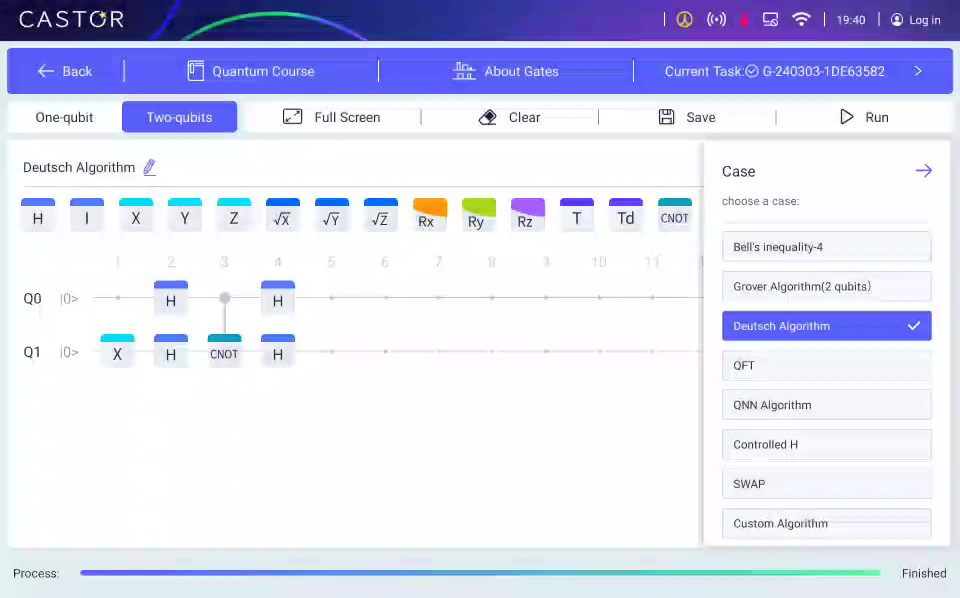}
    \caption{\label{e_12}The case of Deutsch algorithm circuit on a real quantum computer}
\end{figure}

This process is illustrated in Figure~\ref{e_12}, which shows how the circuit for Deutsch's Algorithm is set up and run on a quantum computing platform. Running this simulation on a real quantum computer allows us to observe firsthand the workings of Deutsch's Algorithm and its ability to efficiently determine properties of functions, thereby reinforcing its significance in the realm of quantum computation.

Once the simulation of Deutsch's Algorithm is run on the quantum computer, the next step is to observe and record the projection probabilities of the four quantum states \(\ket{00}\), \(\ket{01}\), \(\ket{10}\), and \(\ket{11}\). These probabilities provide crucial insights into the behavior of the algorithm.

\begin{figure}[H]
    \centering
    \includegraphics[scale=0.5]{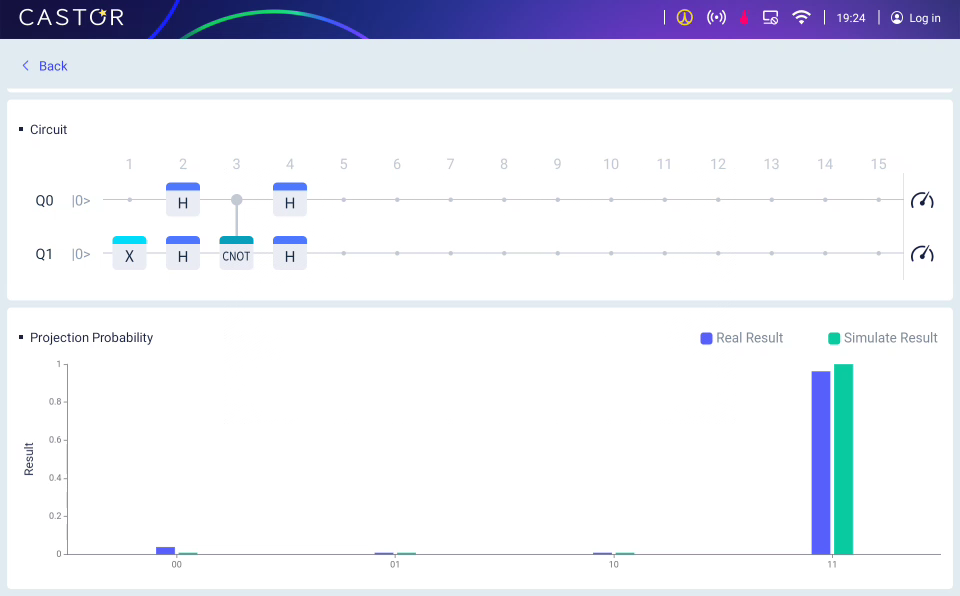}
    \caption{\label{e_13}The Deutsh algorithm experiment result}
\end{figure}

The recorded probabilities are displayed in Figure~\ref{e_13}. By analyzing these results, we can discern the relationship between \(f(0)\) and \(f(1)\). This analysis is pivotal in determining whether the quantum circuit correctly implements Deutsch's Algorithm. Specifically, the relative probabilities of these four states indicate whether the function \(f\) used in the experiment is balanced or constant, thus validating the algorithm's performance.

From the observed results, if the first qubit consistently outputs \(1\), it indicates that the function \(f\) is balanced, meaning \(f(1) \neq f(0)\). This conclusion is drawn from the principle that Deutsch's Algorithm determines whether the function \(f\) applied to the qubits is constant (the same output for both inputs) or balanced (different outputs for each input). Therefore, the outcome where the first qubit is in the state \(1\) directly reflects the nature of \(f\) as a balanced function.

\newpage

\chapter{Week 3: Matrix Representation of Quantum Gates}

Last week, we dipped our toes into the intriguing world of quantum computing with a brief look at Deutsch's algorithm, exploring it through the lens of counting rather than delving into matrices. This week, we're ready to take a step further. We'll begin by introducing two key concepts: vectors, which we'll use to describe quantum states, and matrices, serving as the tools to manipulate these states. Think of this as learning the alphabet and grammar of the quantum computing language. As we grow more comfortable with these ideas, we'll expand our exploration to multi-qubit systems, where the real magic of quantum computing begins to unfold. This journey will lay the groundwork for understanding how quantum computers can perform complex tasks that are currently beyond the capabilities of classical computers. By the end of this chapter, you’ll have a foundational understanding of quantum states and operations, setting you up perfectly for the exciting world of quantum algorithms.

\section{Vectors and matrices}
\textbf{Bits and vectors}

In our previous discussions, we saw how classical bits are the simplest units in computing, representing either 0 or 1. In quantum computing, we use something similar but more powerful: qubits \cite{2_1}. Let's see how we can represent both classical bits and qubits using vectors, which are just lists of numbers arranged in a certain way.

In the world of classical computing, we can represent a bit using a 2D column vector \cite{2_3}. A bit with a value of 0 is shown as $\begin{bmatrix} 1\\0 \end{bmatrix}$, and a bit with a value of 1 is shown as $\begin{bmatrix} 0\\1 \end{bmatrix}$.

Quantum bits, or qubits, have a similar representation \cite{2_2}. The quantum state $|0\rangle$, which is analogous to the classical 0 bit, is represented as $\begin{bmatrix} 1\\0 \end{bmatrix}$. Similarly, the quantum state $|1\rangle$, like the classical 1 bit, is represented as $\begin{bmatrix} 0\\1 \end{bmatrix}$. 

But there's something unique about qubits: they can be in a state called superposition \cite{4_sx}, which means they can be a bit of both 0 and 1 at the same time! We can express a general qubit state $|\varphi\rangle$ as a combination (or superposition) of $|0\rangle$ and $|1\rangle$, which looks like this: $|\varphi\rangle=a|0\rangle+b|1\rangle$. When we use the vector representation, this becomes:
\begin{equation}
  |\varphi\rangle=a|0\rangle+b|1\rangle = a\begin{bmatrix} 1  \\ 0 \end{bmatrix}+b\begin{bmatrix} 0  \\ 1 \end{bmatrix}=\begin{bmatrix} a  \\ b \end{bmatrix}.  
\end{equation}

In this vector, $a$ and $b$ are numbers that tell us how much of $|0\rangle$ and $|1\rangle$ are in the superposition. For example, if we have an equal mix of $|0\rangle$ and $|1\rangle$, it looks like this:
\begin{equation}
  |\varphi\rangle=\frac{1}{\sqrt{2}}|0\rangle+\frac{1}{\sqrt{2}}|1\rangle \to \begin{bmatrix} \frac{1}{\sqrt{2}}  \\ \frac{1}{\sqrt{2}} \end{bmatrix}.  
\end{equation}

\textbf{Gates and matrices}~{}

Now, let's talk about how we can use matrices to represent operations in quantum computing, similar to how we use commands in a computer program. In classical computing, there's a simple operation called the NOT gate, which flips a bit \cite{2_3}. If the bit is 0, it turns it to 1, and if it's 1, it turns it to 0. We already know that bits can be represented as vectors: 0 as $\begin{bmatrix} 1\\0 \end{bmatrix}$ and 1 as $\begin{bmatrix} 0\\1 \end{bmatrix}$. The NOT gate's function, which inverts a bit, can be represented using a matrix. The matrix for a NOT gate looks like this \cite{4_sx}:
\begin{equation}
    \begin{bmatrix}
        0&1\\1&0
    \end{bmatrix}.\nonumber
\end{equation}

In this matrix, the first row tells us what happens when the input is 0, and the second row tells us what happens when the input is 1. Basically, the matrix captures the entire function of the NOT gate. 

When we apply this matrix to a bit, it's like giving a command to the bit. For example, let's apply the NOT matrix to the bit 0 (represented as $\begin{bmatrix} 1\\0 \end{bmatrix}$):
\begin{equation}
\text{NOT} \times \begin{bmatrix} 1\\0 \end{bmatrix} =
   \begin{bmatrix}
        0&1\\1&0
    \end{bmatrix} \times \begin{bmatrix}
    1\\0
\end{bmatrix} =\begin{bmatrix}
    0\\1
\end{bmatrix},
\end{equation}
which is the vector representation of 1, showing that the input 0 has been flipped to 1. Similarly, if we apply the NOT gate to 1 (represented as $\begin{bmatrix} 0\\1 \end{bmatrix}$), it will flip it to 0. This way, matrices provide us with a neat and powerful way to represent and understand the operations of gates in quantum computing.

\textbf{The rule of matrix arithmetic}~{}

In this section, we'll look at some basic rules for doing math with matrices. A matrix is like a big box filled with numbers, arranged in rows and columns. These numbers can be simple numbers (like the ones we use every day) or more complex ones.

Let's consider a matrix called \textbf{A}:
\begin{equation}
    \begin{bmatrix}
a_{11} & a_{12} & \cdots & a_{1n} \\
a_{21} & a_{22} & \cdots & a_{2n} \\
\vdots & \vdots & \ddots & \vdots \\
a_{m1} & a_{m2} & \cdots & a_{mn}
\end{bmatrix}.
\end{equation}
Here, each number in the matrix is called an element. The element in row $j$ and column $k$ is denoted as $a_{jk}$, where $j$ is the number of rows and $k$ is the number of columns.

Now, consider another matrix B, also with dimensions $m \times n$. We can define arithmetic operations between these matrices \cite{2_4,2_5}:
\begin{itemize}
\item \textbf{Addition}: $(A + B)[j,k] = A[j,k] + B[j,k]$
\item \textbf{Scalar Multiplication}: $(c \cdot A)[j,k] = c \times A[j,k]$
\item \textbf{Inverse}: $(-A)[j,k] = -(A[j,k])$
\item \textbf{Matrix Multiplication}:
\end{itemize}

Consider two matrices: {A} (dimensions $m \times n$) and {B} (dimensions $n \times s$). Their matrix product {AB} will be a new matrix with dimensions $m \times s$.

The multiplication rule is:
\begin{equation}
(AB)[j,k] = \sum_{h=0}^{n-1} (A[j,h] \times B[h,k]).
\end{equation}
Where we sum the products of entries from row $j$ of {A} and column $k$ of {B}.

\begin{example}
\begin{equation}
\begin{bmatrix} 0 & 1\\1 & 0 \end{bmatrix}
\times \begin{bmatrix} 1\\0 \end{bmatrix} =
\begin{bmatrix} 0\\1 \end{bmatrix},\begin{bmatrix} 0 & 1\\1 & 0 \end{bmatrix}
\times \begin{bmatrix} 0\\1 \end{bmatrix} =
\begin{bmatrix} 1\\0 \end{bmatrix}.
\end{equation}
\end{example}

Here's an important thing to remember: In matrix multiplication, the order matters! Unlike regular numbers, where \(2 \times 3\) is the same as \(3 \times 2\), with matrices, {A} multiplied by {B} (\({A}{B}\)) can be different from {B} multiplied by {A} (\({B}{A}\)) \cite{2_4}. Generally:
\begin{equation}
    AB \neq BA. 
\end{equation}

\begin{example}
Let's see this with a simple example. Consider two matrices:
\begin{equation}
{A} = \begin{bmatrix} 1 & 2 \\ 3 & 4 \end{bmatrix}, \quad {B} = \begin{bmatrix} 2 & 0 \\ 1 & 2 \end{bmatrix}.
\end{equation}
Now, if we multiply {A} by {B}, we get:
\begin{equation}
AB = \begin{bmatrix} 1 & 2 \\ 3 & 4 \end{bmatrix} \times \begin{bmatrix} 2 & 0 \\ 1 & 2 \end{bmatrix} = \begin{bmatrix} 1\cdot2 + 2\cdot1 & 1\cdot0 + 2\cdot2 \\ 3\cdot2 + 4\cdot1 & 3\cdot0 + 4\cdot2 \end{bmatrix} = \begin{bmatrix} 4 & 4 \\ 10 & 8 \end{bmatrix}.
\end{equation}
But, if we flip the order and multiply {B} by {A}, we get a different result:
\begin{equation}
BA = \begin{bmatrix} 2 & 0 \\ 1 & 2 \end{bmatrix} \times \begin{bmatrix} 1 & 2 \\ 3 & 4 \end{bmatrix} = \begin{bmatrix} 2\cdot1 + 0\cdot3 & 2\cdot2 + 0\cdot4 \\ 1\cdot1 + 2\cdot3 & 1\cdot2 + 2\cdot4 \end{bmatrix} = \begin{bmatrix} 2 & 4 \\ 7 & 10 \end{bmatrix}.
\end{equation}
So, in this case, \(AB \neq BA\).
\end{example}

This rule is important in quantum computing, as the order of operations can change the outcome of a computation!

Matrix multiplication isn't just a cool math trick; it's super useful, especially in quantum computing. To make these calculations easier, we can use software like Matlab \cite{2_20,2_34} and Octave \cite{2_32,2_33}. These tools are like calculators for matrices, helping us multiply and manipulate them without getting lost in all the numbers.

When we get the hang of multiplying matrices, we can start to represent a bunch of operations all together in a really neat and tidy way. It's like having a recipe where each step is clearly laid out, but instead of cooking, we're doing quantum computing!

The rules for working with matrices are pretty straightforward. Once we understand them, we can mix and match matrices to represent complex processes or rules in a simple and clear manner. It's a bit like putting together LEGO bricks to build something big and intricate – each brick (or matrix) has its place and role.

Mastering matrix arithmetic is a key step in unlocking the full potential of matrix computation. It's not just about doing math; it's about gaining a powerful tool to understand and analyze complex operations in quantum computing and beyond. So, diving into the world of matrices is definitely worth the effort!

\section{One bit/qubit vector and gates}

\textbf{Representations of the NOT Gate}~{}

In the world of digital circuits, logic gates are like the basic building blocks. One of these is the NOT gate, which is pretty straightforward but super important. In Figure \ref{Z_1}, we have a classic diagram of a NOT gate. It might look simple, but there are different ways to understand what it does \cite{2_12}:

\begin{figure}[H]
\centering
\includegraphics[scale=0.23]{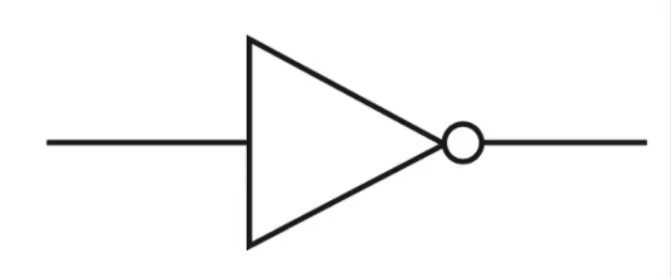}
 \caption{\label{Z_1} Classic NOT-gate logic diagram.}
\end{figure}

\begin{itemize}
\item \textbf{Functional View}: Think of the NOT gate as a function, which we'll call $f(x)$. This function has a simple job: if it gets a 0, it gives back a 1, and if it gets a 1, it gives back a 0. So, $f(0) = 1$ and $f(1) = 0$.

\item \textbf{Truth Table}: To see this in a more organized way, we use a truth table. This table is like a quick reference guide that shows what the output will be for each possible input:
\begin{center}
\setlength{\tabcolsep}{5mm}
\renewcommand\arraystretch{1.5}
\begin{tabular}{c|c}

Input & Output\\
\hline
0 & 1\\

1 & 0
\end{tabular}
\end{center}

\item \textbf{Matrix Representation}: Now, let's talk matrices. We can represent the NOT gate as a matrix too! The matrix for a NOT gate looks like this \cite{2_8}:
\begin{equation}
 \mathrm{NOT} =
\begin{bmatrix}
0 & 1\\
1 & 0
\end{bmatrix}.   
\end{equation}

When we apply this matrix to the vectors $\begin{bmatrix}1\end{bmatrix}$ and $\begin{bmatrix}0\end{bmatrix}$ (which represent the bits 0 and 1), it flips them, just like the NOT function does \cite{2_8}:
\begin{equation}
\mathrm{NOT} \times \begin{bmatrix}1\end{bmatrix} = \begin{bmatrix}0\end{bmatrix},
\mathrm{NOT} \times \begin{bmatrix}0\end{bmatrix} = \begin{bmatrix}1\end{bmatrix}.
\end{equation}
\end{itemize}
These three ways of looking at the NOT gate help us understand it from different angles. Whether it's through a function, a table, or a matrix, each perspective adds to our understanding of how we can use logic gates to control and manipulate information.

\textbf{Single qubit gates and matrices}~{}

In the quantum computing world, we encounter fascinating tools known as quantum gates. These gates are fundamental in building quantum circuits, where they manipulate the states of qubits in specific ways. One such gate is the NOT gate, known in quantum computing as the $X$ gate. This gate is crucial for flipping the state of a qubit \cite{4_sx}:
\begin{align}
    X|0\rangle = |1\rangle,\nonumber\\  
    X|1\rangle = |0\rangle.
\end{align}

Transformations like these are essential in quantum computations, enabling unique processing and manipulation of information.

Each quantum gate corresponds to a specific matrix. For the $X$ gate, the matrix representation is:
\begin{equation}
 X =
\begin{bmatrix}
0 & 1\\
1 & 0
\end{bmatrix}.   
\end{equation}

When this matrix multiplies a qubit state vector, it executes the NOT operation:
\begin{equation}
 \begin{bmatrix}
0 & 1\\
1 & 0
\end{bmatrix} \times \begin{bmatrix}
1\\ 0
\end{bmatrix}=\begin{bmatrix}
0\\ 1
\end{bmatrix}, 
 \begin{bmatrix}
0 & 1\\
1 & 0
\end{bmatrix} \times \begin{bmatrix}
0\\ 1
\end{bmatrix}=\begin{bmatrix}
1\\ 0
\end{bmatrix}.
\end{equation}

The $X$ gate, however, is just one member of a diverse family of quantum gates. For instance, the Hadamard gate, represented as $H$, is notable for creating superpositions of $|0\rangle$ and $|1\rangle$. The matrix for $H$ is \cite{2_10}:
\begin{equation}
 H =
\frac{1}{\sqrt{2}}\begin{bmatrix}
1 & 1\\
1 & -1
\end{bmatrix}.   
\end{equation}

Applying $H$ to $|0\rangle$ results in $\frac{(|0\rangle+|1\rangle)}{\sqrt{2}}$, and applying it to $|1\rangle$ yields $\frac{(|0\rangle-|1\rangle)}{\sqrt{2}}$.

Another vital gate in single-qubit operations is the Phase Gate, denoted as $Z$. The $Z$ gate is unique in that it doesn't change the $|0\rangle$ state but flips the sign of the $|1\rangle$ state:
\begin{align}
    Z|0\rangle &= |0\rangle,\nonumber\\
    Z|1\rangle &= -|1\rangle.
\end{align}
The matrix for $Z$ is:
\begin{equation}
Z =
\begin{bmatrix}
    1&0 \\0&-1\end{bmatrix}.
\end{equation} 

When applied, the $Z$ gate leaves the $|0\rangle$ state unchanged, while it flips the sign of the amplitude for the $|1\rangle$ state:
\begin{equation}
\begin{bmatrix}
    1&0 \\0&-1\end{bmatrix} \times \begin{bmatrix}
    1\\0\end{bmatrix} = \begin{bmatrix}
    1\\0\end{bmatrix},\begin{bmatrix}
    1&0 \\0&-1\end{bmatrix} \times \begin{bmatrix}
    0\\1\end{bmatrix} = -\begin{bmatrix}
    0\\1\end{bmatrix}.
\end{equation} 

Gates like $X$, $H$, and $Z$ are crucial in quantum computing, enabling control and manipulation of qubit states \cite{2_9}. Understanding their matrix representations allows us to analyze and simulate quantum circuits and algorithms, demonstrating the transformative potential of quantum computing for solving complex problems.

In quantum computing, we can construct some quantum gates by cleverly combining other gates. A fascinating example is how we can create the phase gate $Z$ using the NOT gate $X$ and the Hadamard gate $H$ \cite{4_sx}:
\begin{align}
  HZH = X,\nonumber\\
  HXH = Z.
\end{align}

To see how this works, let's examine their matrix representations. The matrix for the Hadamard gate $H$ is defined as:
\begin{equation}
 H = \frac{1}{\sqrt{2}}\begin{bmatrix}
1 & 1\\
1 & -1
\end{bmatrix}.   
\end{equation}
By performing the matrix multiplication, we can validate the relationships:
\begin{equation}
  HZH=\frac{1}{\sqrt{2}}\begin{bmatrix} 1 & 1 \\ 1 & -1 \end{bmatrix}\times\begin{bmatrix} 1 & 0 \\ 0 & -1 \end{bmatrix}\times\frac{1}{\sqrt{2}}\begin{bmatrix} 1 & 1 \\ 1 & -1 \end{bmatrix}=\begin{bmatrix} 0 & 1 \\ 1 & 0 \end{bmatrix}=X. 
\end{equation}

Similarly, for $HXH = Z$:
\begin{equation}
 HXH=\frac{1}{\sqrt{2}}\begin{bmatrix} 1 & 1 \\ 1 & -1 \end{bmatrix}\times\begin{bmatrix} 0 & 1 \\ 1 & 0 \end{bmatrix}\times\frac{1}{\sqrt{2}}\begin{bmatrix} 1 & 1 \\ 1 & -1 \end{bmatrix}=\begin{bmatrix} 1 & 0 \\ 0 & -1 \end{bmatrix}=Z.   
\end{equation}

These calculations show that the Hadamard gate $H$ can transform between the $X$ and $Z$ gates. This example illustrates the power of combining elementary quantum gates to create more complex operations on qubits, a fundamental aspect of quantum computing.

\section{Two bits/qubits vectors and gates}

\textbf{Bits and vectors}~{}

In quantum computing, it's fascinating how we can represent bitstrings, or sequences of bits, as vectors. This method becomes particularly handy when dealing with more complex systems, like those with two bits or qubits. Let's dive into how this works for a two-bit system.

Consider all the possible combinations in a two-bit system: $00$, $01$, $10$, and $11$. Each of these bitstrings can be represented as a unique vector, with the position of the `1' in the vector indicating the value of the bitstring. Here's how we can depict each bitstring as a vector \cite{2_1}:
\begin{center}
 $00 \to \begin{bmatrix} 1 \\0 \\0\\0\end{bmatrix},\quad 01 \to \begin{bmatrix} 0 \\1 \\0\\0\end{bmatrix},\quad 10 \to \begin{bmatrix} 0 \\0 \\1\\0\end{bmatrix},\quad 11 \to \begin{bmatrix} 0 \\0 \\0\\1\end{bmatrix}$.   
\end{center}

In these representations, the position of the `1' in each vector corresponds to the specific bitstring it represents. For instance, the vector $\begin{bmatrix} 0 \\ 0 \\ 1 \\ 0 \end{bmatrix}$ represents the bitstring `10' because the `1' is in the third position, corresponding to the third possibility in our two-bit system.

This vector representation becomes even more powerful in quantum computing. In quantum systems, the elements of these vectors (known as amplitudes) can represent superpositions of multiple states. This means a quantum state can be a combination of several different bitstrings at the same time, unlike in classical computing where a bit is definitely either 0 or 1. Understanding this concept is key to grasping the fundamentals of quantum computing and how it differs from classical computing.

\textbf{Gates and matrices}~{}

In both classical and quantum computing, representing bitstrings as vectors provides a consistent mathematical framework for mapping bitstrings onto states. Operations on bitstrings, such as \text{AND}, \text{OR}, \text{NAND} gates, correspond to specific manipulations of these state vectors.

\begin{figure}[H]
\centering
\includegraphics[scale=0.23]{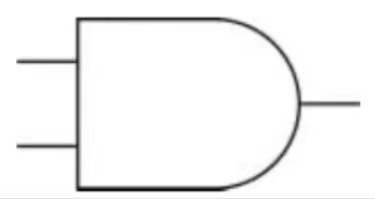}
 \caption{\label{AND_1} Classic AND-gate logic diagram.}
\end{figure}

Consider the \text{AND} gate, as shown in Figure \ref{AND_1}. This gate performs a logical \text{AND} operation on two input bits \cite{2_12}.

Functionally, the \text{AND} gate outputs 1 only if both its input bits are 1. This can be expressed as:
\begin{equation}
  f(00) = 0, f(01) = 0, f(10) = 0, f(11) = 1. \nonumber
\end{equation}

The truth table for the \text{AND} gate is:
\begin{center}
    $00 \to 0$
    
    $01 \to 0$

    $10 \to 0$

    $11 \to 1$.
\end{center}

We can represent the \text{AND} operation using a matrix:
\begin{equation}
  \text{AND} = \begin{bmatrix}
      1&1&1&0\\0&0&0&1
  \end{bmatrix}.
\end{equation}

This matrix, when multiplied by an input vector, performs the \text{AND} operation on the bits. For example, applying the \text{AND} gate to the input $00$:
\begin{equation}
  \text{AND}(00) = 0.  
\end{equation}

The multiplication of the \text{AND} matrix with the vector representation of $00$ is:
\begin{equation}
  \text{AND} = \begin{bmatrix}
      1&1&1&0\\0&0&0&1
  \end{bmatrix}, 00=\begin{bmatrix}
      1\\0\\0\\0
  \end{bmatrix}.
\end{equation}

This results in:
\begin{equation}
  \begin{bmatrix}
      1&1&1&0\\0&0&0&1
  \end{bmatrix} \times \begin{bmatrix}
      1\\0\\0\\0
  \end{bmatrix}  =\begin{bmatrix}
      1\\0
  \end{bmatrix}= 0. 
\end{equation}

Hence, the \text{AND} gate maps the input $00$ to the state vector corresponding to 0, as expected. In general, multiplying the \text{AND} matrix with any input vector performs the \text{AND} logic operation on the bit values. 

\begin{figure}[H]
\centering
\includegraphics[scale=0.23]{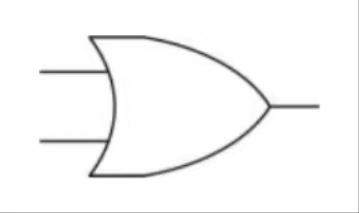}
 \caption{\label{OR_1} Classic OR-gate logic diagram.}
\end{figure}

As depicted in Figure \ref{OR_1}, we see a classic \text{OR}-gate logic diagram. The \text{OR} gate is another fundamental component in digital logic \cite{2_12}.

Now, let's examine the operation of the \text{OR} gate when it acts on the input $00$:
\begin{equation}
\text{OR}(00) = 0.
\end{equation}

To understand this operation in terms of matrix multiplication, we represent the \text{OR} operation as a matrix and multiply it with the vector representation of $00$ \cite{4_sx}:
\begin{equation}
  \text{OR} = \begin{bmatrix}
      1&0&0&0\\0&1&1&1
  \end{bmatrix}, \quad 00=\begin{bmatrix}
      1\\0\\0\\0
  \end{bmatrix}.
\end{equation}

The result of this multiplication is:
\begin{equation}
  \begin{bmatrix}
     1&0&0&0\\0&1&1&1
  \end{bmatrix} \times \begin{bmatrix}
      1\\0\\0\\0
  \end{bmatrix}  =\begin{bmatrix}
      1\\0
  \end{bmatrix}= 0. 
\end{equation}

As anticipated, the \text{OR} gate maps the input $00$ to the state vector corresponding to 0.

More generally, when we multiply the \text{OR} matrix with any input vector, it effectively carries out the logical \text{OR} operation on the bit values. Just like the \text{AND} gate, this matrix formulation allows us to conceptualize the \text{OR} operation as a linear transformation applied to bit vectors. This approach will be extended to understanding how quantum gates operate on qubit state vectors in quantum computing.

\begin{figure}[H]
\centering
\includegraphics[scale=0.23]{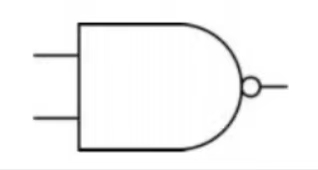}
 \caption{\label{NAND_1} Classic NAND-gate logic diagram.}
\end{figure}

The \text{NAND} gate performs a NOT-AND operation on two bits. Figure \ref{NAND_1} illustrates a classic \text{NAND}-gate logic diagram.

Functionally, the \text{NAND} gate can be described as \cite{2_13}:
\begin{equation}
    f(00) = 1, f(01) = 1, f(10) = 1, f(11) = 0.\nonumber
\end{equation} 

It essentially inverts the output of an \text{AND} gate. The truth table for \text{NAND} is:
\begin{center}
  $00 \to 1$
  
$01 \to  1$

$10 \to  1$

$11 \to  0 $
\end{center}

The matrix representation of \text{NAND} is:
\begin{equation}
    \text{NAND} = \begin{bmatrix}
        0&0&0&1\\1&1&1&0
    \end{bmatrix}.
\end{equation}

When this matrix multiplies input vectors, it effectively flips the output of an \text{AND} operation.

The \text{NAND} gate demonstrates how basic logic gates can be combined to create more complex operations. The matrix formulation provides a systematic way to manipulate these composite gates \cite{2_14}.

\begin{figure}[H]
\centering
\includegraphics[scale=0.23]{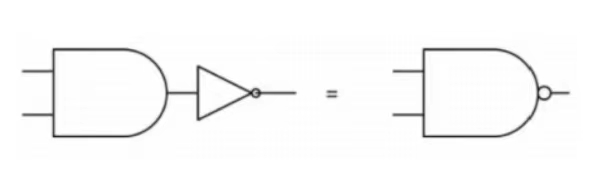}
 \caption{\label{NAND_2} NAND-gate construction.}
\end{figure}

Figure \ref{NAND_2} shows the construction of a \text{NAND} gate from \text{AND} and \text{NOT} gates:
\begin{equation}
   \text{NAND} = \text{NOT} \times \text{AND}. 
\end{equation}
The matrix for the \text{NOT} gate is:
\begin{equation}
   \text{NOT} = \begin{bmatrix}
    0&1\\1&0
\end{bmatrix}. 
\end{equation}
And the matrix for the \text{AND} gate is:
\begin{equation}
   \text{AND} = \begin{bmatrix}
    1&1&1&0\\0&0&0&1
\end{bmatrix}.
\end{equation}
Multiplying these matrices gives:
\begin{equation}
    \text{NOT} \times \text{AND} = \begin{bmatrix}
        0&1\\1&0
    \end{bmatrix} \times \begin{bmatrix}
         1&1&1&0\\0&0&0&1
    \end{bmatrix}=\begin{bmatrix}
        0&0&0&1\\1&1&1&0
    \end{bmatrix}.
\end{equation}

This results in the \text{NAND} matrix, illustrating how we can construct complex gates like \text{NAND} from simpler gates such as \text{AND} and \text{NOT} using matrix operations. This approach is extendable to the construction of complex quantum operations from basic quantum gates.

\textbf{Two-qubit gates}~{}

An important two-qubit gate is the controlled-NOT (CNOT) gate. This flips the second qubit if the first qubit is $|1\rangle$:
\begin{equation}
  |x\rangle \otimes |y\rangle \to  |x\rangle \otimes |y \oplus x\rangle.  
\end{equation}
Where $\oplus$ is addition modulo 2.

In the computational basis ${|00\rangle, |01\rangle, |10\rangle, |11\rangle}$, the CNOT matrix is \cite{4_sx}:
\begin{equation}
    \text{CNOT} = \begin{pmatrix}1 &0& 0& 0\\
0 &1 &0 &0\\0& 0& 0 &1\\
0& 0& 1 &0  
    \end{pmatrix}.
\end{equation}

\begin{figure}[H]
	\centering
	\[\Qcircuit @!C @!R {
        &\lstick{|x\rangle} &\ctrl{1}  & \rstick{|x\rangle}\qw  & &  &\lstick{|x\rangle} &\targ    & \rstick{|x \oplus y\rangle}\qw \\
        &\lstick{|y\rangle} &\targ    & \rstick{|y \oplus x\rangle}\qw &   & & \lstick{|y\rangle} &\ctrl{-1}  & \rstick{|y\rangle}\qw
        }\]
	 \caption{\label{CNOT} A CNOT with the second qubit as control.}
\end{figure}
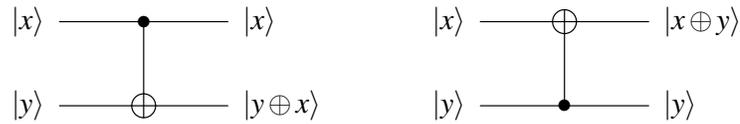

We can also construct a CNOT with the second qubit as control, taking $|x\rangle \otimes |y\rangle$ to $|x\oplus y\rangle \otimes|y\rangle$. As shown in Figure \ref{CNOT}. In the left diagram of Figure \ref{CNOT}, flips $|y\rangle$ if $|x\rangle = |1\rangle$. In the right diagram of Figure \ref{CNOT}, flips $|x\rangle$ if $|y\rangle = |1\rangle$.

CNOT allows qubit-qubit interactions and entanglement in multi-qubit systems.

Another two-qubit gate is the controlled-Z (CZ) gate. This applies a phase flip to the second qubit if the first qubit is $|1\rangle$:
\begin{center}
    $|00\rangle \to |00\rangle$
    
$|01\rangle \to |01\rangle$

$|10\rangle \to |10\rangle$

$\quad|11\rangle \to -|11\rangle$.
\end{center}

The CZ gate is symmetric so the control/target roles are interchangeable. In the computational basis, the CZ matrix is \cite{4_sx,2_3}:
\begin{equation}
   CZ =  \begin{pmatrix}
        1& 0 &0 &0\\
0 &1& 0& 0\\
0& 0& 1 &0\\
0 &0 &0& -1
    \end{pmatrix}.
\end{equation}

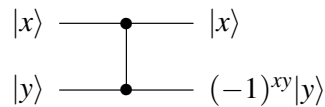
\begin{figure}[H]
	\centering
	\[\Qcircuit @!C @!R {
        &\lstick{|x\rangle} &\ctrl{1}  & \rstick{|x\rangle}\qw \\
        &\lstick{|y\rangle} &\control\qw & \rstick{(-1)^{xy}|y\rangle }\qw}\]
	\caption{\label{Cz} Circuit representation of the CZ gate.}
\end{figure}

Figure \ref{Cz} depicts the graphical representation of the CZ gate.

Applies phase $(-1)^{xy}$ to $|y\rangle$ if $|x\rangle = |1\rangle$.

CZ creates entanglement between the two qubits. Multi-qubit gates like CZ and CNOT are key to quantum algorithms.

\section{Tensor Products}

In quantum computing, single-qubit states $|0\rangle$ and $|1\rangle$ can be expressed as vectors 
$\begin{bmatrix}
1\\ 0
\end{bmatrix}$
and
$\begin{bmatrix}
0\\ 1
\end{bmatrix}$, respectively. Similarly, two-qubit states like $|00\rangle$, $|01\rangle$, $|10\rangle$, and $|11\rangle$ can be expressed as the vectors 
$\begin{bmatrix}
1\\ 0\\ 0\\ 0
\end{bmatrix}$,
$\begin{bmatrix}
0\\ 1\\ 0\\ 0
\end{bmatrix}$,
$\begin{bmatrix}
0\\ 0\\ 1\\ 0
\end{bmatrix}$,
and
$\begin{bmatrix}
0\\ 0\\ 0\\ 1
\end{bmatrix}$.

To combine two single qubits into a two-qubit system, we use a mathematical operation known as the tensor product. The tensor product, denoted by $\otimes$, allows us to calculate the vector representation of the combined state.

\begin{example}
Calculate the tensor product of $|1\rangle \otimes |1\rangle$.

To find the tensor product of $|1\rangle \otimes |1\rangle$, we first express each qubit as its vector representation:
$|1\rangle = \begin{bmatrix} 0 \\ 1 \end{bmatrix}$.

The tensor product is calculated by multiplying each element of the first vector by the entire second vector:
\begin{align*}
|1\rangle \otimes |1\rangle & = \begin{bmatrix} 0 \\ 1 \end{bmatrix} \otimes \begin{bmatrix} 0 \\ 1 \end{bmatrix} \\
& = \begin{bmatrix} 0 \times \begin{bmatrix} 0 \\ 1 \end{bmatrix} \\ 1 \times \begin{bmatrix} 0 \\ 1 \end{bmatrix} \end{bmatrix} \\
& = \begin{bmatrix} 0 \times 0 \\ 0 \times 1 \\ 1 \times 0 \\ 1 \times 1 \end{bmatrix} \\
& = \begin{bmatrix} 0 \\ 0 \\ 0 \\ 1 \end{bmatrix}.
\end{align*}
Thus, the tensor product of $|1\rangle \otimes |1\rangle$ results in the vector representation of the two-qubit state $|11\rangle$, which is $\begin{bmatrix} 0 \\ 0 \\ 0 \\ 1 \end{bmatrix}$.
\end{example}

This process of using tensor products is fundamental in building multi-qubit systems from single-qubit states in quantum computing.

The symbol for the tensor product in mathematics and quantum computing is $\otimes$. This operation is particularly important in quantum computing for combining states. For example, the tensor product of two vectors is expressed as follows \cite{4_sx}:    
$\begin{bmatrix}
	a_0\\ a_1\\ a_2\\ a_3
\end{bmatrix}$ $\otimes$
$\begin{bmatrix}
	b_0\\ b_1
\end{bmatrix}$.

The rules for the tensor product of vectors are as follows: 
\begin{enumerate}
    \item If vector A has $m$ elements and vector B has $n$ elements, their tensor product results in a new vector with $m \times n$ elements.
    \item The $i$th element of vector A is multiplied by the $j$th element of vector B to form the $(i \times n + j)$th element of the new vector.
\end{enumerate}

For example, the tensor product of two vectors A and B is calculated as:
\begin{equation}
 \begin{bmatrix}
	a_0\\ a_1\\ a_2\\ a_3
\end{bmatrix} \otimes
\begin{bmatrix}
	b_0\\ b_1
\end{bmatrix} =
\begin{bmatrix}
	a_0b_0\\ a_0b_1\\a_1b_0\\a_1b_1\\a_2b_0\\a_2b_1\\a_3b_0\\a_3b_1
\end{bmatrix},
\end{equation}
and the general formula can be expressed as: 
\begin{equation}
 \begin{bmatrix}
	a_0\\ a_1\\ \vdots\\ a_{m-1}
\end{bmatrix} \otimes
\begin{bmatrix}
	b_0\\ b_1\\ \vdots \\b_{n-1}
\end{bmatrix} =
\begin{bmatrix}
	a_0b_0\\ \vdots\\ a_0b_{n-1}\\a_1b_0\\ \vdots\\ a_1b_{n-1}\\ \vdots\\a_{m-1}b_{n-1}
\end{bmatrix}.    
\end{equation}

\begin{example}
For a qubit, the tensor product can be expressed as: 
\begin{equation}
   |0\rangle \otimes |0\rangle \otimes |0\rangle 
= \begin{bmatrix} 1\\ 0 \end{bmatrix} \otimes \begin{bmatrix} 1\\ 0 \end{bmatrix} \otimes \begin{bmatrix} 1\\ 0 \end{bmatrix} 
= \begin{bmatrix} 1\\ 0 \\0 \\0 \end{bmatrix} \otimes \begin{bmatrix} 1\\ 0 \end{bmatrix}
= \begin{bmatrix} 1\\ 0 \\0 \\0 \\0 \\0 \\0 \\0 \end{bmatrix} = |000\rangle. 
\end{equation}
\end{example}

\begin{exercise}
Similarly, calculate the tensor product of $|010\rangle$.    
\end{exercise}

The tensor product is a fundamental operation in quantum computing with several important properties \cite{2_3}:

1. A tensor product of a vector A with \( m \) elements and a vector B with \( n \) elements results in a new vector C with \( m \times n \) elements. However, a vector with \( m \times n \) elements is not necessarily the result of a tensor product of two vectors with \( m \) and \( n \) elements. It can also be a sum of tensor products of several vectors.

\begin{example}
Consider the vector 
\begin{equation}
 \begin{bmatrix} 1\\ 0\\ 0\\1 \end{bmatrix} \neq \begin{bmatrix} x\\ y \end{bmatrix} \otimes \begin{bmatrix} a\\ b \end{bmatrix} = \begin{bmatrix} xa\\ xb\\ya \\yb \end{bmatrix}, 
\end{equation}
but
\begin{equation}
  \begin{bmatrix} 1\\ 0\\ 0\\1 \end{bmatrix} = \begin{bmatrix} 1\\ 0 \end{bmatrix} \otimes \begin{bmatrix} 1\\ 0 \end{bmatrix} + \begin{bmatrix} 0\\ 1 \end{bmatrix} \otimes \begin{bmatrix} 0\\ 1 \end{bmatrix}.    
\end{equation}    
\end{example}

2. The tensor product does not satisfy the commutative law, meaning \( A \otimes B \neq B \otimes A \).

\begin{example}
For instance:
 \begin{equation}
\begin{bmatrix} 1\\ 0 \end{bmatrix} \otimes \begin{bmatrix} 0\\ 1 \end{bmatrix} \neq \begin{bmatrix} 0\\ 1 \end{bmatrix} \otimes \begin{bmatrix} 1\\ 0 \end{bmatrix},
\begin{bmatrix} 0\\ 1 \end{bmatrix} \otimes \begin{bmatrix} 8\\ 0 \\0 \end{bmatrix} \neq \begin{bmatrix} 8\\ 0 \\0 \end{bmatrix} \otimes \begin{bmatrix} 0\\ 1 \end{bmatrix}.
\end{equation}     
\end{example}

3. The tensor product operation satisfies the associative law, such that \((A \otimes B) \otimes C = A \otimes (B \otimes C)\).

\begin{example}
Here is an example:
 \begin{equation}
\left(
\begin{bmatrix} a_0\\ a_1 \end{bmatrix} \otimes \begin{bmatrix} b_0\\ b_1  \end{bmatrix}\right) \otimes \begin{bmatrix} c_0\\ c_1 \end{bmatrix}  =
\begin{bmatrix} a_0b_0c_0\\ a_0b_0c_1\\a_0b_1c_0\\a_0b_1c_1\\a_1b_0c_0\\a_1b_0c_1\\a_1b_1c_0\\a_1b_1c_1 \end{bmatrix} = \begin{bmatrix} a_0\\ a_1 \end{bmatrix} \otimes \left(
\begin{bmatrix} b_0\\ b_1 \end{bmatrix} \otimes \begin{bmatrix} c_0\\ c_1  \end{bmatrix}\right).
\end{equation}    
\end{example}

4. Expressing a 3-bit string (e.g., 101) or 3 qubits (e.g., \( |1 \rangle \otimes |0 \rangle \otimes |1\rangle \)) results in a vector with \( 2^3 = 8 \) elements. Generally, expressing \( n \) qubits yields a vector with \( 2^n \) elements.

\begin{example}
For instance:
\begin{equation}
  |1\rangle \otimes |0\rangle \otimes |1\rangle 
= \begin{bmatrix} 0\\ 1 \end{bmatrix} \otimes \begin{bmatrix} 1\\ 0 \end{bmatrix} \otimes \begin{bmatrix} 0\\ 1 \end{bmatrix} 
= \begin{bmatrix} 0\\ 0 \\1 \\0 \end{bmatrix} \otimes \begin{bmatrix} 0\\ 1 \end{bmatrix}
= \begin{bmatrix} 0\\ 0 \\0 \\0 \\0 \\1 \\0 \\0 \end{bmatrix} = |101\rangle.  
\end{equation}
\end{example}

The tensor product is an essential operation in quantum computing, particularly useful for constructing larger matrices from smaller ones. Let's consider two sample matrices $A$ and $B$, where $A$ is a $2\times 2$ matrix and $B$ is a $3\times 3$ matrix. The tensor product $A\otimes B$ involves multiplying each element of $A$ by the entire matrix $B$. This process results in a new $6\times 6$ matrix.

Let's define matrices $A$ and $B$ as follows:
\begin{align}
	A = \begin{bmatrix}
		a_{00} & a_{01}\\
		a_{10} & a_{11}
	\end{bmatrix},
	B = \begin{bmatrix}
		b_{00} & b_{01} & b_{02}\\
		b_{10} & b_{11} & b_{12}\\
		b_{20} & b_{21} & b_{22}
	\end{bmatrix}.
\end{align}
Then, the tensor product $A \otimes B$ is calculated as:
\begin{align}
	A \otimes B = \begin{bmatrix}
		a_{00}B & a_{01}B\\
		a_{10}B & a_{11}B
	\end{bmatrix}
	= \begin{bmatrix}
		a_{00}\begin{bmatrix}
			b_{00} & b_{01} & b_{02}\\
			b_{10} & b_{11} & b_{12}\\
			b_{20} & b_{21} & b_{22}
		\end{bmatrix} &  a_{01}\begin{bmatrix}
		b_{00} & b_{01} & b_{02}\\
		b_{10} & b_{11} & b_{12}\\
		b_{20} & b_{21} & b_{22}
	\end{bmatrix} \\
	 a_{10}\begin{bmatrix}
		b_{00} & b_{01} & b_{02}\\
		b_{10} & b_{11} & b_{12}\\
		b_{20} & b_{21} & b_{22}
	\end{bmatrix} &  a_{11}\begin{bmatrix}
	b_{00} & b_{01} & b_{02}\\
	b_{10} & b_{11} & b_{12}\\
	b_{20} & b_{21} & b_{22}
	\end{bmatrix} \end{bmatrix}.
\end{align}

This result showcases the new $6\times 6$ matrix generated from the tensor product of $A$ and $B$. Each block of this matrix is a scaled version of $B$, with the scaling factor being the corresponding element from matrix $A$.

\begin{example}
Calculate the tensor product of 
$\begin{bmatrix}
	0 & 1\\
	1 & 0
\end{bmatrix}$ 
and 
$\begin{bmatrix}
1 & 1 \\
1 & -1 
\end{bmatrix}$.

First, let's define the matrices as $A$ and $B$:
$A = \begin{bmatrix} 0 & 1\\ 1 & 0 \end{bmatrix}$,
$B = \begin{bmatrix} 1 & 1 \\ 1 & -1 \end{bmatrix}$.

The tensor product $A \otimes B$ is calculated by multiplying each element of $A$ by the entire matrix $B$. This results in a $4 \times 4$ matrix, obtained as follows:

\begin{align*}
A \otimes B &= \begin{bmatrix}
	0B & 1B\\
	1B & 0B
\end{bmatrix} \\
&= \begin{bmatrix}
	0 \cdot \begin{bmatrix} 1 & 1 \\ 1 & -1 \end{bmatrix} & 1 \cdot \begin{bmatrix} 1 & 1 \\ 1 & -1 \end{bmatrix} \\
	1 \cdot \begin{bmatrix} 1 & 1 \\ 1 & -1 \end{bmatrix} & 0 \cdot \begin{bmatrix} 1 & 1 \\ 1 & -1 \end{bmatrix}
\end{bmatrix} \\
&= \begin{bmatrix}
	\begin{bmatrix} 0 & 0 \\ 0 & 0 \end{bmatrix} & \begin{bmatrix} 1 & 1 \\ 1 & -1 \end{bmatrix} \\
	\begin{bmatrix} 1 & 1 \\ 1 & -1 \end{bmatrix} & \begin{bmatrix} 0 & 0 \\ 0 & 0 \end{bmatrix}
\end{bmatrix} \\
&= \begin{bmatrix}
	0 & 0 & 1 & 1 \\
	0 & 0 & 1 & -1 \\
	1 & 1 & 0 & 0 \\
	1 & -1 & 0 & 0
\end{bmatrix}.
\end{align*}

Therefore, the tensor product of the matrices $A$ and $B$ is a $4 \times 4$ matrix:
$\begin{bmatrix}
	0 & 0 & 1 & 1 \\
	0 & 0 & 1 & -1 \\
	1 & 1 & 0 & 0 \\
	1 & -1 & 0 & 0
\end{bmatrix}$.
\end{example}

The tensor product is a vital operation in quantum computing, especially for combining matrices. However, it behaves differently from standard matrix multiplication. Here are some key properties of the tensor product \cite{4_sx}:

1. \textbf{Non-Commutative Property}: The tensor product does not satisfy the commutative law. This means the order of the matrices is important. Generally, for two matrices \( A \) and \( B \), \( A \otimes B \neq B \otimes A \).

2.  \textbf{Associative Property}: The tensor product is associative. This has a couple of implications:

    2.1. When taking tensor products of three or more matrices, the grouping of matrices does not affect the result: \( (A \otimes B) \otimes C = A \otimes (B \otimes C) \).

    2.2. For matrices that satisfy the conditions for matrix multiplication, the tensor product of products is the product of tensor products: \( (A \times A')\otimes(B \times B') = (A\otimes B) \times (A' \otimes B') \).

    2.3. If matrices \( A \), \( B \) and vectors \( V \), \( V' \) satisfy the conditions for matrix multiplication, then \( (A \otimes B) \times (V \otimes V') = (A \times V) \otimes (B \times V') \).

These properties illustrate how the tensor product, while similar to regular matrix multiplication in some respects, has unique characteristics that are crucial for operations in quantum computing.

\begin{exercise}
 For
 \begin{align}
	A = \begin{pmatrix}
		1 & 0\\
		0 & 1
	\end{pmatrix},
	B = \begin{pmatrix}
		0 & 1\\
		1 & 0
	\end{pmatrix},
        C = \begin{pmatrix}
		1 & 0\\
		0 & -1
	\end{pmatrix},
	V = \begin{pmatrix}
		1\\
		0
	\end{pmatrix},
	V' = \begin{pmatrix}
		0\\
		1
	\end{pmatrix},
\end{align} 
Calculate 
\[ (A \otimes B) \otimes C,\]
and
\[(A \times V) \otimes (B \times V'). \]
\end{exercise}

\newpage

\chapter{Week 4: Quantum Circuits and Search Algorithm}
\section{Quantum circuits}

Building on our exploration of matrix representation in Week 3, this week, we delve into the realm of quantum circuits. A quantum circuit is a schematic that effectively illustrates how quantum gates are arranged to execute complex computational operations. There are chiefly two methodologies for arranging these gates: sequential and parallel. In a sequential arrangement, gates are applied consecutively, and this process equates to performing a series of matrix multiplications. Here, each quantum gate is depicted by its corresponding matrix, and the cumulative effect of the circuit is determined by the sequential multiplication of these matrices, mirroring the order of gate application \cite{2_3,4_sx}. Conversely, a parallel arrangement entails the simultaneous application of gates to distinct qubits. This type of arrangement is quantified mathematically through the tensor product of the matrices representing the individual gates \cite{4_sx}. Grasping these two key arrangements—sequential and parallel—is essential for a comprehensive understanding of how quantum circuits manipulate and process information at the quantum level.

\textbf{Sequential Quantum Circuits}

\begin{figure}[H]
	\centering
	\[\Qcircuit @!C @!R {
        &\gate{A}  & \gate{B} &\qw}\]
	\caption{Sequential gate application in a single-qubit circuit.}
	\label{1qubit}
\end{figure}
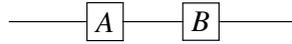

In quantum computing, understanding the sequential arrangement of circuits is crucial, especially when dealing with single qubit operations. Consider single qubit circuits with $2 \times 2$ matrices A and B. The sequence in which these matrices are applied is pivotal. For instance, if matrix A represents the first gate acting on a qubit, it appears on the left in the matrix multiplication. Therefore, if gate A acts first, followed by gate B, the corresponding matrix operation is $B \times A$, as depicted in Figure \ref{1qubit}.

\begin{figure}[H]
	\centering
        \[\Qcircuit @C=2em @R=0.25em {
        & & & & & &\qw& \multigate{4}{A}  & \multigate{4}{B} &\qw&\qw\\
        & &  &  & & &\qw& \ghost{A} & \ghost{B}&\qw&\qw\\
        &{/^{m}}\qw&\gate{A}  & \gate{B} &\qw & &\qw& \ghost{A} & \ghost{B}&\qw &\qw\\
        & &   &  & & &\cdots& \nghost{A} & \nghost{B}&\cdots & \\
        & & & & & & \qw&\ghost{A} &\ghost{B}&\qw&\qw
        }\]
	\caption{Sequential gate application in an m-qubit circuit.}
	\label{m_qubit}
\end{figure}
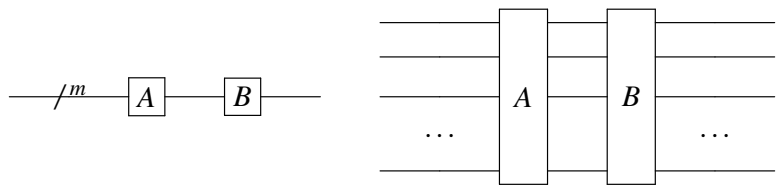

When considering multiple qubits, the operations are represented by tensor products \cite{4_sx} of the matrices. In an m-qubit circuit, the dimensionality of matrices A and B increases to $2^m \times 2^m$, as illustrated in Figure \ref{m_qubit}.

\begin{figure}[H]
	\centering
	\[\Qcircuit @!C @!R {
        &\gate{H}  & \gate{X} & \gate{H} &\qw}\]
	\caption{Matrix representation of operations in a quantum circuit.}
	\label{HXH}
\end{figure}
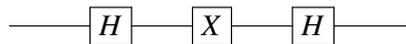

\begin{example}
Consider the circuit shown in Figure \ref{HXH}. It can be expressed as:
\begin{align}
	G &= HXH \nonumber\\
	&= \frac{1}{\sqrt{2}} \begin{bmatrix}
		1 & 1\\
		1 & -1
	\end{bmatrix}
	\begin{bmatrix}
		0 & 1\\
		1 & 0
	\end{bmatrix}
	\frac{1}{\sqrt{2}} \begin{bmatrix}
		1 & 1\\
		1 & -1
	\end{bmatrix}\nonumber\\
	&= \frac{1}{2} \begin{bmatrix}
		1 & 1\\
		-1 & 1
	\end{bmatrix}
	\begin{bmatrix}
		1 & 1\\
		1 & -1
	\end{bmatrix}\nonumber\\
&= \frac{1}{2} \begin{bmatrix}
	2 & 0\\
	0 & -2
\end{bmatrix}
=\begin{bmatrix}
	1 & 0\\
	0 & -1
\end{bmatrix}
	= Z.
\end{align}
\end{example}

\begin{exercise}
Using the principles discussed, calculate the resultant matrix for the operation $G = HZH$.
\end{exercise}

\textbf{Parallel Quantum Circuits}

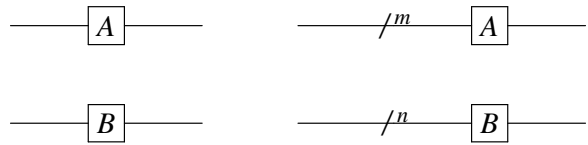
\begin{figure}[H]
	\centering
	\[\Qcircuit @!C @!R {
        &\gate{A}&\qw   &  &{/^{m}}\qw&\gate{A} &\qw\\
        &\gate{B}&\qw    &  &{/^{n}}\qw&\gate{B} &\qw}\]
	\caption{\label{mn_qubit} The gate sequence representation on the multi-qubit  circuit.}
\end{figure}

Quantum circuits often involve operations occurring simultaneously on different qubits, known as parallel operations. To represent these parallel operations mathematically, we employ the tensor product, denoted as $A \otimes B$. This concept is illustrated in Figure \ref{mn_qubit} for a circuit comprising $m$ qubits in the top wire and $n$ qubits in the bottom wire, with corresponding matrices $A$ (of size $2^m \times 2^m$) and $B$ (of size $2^n \times 2^n$).

Here are some examples:
\begin{example}
As shown on the left in Figure \ref{exampl} can be expressed as: 
\begin{align}
	G &= H \otimes H 
	= \frac{1}{\sqrt{2}} \begin{bmatrix}
		1 & 1\\
		1 & -1
	\end{bmatrix}
	\otimes
	\frac{1}{\sqrt{2}} \begin{bmatrix}
		1 & 1\\
		1 & -1
	\end{bmatrix}
	= \frac{1}{2}\begin{bmatrix}
		1 & 1 & 1 & 1\\
		1 & -1 & 1 & -1\\
		1 & 1 & -1 & -1\\
		-1 & -1 & -1 & 1
	\end{bmatrix}.
\end{align}  
\end{example}
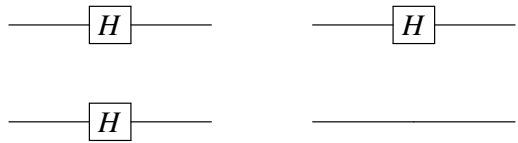
\begin{figure}[H]
	\centering
	\[\Qcircuit @!C @!R {
        &\gate{H}&\qw   &  &\gate{H} &\qw\\
        &\gate{H}&\qw    &  &\qw  &\qw}\]
	\caption{Calculation methods of different combinations of gates.}\label{exampl}
\end{figure}
\begin{example}
    As shown on the right in Figure \ref{exampl}, if a circuit is empty, it will automatically fill the $``I"$ gate , and it can be expressed as: 
\begin{align}
	G &= H \otimes I
	= \frac{\sqrt{2}}{2} \begin{bmatrix}
		1 & 1\\
		1 & -1
	\end{bmatrix}
	\otimes
	\begin{bmatrix}
		1 & 0\\
		0 & 1
	\end{bmatrix}
	= \frac{\sqrt{2}}{2}\begin{bmatrix}
		1 & 0 & 1 & 0\\
		0 & 1 & 0 & 1\\
		1 & 0 & -1 & 0\\
		0 & 1 & 0 & -1
	\end{bmatrix}.
\end{align}
\end{example}

\begin{example}
   In Figure \ref{example2} can be expressed as:
\begin{align}
	G &= CNOT \times (H \otimes I)\nonumber\\
	&= \begin{bmatrix}
		1 & 0 & 0 & 0\\
		0 & 1 & 0 & 0\\
		0 & 0 & 0 & 1\\
		0 & 0 & 1 & 0
	\end{bmatrix}
	\left( \frac{\sqrt{2}}{2} \begin{bmatrix} 1 & 1\\ 1 & -1 \end{bmatrix} \otimes \begin{bmatrix} 1 & 0\\ 0 & 1 \end{bmatrix} \right)\nonumber\\
	&= \begin{bmatrix} 1 & 0 & 0 & 0\\ 0 & 1 & 0 & 0\\ 0 & 0 & 0 & 1\\ 0 & 0 & 1 & 0 \end{bmatrix} \times \frac{\sqrt{2}}{2}  \begin{bmatrix} 1 & 0 & 1 & 0\\ 0 & 1 & 0 & 1\\ 1 & 0 & -1 & 0\\ 0 & 1 & 0 & -1 \end{bmatrix}\nonumber\\
	&= \frac{\sqrt{2}}{2} \begin{bmatrix} 1 & 0 & 1 & 0\\ 0 & 1 & 0 & 1\\ 0 & 1 & 0 & -1\\ 1 & 0 & -1 & 0 \end{bmatrix}. 
 \end{align} 
\end{example}

\begin{figure}[H]
	\centering
	\[\Qcircuit @C=1.5em @R=1em {
         &\lstick{}&\gate{H} &\ctrl{2} &\qw & & &\multigate{2}{G} &\qw\\
         &\lstick{}& & & &= & & &  & & &\\
         &\lstick{}&\qw &\targ &\qw& & &\ghost{G} &\qw
        }\]
	\caption{\label{example2} The circuit diagram on the left of the equal sign can be combined from H gate and CNOT gate.}
\end{figure}

\begin{figure}[H]
	\centering
	\[\Qcircuit @C=1.5em @R=1em {
         &\lstick{}& \qw &\ctrl{2} &\qw &\qw & & &\multigate{2}{G} &\qw\\
         & &  & & & & =& & & \\
        &\lstick{}& \gate{H} & \control\qw &\gate{H} &\qw & & &\ghost{G} &\qw
        }\]
	\caption{\label{example3} The circuit diagram on the left of the equal sign can be combined from two H gates and CZ gate.}
\end{figure}
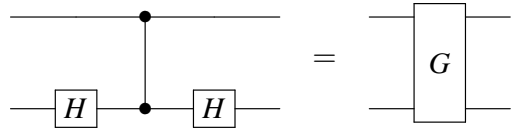

\begin{example}
    In Figure \ref{example3} can be expressed as:
\begin{align}
	G &= (I \otimes H) \times  CZ \times (I \otimes H)\nonumber\\
	&= \left( \begin{bmatrix}
		1 & 0\\
		0 & 1
	\end{bmatrix}
	\otimes
	\frac{\sqrt{2}}{2} \begin{bmatrix}
		1 & 1\\
		1 & -1
	\end{bmatrix} \right) 
\times  \begin{bmatrix}
		1 & 0 & 0 & 0\\
		0 & 1 & 0 & 0\\
		0 & 0 & 1 & 0\\
		0 & 0 & 0 & -1
	\end{bmatrix} \times  \left( \begin{bmatrix}
		1 & 0\\
		0 & 1
	\end{bmatrix}
	\otimes
	\frac{\sqrt{2}}{2} \begin{bmatrix}
		1 & 1\\
		1 & -1
	\end{bmatrix} \right)\nonumber\\
&= \frac{\sqrt{2}}{2}\begin{bmatrix}
	1 & 1 & 0 & 0\\
	1 & -1 & 0 & 0\\
	0 & 0 & 1 & 1\\
	0 & 0 & 1 & -1
\end{bmatrix} \times 
\begin{bmatrix}
	1 & 0 & 0 & 0\\
	0 & 1 & 0 & 0\\
	0 & 0 & 1 & 0\\
	0 & 0 & 0 & -1
\end{bmatrix} 
\times \frac{\sqrt{2}}{2}\begin{bmatrix}
	1 & 1 & 0 & 0\\
	1 & -1 & 0 & 0\\
	0 & 0 & 1 & 1\\
	0 & 0 & 1 & -1
\end{bmatrix}\nonumber\\
	&= \begin{bmatrix}
		1 & 0 & 0 & 0\\
		0 & 1 & 0 & 0\\
		0 & 0 & 0 & 1\\
		0 & 0 & 1 & 0
	\end{bmatrix}
	= \text{CNOT}
\end{align}
\end{example}

\textbf{Reversible Circuits}

Quantum gates differ from traditional logic gates in a critical aspect: reversibility. They have the unique ability to ``undo" their operations, allowing them to revert to their original state. This property of reversibility is crucial for constructing quantum circuits that can maintain the delicate superposition states of qubits.

\begin{figure}[H]
	\centering
	\[\Qcircuit @C=3em @R=2.5em {
        &\lstick{x} &\ctrl{1}  & \rstick{x}\qw \\
        &\lstick{y} &\gate{X}  & \rstick{y \oplus x}\qw}\]
	\caption{CNOT gate circuit.}
	\label{18_1}
\end{figure}
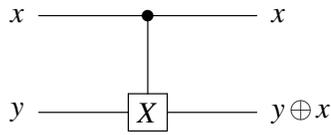

One of the most versatile reversible gates in quantum computing is the controlled-NOT (CNOT) gate. The CNOT gate, depicted in Figure \ref{18_1}, takes two qubit inputs: a control qubit and a target qubit \cite{2_3}. When the control qubit is set to 1, the CNOT gate flips the state of the target qubit. However, if the control qubit is 0, the target qubit remains unchanged.

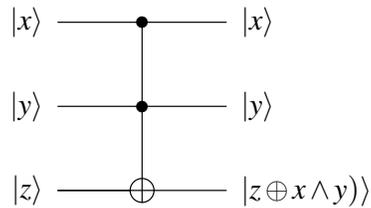
\begin{figure}[H]
	\centering
	 \[\Qcircuit @!C @!R {
        &\lstick{|x\rangle} &\ctrl{1}  & \rstick{|x\rangle}\qw \\
         &\lstick{|y\rangle} &\ctrl{1}  & \rstick{|y\rangle}\qw \\
        &\lstick{|z\rangle} &\targ \qw & \rstick{|z\oplus x\wedge y)\rangle }\qw}\]
	\caption{Toffoli gate circuit.}
	\label{toffoli_fig}
\end{figure}

Another significant reversible gate is the Toffoli gate, also known as the CCNOT gate \cite{4_sx,2_1}. Illustrated in Figure \ref{toffoli_fig}, the Toffoli gate operates on three qubits - two control qubits and one target qubit. The target qubit flips its state only when both control qubits are set to 1.

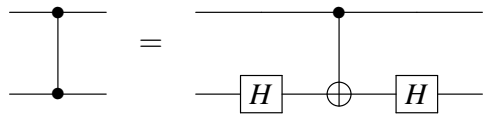
\begin{figure}[H]
	\centering
       \[\Qcircuit @C=1.5em @R=1em {
        &\ctrl{2} &\qw &  &\lstick{}& \qw &\ctrl{2} &\qw &\qw\\
        & & &= & &  & & &\\
        &\control\qw &\qw &  &\lstick{}& \gate{H} & \targ &\gate{H} &\qw
        }\]
	\caption{Constructing a CZ gate from a CNOT gate.}
	\label{cz_construction}
\end{figure}

The controlled-Z (CZ) gate is another vital component in quantum circuits \cite{4_sx}. It can be constructed using a CNOT gate combined with two Hadamard gates, as shown in Figure \ref{cz_construction}. The Hadamard gates change the basis states of the qubits ($H|0/1\rangle = \frac{1}{\sqrt{2}}(0 \pm 1)$), effectively transforming X operations into Z operations and vice versa. This transformation allows the CNOT gate to simulate the functionality of a CZ gate.

These multi-qubit reversible gates are the foundational elements for constructing more complex quantum circuits and algorithms, enabling conditional flipping, phase flipping, or entangling of qubits.

\textbf{Summary of Quantum Circuits}

Quantum circuits form the core framework of quantum computing, enabling the manipulation and processing of quantum information. This summary encapsulates the essence of quantum circuits, delving into their building blocks, the arrangement of these blocks, and the vital role of measurement. Below are the key elements that define quantum circuits:

\begin{itemize}
  \item \textbf{Building Blocks:} The fundamental components of quantum circuits are quantum gates \cite{4_sx}. Distinct from classical logic gates, quantum gates are reversible and can operate on qubits in superposition states. Essential examples include single-qubit gates such as the Hadamard (H) and Pauli (X, Y, Z) gates, along with multi-qubit gates like the Controlled-NOT (CNOT), Toffoli, and Controlled-Z (CZ) gates. These gates manipulate qubit states to facilitate quantum computations.

  \item \textbf{Arranging Quantum Gates:} In quantum circuits, gates can be arranged sequentially or in parallel. Sequential arrangements involve gates applied in succession, with their operations equating to matrix multiplications. The sequence of gates is crucial, as it affects the qubits' final state. Parallel arrangements involve simultaneous operations of gates on different qubits, represented mathematically by the tensor product of the gates' matrices, enabling intricate multi-qubit operations \cite{2_1}.

  \item \textbf{Measurement:} Measurement \cite{2_2} is the process of deriving classical information from quantum states in a quantum circuit. It causes the quantum state to collapse into one of its basis states, yielding a specific outcome. Measurements, usually conducted at the circuit's end, are integral for acquiring the results of quantum computations. The probabilistic nature of quantum measurement leads to varied outcomes based on the qubit's state before measurement.

  \item \textbf{Integration of Elements:} Quantum circuits integrate quantum gates, their arrangements, and measurements to execute complex computational tasks \cite{2_9}. The selection and configuration of gates define the functionality of the algorithm, while measurement delivers the results. Through thoughtful design of these circuits, the principles of quantum mechanics are harnessed for computational applications.

\end{itemize}

In the following we provide an example that brings together these elements, forming a comprehensive quantum circuit and illustrating the concepts discussed in this section.

\begin{example}
Find the output state $|\phi\rangle$ of the circuit shown in Figure \ref{99}:
\begin{figure}[H]
	\centering
	\[\Qcircuit @!C @!R {
        &\lstick{} &\ctrl{1} &\qw &\qw &\meter &\rstick{|0\rangle}\cw\\
        &\lstick{\frac{1}{\sqrt{2}}(|000\rangle+|111\rangle)} &\ctrl{1} &\gate{H} &\targ &\meter &\rstick{|1\rangle}\cw\\
        &\lstick{} &\targ &\qw &\ctrl{-1} &\qw &\rstick{|\phi\rangle}\qw
        }\]
	\caption{\label{99} An example of how quantum gates work on qubits in a quantum circuit.}
\end{figure}
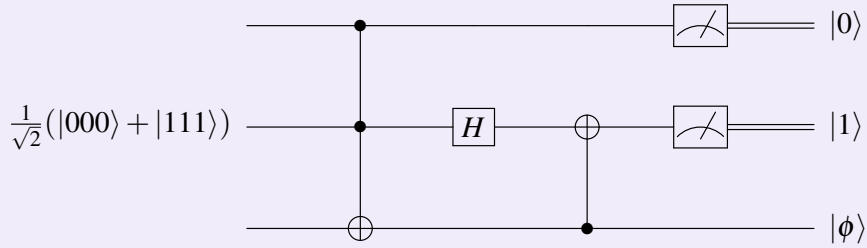

It starts with three qubits in the state $\frac{1}{\sqrt{2}}(|000\rangle+|111\rangle)$. A Toffoli gate is applied, which flips the third qubit if both the first two are 1. This results in the state $\frac{1}{\sqrt{2}}(|000\rangle + |110\rangle)$.

Next, a Hadamard gate H is applied to the second qubit. This puts the qubit into a superposition of 0 and 1. The state becomes $\frac{1}{2}(|000\rangle + |010\rangle + |100\rangle - |110\rangle)$.

Finally, a CNOT gate with qubit 3 as control and qubit 2 as target is applied. This flips qubit 2 if qubit 3 is $|1\rangle$. The final state is $\frac{1}{2}(|000\rangle + |010\rangle + |100\rangle - |110\rangle)$.

The specific process is as follows:
\begin{align}
 \frac{1}{\sqrt{2}}(|000\rangle + |111\rangle&\stackrel{\text{Toffoli}}{\longrightarrow}\frac{1}{\sqrt{2}}(|000\rangle + |110\rangle)\nonumber\\
&\stackrel{\text{H}_2}{\longrightarrow}\frac{1}{\sqrt{2}}[|0\rangle\frac{1}{\sqrt{2}}(|0\rangle+|1\rangle)+|1\rangle\frac{1}{\sqrt{2}}(|0\rangle-|1\rangle)|0\rangle]\nonumber \\ & \qquad=\frac{1}{2}(|000\rangle + |010\rangle + |100\rangle - |110\rangle)\nonumber\\ 
&\stackrel{\text{CNOT}_{32}}{\longrightarrow}\frac{1}{2}(|000\rangle + |010\rangle + |100\rangle - |110\rangle) \to |\phi\rangle=|0\rangle.   
\end{align}

\textbf{Note:} $H_2$ means to perform $H$ gate operation on the second qubit; $CNOT_{32}$ means that the third qubit is the control qubit and the second qubit is the controlled qubit.\\
The matrix representation of this circuit can be written as:
\begin{align}
G &= (I \otimes CNOT_{32}) \times (I \otimes H_2 \otimes I) \times Toffoli  \nonumber \\ 
  &= 
  \left( \begin{bmatrix} 
  1 & 0 \\
  0 & 1 
  \end{bmatrix} \otimes 
  \begin{bmatrix} 
  1 & 0 & 0 & 0\\ 
  0& 0& 0 & 1 \\ 
  0& 0& 1& 0\\ 
  0& 1& 0& 0 
  \end{bmatrix} \right) \times 
  \left( \begin{bmatrix} 
  1 & 0 \\ 
  0 & 1 
  \end{bmatrix} \otimes 
  \frac{\sqrt{2}}{2} 
  \begin{bmatrix}
		1 & 1\\
		1 & -1
  \end{bmatrix} \otimes 
 \begin{bmatrix} 
 1 & 0 \\ 
 0 & 1 
 \end{bmatrix} \right) \nonumber \\& \qquad \times
\begin{bmatrix}
1& 0 &0 &0 &0 &0 &0 &0 \\
0& 1 &0 &0 &0 &0 &0 &0 \\
0& 0 &1 &0 &0 &0 &0 &0 \\
0& 0 &0 &1 &0 &0 &0 &0 \\
0& 0 &0 &0 &1 &0 &0 &0 \\
0& 0 &0 &0 &0 &1 &0 &0 \\
0& 0 &0 &0 &0 &0 &0 &1 \\
0& 0 &0 &0 &0 &0 &1 &0 
\end{bmatrix} \nonumber \\ 
&= \frac{\sqrt{2}}{2}
\begin{bmatrix}
1& 0& 1& 0& 0& 0& 0& 0\\
0& 1& 0& -1& 0& 0& 0& 0\\
1& 0& -1& 0& 0& 0& 0& 0\\
0& 1& 0& 1& 0& 0& 0& 0\\
0& 0& 0& 0& 1& 0& 0& 1\\
0& 0& 0& 0& 0& 1& -1& 0\\
0& 0& 0& 0& 1& 0& 0& -1\\
0& 0& 0& 0& 0& 1& 1& 0
\end{bmatrix}.
\end{align}
For the initial quantum state can be written as:
\begin{align}
 \frac{1}{\sqrt{2}}(|000\rangle + |111\rangle = \frac{1}{\sqrt{2}} \times
\begin{bmatrix}
1\\
0\\
0\\
0\\
0\\
0\\
0\\
1
\end{bmatrix}.
\end{align}
Therefore, the final state obtained after the initial quantum state passes through the circuit G is:
\begin{align}
\frac{1}{2} \times
\begin{bmatrix}
1\\
0\\
1\\
0\\
1\\
0\\
-1\\
0
\end{bmatrix},
\end{align}
The quantum state represented by this vector is: $\frac{1}{2}(|000\rangle + |010\rangle + |100\rangle - |110\rangle)$.

It is worth noting that for the operations of these matrices, we can fully utilize MATLAB for calculations, and there will be specific examples in the later courses.
\end{example}

\section{Grover’s algorithm}

\textbf{Grover's Algorithm: Background}

Grover's algorithm is designed to efficiently solve a specific type of search problem: locating a target element in a large, unstructured search space. A typical example of this problem is searching through a database. Imagine a database containing \( N = 2^n \) entries, among which only one entry (e.g., a red ball) satisfies our search criteria. We define a function \( f \): \( \{0, 1\}^n \rightarrow \{0, 1\} \), where \( n \) is a positive integer. This function \( f \) checks whether each entry is the target red ball \( x_0 \) \cite{2_15}:
\begin{equation}
    f(x) = \begin{cases}
    1, & \text{if } x = x_0, \\
    0, & \text{otherwise}.
    \end{cases}
\end{equation}

The objective is to find this specific red ball \( x_0 \), meaning we need to locate the unique input \( x_0 \) for which \( f(x) \) maps to 1.

Consider an example with \( n = 2 \). There exists a function \( f \) such that:
\begin{equation}
  f(00) = 0, \ f(01) = 0, \ f(10) = 0, \ f(11) = 1.
\end{equation}
Here, \( x_0 = 11 \) is the particular string we are trying to find.

\begin{figure}[H]
	\centering
	\subfigure[classical solution]{
		\includegraphics[scale=0.35]{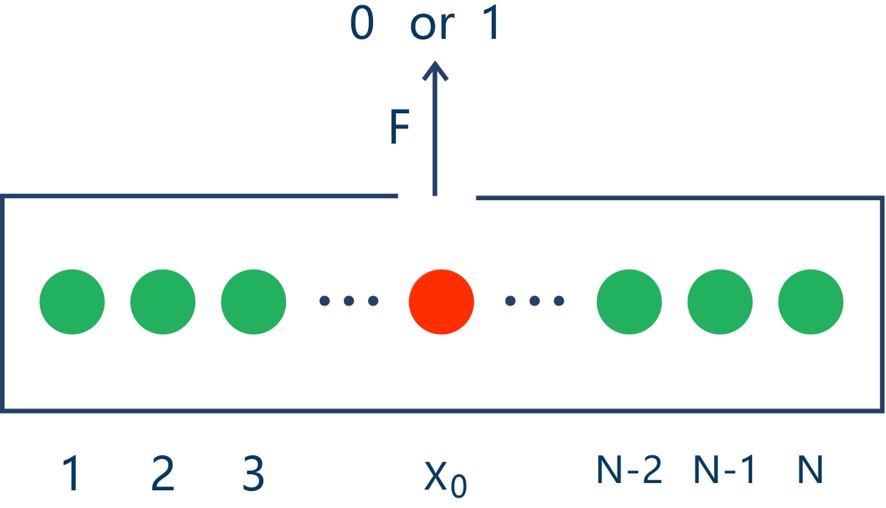}}
	\quad
	\subfigure[quantum solution]{
		\includegraphics[scale=0.35]{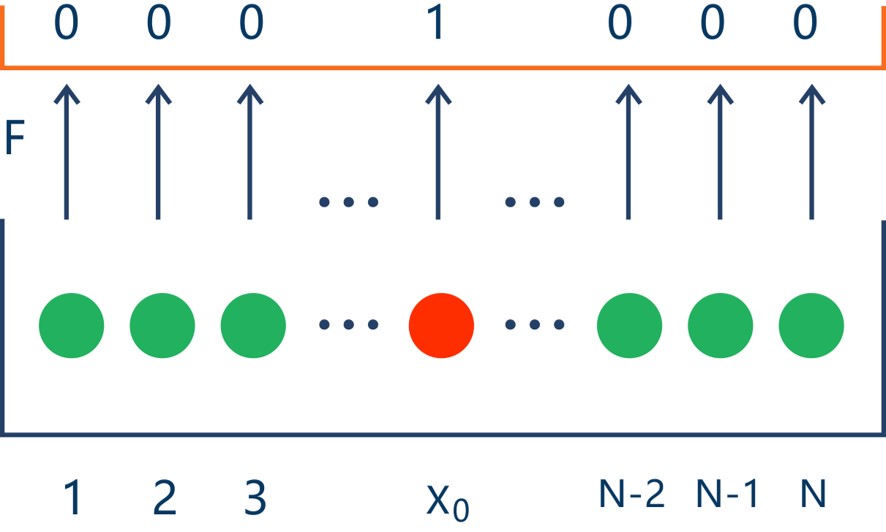}}
	\caption{Comparative illustration of solving an unstructured database search problem.}
	\label{solution}
\end{figure}

On a classical computer, the typical approach is to search through all \( N \) entries linearly, requiring, on average, about \( \frac{N}{2} \) queries to find the red ball. However, Grover's algorithm on a quantum computer provides a significant advantage \cite{2_16}. It offers a quadratic speedup for this kind of unstructured database search, needing only approximately \( \sqrt{N} \) queries to locate \( x_0 \), as illustrated in Figure \ref{solution}.

\textbf{Grover’s Algorithm: Separation of phase}

In the field of quantum computing, phase separation \cite{2_2,4_sx} is a key concept that plays a central role in understanding and implementing quantum algorithms. Phase separation involves the phase changes of a quantum system's state after undergoing a series of operations, which is crucial for revealing the workings of quantum algorithms.

\begin{figure}[H]
    \centering
        \[\Qcircuit @!C @!R{
        &\lstick{\mid 0\rangle}&{/^{n}}\qw&\gate{H^{\otimes n}}\qw&{/^{n}}\qw&\multigate{1}{U_{f}}&{/^{n}}\qw &\meter\\
        &\lstick{\mid 0\rangle}&\qw    &\qw     &\qw    &\ghost{U_{f}}       &\qw&\qw\\
        &                      &\dstick{\mid\phi_{0}\rangle}& &\dstick{\mid\phi_{1}\rangle}& &\dstick{\mid\phi_{2}\rangle}&&
        }\]
    \caption{\label{Grover1} Quantum phase separation evolution circuit diagram.}
\end{figure}
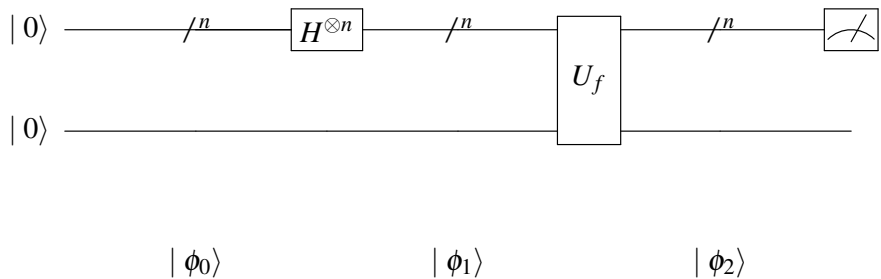

Let us examine this concept through two steps. As shown in Figure \ref{Grover1}. 

First, imagine an initial state composed of two qubits, each initialized to $\ket{0}$, so that the entire system's state can be represented as $\ket{\phi_0} = \ket{00}$. Next, by performing a Hadamard transform on the first qubit, we put it into a superposition state, producing a new state $\ket{\phi_1}$ that contains a superposition of all possible $n$-bit strings $x$, which can be represented as :
\[
\ket{\phi_1} = \left( \sum_{x \in \{0,1\}^n} \ket{x} \right) \ket{0},
\]
this superposition state demonstrates the powerful capability of quantum computing, as it allows us to examine all possible inputs in a single quantum operation.

Further, we consider a state $\ket{\phi_2}$ that includes the results of applying a certain function $f$ to all possible inputs $x$. By applying a special quantum gate $U_f$, we encode the output of the function $f$ into the phase of the qubits, thus obtaining the new state 
\[
\ket{\phi_2} = \sum_{x \in \{0,1\}^n} \ket{x, f(x)},
\] 
this step effectively showcases the efficiency of quantum algorithms, theoretically allowing us to simultaneously compute the value of the function $f$ on all possible inputs.

\begin{figure}[H]
    \centering
        \[\Qcircuit @!C @!R{
        &\lstick{\mid x\rangle}&{/^{n}}\qw&\qw         &\qw&\multigate{1}{U_{f}}&{/^{n}}\qw &\meter\\
        &\lstick{\mid 1\rangle}&\qw&\gate{H}    &\qw&\ghost{U_{f}}       &\qw&\qw\\
        &                      &\dstick{\mid\phi_{0}\rangle}& &\dstick{\mid\phi_{1}\rangle}& &\dstick{\mid\phi_{2}\rangle}&&
        }\]
    \caption{\label{Grover2} Quantum phase separation and controlled phase flip circuit diagram}
\end{figure}
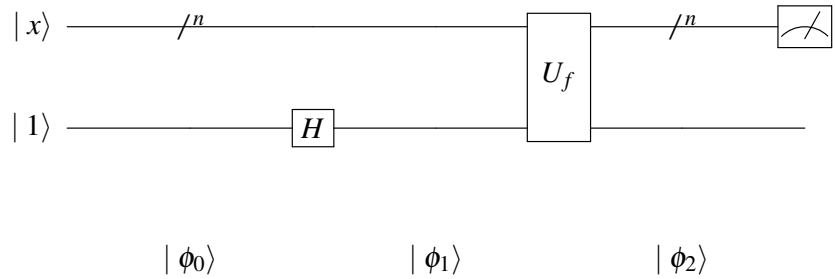

Now, let us look specifically at how phase separation is implemented in quantum computing. As shown in Figure \ref{Grover2}. 
Starting with an initial state composed of a determined qubit $x$ and another qubit set to $1$, i.e., $\ket{\phi_0} = \ket{x,1}$, we then apply a Hadamard gate (H gate) to the second qubit, transforming $\ket{1}$ into a superposition state $\ket{0} - \ket{1}$. The result of this operation leads to the state $\ket{\phi_1}$:
\[
\ket{\phi_1} = \ket{x}(\ket{0} - \ket{1}) = (\ket{x,0} - \ket{x,1}).
\]
Subsequently, we apply a special quantum gate $U_f$ to this superposition state, which operates on the qubits based on the value of the function $f$. This causes our system to transition to the state $\ket{\phi_2}$, which can be represented as:
\[
\ket{\phi_2} = \ket{x}(\ket{f(x) \oplus 0} - \ket{f(x) \oplus 1})=\ket{x}(\ket{f(x)}-\overline{\ket{f(x)}}).
\]
Given that the phase of a qubit is a global property, we can further simplify the state $\ket{\phi_2}$ based on the value of $x$. If $x=x_0$, we get:
\[
\ket{\phi_2} = -\ket{x}(\ket{0} - \ket{1}),
\]
otherwise, when $x\neq x_0$, we have:
\[
\ket{\phi_2} = \ket{x}(\ket{0} - \ket{1}).
\]

This shows that we can control the phase of the qubits through the value of the function $f(x)$, a feature that allows us to encode the information of the function $f$ in the output state $\ket{\phi_2}$. Ultimately, the state $\ket{\phi_2}$ can be expressed as:
\[
\ket{\phi_2} = (-1)^{f(x)}\ket{x}(\ket{0} - \ket{1}).
\]
In this expression, we can clearly see how the phase of the quantum state is modulated by the value of the function $f(x)$. This phase modulation is a unique characteristic of quantum algorithms, reflecting the unique advantages of quantum computing in information processing.

\textbf{Operator R1: From Function \( f \) to Operator \( U_f \)}

In Grover's algorithm, the first key operator, R1, is derived from the function \( f \). This function \( f \) is defined as \( f(x) = 1 \) if \( x \) is the target element (our `red ball') and \( f(x) = 0 \) otherwise. The role of R1 is to apply a phase inversion to the target state, and it is realized through the unitary operator \( U_f \) \cite{2_15}.

The operator \( U_f \) is defined as follows:
\begin{equation}
    U_f|x\rangle = (-1)^{f(x)}|x\rangle.
\end{equation}
This operation leaves the non-target states (where \( f(x) = 0 \)) unchanged, while it multiplies the target state (where \( f(x) = 1 \)) by -1, effectively applying a phase shift of \(\pi\) to it.

The geometrical interpretation of this operation is that it reflects the quantum state vector about the origin in the subspace defined by the target state. If \( |x_0\rangle \) is the target state, then \( U_f \) acts as:
\begin{equation}
    U_f|x_0\rangle = -|x_0\rangle,
\end{equation}
and for all other states \( |x\rangle \) where \( x \neq x_0 \),
\begin{equation}
    U_f|x\rangle = |x\rangle.
\end{equation}

This selective phase inversion is crucial in Grover's algorithm as it sets up the state for the subsequent operation R2, which is the inversion about the mean. Together, these operations amplify the probability amplitude of the target state, leading to its eventual domination in the quantum state vector after repeated iterations of the algorithm.

\textbf{Operator R2: Inversion About the Average}

\begin{figure}[H]
	\centering
	\subfigure[Before inverting]{
		\includegraphics[scale=0.4]{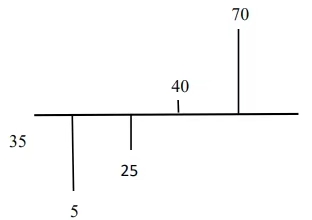}}
	\quad
	\subfigure[After inverting]{
		\includegraphics[scale=0.4]{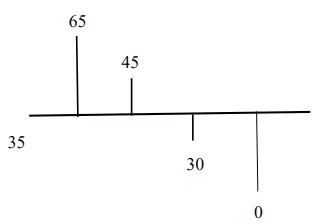}}
	\caption{Invert Diagram.}\label{inverting}
\end{figure}

To better understand the R2 operation in Grover's algorithm, consider the following example with a smaller set of numbers:

\begin{example}
As shown in figure \ref{inverting}. Consider an array of four numbers: [5, 25, 40, 70]. The first step is to calculate their average, which is \( a = \frac{5 + 25 + 40 + 70}{4} = 35 \).

Next, we invert each number about this average. For instance, inverting the number 5, we compute \( 2 \times 35 - 5 = 65 \). Applying this inversion to all the numbers in the array, we get the following results:
\begin{align*}
    \text{Invert 5:} & \quad 2 \times 35 - 5 = 65, \\
    \text{Invert 25:} & \quad 2 \times 35 - 25 = 45, \\
    \text{Invert 40:} & \quad 2 \times 35 - 40 = 30, \\
    \text{Invert 70:} & \quad 2 \times 35 - 70 = 0.
\end{align*}
Thus, the inverted array becomes: [65, 45, 30, 0].
\end{example}

This operation amplifies the amplitudes of values that are further away from the mean while keeping the average itself unchanged. In Grover's algorithm, this is crucial for enhancing the amplitude of the target state \cite{4_sx}.

In summary, the formula for inverting about the average is \( v' = -v + 2a \), where \( a \) is the average. This inversion process is essential for improving the probability amplitude of the searched values in Grover's algorithm.

\begin{example}
Consider a two-qubit system where we aim to find a particular state using Grover's algorithm. Let's say our target state is \( |10\rangle \). The quantum state is initially prepared in an equal superposition of all basis states, denoted as \( |\varphi_0\rangle \). For two qubits, this state is:
\[ |\varphi_0\rangle = \frac{1}{2}\begin{bmatrix} 1 \\ 1 \\ 1 \\ 1 \end{bmatrix}. \]

The first step of Grover's algorithm is to apply the oracle operation specific to our target state \( |10\rangle \). The oracle flips the sign of the amplitude of the target state, resulting in:
\[ |\varphi_1\rangle = \begin{bmatrix} \frac{1}{2} \\ \frac{1}{2} \\ -\frac{1}{2} \\ \frac{1}{2} \end{bmatrix}. \]

Next, we perform the inversion about the average. The average amplitude \( a \) of the state \( |\varphi_1\rangle \) is \( \frac{1}{4} \). The inversion about the average operation reflects each amplitude about this average, resulting in the state \( |\varphi_2\rangle \):
\[ |\varphi_2\rangle = \begin{bmatrix} 0 \\ 0 \\ 1 \\ 0 \end{bmatrix}. \]

Finally, measuring this quantum state will collapse it to the state \( |10\rangle \) with \( 100\% \) probability, as this is now the only state with a non-zero amplitude.
\end{example}

\begin{example}
Consider a three-qubit system in Grover's algorithm. We start with the initial state in an equal superposition of all basis states (we use row vector for convenience):
\begin{equation}
    |\varphi_0\rangle = \frac{1}{\sqrt{8}}\begin{bmatrix}
    1 & 1 & 1 & 1 & 1 & 1 & 1 & 1
    \end{bmatrix}^T.
\end{equation}

After applying the oracle operation \( U_f \) for the function \( f \) where \( f(100) = -1 \), the state becomes:
\begin{equation}
    |\varphi_1\rangle = U_f|\varphi_0\rangle = \frac{1}{\sqrt{8}}\begin{bmatrix}
    1 & 1 & 1 & 1 & -1 & 1 & 1 & 1
    \end{bmatrix}^T.
\end{equation}

The average amplitude \( a \) after the oracle operation is \( \frac{3}{4\sqrt{8}} \). Inverting about this average gives:
\begin{equation}
    |\varphi_2\rangle = \frac{1}{2\sqrt{8}}\begin{bmatrix}
    1 & 1 & 1 & 1 & 5 & 1 & 1 & 1
    \end{bmatrix}^T.
\end{equation}

At this stage, the probability of measuring the state \( |100\rangle \) is \( \left(\frac{5\sqrt{2}}{8}\right)^2 = \frac{50}{64} \approx 78.1\% \).

For the second iteration, we again apply \( U_f \) to \( |\varphi_2\rangle \) and invert about the new average:
\begin{equation}
    |\varphi_3\rangle = U_f|\varphi_2\rangle = \frac{1}{2\sqrt{8}}\begin{bmatrix}
    1 & 1 & 1 & 1 & -5 & 1 & 1 & 1
    \end{bmatrix}^T,
\end{equation}
\begin{equation}
    \text{After inversion, } |\varphi_4\rangle = \frac{1}{4\sqrt{8}}\begin{bmatrix}
    -1 & -1 & -1 & -1 & 11 & -1 & -1 & -1
    \end{bmatrix}^T.
\end{equation}

Now, measuring \( |\varphi_4\rangle \), the probability of observing \( |100\rangle \) is \( \left(\frac{11\sqrt{2}}{16}\right)^2 \approx 94.5\% \), which is much higher than the \( \frac{2}{64} \approx 3.1\% \) for the other states.
\end{example}

In arithmetic, inverting the average value is straightforward, but for a quantum computer, this operation must be expressed as a matrix operation, as quantum computing fundamentally relies on matrix transformations. 

\begin{example}
First, let's construct a matrix that computes the average for a two-qubit system (4 quantum states). The mean matrix \( A \) is:
\begin{equation}
A = \frac{1}{4}\begin{bmatrix}
1 & 1 & 1 & 1 \\
1 & 1 & 1 & 1 \\
1 & 1 & 1 & 1 \\
1 & 1 & 1 & 1
\end{bmatrix}.
\end{equation}

The inversion about the average can then be written as \( V' = -V + 2AV \), which simplifies to \( V' = (-I + 2A)V \). For instance, if we have \( V = \frac{1}{2}\begin{bmatrix} 1 & 1 & -1 & 1 \end{bmatrix}^T \), then:
\begin{align}
V' &= \left(-I + 2A\right)V \nonumber \\
&= \left(-\begin{bmatrix}
1 & 0 & 0 & 0 \\
0 & 1 & 0 & 0 \\
0 & 0 & 1 & 0 \\
0 & 0 & 0 & 1
\end{bmatrix} + 2\begin{bmatrix}
\frac{1}{4} & \frac{1}{4} & \frac{1}{4} & \frac{1}{4} \\
\frac{1}{4} & \frac{1}{4} & \frac{1}{4} & \frac{1}{4} \\
\frac{1}{4} & \frac{1}{4} & \frac{1}{4} & \frac{1}{4} \\
\frac{1}{4} & \frac{1}{4} & \frac{1}{4} & \frac{1}{4}
\end{bmatrix} \right)\begin{bmatrix}
\frac{1}{2} \\
\frac{1}{2} \\
-\frac{1}{2} \\
\frac{1}{2}
\end{bmatrix} \nonumber \\
&= \begin{bmatrix}
0 \\
0 \\
1 \\
0
\end{bmatrix}.
\end{align}
\end{example}

Observing the operation \( V' = -V + 2AV \), \( 2AV \) returns a vector of averages \( \frac{1}{2}\begin{bmatrix} 1 & 1 & 1 & 1 \end{bmatrix}^T \). The vector \( -V \) calculates the negative of \( V \), so \( -V + 2AV \) effectively inverts each element of \( V \) about the average.

For \( n \) qubits, we construct an \( n \)-qubit matrix that computes the average:
\begin{equation}
A = \frac{1}{2^n}\begin{bmatrix}
1 & 1 & \cdots & 1 \\
1 & 1 & \cdots & 1 \\
\vdots & \vdots & \ddots & \vdots \\
1 & 1 & \cdots & 1
\end{bmatrix}.
\end{equation}

The matrix for inverting about the average becomes:
\begin{equation}
V' = (-I +2A)V  = \frac{1}{2^n}\begin{bmatrix}
-2^{n}+2 & 2 & \cdots & 2 \\
2 & -2^{n}+2 & \cdots & 2 \\
\vdots & \vdots & \ddots & \vdots \\
2 & 2 & \cdots & -2^{n}+2
\end{bmatrix}V.
\end{equation}

This change does not affect the final quantum state normalization or the increased probability of finding the searched quantum state.

\begin{example}
In the 2-qubit case, searching for the \( |10\rangle \) state, \( V = \frac{1}{2}\begin{bmatrix} 1 & 1 & -1 & 1 \end{bmatrix}^T \), the transformed vector \( V' \) will be \( \begin{bmatrix} 0 & 0 & -1 & 0 \end{bmatrix}^T \), leading to a \( 100\% \) probability of measuring the quantum state \( |10\rangle \).
\end{example}

\textbf{Grover's Algorithm: A Geometrical View}

We have previously introduced phase separation as the mathematical description of the core steps in Grover's algorithm. Similarly, the geometrical perspective also offers a graphical representation that provides an intuitive understanding of the algorithm's behavior. Next, we will see how Grover's algorithm incrementally rotates the quantum state to approach the target state with high probability.

\begin{figure}[htbp]
\centering
\includegraphics[scale=0.23]{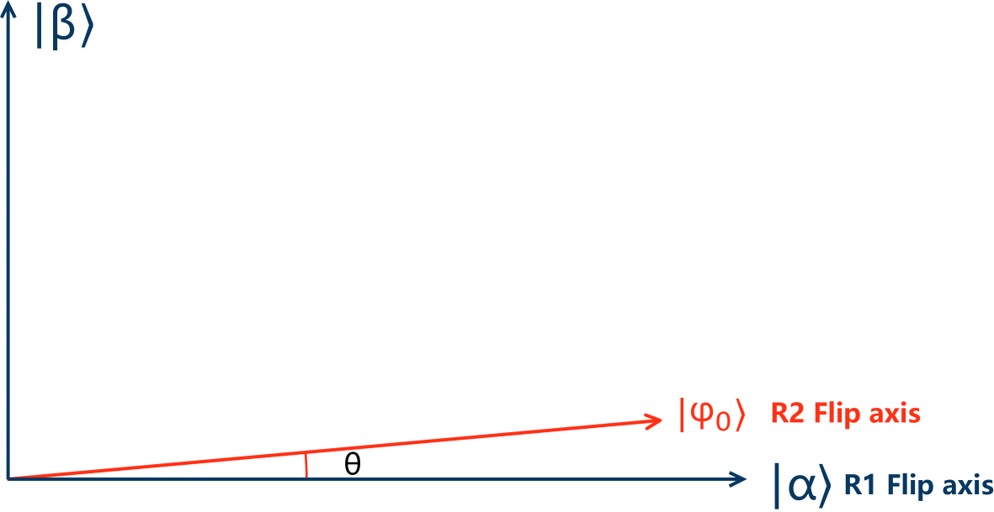}
\caption{Initial state representation: \( |\varphi_0\rangle = \frac{\sqrt{N-1}}{\sqrt{N}}|\alpha\rangle + \frac{1}{\sqrt{N}}|\beta\rangle \).}
\label{8_4}
\end{figure}

\begin{figure}[htbp]
\centering
\includegraphics[scale=0.23]{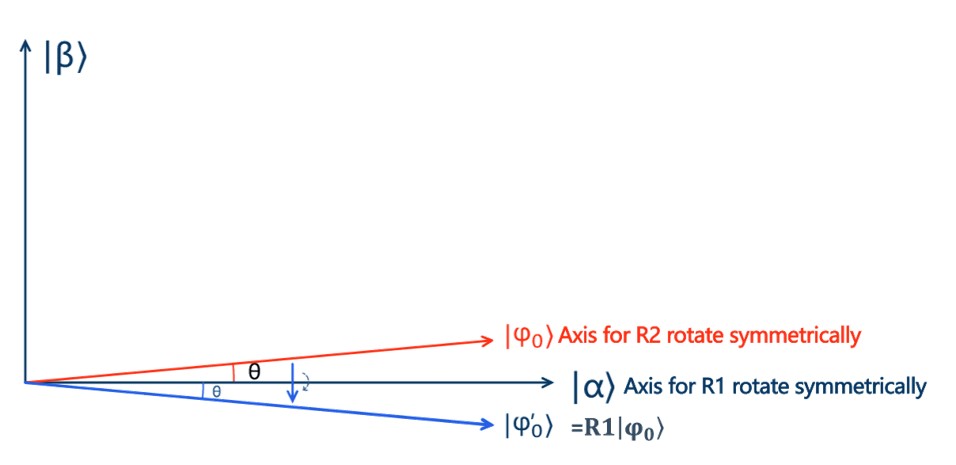}
\caption{Effect of R1 operation: Equivalent to flipping the vector \( |\varphi'\rangle = R1|\varphi_0\rangle \).}
\label{8_5}
\end{figure}

\begin{figure}[htbp]
\centering
\includegraphics[scale=0.23]{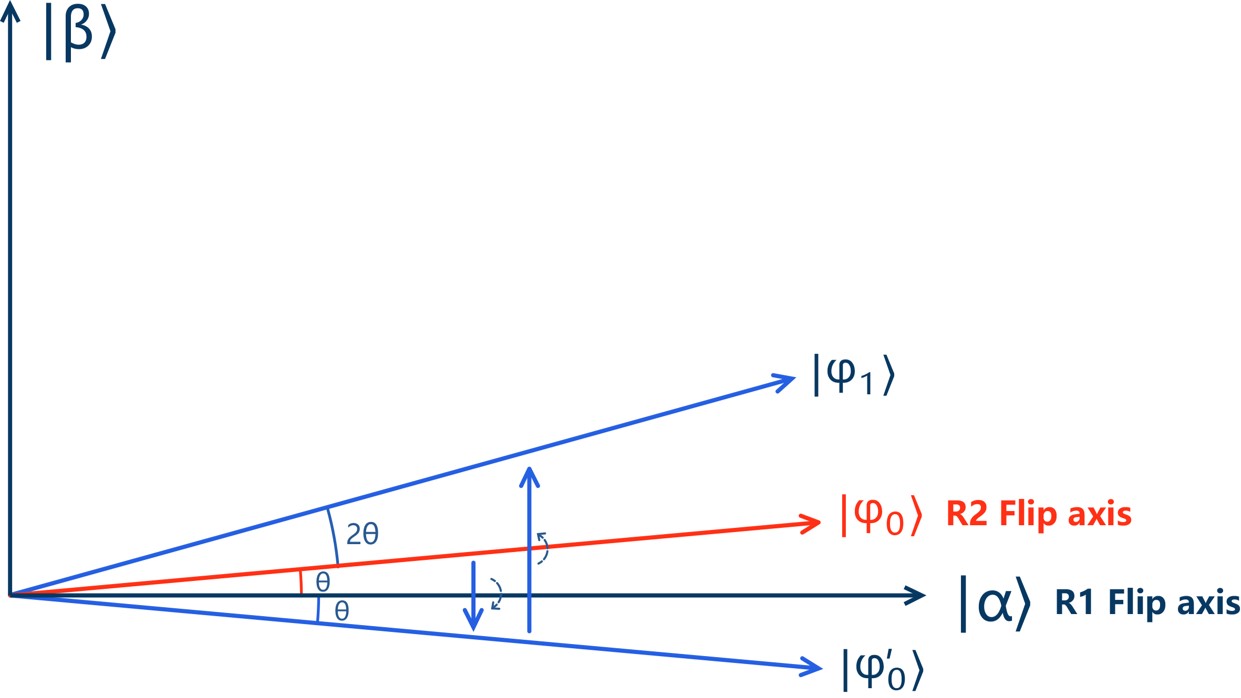}
\caption{Effect of R2 operation: Flipping the vector \( |\varphi_1\rangle = R2|\varphi'_0\rangle \).}
\label{8_6}
\end{figure}

\begin{figure}[htbp]
\centering
\includegraphics[scale=0.23]{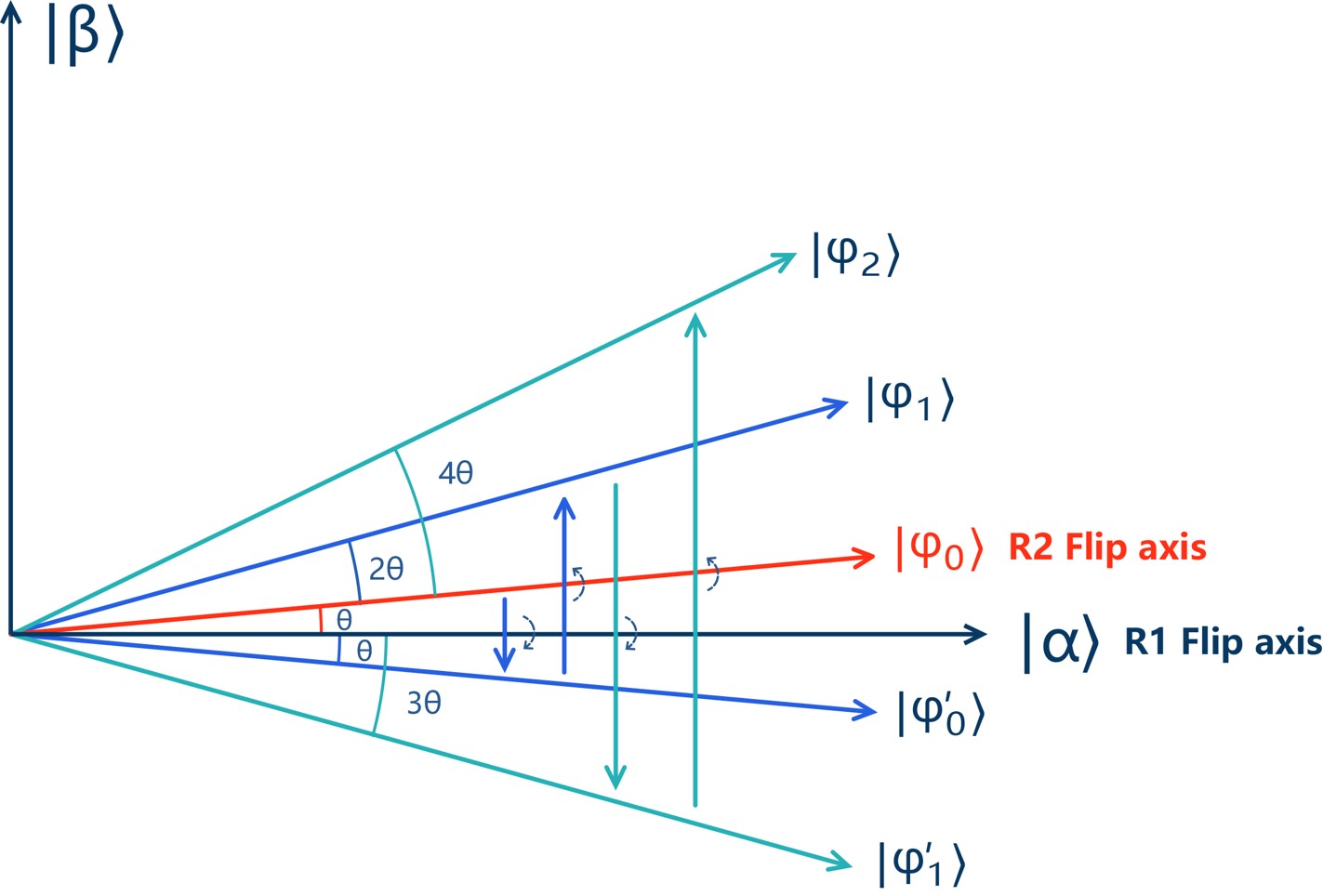}
\caption{Subsequent iterations: Each iteration rotates the state vector closer to \( |\beta\rangle \) by \( 2\theta \).}
\label{8_7}
\end{figure}

Grover's Algorithm can be intuitively understood through a geometrical lens, as illustrated in Figures \ref{8_4}, \ref{8_5}, \ref{8_6}, and \ref{8_7}. We consider the quantum state space spanned by two state vectors \( |\alpha\rangle \) and \( |\beta\rangle \). Here, \( |\beta\rangle \) represents the target state \( |x_0\rangle \), and \( |\alpha\rangle \) is the superposition of all other states, i.e., \( |\alpha\rangle = \frac{1}{\sqrt{N-1}}\sum_{x \neq x_0} |x\rangle \). The initial state \( |\varphi\rangle \) can then be expressed as \( |\varphi\rangle = \frac{\sqrt{N-1}}{\sqrt{N}}|\alpha\rangle + \frac{1}{\sqrt{N}}|\beta\rangle \).

The first operation, R1, applies a phase inversion to the target state \( |\beta\rangle \), effectively flipping its sign. In this operation, \( |\beta\rangle = |x_0\rangle \) becomes \( -|x_0\rangle \), while \( |\alpha\rangle \) remains unchanged. As a result, the state \( |\varphi_0\rangle \) is transformed to \( |\varphi'\rangle = \frac{\sqrt{N-1}}{\sqrt{N}}|\alpha\rangle - \frac{1}{\sqrt{N}}|\beta\rangle \).

The second operation, R2, is a reflection through the average amplitude, flipping the vector \( |\varphi'\rangle \) around the initial state \( |\varphi_0\rangle \) to become \( |\varphi_1\rangle \).

In this geometric representation, \( \sin\theta = \frac{1}{\sqrt{N}} \) and \( \cos\theta = \frac{\sqrt{N-1}}{\sqrt{N}} \), leading to \( \sin2\theta = 2\sin\theta\cos\theta = \frac{2\sqrt{N-1}}{N} \). Each iteration of the Grover operation (R1 followed by R2) rotates the state vector towards \( |\beta\rangle \) by an angle of \( 2\theta \).

For large \( N \), where \( N \gg 1 \), \( 2\theta \) can be approximated as \( \frac{2}{\sqrt{N}} \). The algorithm, therefore, requires approximately \( \frac{\pi\sqrt{N}}{4} \) iterations to rotate the state vector close to \( |\beta\rangle \) with high probability, and the error probability is less than \( \frac{1}{N} \) \cite{4_sx}.

\section{Circuits for Grover's algorithm}

\begin{figure}[H]
	\centering
	\[\Qcircuit @C=2em @!R {
        &\lstick{|0\rangle} &\multigate{2}{H^{\otimes n}} &\multigate{2}{U_{f}} &\multigate{2}{I-2A}&\qw &\cdots & &\multigate{2}{U_{f}} &\multigate{2}{I-2A}&\qw\\
        &\lstick{|0\rangle} &\ghost{H^{\otimes n}} &\ghost{U_{f}} &\ghost{I-2A}&\qw &\cdots & &\ghost{U_{f}} &\ghost{I-2A}&\qw\\
        &\lstick{|0\rangle} &\ghost{H^{\otimes n}} &\ghost{U_{f}} &\ghost{I-2A} &\qw &\cdots & &\ghost{U_{f}} &\ghost{I-2A}&\qw
        \inputgroupv{1}{3}{.1em}{2.5em}{\text{$n$ \ qubits\quad}}
        }\]
	\caption{\label{grover_circuit} The circuit of Grover's algorithm, here should repeat $U_{f}$ and $I-2A$ $\sqrt{2^{n}}$ times.}
\end{figure}
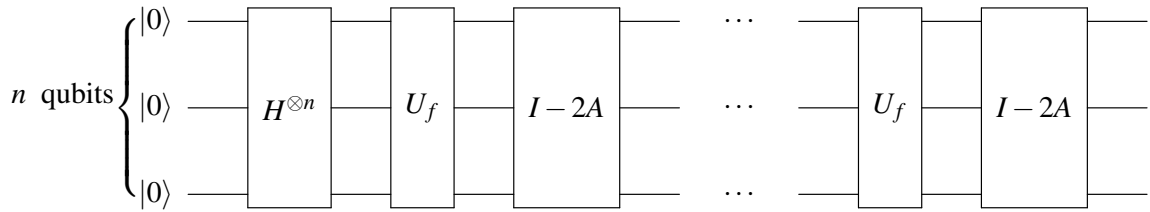

The circuit diagram for Grover's algorithm, as shown in Figure \ref{grover_circuit}, illustrates the algorithm's operational steps. Initially, all qubits are set to the \( |0\rangle \) state. This is achieved by applying Hadamard gates (H gates) to each qubit, denoted as \( H^{\otimes n} \), which creates a superposition of all possible states.

In the circuit, Grover's algorithm involves repeating a sequence of operations several times. These operations consist of two main parts: the phase inversion and the amplitude amplification. The phase inversion, which flips the phase of the target state, is implemented by the oracle gate \( U_f \) \cite{4_sx}. The amplitude amplification, which increases the probability amplitude of the target state relative to the other states, is represented by the gate \( I - 2A \).

After approximately \( \sqrt{2^n} \) iterations of these steps, the algorithm significantly amplifies the amplitude of the target state. The final step is to measure the quantum state, which, with high probability, collapses to the target state.

Grover's algorithm leverages quantum parallelism and amplitude amplification, enabling a quadratic speedup with a complexity of \( O(\sqrt{2^n}) \) compared to classical search algorithms \cite{2_16}. This makes it a powerful demonstration of the potential advantages of quantum computing in solving specific types of problems more efficiently than classical methods.

\textbf{Circuit for R1}

In Grover's algorithm, constructing the oracle function \( U_f \) is crucial \cite{2_15}. The choice of \( U_f \) depends on the target state we wish to find. Key quantum gates such as Pauli \( X \), \( Z \), Hadamard (\( H \)), controlled-\( Z \) (CZ), and controlled-NOT (CNOT) are used in this process.

Let's consider some examples:

\begin{example}
In a single-qubit case if we want to find state $|1\rangle$, we could use a Z gate as our $U_f$. This is because Z flips the phase of $|1\rangle$ but leaves $|0\rangle$ unchanged. So applying $Z$ to a superposition $a|0\rangle + b|1\rangle$ would give $a|0\rangle - b|1\rangle$, inverting the amplitude of the desired state shown in Table \ref{Truth_table}.
\begin{table}[H]
	\centering
        \setlength{\tabcolsep}{5mm}
        \renewcommand\arraystretch{1.5}
	\begin{tabular}{|l|r|}
            \hline
		Input & Output \\\hline
		$|0\rangle$ &  $|0\rangle$\\\hline
		$|1\rangle$ & $-|1\rangle$\\\hline
	\end{tabular}
	\caption{\label{Truth_table}Truth table of Z gate.}
\end{table}
\end{example}

\begin{example}
For more example, in the single qubit case,  we need to find $|0\rangle$, the truth table of $U_f$ is in Table \ref{Truth_table1}, we can realize this truth table by $XZX$ combination gate.
\begin{table}[H]
	\centering
        \setlength{\tabcolsep}{5mm}
        \renewcommand\arraystretch{1.5}
	\begin{tabular}{|l|r|}
            \hline
		Input & Output \\\hline
		$|0\rangle$ &  $-|0\rangle$\\\hline
		$|1\rangle$ & $|1\rangle$\\\hline
	\end{tabular}
	\caption{\label{Truth_table1}Truth table of XZX gate.}
\end{table}
\end{example}

Now we look at some two-qubit examples.

\begin{example}
For example, we need to find $|11\rangle$, the truth table of $U_f$ is in Tabel \ref{CZ_truth}, we can realize this truth table by CZ combination gates.
\begin{table}[H]
	\centering
        \setlength{\tabcolsep}{5mm}
        \renewcommand\arraystretch{1.5}
	\begin{tabular}{|l|r|}
            \hline
		Input & Output \\\hline
		$|00\rangle$ &  $|00\rangle$\\\hline
		$|01\rangle$ & $|01\rangle$\\\hline
		$|10\rangle$ & $|10\rangle$\\\hline
		$|11\rangle$ & $-|11\rangle$\\\hline
	\end{tabular}
	\caption{\label{CZ_truth}Truth table of $CZ$ gate.}
\end{table}
\end{example}

\begin{example}
One more example, we need to find $|10\rangle$, then the truth table of $U_f$ is as follows, we can use the following combination gate of CZ and X to realize this truth table in Tabel \ref{XCZX}, here $U_f = (I \otimes X)CZ(I \otimes X)$.
\begin{table}[H]
	\centering
        \setlength{\tabcolsep}{5mm}
        \renewcommand\arraystretch{1.5}
	\begin{tabular}{|l|r|}
            \hline
		Input & Output \\\hline
		$|00\rangle$ &  $|00\rangle$\\\hline
		$|01\rangle$ & $|01\rangle$\\\hline
		$|10\rangle$ & $-|10\rangle$\\\hline
		$|11\rangle$ & $|11\rangle$\\\hline
	\end{tabular}
	\caption{\label{XCZX}Truth table of $(I\otimes X)CZ(I\otimes X)$.}
\end{table}
\end{example}

Now we look at a three-qubit example.

\begin{example}
In the case of 3 qubits, for example, we need to find $|111\rangle$ , the truth table of $U_f$ is shown in Table \ref{HCCXH}, and the corresponding circuit shown in Figure \ref{HCCXH1} can be built with Toffoli gate and H gate.
\begin{table}[H]
	\centering
        \setlength{\tabcolsep}{5mm}
        \renewcommand\arraystretch{1.5}
	\begin{tabular}{|l|r|}
            \hline
		Input & Output \\\hline
		$|000\rangle$ &  $|000\rangle$\\\hline
		$|001\rangle$ & $|001\rangle$\\\hline
		$|010\rangle$ & $|010\rangle$\\\hline
		$|011\rangle$ & $|011\rangle$\\\hline
		$|100\rangle$ &  $|100\rangle$\\\hline
		$|101\rangle$ & $|101\rangle$\\\hline
		$|110\rangle$ & $|110\rangle$\\\hline
		$|111\rangle$ & $-|111\rangle$\\\hline
	\end{tabular}
	\caption{\label{HCCXH}Truth table of circuit in Figure \ref{HCCXH1}.}
\end{table}

\begin{figure}[H]
	\centering
	\[\Qcircuit @!C @!R {
        &\lstick{a} &\qw &\ctrl{1} &\qw   &\rstick{a^{'}}\qw \\
        &\lstick{b}  &\qw &\ctrl{1} &\qw &\rstick{b^{'}}\qw\\
        &\lstick{c}  &\gate{H} &\targ &\gate{H} &\rstick{c^{'}}\qw
        }\]
	\caption{\label{HCCXH1} Circuit of gates $(I\otimes I \otimes H)CCX(I\otimes I\otimes H)$.}
\end{figure}
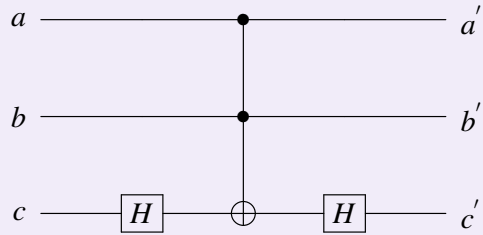
\end{example}

\begin{example}
In the 3 qubit case, for example, we need to find $|101\rangle$, the truth table of $U_f$ is shown in Table \ref{XHCCXHX}. It is only one qubit different from the truth table of $U_f$ for finding $|111\rangle$, so we can imagine that if we modify the 0 of the second qubit to 1, then it will be exactly the same as $|111\rangle$, so we only need to add an X gate to the second qubit of $U_f$ of $|111\rangle$, then we can get the $U_f$ circuit shown in Figure \ref{XHCCXHX1} of search $|101\rangle$.
\begin{table}[H]
	\centering
        \setlength{\tabcolsep}{5mm}
        \renewcommand\arraystretch{1.5}
	\begin{tabular}{|l|r|}
            \hline
		Input & Output \\
            \hline
		$|000\rangle$ &  $|000\rangle$\\
            \hline
		$|001\rangle$ & $|001\rangle$\\
            \hline
		$|010\rangle$ & $|010\rangle$\\
            \hline
		$|011\rangle$ & $|011\rangle$\\
            \hline
		$|100\rangle$ &  $|100\rangle$\\
            \hline
		$|101\rangle$ & $-|101\rangle$\\
            \hline
		$|110\rangle$ & $|110\rangle$\\
            \hline
		$|111\rangle$ & $|111\rangle$\\
            \hline
	\end{tabular}
	\caption{\label{XHCCXHX}Truth table of circuit in Figure \ref{XHCCXHX1}.}
\end{table}

Correspondingly, for the 3 qubit case, we can easily build gates that search for other values (quantum states). 
We need to search $|110\rangle$, just add an X gate to the 3rd qubit of the $|111\rangle$ $U_f$ gate to get the $U_f$ gate of $|110\rangle$.
We need to search $|100\rangle$, just add an X gate to the 2nd qubit of the $|110\rangle$ $U_f$ gate to get the $U_f$ gate of $|100\rangle$.
We need to search $|000\rangle$, just add an X gate to the 1st qubit of the $|100\rangle$ $U_f$ gate to get the $U_f$ gate of $|000\rangle$.
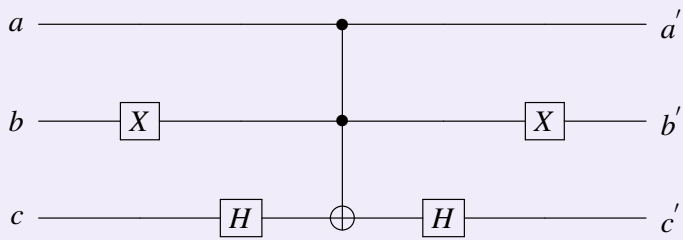
\begin{figure}[H]
	\centering
	\[\Qcircuit @!C @!R {
        &\lstick{a} &\qw &\qw &\ctrl{1} &\qw &\qw  &\rstick{a^{'}}\qw \\
        &\lstick{b} &\gate{X} &\qw &\ctrl{1} &\qw &\gate{X}&\rstick{b^{'}}\qw\\
        &\lstick{c} &\qw &\gate{H} &\targ &\gate{H} &\qw&\rstick{c^{'}}\qw
        }\]
	\caption{\label{XHCCXHX1} Circuit of gate $(I\otimes X \otimes I)(I\otimes I\otimes H)CCX(I\otimes I\otimes H)(I\otimes X \otimes I)$.}
\end{figure}
\end{example}

\textbf{Circuits for R2}

We have examined Grover's algorithm, a pivotal quantum search algorithm. An integral part of this algorithm is the matrix \( I - 2A \) \cite{4_sx}, which is essential for amplitude amplification. This matrix is implemented as quantum gate operations on the qubits being searched. Let's consider its properties:

\begin{enumerate}
    \item The matrix is independent of the specific state being searched. Its structure remains constant regardless of which state we target.
    \item The matrix does depend on the number of qubits in the system. For a single-qubit system, the matrix can be implemented using an \( HXZXH \) gate sequence. For systems with more qubits, different gate sequences are required.
\end{enumerate}

To better understand the matrix for a single-qubit system, consider the truth table shown in Table \ref{HXZXH}, which corresponds to the \( HXZXH \) gate sequence. This sequence effectively realizes the \( I - 2A \) operation.

\begin{table}[H]
    \centering
    \setlength{\tabcolsep}{5mm}
    \renewcommand\arraystretch{1.5}
    \begin{tabular}{|l|r|}
        \hline
        Input & Output \\
        \hline
        \( |0\rangle \) & \( -|1\rangle \) \\
        \hline
        \( |1\rangle \) & \( -|0\rangle \) \\
        \hline
    \end{tabular}
    \caption{Truth table for the \( HXZXH \) gate sequence.}
    \label{HXZXH}
\end{table}

\begin{figure}[H]
    \centering
    \[\Qcircuit @!C @!R {
        &\lstick{b} &\gate{H} &\gate{X} &\gate{Z} &\gate{X} &\gate{H} &\rstick{b'}\qw
    }\]
    \caption{Circuit for the \( HXZXH \) gate sequence.}
    \label{HXZXH1}
\end{figure}
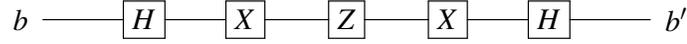

Figure \ref{HXZXH1} illustrates the quantum circuit for the \( HXZXH \) sequence, demonstrating how the matrix operation is translated into quantum gates.

Through the analysis of this matrix and its implementation in quantum circuits, we gain valuable insights into Grover's algorithm. Understanding the translation of mathematical operations into quantum circuits is essential for implementing the algorithm on quantum computers.

With two qubits, the expression for the matrix $I-2A$ can be written as:
\begin{align}
   &I - 2A = \begin{bmatrix}
	1 & 0 & 0 & 0\\
	0 & 1 & 0 & 0\\
	0 & 0 & 1 & 0\\
	0 & 0 & 0 & 1
\end{bmatrix} - 2\begin{bmatrix}
\frac{1}{4} & \frac{1}{4} & \frac{1}{4} & \frac{1}{4}\\
\frac{1}{4} & \frac{1}{4} & \frac{1}{4} & \frac{1}{4}\\
\frac{1}{4} & \frac{1}{4} & \frac{1}{4} & \frac{1}{4}\\
\frac{1}{4} & \frac{1}{4} & \frac{1}{4} & \frac{1}{4}
\end{bmatrix}
= \begin{bmatrix}
	\frac{1}{2} & -\frac{1}{2} & -\frac{1}{2} & -\frac{1}{2}\\
	-\frac{1}{2} & \frac{1}{2} & -\frac{1}{2} & -\frac{1}{2}\\
	-\frac{1}{2} & -\frac{1}{2} & \frac{1}{2} & -\frac{1}{2}\\
	-\frac{1}{2} & -\frac{1}{2} & -\frac{1}{2} & \frac{1}{2}
\end{bmatrix}. 
\end{align}
Decompose the matrix to get: 
\begin{align}
 I - 2A  
&= \frac{1}{2}\begin{bmatrix}
1 & 1 & 1 & 1\\
1 & -1 & 1 & -1\\
1 & 1 & -1 & -1\\
1 & -1 & -1 & 1
\end{bmatrix} \begin{bmatrix}
1 & 0 & 0 & 0\\
0 & -1 & 0 & 0\\
0 & 0 & -1 & 0\\
0 & 0 & 0 & -1
\end{bmatrix} 
\frac{1}{2}\begin{bmatrix}
1 & 1 & 1 & 1\\
1 & -1 & 1 & -1\\
1 & 1 & -1 & -1\\
1 & -1 & -1 & 1
\end{bmatrix} \nonumber\\
&= (H \otimes H) \begin{bmatrix}
	1 & 0 & 0 & 0\\
	0 & -1 & 0 & 0\\
	0 & 0 & -1 & 0\\
	0 & 0 & 0 & -1
\end{bmatrix}  (H \otimes H)\nonumber\\
&= (H \otimes H) \begin{bmatrix}
	0 & 0 & 0 & 1\\
	0 & 0 & 1 & 0\\
	0 & 1 & 0 & 0\\
	1 & 0 & 0 & 0
\end{bmatrix} \begin{bmatrix}
-1 & 0 & 0 & 0\\
0 & -1 & 0 & 0\\
0 & 0 & -1 & 0\\
0 & 0 & 0 & 1
\end{bmatrix} 
\begin{bmatrix}
0 & 0 & 0 & 1\\
0 & 0 & 1 & 0\\
0 & 1 & 0 & 0\\
1 & 0 & 0 & 0\end{bmatrix} \begin{bmatrix}
0 & 0 & 0 & 1\\
0 & 0 & 1 & 0\\
0 & 1 & 0 & 0\\
1 & 0 & 0 & 0
\end{bmatrix} (H \otimes H)\nonumber\\
&= (H\otimes H)(X\otimes X)\text{CZ}(X\otimes X)(H\otimes H) \nonumber\\
&= H^{\otimes 2} X^{\otimes 2}\text{CZ} X^{\otimes 2} H^{\otimes 2}.    
\end{align}
Where CZ is a controlled-Z gate.

Therefore for two qubits,
\[I - 2A = (H\otimes H)(X\otimes X)\text{CZ}(X\otimes X)(H\otimes H),\] 
so the circuit for inversion by the average is shown in Figure \ref{oracle}.

\begin{figure}[H]
	\centering
	\[\Qcircuit @!C @!R {
        &\lstick{|0\rangle} &\gate{H} &\gate{X} &\ctrl{1} &\gate{X} &\gate{H}&\qw\\
        &\lstick{|0\rangle} &\gate{H} &\gate{X} &\control\qw &\gate{X} &\gate{H}&\qw
        }\]
	\caption{\label{oracle} $I-2A$ circuit of 2 qubits Grover Algorithm.}
\end{figure}

We can also given an alternative proof of this two-qubit circuit:
\begin{align}
H^{\otimes 2} X^{\otimes 2}\text{CZ} X^{\otimes 2} H^{\otimes 2}
&= (I - 2A)_{4 \times 4} \nonumber\\
&= \frac{\sqrt{2}}{2}\begin{bmatrix}
	H & H  \\
	H & -H 
\end{bmatrix}
\begin{bmatrix}
	 0 & X\\
	 X & 0 
\end{bmatrix}\begin{bmatrix}
	I_{2 \times 2} & 0 \\
	0 & Z 
\end{bmatrix}\begin{bmatrix}
0 & X\\
X & 0 
\end{bmatrix}
\frac{\sqrt{2}}{2}\begin{bmatrix}
H & H  \\
H & -H 
\end{bmatrix}\nonumber\\
&= \frac{1}{2}\begin{bmatrix}
	HX & HX \\
	-HX & HX 
\end{bmatrix}\begin{bmatrix}
I_{2 \times 2} & 0 \\
0 & Z 
\end{bmatrix}\begin{bmatrix}
XH & -XH \\
XH & XH 
\end{bmatrix} \nonumber\\
&= \frac{1}{2}\begin{bmatrix}
HX  &  HXZ\\
-HX &  HXZ
\end{bmatrix} \begin{bmatrix}
XH & -XH \\
XH & XH 
\end{bmatrix}\nonumber\\
&= \frac{1}{2}\begin{bmatrix}
	HXXH + HXZXH & -HXXH + HXZXH\\
	-HXXH + HXZXH & HXXH + HXZXH
\end{bmatrix}. 
\end{align}

We kown: $HXXH= I_{2 \times 2}, HXZXH = (1-2A)_{2 \times 2}$,\\
so:
\begin{align}
H^{\otimes 2} X^{\otimes 2}\text{CZ} X^{\otimes 2} H^{\otimes 2} 
&= \frac{1}{2}\begin{bmatrix}
	I_{2 \times 2}+(I-2A)_{2 \times 2} & -I_{2 \times 2}+(I-2A)_{2 \times 2}\\
	-I_{2 \times 2}+(I-2A)_{2 \times 2} & -I_{2 \times 2}+(I-2A)_{2 \times 2}
\end{bmatrix}\nonumber\\
&= \frac{1}{2}\begin{bmatrix}
	2(I-A)_{2 \times 2} & -(2A)_{2 \times 2}\\
	-(2A)_{2 \times 2} & 2(I-A)_{2 \times 2}
\end{bmatrix}\nonumber\\
&= I_{4 \times 4}-\begin{bmatrix}
	A_{2 \times 2} & A_{2 \times 2}\\
	A_{2 \times 2} & A_{2 \times 2}
\end{bmatrix}\nonumber\\
&= I_{4 \times 4}-2\begin{bmatrix}
	A_{2 \times 2}/2 & A_{2 \times 2}/2\\
	A_{2 \times 2}/2 & A_{2 \times 2}/2
\end{bmatrix}\nonumber\\
&= I_{4 \times 4} -2A_{4 \times 4} = (I-2A)_{4 \times 4}.
\end{align}

\begin{figure}[H]
	\centering
	\[\Qcircuit @!C @!R {
        &\lstick{} &\gate{H} &\gate{X} &\ctrl{1} &\gate{X} &\gate{H}&\qw\\
        &\lstick{} &\gate{H} &\gate{X} &\ctrl{1} &\gate{X} &\gate{H}&\qw\\
        &\lstick{} &\gate{H} &\gate{X} &\gate{CZ} &\gate{X} &\gate{H}&\qw
        }\]
	\caption{\label{3_qubit_oracle} $I-2A$ circuit of 3 qubits Grover Algorithm.}
\end{figure}
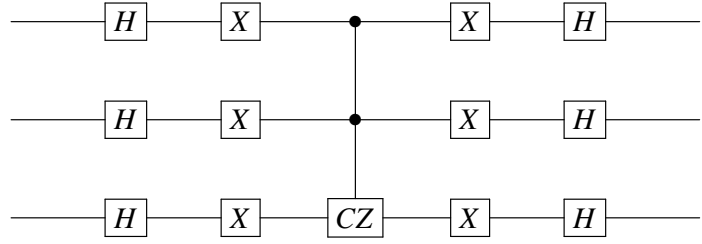

For three qubits, the circuit for inversion by the average is shown in Figure \ref{3_qubit_oracle},

To show this, notice that:
\[H^{\otimes3} X^{\otimes3}\begin{bmatrix}
	1 & 0 & \cdots & 0\\
	0 & 1 & \cdots & 0\\
	\vdots & \vdots & \ddots & \vdots\\
	0 & 0 & \cdots & -1
\end{bmatrix} X^{\otimes 3}H^{\otimes 3} = I - 2A,\] $A$ is the matrix of $2^3\times2^3$, each element is $1/2^3$.

The controlled-$Z$(CZ) gate in the Figure \ref{3_qubit_oracle} is similar to the previous Toffoli gate, in order to distinguish it from the 2 qubits CZ gate, we denote it by $CZ_3$, whose matrix is the elements on the diagonal are 1 except the last $CZ_3$[7][7] is $-1$, the off-diagonal element is $0$, easy to verify $CZ_3 = \begin{bmatrix}
I_{4 \times 4} & 0\\
0 & CZ
\end{bmatrix}$.

\begin{align}
H^{\otimes 3} X^{\otimes 3}CZ_3 X^{\otimes 3} H^{\otimes 3} 
=& (I - 2A)_{8 \times 8} \nonumber\\
=& \frac{\sqrt{2}}{2}\begin{bmatrix}
	H^{\otimes 2} & H^{\otimes 2}  \\
	H^{\otimes 2} & -H^{\otimes 2} 
\end{bmatrix}
\begin{bmatrix}
	0 & X^{\otimes 2}\\
	X^{\otimes 2} & 0 
\end{bmatrix}
\begin{bmatrix}
	I_{4 \times 4} & 0 \\
	0 & CZ 
\end{bmatrix}\begin{bmatrix}
	0 & X^{\otimes 2}\\
	X^{\otimes 2} & 0 
\end{bmatrix}
\frac{\sqrt{2}}{2}\begin{bmatrix}
	H^{\otimes 2} & H^{\otimes 2}  \\
	H^{\otimes 2} & -H^{\otimes 2} 
\end{bmatrix}\nonumber\\
=& \frac{1}{2}\begin{bmatrix}
	H^{\otimes 2}X^{\otimes 2} & H^{\otimes 2}X^{\otimes 2} \\
	-H^{\otimes 2}X^{\otimes 2} & H^{\otimes 2}X^{\otimes 2} 
\end{bmatrix}\begin{bmatrix}
	I_{4 \times 4} & 0 \\
	0 & CZ 
\end{bmatrix}
\begin{bmatrix}
	X^{\otimes 2}H^{\otimes 2} & -X^{\otimes 2}H^{\otimes 2} \\
	X^{\otimes 2}H^{\otimes 2} & X^{\otimes 2}H^{\otimes 2} 
\end{bmatrix}\nonumber \\
=& \frac{1}{2}\begin{bmatrix}
	H^{\otimes 2}X^{\otimes 2}  &  H^{\otimes 2}X^{\otimes 2}CZ\\
	-H^{\otimes 2}X^{\otimes 2} &  H^{\otimes 2}X^{\otimes 2}CZ
\end{bmatrix}
 \begin{bmatrix}
	X^{\otimes 2}H^{\otimes 2} & -X^{\otimes 2}H^{\otimes 2} \\
	X^{\otimes 2}H^{\otimes 2} & X^{\otimes 2}H^{\otimes 2} 
\end{bmatrix}\nonumber\\
=& 
\frac{1}{2} \begin{bmatrix}
	 B & -B\\
	 -B & B
\end{bmatrix},
\end{align}
where 
$B = H^{\otimes 2}X^{\otimes 2}X^{\otimes 2}H^{\otimes 2} + H^{\otimes 2}X^{\otimes 2}CZ_2X^{\otimes 2}H^{\otimes 2}.$

We kown:
\begin{align}
H^{\otimes 2}X^{\otimes 2}X^{\otimes 2}H^{\otimes 2}&= I_{4 \times 4},\nonumber \\
H^{\otimes 2}X^{\otimes 2}CZX^{\otimes 2}H^{\otimes 2} &= (1-2A)_{4 \times 4}, 
\end{align}
so:
\begin{align}
H^{\otimes 3} X^{\otimes 3}CZ_3 X^{\otimes 3} H^{\otimes 3} &= \frac{1}{2}\begin{bmatrix}
	I_{4 \times 4}+(I-2A)_{4 \times 4} & -I_{4 \times 4}+(I-2A)_{4 \times 4}\\
	-I_{4 \times 4}+(I-2A)_{4 \times 4} & -I_{4 \times 4}+(I-2A)_{4 \times 4}
\end{bmatrix}\nonumber\\
&= \frac{1}{2}\begin{bmatrix}
	2(I-A)_{4 \times 4} & -(2A)_{4 \times 4}\\
	-(2A)_{4 \times 4} & 2(I-A)_{4 \times 4}
\end{bmatrix}\nonumber\\
&= I_{8 \times 8}-\begin{bmatrix}
	A_{4 \times 4} & A_{4 \times 4}\\
	A_{4 \times 4} & A_{4 \times 4}
\end{bmatrix}\nonumber\\
&= I_{8 \times 8}-2\begin{bmatrix}
	A_{4 \times 4}/2 & A_{4 \times 4}/2\\
	A_{4 \times 4}/2 & A_{4 \times 4}/2
\end{bmatrix}\nonumber\\
&= I_{8 \times 8} -2A_{8 \times 8} = (I-2A)_{8 \times 8}.
\end{align}



For the amplitude amplification step in Grover's algorithm for an \( n \)-qubit system, the key operation is the inversion about the average, which can be implemented using a specific sequence of quantum gates. This sequence is \( H^{\otimes n}X^{\otimes n}CZ_nX^{\otimes n}H^{\otimes n} \), where \( CZ_n \) represents an \( n \)-qubit controlled \( Z \) gate.

In this context, \( CZ_n \) is a gate where the first \( n-1 \) qubits are control qubits, and the \( n \)-th qubit is the target. The \( CZ_n \) gate performs a phase inversion only if all qubits are in the \( |1\rangle \) state. In matrix terms, \( CZ_n \) is a \( 2^n \times 2^n \) matrix where all off-diagonal elements are 0, and all diagonal elements are 1 except for the last element, which is -1, i.e., \( CZ_n[2^n-1][2^n-1] = -1 \).

Using this gate sequence, we can effectively implement the \( (I - 2A) \) operation required in Grover's algorithm. The Hadamard gates \( H^{\otimes n} \) create a superposition, the \( X^{\otimes n} \) gates prepare the state for the controlled phase flip, and the \( CZ_n \) gate applies the phase flip \cite{4_sx}. Finally, the \( X^{\otimes n} \) and \( H^{\otimes n} \) gates reverse the initial state preparation. This can be mathematically verified and shown to be equivalent to the \( (I - 2A)_{2^n \times 2^n} \) matrix.

\section{Experiment of Grover's algorithm}

The real-machine demonstration experiment of the Grover algorithm is designed to provide students with a hands-on and intuitive understanding of the operational mechanisms of parallelism and coherence on quantum bits in the Grover algorithm. By observing the actual execution process, students can gain a more concrete insight into why quantum computing exhibits extraordinary computational speed in certain problems.

Furthermore, the experiment offers students an interactive opportunity to actively participate in the real-time operation of the Grover algorithm. This participatory experience not only enhances the enjoyment of learning but also deepens students' comprehension of the principles of quantum computing.

\begin{figure}[H]
	\centering
	\includegraphics[scale=0.4]{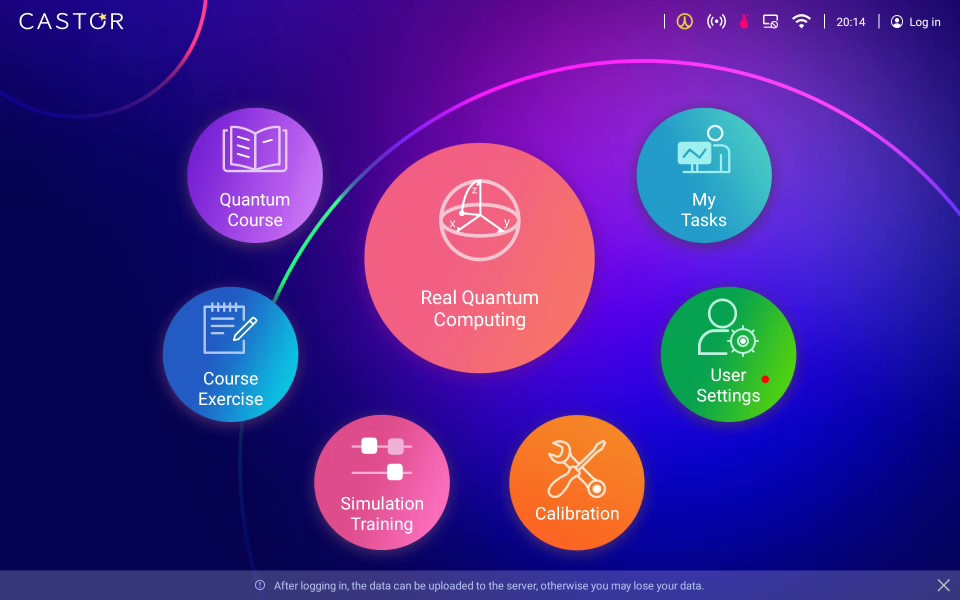}
	\caption{\label{6_21} Display interface of a real quantum computer.}
\end{figure}

As shown in Figure \ref{6_21}. Here describes entering the “Real Quantum Computing" section of the quantum circuit simulator. This allows running circuits on real quantum processors.

The ``Real Quantum Computing" feature allows executing quantum circuits on actual quantum processing units. It provides a list of available hardware backends to select as the target.

The processors have varying specifications like number of qubits, connectivity, and gate error rates \cite{4_sx}. More advanced processors can run larger circuits with higher fidelities. The compiler optimizes circuits to run on the selected processor.

After choosing a processor, users can execute pre-built circuits or their own designed circuits. The results reflect real quantum noise, interference, and errors. Debugging tools help characterize hardware performance and improve circuits.

Access to real hardware is critical for quantum programmers and researchers. Real-world experience identifies practical challenges not seen in simulators. It builds intuition for classical-quantum interactions. As quantum processors advance, hands-on access will catalyze practical quantum applications.

\begin{figure}[H]
	\centering
	\includegraphics[scale=0.4]{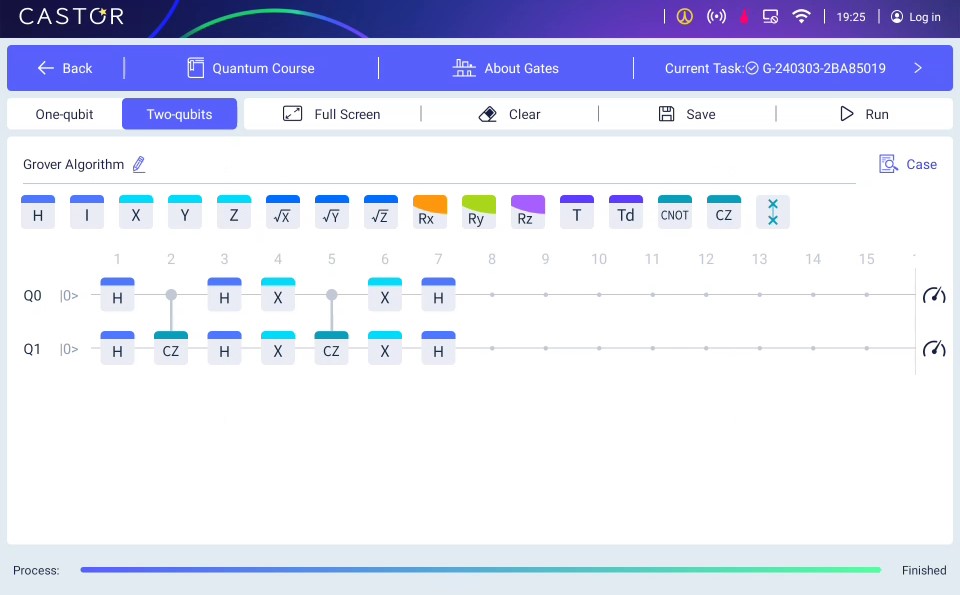}
	\caption{\label{6_22} Select the experimental case of Grover Algorithm.}
\end{figure}

\begin{figure}[H]
	\centering
	\includegraphics[scale=0.7]{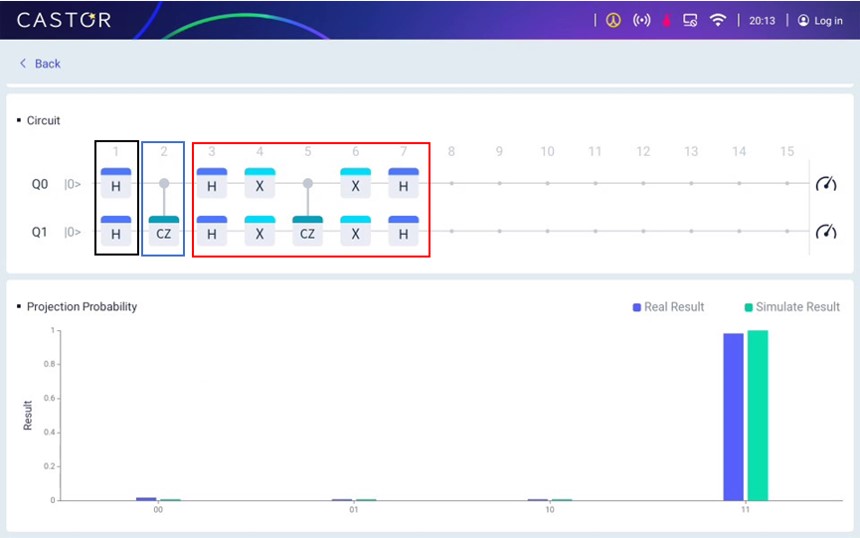}
	\caption{\label{66} The implementation of the Grover algorithm on SpinQ Gemini mini.}
\end{figure}

As shown in Figure \ref{6_22}. It describes selecting a two-qubit quantum circuit like Grover's algorithm \cite{4_sx,2_15} in the simulator, clicking ``Run" to execute it on a real quantum processor, and observing the results.
In Figure \ref{66}, the algorithm utilizes a quantum system with three main components, represented by boxes in the circuit diagram:
\begin{itemize}
    \item $H^{\otimes 2}$(Black box): This represents the preparation of superposition for data input.
    \item $U_f$(Blue box): This represents the Oracle or function that identifies the desired solution. It is the diffusion operator that amplifies the amplitude of the marked state.
    \item $-I + 2A$(Red box): This inverts the amplitudes about their average.
\end{itemize}

The gates in the circuit diagram correspond to these key algorithmic components. By wiring together quantum logic gates to realize these operators, the quantum computer can exploit superposition and interference to search the state space efficiently. However, it is worth noting that due to the presence of noise, experiments conducted on actual quantum computers often yield results that differ from those obtained on simulators.

The objective of the 3-qubit Grover's algorithm experiment is to demonstrate on a quantum computer that Grover's algorithm can effectively locate a predetermined target state $\ket{111}$. The experiment begins with the initialization of three qubits into a uniform superposition state, achieved by applying a Hadamard gate to each qubit. Following this, an Oracle(CCZ, the blue part) \cite{4_sx} operation is used to perform a phase flip on the target state $\ket{111}$, while leaving the phases of other states unchanged. Subsequently, a Grover diffusion transform($-I+2A$, the red part) is applied, designed to amplify the probability amplitude of the target state. In a 3-qubit system, after one iteration of Grover's algorithm (which includes one Oracle operation and one diffusion transform) the probability of the target state is already expected to increase. Theoretically, after two iterations, the probability of the target state $\ket{111}$ should be near its maximum. 
\begin{figure}[H]
	\centering
	\includegraphics[scale=0.3]{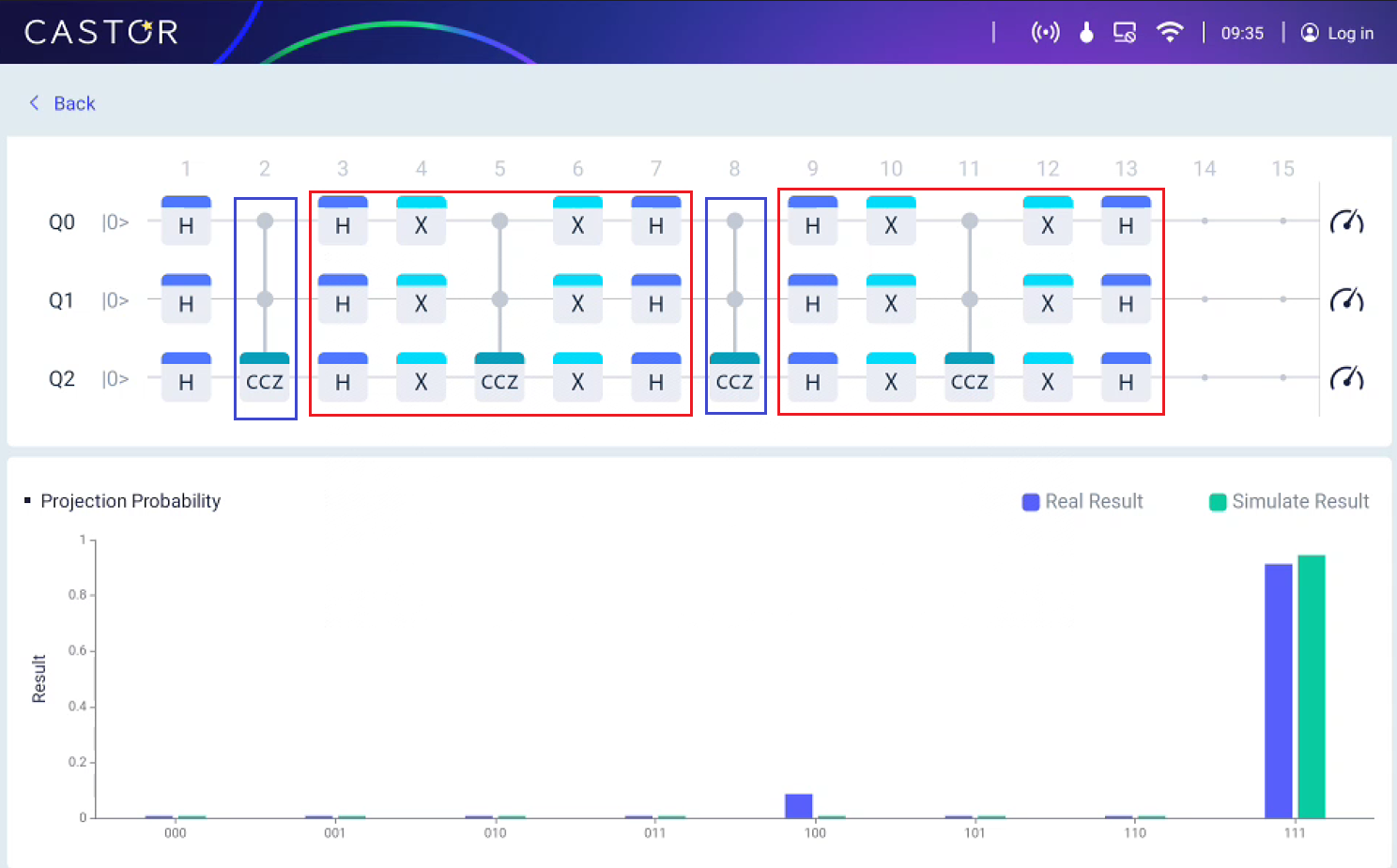}
	\caption{\label{grover3} The implementation of the 3-qubit Grover algorithm on SpinQ Triangulum mini.}
\end{figure}

In the experiment, as shown in Figure~\ref{grover3}, we directly performed two Grover iterations on the SpinQ Triangulum mini, and even in the presence of noise, measurements indicated that the probability of the target state $\ket{111}$ remained significantly higher than that of the other states, confirming that Grover's algorithm can rapidly locate the target solution.

\newpage
\chapter{Week 5: Complex Numbers and Single-Qubit Gates}

Our previous exploration of quantum computing has primarily focused on topics that do not involve complex numbers, discussing the fundamental principles and algorithms of quantum computing based on real numbers. However, in order to gain a deeper understanding of advanced topics such as universal quantum computing and Shor's algorithm, we must introduce and master the concept of complex numbers. Complex numbers are the essential foundation for delving into the deeper and broader realms of quantum computing.

\section{Complex numbers}

\textbf{Basic definitions}

Within the realm of real numbers, the equation \( x^2 + 1 = 0 \) has no solution because no real number squared equals \( -1 \). To address this issue, we rewrite the equation as \( x^2 = -1 \). and introduce the imaginary unit $i$, defined as \(\sqrt{-1}\). With this definition, we find the solutions to the equation are \( x = \pm i \), where 
\( i^2 = -1 \) \cite{2_3, 2_5}.

The introduction of the imaginary unit allows us to extend the system of real numbers to include complex numbers, which can provide solutions to equations like \( x^2 + 1 = 0 \) that lack real solutions.

A complex number is typically expressed as \( c = a + bi \), where $a$ is the real part and $b$ is the imaginary part, both of which are real numbers.

\textbf{The Fundamental Theorem of Algebra} 

This theorem asserts that every polynomial equation with complex coefficients has at least one complex solution \cite{2_17}. It establishes the fundamental role of complex numbers in solving algebraic equations and underscores their significance and wide-ranging applications in mathematical and physical theories, including quantum mechanics.

\begin{example}
    Verify that the complex number \( 2 + 3i \) is a solution for the polynomial equation \( x^2 - 4x + 13 = 0 \).

    \textbf{Solution:}
Substitute \( x = 2 + 3i \) into the equation:
\begin{align*}
&(2 + 3i)^2 - 4(2 + 3i) + 13 \\
&= (4 + 12i + 9i^2) - 8 - 12i + 13 \\
&= 4 + 12i - 9 - 8 - 12i + 13 \\
&= (4 - 9 - 8 + 13) + (12i - 12i) \\
&= 0 + 0i \\
&= 0.
\end{align*}
Since the equation is satisfied, \( 2 + 3i \) is indeed a solution to the polynomial equation \( x^2 - 4x + 13 = 0 \).
\end{example}

\textbf{The Structure of Complex Number Algebra}

Complex numbers can be expressed as ordered pairs \( (a, b) \), where $a$ represents the real part and $b$ represents the imaginary part \cite{2_4}. This representation not only facilitates an intuitive understanding of complex numbers but also simplifies the process of performing arithmetic operations with them.

\textbf{Addition of Complex Numbers:}

When we add two complex numbers, the rule is to add their real and imaginary parts separately \cite{2_18}. For instance, if we have two complex numbers \( (a_1, b_1) \) and \( (a_2, b_2) \),  their sum is calculated as: \[(a_1 + a_2, b_1 + b_2). \]

\textbf{Multiplication of Complex Numbers:}

The multiplication of complex numbers is based on the distributive law \cite{2_3,2_18}. Given two complex numbers \( (a_1, b_1) \) and \( (a_2, b_2) \),  their product is calculated by the following formula: \[ (a_1a_2 - b_1b_2, a_1b_2 + a_2b_1). \]

This operation takes into account the property that the square of the imaginary unit $i$ is $-1$, ensuring that the outcome remains in the form of a complex number.

\begin{example}
   Consider two complex numbers \( c_1 = (4, -3) \) and \( c_2 = (2, 1) \). Compute the product \( c_1 \times c_2 \).

\textbf{Solution:}
Using the multiplication formula:
\begin{align*}
    c_1 \times c_2 &= (4, -3) \times (2, 1) \\
                  &= (4 \times 2 - (-3) \times 1, 4 \times 1 + 2 \times (-3)) \\
                  &= (8 + 3, 4 - 6) \\
                  &= (11, -2).
\end{align*}
Therefore, the product of \( c_1 \) and \( c_2 \) is \( (11, -2) \).
\end{example}

We continue our exploration of complex numbers in ordered pair notation, focusing on their key properties and operations \cite{2_4,2_5,2_18}.

\begin{enumerate}
    \item \textit{Commutativity:} Both addition and multiplication of complex numbers are commutative, meaning that the order of the numbers does not affect the result:
    \[ c_1 + c_2 = c_2 + c_1, \quad c_1 \times c_2 = c_2 \times c_1. \]

    \item \textit{Associativity:} The operations of addition and multiplication are associative:
    \[ (c_1 + c_2) + c_3 = c_1 + (c_2 + c_3), \quad (c_1 \times c_2) \times c_3 = c_1 \times (c_2 \times c_3). \]

    \item \textit{Distributivity:} The distributive property holds for multiplication over addition:
    \[ c_1 \times (c_2 + c_3) = (c_1 \times c_2) + (c_1 \times c_3). \]
\end{enumerate}

Now we look at subtraction and division of complex numbers \cite{2_3}.
\begin{itemize}
    \item \textit{Subtraction:} Subtraction is carried out by separately subtracting the real and imaginary parts:
    \[ (a_1, b_1) - (a_2, b_2) = (a_1 - a_2, b_1 - b_2). \]

    \item \textit{Division:} Division involves rationalizing the denominator and separating the result into real and imaginary parts. For complex numbers \( (a_1, b_1) \) and \( (a_2, b_2) \), division is performed as follows:
    \[ \frac{(a_1, b_1)}{(a_2, b_2)} = \frac{a_1a_2 + b_1b_2}{a_2^2 + b_2^2} + \frac{a_2b_1 - a_1b_2}{a_2^2 + b_2^2}i. \]
\end{itemize}

\begin{example}
   Let \( c_1 = -2 + 3i \) and \( c_2 = 2 - i \). Compute \( \frac{c_1}{c_2} \).

\textbf{Solution:}
Using the division formula:
\begin{align*}
    \frac{c_1}{c_2} &= \frac{(-2 + 3i)}{(2 - i)} \\
                    &= \frac{(-2)(2) + (3)(-1) + ((3)(2) - (-2)(-1))i}{(2)^2 + (-1)^2} \\
                    &= \frac{-4 - 3 + (6 - 2)i}{4 + 1} \\
                    &= \frac{-7 + 4i}{5} \\
                    &= -\frac{7}{5} + \frac{4}{5}i.
\end{align*}
Therefore, \( \frac{c_1}{c_2} = -\frac{7}{5} + \frac{4}{5}i \).
\end{example}

The \textbf{modulus} of a complex number \( c = a + bi \) is a crucial concept for understanding its properties. It is defined as \( |c| = |a + bi| = \sqrt{a^2 + b^2} \), measuring the `size' or `length' of the complex number from the origin in the complex plane. For a real number \( a \), \( |a| \) represents its absolute value.

\begin{example}
Consider the complex number \( c = 3 - 2i \). The modulus of \( c \) is calculated as:
\[ |c| = \sqrt{3^2 + (-2)^2} = \sqrt{9 + 4} = \sqrt{13}. \]
Thus, the modulus of \( c \) is \( \sqrt{13} \).
\end{example}

\textbf{Complex conjugation} is another fundamental concept. The complex conjugate of a complex number \( c = a + bi \) is denoted as \( \overline{c} \) and is defined as \( \overline{c} = a - bi \), involving only the change in sign of the imaginary part \cite{2_18}.

\begin{example}
Let \( c_1 = -3 + 2i \). The complex conjugate of \( c_1 \) is:
\[ \overline{c_1} = -3 - 2i. \]
Hence, the complex conjugate of \( c_1 \) is \( -3 - 2i \).
\end{example}

 The modulus and complex conjugation is vital in complex number theory and they play a significant role in the field of quantum computing.

\textbf{The geometry of complex numbers}

The geometry of complex numbers allows for a visual representation of these numbers in the complex plane, where they can be depicted as vectors. The modulus of a complex number corresponds to the length of this vector \cite{2_5}.

\begin{example}
Consider the complex number \( c = 9 + 4i \). This number can be represented as a vector from the origin to the point (9, 4) in the complex plane. Using the Pythagorean theorem, the length of this vector, which is the modulus of \( c \), is calculated as:
\[ |c| = \sqrt{9^2 + 4^2} = \sqrt{97}. \]
This result aligns with the algebraic calculation of the modulus:
\[ |c| = |9 + 4i| = \sqrt{9^2 + 4^2} = \sqrt{97}. \]
\end{example}

\begin{figure}[H]
    \centering
    \includegraphics[scale=0.6]{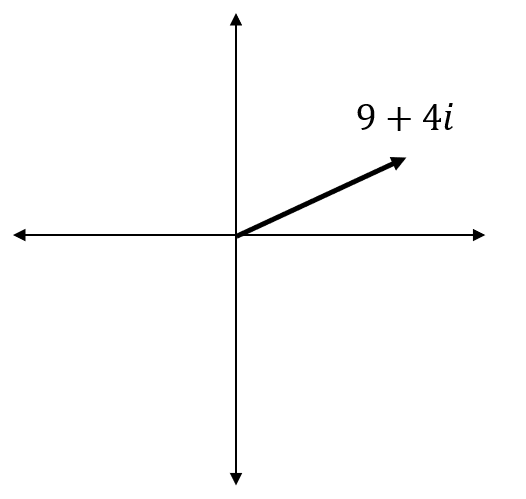}
    \caption{Representation of the vector \( 9 + 4i \) on coordinate axes.}
    \label{17_vector_length}
\end{figure}

A complex number can be uniquely represented by its modulus \( \rho \) and angle \( \theta \). For a complex number in ordered pair form \( (a, b) \), the modulus \( \rho \) is:
\begin{equation}
    \rho = \sqrt{a^2 + b^2}.
\end{equation}
The angle \( \theta \) is given by:
\begin{equation}
    \theta = \tan^{-1}\left(\frac{b}{a}\right), \quad \text{for } a > 0.
\end{equation}

\begin{example}
Let \( c = 2 - 2i \). Find its polar representation.

\textbf{Solution:}
The modulus of \( c \) is 
\[
\rho = \sqrt{2^2 + (-2)^2} = \sqrt{8} = 2\sqrt{2}.
\]
The angle \( \theta \) is 
\[
\theta = \tan^{-1}\left(\frac{-2}{2}\right) = \tan^{-1}(-1) = \frac{7 \pi}{4} \quad (\text{since the angle is in the fourth quadrant}).
\]
Thus, the polar representation of \( c \) is 
\[
(\rho, \theta) = \left(2\sqrt{2}, \frac{7 \pi}{4}\right).
\]
\end{example}

The polar representation of complex numbers is defined by a magnitude \( \rho \geq 0 \) and a phase \( \theta \), where \( 0 \leq \theta < 2\pi \). The phase angle \( \theta \) is periodic with a period of \( 2\pi \), meaning \( \theta_1 = \theta_2 \) if and only if \( \theta_2 = \theta_1 + 2\pi k \) for some integer \( k \).

\begin{example}
Are the numbers \( (5,\pi) \) and \( (5,-\pi) \) the same?

\textbf{Solution:}
Since \( -\pi \) and \( \pi \) differ by \( 2\pi \) (one full rotation in the complex plane), they represent the same angle. Therefore, \( (5,\pi) \) and \( (5,-\pi) \) are the same in polar coordinates.
\end{example}

Multiplying complex numbers in polar form is straightforward \cite{2_18}. Given two complex numbers in polar coordinates \( (\rho_1, \theta_1) \) and \( (\rho_2, \theta_2) \), their product is given by:
\begin{equation}
(\rho_1, \theta_1) \times (\rho_2, \theta_2) = (\rho_1\rho_2, \theta_1 + \theta_2).
\end{equation}

\begin{example}
Let \( c_1 = \sqrt{3} + i \) and \( c_2 = -\sqrt{3} + i \). Compute \( c_1c_2 \) using polar representation.

\textbf{Solution:}
First, convert \( c_1 \) and \( c_2 \) to polar form. \( c_1 = \sqrt{3} + i \) has \( \rho_1 = \sqrt{(\sqrt{3})^2 + 1^2} = 2 \) and \( \theta_1 = \tan^{-1}\left(\frac{1}{\sqrt{3}}\right) = \frac{\pi}{6} \). Similarly, \( c_2 = -\sqrt{3} + i \) has \( \rho_2 = \sqrt{(-\sqrt{3})^2 + 1^2} = 2 \) and \( \theta_2 = \tan^{-1}\left(\frac{1}{-\sqrt{3}}\right) =\frac{5\pi}{6} \) since it is in the second quadrant. The product in polar form is:
\[
(\rho_1\rho_2, \theta_1 + \theta_2) = (2 \times 2, \frac{\pi}{6} + \frac{5\pi}{6}) = (4, \pi).
\]
Hence, the product \( c_1c_2 \) in polar representation is \( (4, \pi) \).
\end{example}

Dividing complex numbers in polar form is similar. Given \( c_1 = (\rho_1, \theta_1) \) and \( c_2 = (\rho_2, \theta_2) \), their quotient is:
\begin{equation}
\frac{c_1}{c_2} = \left(\frac{\rho_1}{\rho_2}, \theta_1 - \theta_2\right).
\end{equation}

\begin{example}
Let \( c_1 = -1 + i \) and \( c_2 = -1 - i \). Compute \( \frac{c_1}{c_2} \) using polar representation.

\textbf{Solution:}
Convert \( c_1 \) and \( c_2 \) to polar form. \( c_1 = -1 + i \) has \( \rho_1 = \sqrt{(-1)^2 + 1^2} = \sqrt{2} \) and \( \theta_1 = \tan^{-1}\left(\frac{1}{-1}\right) = \frac{3\pi}{4} \). For \( c_2 = -1 - i \), \( \rho_2 = \sqrt{(-1)^2 + (-1)^2} = \sqrt{2} \) and \( \theta_2 = \tan^{-1}\left(\frac{-1}{-1}\right) = \frac{5 \pi}{4} \). The quotient is:
\[ \left(\frac{\rho_1}{\rho_2}, \theta_1 - \theta_2\right) = \left(\frac{\sqrt{2}}{\sqrt{2}}, \frac{3\pi}{4} - \frac{5 \pi}{4}\right) = \left(1, -\frac{\pi}{2}\right). \]
\end{example}

Finding the roots of a complex number in polar form involves determining the nth roots. If \( c = (\rho, \theta) \), then the \( n^{th} \) roots of \( c \) are given by:
\[ c^{\frac{1}{n}} = \left(\rho^{\frac{1}{n}}, \frac{1}{n}(\theta + k2\pi)\right), \ k=0,1,2,...,n-1. \]

\begin{example}
Find all the cube roots of \( c = \sqrt{3} + i \).

\textbf{Solution:}
Convert \( c \) to polar form: \( \rho = \sqrt{\sqrt{3}^2 + 1^2} = 2 \), and \( \theta = \tan^{-1}\left(\frac{1}{\sqrt{3}}\right) = \frac{\pi}{6} \). The cube roots of \( c \) are given by \( c^{\frac{1}{3}} \):
\[ c^{\frac{1}{3}} = \left(2^{\frac{1}{3}}, \frac{1}{3}\left(\frac{\pi}{6}  + 2\pi k\right)\right), \ k=0,1,2. \]
\end{example}

To understand the exponential form of complex numbers, let's start with their polar form representation:
\begin{equation}
c = \rho(\cos(\theta) + i\sin(\theta)).
\end{equation}
Euler's formula, a powerful and elegant link between trigonometry and complex exponentials, is expressed as:
\begin{equation}
e^{i\theta} = \cos(\theta) + i\sin(\theta).
\end{equation}

This formula, discovered by Leonhard Euler, shows that complex exponentials can be represented using sine and cosine functions. Even without a background in calculus, you can appreciate its beauty and utility in simplifying complex number calculations.

Using Euler's formula, we can rewrite the complex number in its exponential form:
\begin{equation}
c = \rho e^{i\theta}.
\end{equation}
This form is particularly useful because it compactly encapsulates both the magnitude (\(\rho\)) and the phase (\(\theta\)) of the complex number.

The conjugate of a complex number in exponential form is straightforward. Given \(c = \rho e^{i\theta}\), its conjugate is:
\begin{equation}
\overline{c} = \rho e^{-i\theta}.
\end{equation}
The negative sign in the exponent reflects the change in the direction of the angle, which is characteristic of conjugation in complex numbers.

Multiplication of complex numbers in exponential form is also simplified. Given two complex numbers \(c_1 = \rho_1 e^{i\theta_1}\) and \(c_2 = \rho_2 e^{i\theta_2}\), their product is:
\begin{equation}
c_1c_2 = \rho_1 e^{i\theta_1}\rho_2 e^{i\theta_2} = \rho_1\rho_2 e^{i(\theta_1 + \theta_2)}.
\end{equation}
This formula shows how the magnitudes multiply while the angles add, making computations more intuitive and straightforward.

The \( n^{th} \) roots of unity are solutions to the equation \( x^n = 1 \). These are the \( n \) distinct complex numbers which, when raised to the \( n^{th} \) power, equal one. 

Starting with the complex number \( c = (1, 0) \) in polar form, its \( n^{th} \) roots are given by:
\begin{equation}
c^{\frac{1}{n}} = (1, \frac{2\pi k}{n}), \quad k=0,1,2,...,n-1.
\end{equation}
In exponential form, the \( k^{th} \) root of unity is expressed as:
\begin{equation}
\omega_k = e^{\frac{2\pi i k}{n}}.
\end{equation}
The \( n \) roots of unity are thus denoted as:
\begin{equation}
\omega_0 = 1, \omega_1, \omega_2,...,\omega_{n-1}.
\end{equation}
These roots of unity are evenly spaced around the unit circle in the complex plane, forming the vertices of a regular \( n \)-sided polygon.

\begin{figure}[H]
    \centering
    \includegraphics[scale=0.6]{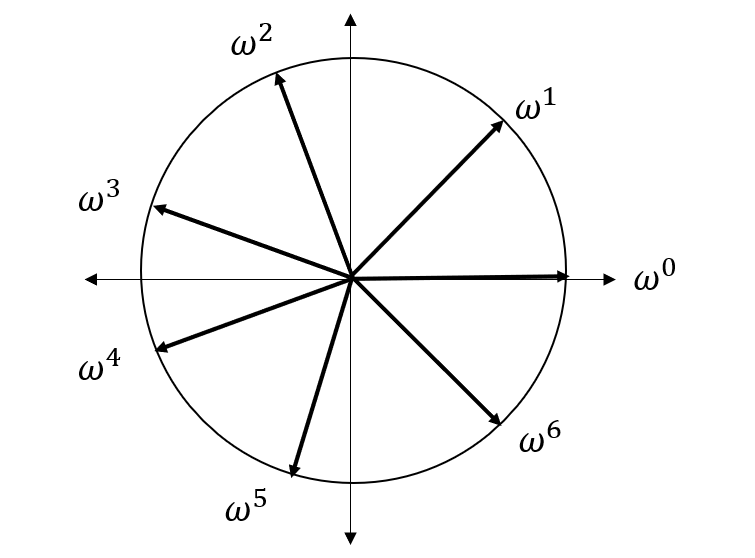}
    \caption{The seventh root of unity and its powers, representing a regular heptagon on the unit circle.}
    \label{17_root_of_unity}
\end{figure}

\begin{example}
Find all the fourth roots of unity.

\textbf{Solution:}
The fourth roots of unity are solutions to \( x^4 = 1 \). Using the formula for the \( k^{th} \) root of unity, we have:
\begin{align*}
\omega_0 &= e^{\frac{2\pi i \times 0}{4}} = 1, \\
\omega_1 &= e^{\frac{2\pi i \times 1}{4}} = i, \\
\omega_2 &= e^{\frac{2\pi i \times 2}{4}} = -1, \\
\omega_3 &= e^{\frac{2\pi i \times 3}{4}} = -i.
\end{align*}
Thus, the fourth roots of unity are \( 1, i, -1, \) and \( -i \).
\end{example}

\section{Bloch sphere}~{}

The Bloch sphere is a powerful tool for visualizing the state of a qubit \cite{4_sx}. A qubit, the fundamental unit of quantum information, can be represented as a linear combination of its basis states \( |0\rangle \) and \( |1\rangle \):
\begin{equation}
|\psi\rangle = a|0\rangle + b|1\rangle,
\end{equation}
where \( a \) and \( b \) are complex coefficients. In quantum mechanics, it's essential that the state of a qubit is normalized, meaning the sum of the probabilities (squared magnitudes of the coefficients) must equal 1:
\begin{equation}
|a|^2 + |b|^2 = 1.
\end{equation}

Normalization ensures that the total probability of finding the qubit in either of the states \( |0\rangle \) or \( |1\rangle \) is 1. The coefficients \( a \) and \( b \) can be parameterized in terms of angles \( \theta \), \( \alpha \), and \( \beta \) as follows: \( a = \cos(\frac{\theta}{2})e^{i\alpha} \) and \( b = \sin(\frac{\theta}{2})e^{i\beta} \). However, we often represent \( b \) using a single phase factor \( \phi \), where \( \phi = \beta - \alpha \), simplifying the representation to \cite{4_sx}:
\begin{equation}
|\psi\rangle = \cos(\frac{\theta}{2})|0\rangle + e^{i\phi}\sin(\frac{\theta}{2})|1\rangle,
\end{equation}
where \( \theta \) and \( \phi \) determine the point on the Bloch sphere representing the qubit state.

\begin{figure}[H]
    \centering
    \includegraphics[scale=0.4]{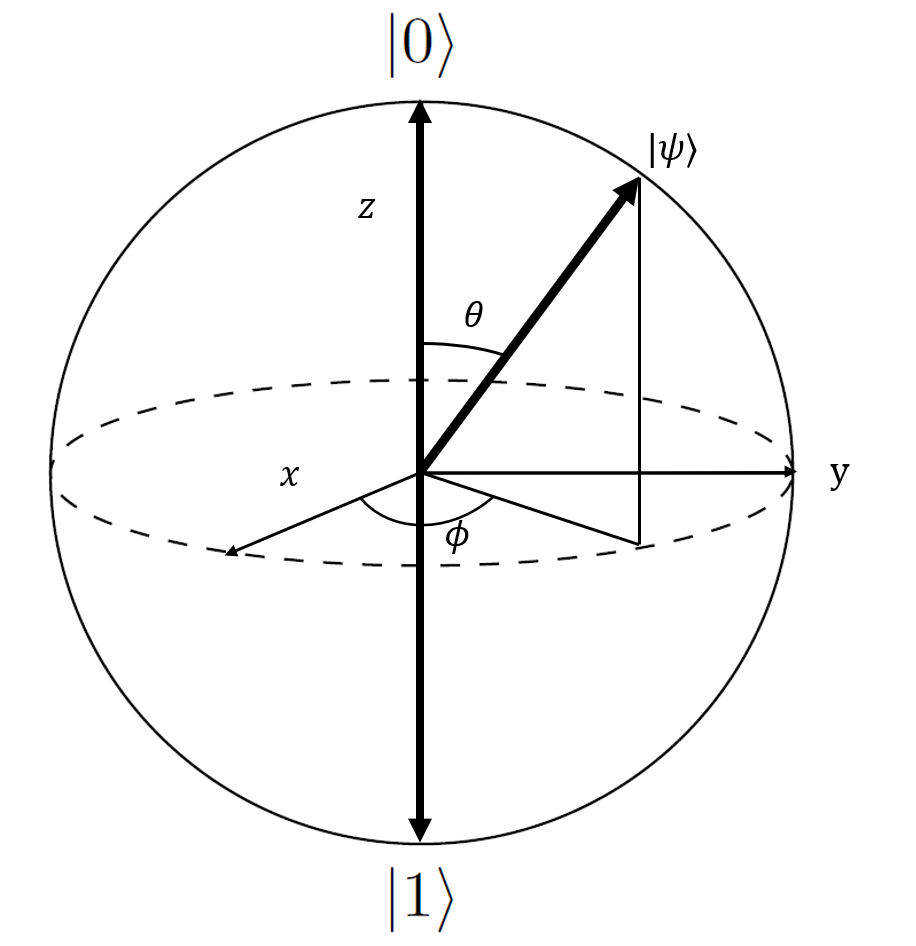}
    \caption{Representation of quantum states on the Bloch sphere.}
    \label{17_bloch_sphere}
\end{figure}

In the Bloch sphere, shown in Figure \ref{17_bloch_sphere}, every point on the surface of the sphere represents a possible state of the qubit. The angles \( \theta \) and \( \phi \) in the expression of \( |\psi\rangle \) correspond to the spherical coordinates on the Bloch sphere, with \( 0 \leq \phi < 2\pi \) and \( 0 \leq \theta \leq \pi \).

The Bloch sphere is a valuable visualization tool in quantum computing, representing the state of a qubit in a way that is easy to conceptualize, especially when discussing quantum gates and their effects on qubit states.

Quantum gates are fundamental in quantum computing as they represent reversible operations that manipulate qubit states \cite{2_2}. Understanding how these gates are represented on the Bloch sphere helps us visualize their effects on qubits.

Consider the phase shift gate \( R(\theta) \), which rotates the state vector about the \( z \)-axis. Its matrix representation is \cite{4_sx}:
\begin{equation}
R(\theta) =
\begin{bmatrix}
    1 & 0 \\
    0 & e^{i\theta}
\end{bmatrix}.
\end{equation}
When this gate acts on a qubit state \( |\psi\rangle = \cos(\frac{\theta'}{2})|0\rangle + e^{i\phi}\sin(\frac{\theta'}{2})|1\rangle \), it rotates the relative phase between the basis states \( |0\rangle \) and \( |1\rangle \):
\begin{align}
R(\theta)|\psi\rangle &= \cos(\frac{\theta'}{2})|0\rangle + e^{i\theta}e^{i\phi}\sin(\frac{\theta'}{2})|1\rangle  \nonumber\\
&= \begin{bmatrix}
    1 & 0 \\
    0 & e^{i\theta}
\end{bmatrix}  
\begin{bmatrix}
    \cos(\frac{\theta'}{2}) \\
    e^{i\phi}\sin(\frac{\theta'}{2})
\end{bmatrix}  
= \begin{bmatrix}
    \cos(\frac{\theta'}{2}) \\
    e^{i\theta}e^{i\phi}\sin(\frac{\theta'}{2})
\end{bmatrix}.
\end{align}

\begin{figure}[H]
    \centering
    \includegraphics[scale=0.3]{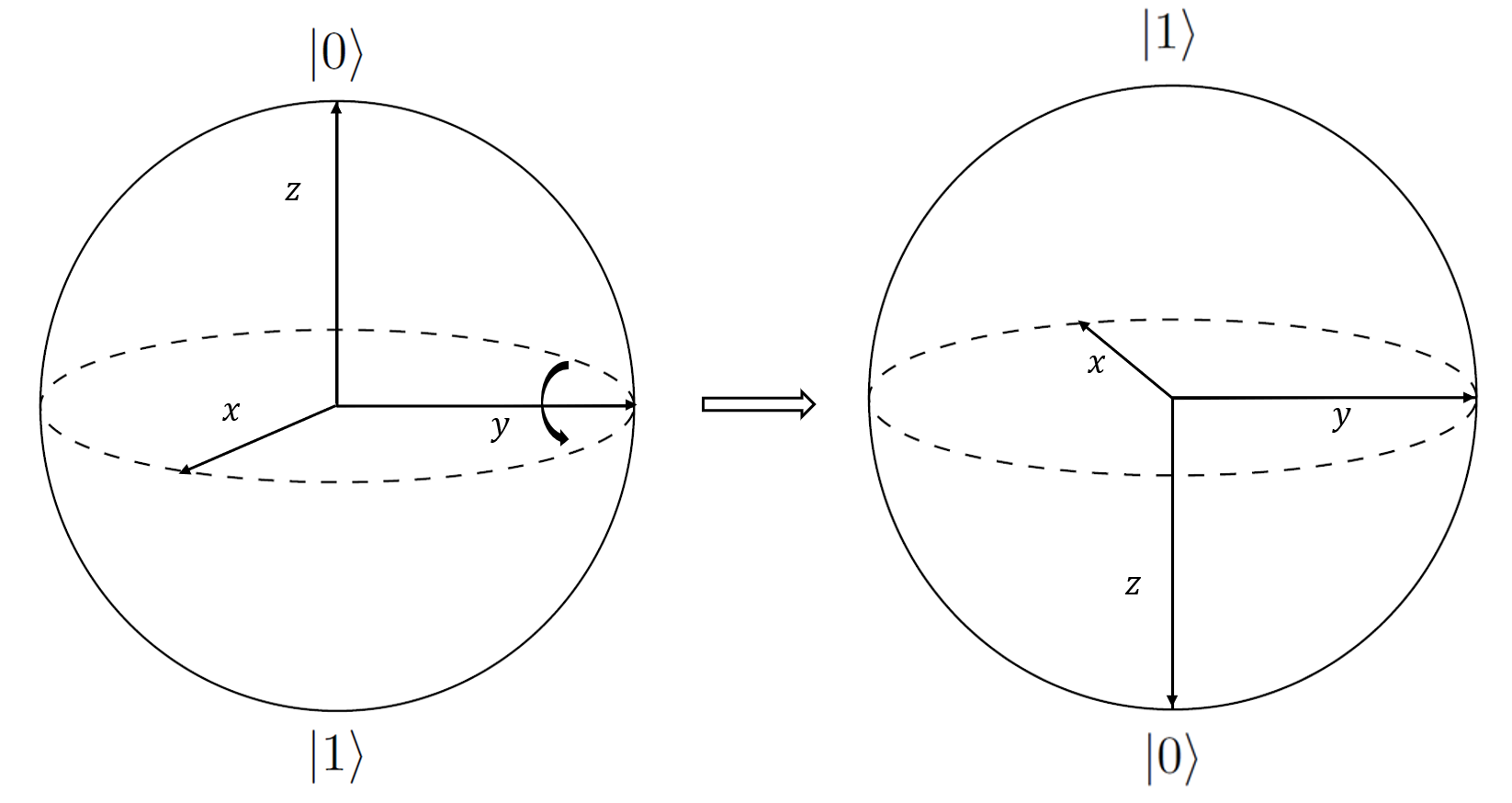}
    \caption{Quantum state rotation around the \( y \) axis on the Bloch sphere.}
    \label{17_rotation_y}
\end{figure}

Next, consider the rotation operator \( R_y(\theta) \) which performs a rotation about the \( y \)-axis, affecting the latitude on the Bloch sphere while keeping the longitude fixed:
\begin{align}
R_y(\theta) =
\begin{bmatrix}
    \cos \frac{\theta}{2} & -\sin \frac{\theta}{2} \\
    \sin \frac{\theta}{2} & \cos \frac{\theta}{2}
\end{bmatrix}.
\end{align}
This rotation is depicted in Figure \ref{17_rotation_y}.

The matrix representations of qubit rotations about the \( x \), \( y \) and \( z \) axes on the Bloch sphere can be summarized as \cite{4_sx}:
\begin{align}
R_x(\theta) &= \cos \frac{\theta}{2} I - i\sin \frac{\theta}{2} X, \nonumber\\
R_y(\theta) &= \cos \frac{\theta}{2} I - i\sin \frac{\theta}{2} Y, \nonumber\\
R_z(\theta) &= \cos \frac{\theta}{2} I - i\sin \frac{\theta}{2} Z,
\end{align}
where \( X \), \( Y \) and \( Z \) are the Pauli matrices, representing rotations about the respective axes.

Euler's formula, \( e^{i\theta} = \cos(\theta) + i\sin(\theta) \), plays a crucial role in deriving the matrix representations of rotations in quantum mechanics \cite{2_3}. This formula allows us to express rotations about the \( x \), \( y \), and \( z \) axes on the Bloch sphere in terms of the Pauli matrices. The Pauli matrices are special matrices that represent rotations about these axes in quantum computing.

For rotation about the \( z \)-axis, the rotation operator \( R_z(\theta) \) is given by:
\begin{align}
R_z(\theta) &= e^{-i\frac{\theta}{2} Z}  \nonumber \\
&= \cos \frac{\theta}{2} \begin{bmatrix} 1 & 0 \\ 0 & 1 \end{bmatrix} - i\sin \frac{\theta}{2} \begin{bmatrix} 1 & 0 \\ 0 & -1 \end{bmatrix} \nonumber\\
&= \begin{bmatrix} e^{-i\frac{\theta}{2}} & 0 \\ 0 & e^{i\frac{\theta}{2}} \end{bmatrix}.
\end{align}
This matrix represents a rotation by an angle \( \theta \) around the \( z \)-axis.

Similarly, for the \( y \)-axis, \( R_y(\theta) \) is derived using Euler's formula:
\begin{align}
R_y(\theta) &= e^{-i\frac{\theta}{2} Y} \nonumber\\
&= \cos \frac{\theta}{2} \begin{bmatrix} 1 & 0 \\ 0 & 1 \end{bmatrix} - i\sin \frac{\theta}{2} \begin{bmatrix} 0 & -i \\ i & 0 \end{bmatrix} \nonumber\\
&= \begin{bmatrix} \cos{\frac{\theta}{2}} & -\sin{\frac{\theta}{2}} \\ \sin{\frac{\theta}{2}} & \cos{\frac{\theta}{2}} \end{bmatrix}.
\end{align}

For the \( x \)-axis, the rotation operator \( R_x(\theta) \) is obtained in a similar manner:
\begin{align}
R_x(\theta) &= e^{-i\frac{\theta}{2} X} \nonumber\\
&= \cos \frac{\theta}{2} \begin{bmatrix} 1 & 0 \\ 0 & 1 \end{bmatrix} - i\sin \frac{\theta}{2} \begin{bmatrix} 0 & 1 \\ 1 & 0 \end{bmatrix} \nonumber\\
&= \begin{bmatrix} \cos{\frac{\theta}{2}} & -i\sin{\frac{\theta}{2}} \\ -i\sin{\frac{\theta}{2}} & \cos{\frac{\theta}{2}} \end{bmatrix}.
\end{align}

These matrices \( R_x(\theta) \), \( R_y(\theta) \) and \( R_z(\theta) \) represent rotations of a qubit state around the respective axes on the Bloch sphere. Euler's formula provides a bridge between the trigonometric functions (cosine and sine) and complex exponentials \cite{2_5}, which is fundamental in quantum computing.

\begin{figure}[H]
    \centering
    \includegraphics[scale=0.3]{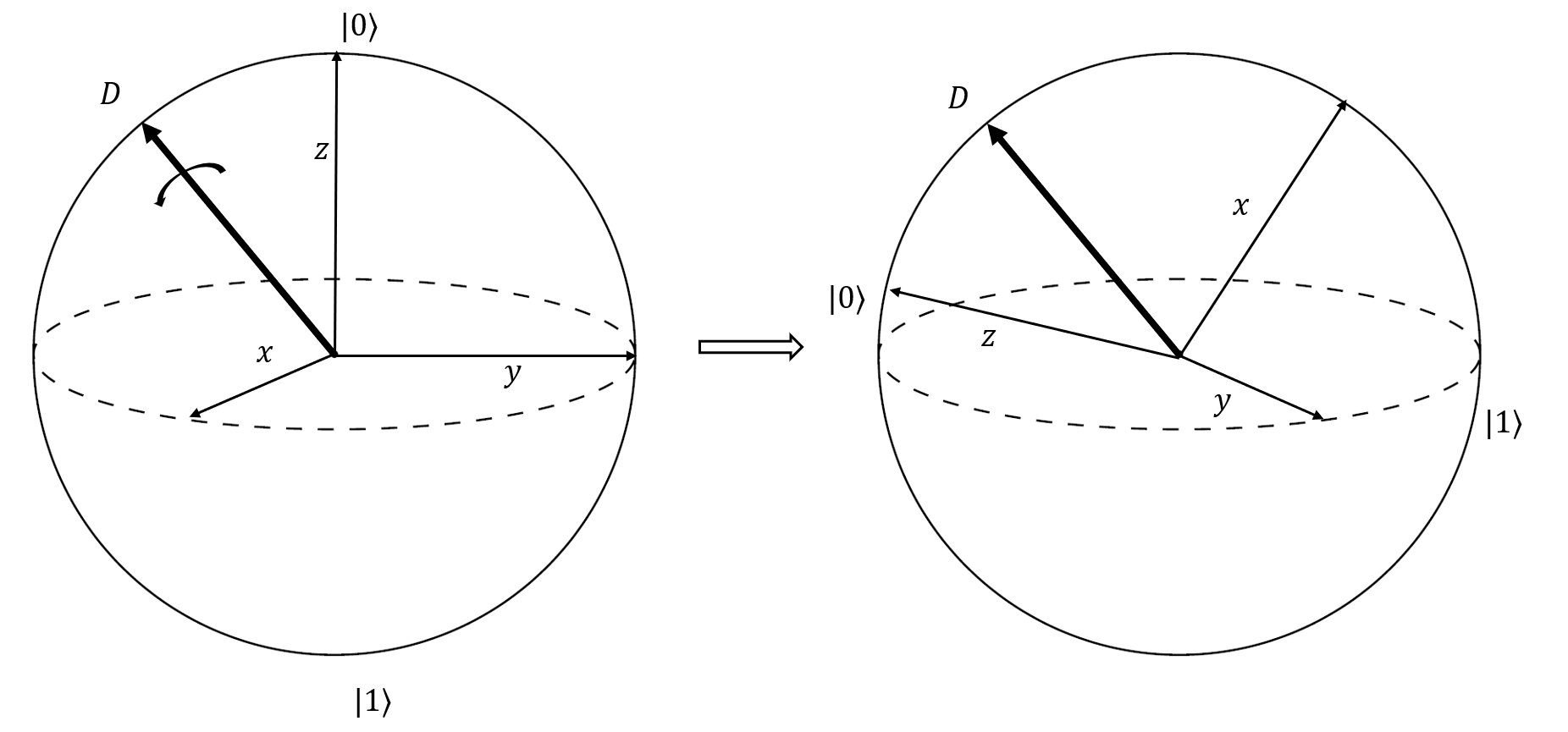}
    \caption{Quantum state rotation around an arbitrary axis \( D \) on the Bloch sphere.}
    \label{17_rotation_D}
\end{figure}

For any arbitrary axis \( D \), characterized by its directional cosines \( D_x \), \( D_y \) and \( D_z \), a rotation \( R_D(\theta) \) can be expressed as a linear combination of these three rotations. This is shown in Figure \ref{17_rotation_D}:
\begin{align}
    R_D(\theta) = \cos \frac{\theta}{2} I - i\sin \frac{\theta}{2} (D_x X + D_y Y + D_z Z).
\end{align}

These rotations on the Bloch sphere provide an intuitive understanding of how quantum gates manipulate the state of qubits in quantum computing.

\section{Single-qubit gates}

Single-qubit quantum gates play a crucial role in quantum computing. A single-qubit gate \( U \) can be expressed using matrix exponentials of the Pauli matrices \( Z \) and \( Y \). Specifically, for some real numbers \( \alpha, \beta, \gamma, \delta \) the gate \( U \) can be written as a combination of rotations around the \( Z \) and \( Y \) axes \cite{4_sx}:

\begin{align}
U = e^{i\alpha} R_z(\beta) R_y(\gamma) R_z(\delta),
\end{align}
where \( R_z(\theta) \) and \( R_y(\theta) \) are the rotation operators about the \( Z \) and \( Y \) axes, respectively. These rotation operators can be represented as:
\begin{align}
R_z(\theta) = e^{-i\frac{\theta}{2} Z} = 
\begin{bmatrix}
e^{-i\frac{\theta}{2}} & 0 \\
0 & e^{i\frac{\theta}{2}}
\end{bmatrix}, \quad
R_y(\theta) = e^{-i\frac{\theta}{2} Y} = 
\begin{bmatrix}
\cos{\frac{\theta}{2}} & -\sin{\frac{\theta}{2}} \\
\sin{\frac{\theta}{2}} & \cos{\frac{\theta}{2}}
\end{bmatrix}.
\end{align}

Using these expressions, we can derive the matrix form of \( U \) as:
\begin{align}
U &= e^{i\alpha} 
\begin{bmatrix}
e^{-i\frac{\beta}{2}} & 0 \\
0 & e^{i\frac{\beta}{2}}
\end{bmatrix}
\begin{bmatrix}
\cos{\frac{\gamma}{2}} & -\sin{\frac{\gamma}{2}} \\
\sin{\frac{\gamma}{2}} & \cos{\frac{\gamma}{2}}
\end{bmatrix}
\begin{bmatrix}
e^{-i\frac{\delta}{2}} & 0 \\
0 & e^{i\frac{\delta}{2}}
\end{bmatrix} \nonumber\\
&= e^{i\alpha} 
\begin{bmatrix}
e^{i(-\beta/2-\delta/2)}\cos{\frac{\gamma}{2}} & -e^{i(-\beta/2+\delta/2)}\sin{\frac{\gamma}{2}} \\
e^{i(\beta/2-\delta/2)}\sin{\frac{\gamma}{2}} & e^{i(\beta/2+\delta/2)}\cos{\frac{\gamma}{2}}
\end{bmatrix},
\end{align}
where \( \alpha, \beta, \gamma \) and \( \delta \) are angles that determine the rotation on the Bloch sphere.

This parametrization of \( U \) demonstrates that any single-qubit unitary transformation can be constructed from rotations about the \( Z \) and \( Y \) axes on the Bloch sphere. The angles \( \alpha, \beta, \gamma \) and \( \delta \) provide a comprehensive description of how the quantum state is transformed by \( U \).

Quantum gates are the building blocks of quantum circuits. Let's introduce some common single-qubit quantum gates and their matrix representations. We start with the three Pauli matrices, which represent the \( X \), \( Y \) and \( Z \) gates \cite{2_1}:

\begin{align}
X=
\begin{bmatrix}
0 & 1\\1&0
\end{bmatrix}, \quad
Y=
\begin{bmatrix}
0 & -i\\i&0
\end{bmatrix}, \quad
Z=
\begin{bmatrix}
1 & 0\\0&-1
\end{bmatrix}.
\end{align}

In addition to these, there are three other important single-qubit gates:
\begin{align}
H=\frac{1}{\sqrt{2}}
\begin{bmatrix}
1 & 1\\1&-1
\end{bmatrix}, \quad
S=
\begin{bmatrix}
1 & 0\\0&i
\end{bmatrix}, \quad
T=
\begin{bmatrix}
1 & 0\\0&e^{i\pi/4}
\end{bmatrix}.
\end{align}

\begin{example}
These quantum gates are closely related to each other. Let's prove the following relationships between these operations:

\begin{itemize}
    \item \( X^2 = Y^2 = Z^2 = I \):
    \begin{align*}
        X^2 &= \begin{bmatrix} 0 & 1 \\ 1 & 0 \end{bmatrix} \begin{bmatrix} 0 & 1 \\ 1 & 0 \end{bmatrix} = \begin{bmatrix} 1 & 0 \\ 0 & 1 \end{bmatrix} = I, \\
        Y^2 &= \begin{bmatrix} 0 & -i \\ i & 0 \end{bmatrix} \begin{bmatrix} 0 & -i \\ i & 0 \end{bmatrix} = \begin{bmatrix} 1 & 0 \\ 0 & 1 \end{bmatrix} = I, \\
        Z^2 &= \begin{bmatrix} 1 & 0 \\ 0 & -1 \end{bmatrix} \begin{bmatrix} 1 & 0 \\ 0 & -1 \end{bmatrix} = \begin{bmatrix} 1 & 0 \\ 0 & 1 \end{bmatrix} = I.
    \end{align*}

    \item \( H = \frac{1}{\sqrt{2}}(X+Z) \):
    \begin{align*}
        \frac{1}{\sqrt{2}}(X+Z) &= \frac{1}{\sqrt{2}}\left( \begin{bmatrix} 0 & 1 \\ 1 & 0 \end{bmatrix} + \begin{bmatrix} 1 & 0 \\ 0 & -1 \end{bmatrix} \right) \\
        &= \frac{1}{\sqrt{2}}\begin{bmatrix} 1 & 1 \\ 1 & -1 \end{bmatrix} = H.
    \end{align*}

    \item \( X = HZH \):
    \begin{align*}
        HZH &= \frac{1}{\sqrt{2}}\begin{bmatrix} 1 & 1 \\ 1 & -1 \end{bmatrix} \begin{bmatrix} 1 & 0 \\ 0 & -1 \end{bmatrix} \frac{1}{\sqrt{2}}\begin{bmatrix} 1 & 1 \\ 1 & -1 \end{bmatrix} \\
        &= \frac{1}{2}\begin{bmatrix} 1 & -1 \\ 1 & 1 \end{bmatrix} \begin{bmatrix} 1 & 1 \\ 1 & -1 \end{bmatrix} = \begin{bmatrix} 0 & 1 \\ 1 & 0 \end{bmatrix} = X.
    \end{align*}

    \item \( Z = HXH \):
    \begin{align*}
        HXH &= \frac{1}{\sqrt{2}}\begin{bmatrix} 1 & 1 \\ 1 & -1 \end{bmatrix} \begin{bmatrix} 0 & 1 \\ 1 & 0 \end{bmatrix} \frac{1}{\sqrt{2}}\begin{bmatrix} 1 & 1 \\ 1 & -1 \end{bmatrix} \\
        &= \frac{1}{2}\begin{bmatrix} 1 & 1 \\ -1 & 1 \end{bmatrix} \begin{bmatrix} 1 & 1 \\ 1 & -1 \end{bmatrix} = \begin{bmatrix} 1 & 0 \\ 0 & -1 \end{bmatrix} = Z.
    \end{align*}

    \item \( -Y = HYH \):
    \begin{align*}
        HYH &= \frac{1}{\sqrt{2}}\begin{bmatrix} 1 & 1 \\ 1 & -1 \end{bmatrix} \begin{bmatrix} 0 & -i \\ i & 0 \end{bmatrix} \frac{1}{\sqrt{2}}\begin{bmatrix} 1 & 1 \\ 1 & -1 \end{bmatrix} \\
        &= \frac{1}{2}\begin{bmatrix} i & -i \\ -i & -i \end{bmatrix} \begin{bmatrix} 1 & 1 \\ 1 & -1 \end{bmatrix} = \begin{bmatrix} 0 & i \\ -i & 0 \end{bmatrix} = -Y.
    \end{align*}

    \item \( S = T^2 \):
    \begin{align*}
        T^2 &= \begin{bmatrix} 1 & 0 \\ 0 & e^{i\pi/4} \end{bmatrix} \begin{bmatrix} 1 & 0 \\ 0 & e^{i\pi/4} \end{bmatrix} = \begin{bmatrix} 1 & 0 \\ 0 & e^{i\pi/2} \end{bmatrix} = S.
    \end{align*}

    \item \( -Y = XYX \):
    \begin{align*}
        XYX &= \begin{bmatrix} 0 & 1 \\ 1 & 0 \end{bmatrix} \begin{bmatrix} 0 & -i \\ i & 0 \end{bmatrix} \begin{bmatrix} 0 & 1 \\ 1 & 0 \end{bmatrix} \\
        &= \begin{bmatrix} i & 0 \\ 0 & -i \end{bmatrix} \begin{bmatrix} 0 & 1 \\ 1 & 0 \end{bmatrix} = \begin{bmatrix} 0 & i \\ -i & 0 \end{bmatrix} = -Y.
    \end{align*}
\end{itemize}
\end{example}

\begin{example}
For implementing a more complex operation on a single qubit, such as an $R_y$ gate, the corresponding MATLAB \cite{2_19,2_20} code is as follows:
\begin{lstlisting}
clear,clc;

% Define the rotation angle theta (for example, pi/3 radians)
theta = pi/3;

% Define the R_y rotation matrix for the given theta
R_y = [cos(theta/2), -sin(theta/2);
sin(theta/2),  cos(theta/2)];

% Define a single qubit state, for example, 0 (you can change this to any state)
qubit_state = [1; 0];  % This represents the 0 state

% Apply the R_y gate to the qubit state to get the new state
new_qubit_state = R_y * qubit_state;

% Display the new qubit state after applying the R_y gate
disp('The new qubit state after R_y rotation is:');
disp(new_qubit_state);
\end{lstlisting}
\end{example}

\section{Experiments of single-qubit gates}

\begin{figure}[H]
    \centering
    \includegraphics[scale=0.5]{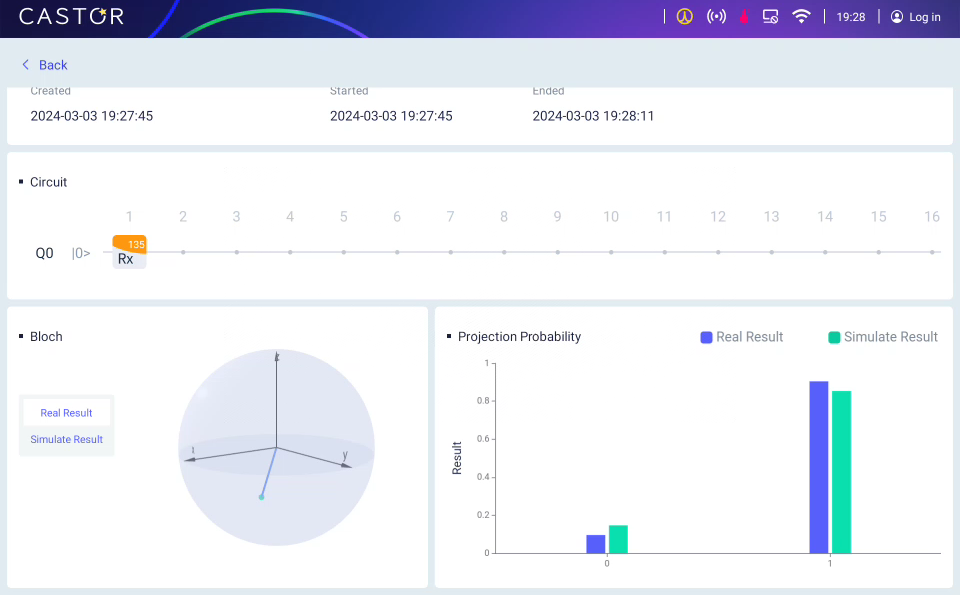}
    \caption{Measurement results of the qubit state after the \( R_x \) rotation gate.}
    \label{rx_result}
\end{figure}

\begin{figure}[H]
    \centering
    \includegraphics[scale=0.5]{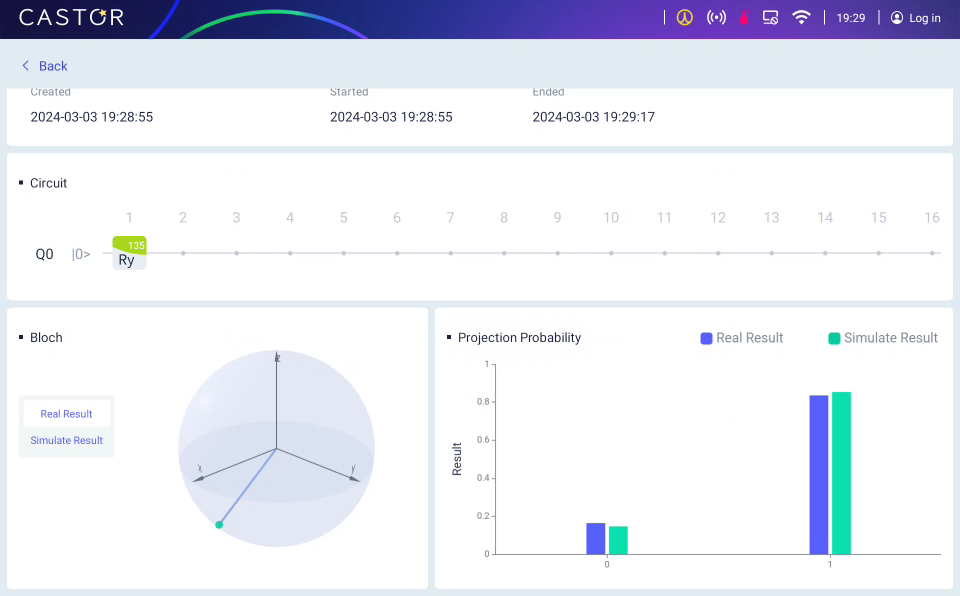}
    \caption{Measurement results of the qubit state after the \( R_y \) rotation gate.}
    \label{ry_result}
\end{figure}

In this quantum computing experiment, we investigated the dynamics of a single qubit under the action of two different quantum logic gates, the \( R_x \) and \( R_y \)  rotation gates. 
The experiment began with the initialization of the qubit to ensure it was in the ground state \( |0\rangle \). Subsequently, we applied parameterized \( R_x(\theta) \) and \( R_y(\phi) \) rotation gates \cite{4_sx} to the qubit in sequence, where \(\theta\) and \(\phi\) represent the angles of rotation. These angles were preset according to our quantum circuit design. After each rotation operation, the state of the qubit was measured to determine its position on the Bloch sphere. The experimental results were recorded as probability distributions, as shown in Figure \ref{rx_result} and Figure \ref{ry_result}, displaying the probability distribution of the qubit state after rotation. 
The experiments provide an intuitive view of quantum state manipulation, which are of great importance for constructing complex quantum algorithms and quantum logic gates. However, it is worth noting that due to the presence of noise, experiments conducted on actual quantum computers often yield results that differ from those obtained on simulators.

\newpage

\chapter{Week 6: Universal Circuits, Quantum Fourier Transform}

\section{Unitary matrices}

We will explore several important linear algebra concepts related to matrices: transpose, conjugate, and adjoint. First, let's introduce the transpose of a matrix. For a matrix $\mathbf{A}$, its transpose is denoted $\mathbf{A}^T$, where the operation involves interchanging the rows and columns of $\mathbf{A}$, meaning for all $j$ and $k$, we have $\mathbf{A}^T[j,k] = \mathbf{A}[k,j]$ \cite{2_4}.

Next is the conjugate of a matrix, represented by $\bar{\mathbf{A}}$. This operation replaces each element in matrix $\mathbf{A}$ with its complex conjugate, that is, $\bar{\mathbf{A}}[j,k] = \overline{\mathbf{A}[j,k]}$ \cite{2_21}.

Finally, the adjoint of a matrix is a combination of conjugation and transposition, denoted as $\mathbf{A}^\dagger$.  It is equivalent to taking the conjugate of matrix $\mathbf{A}$ followed by the transpose of this conjugate matrix, thus $\mathbf{A}^\dagger = \bar{\mathbf{A}}^T$. Therefore, for all $j$ and $k$, the elements of the adjoint matrix can be expressed as $\mathbf{A}^\dagger[j,k] = \overline{\mathbf{A}[k,j]}$ \cite{2_21,2_22}.

\begin{exercise}
Find the transpose, conjugate, and adjoint of
  \begin{align*}
  \begin{bmatrix}
 7-4i & 3+8i & -5i \\
 2 & 6-1.3i & 14 \\
 4-1i & 7+3i & 8-2.5i
\end{bmatrix}.
  \end{align*}
\end{exercise}

Then we introduce the important concept of invertible matrix, which satisfies:
\begin{equation}
\mathbf{A}\mathbf{A}^{-1} = \mathbf{A}^{-1}\mathbf{A} = \mathbf{I}_n,
\end{equation}
for some inverse matrix $\mathbf{A}^{-1}$.

Another important concept of unitary matrix, $\mathbf{U} \in \mathbb{C}^{n\times n}$ satisfies the property:
\begin{equation}
\mathbf{U}\mathbf{U}^\dagger = \mathbf{U}^\dagger\mathbf{U} = \mathbf{I}_n,
\end{equation}
where $\mathbf{U}^\dagger$ is the adjoint (conjugate transpose) of $\mathbf{U}$, and $\mathbf{I}_n$ is the $n \times n$ identity matrix.

The key difference in the definition of unitary is that it involves the adjoint $\mathbf{U}^\dagger$ instead of the typical matrix inverse $\mathbf{U}^{-1}$. Unitary matrices have nice properties related to inner products and norms preservation that make them very useful in areas like quantum computing and signal processing.

\begin{example}
For any $\theta$ the matrix\\
 \begin{align}
 \begin{bmatrix}
   \cos{\theta}&-\sin{\theta}&0\\
   \sin{\theta}&\cos{\theta}&0\\
   0&0&1
 \end{bmatrix}\nonumber
 \end{align}\\
 is a unitary matrix.
\end{example}

\begin{figure}[H]
	\centering
	\includegraphics[scale=0.5]{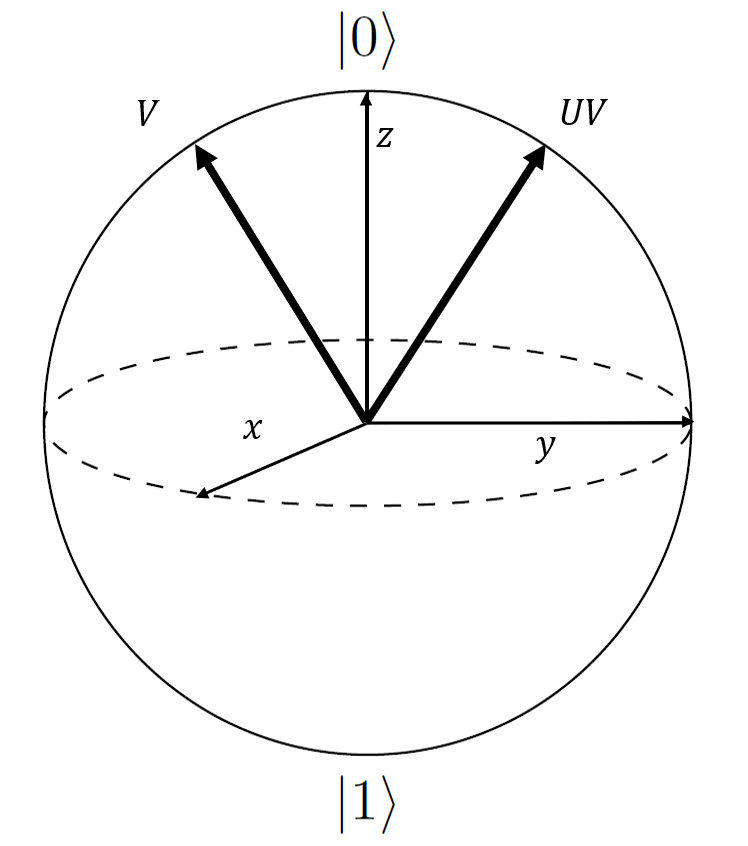}
	\caption{\label{FIG2141} The unit sphere and the action of $U$ on $V$.}
\end{figure}

Let's delve into the concept of the unit sphere in a vector space. The unit sphere is the set of all vectors that have a length (or norm) exactly equal to 1. In the vector space, these vectors are centered around the origin, forming a geometric sphere, as illustrated in Figure \ref{FIG2141}.

Several key points include:
\begin{itemize}
\item The unit sphere is centered around the zero vector (the origin) of the vector space.
\item In three-dimensional space, it can be intuitively visualized as a spherical body enveloping the origin.
\item The concept of the unit sphere is of great importance in mathematics and physics, especially in processes of normalization and in the definition of metric distances.
\end{itemize}

\section{Two-qubit gate}

Here, we provide a definition and some examples of quantum gates. A quantum gate is an operator represented by a unitary matrix that acts on qubits. 

By applying a specific unitary operator to an input quantum state, the quantum state is transformed into an output state. In this manner, quantum gates manipulate and process the quantum information stored in the states of qubits. Common examples include the Pauli matrices X, Y, and Z. The Pauli X gate inverts or performs a bit-flip on a qubit state. The Pauli Y gate rotates the phase by 90 degrees. The Pauli Z gate leaves $|0\rangle$ unchanged but reverses the sign of $|1\rangle$, providing a phase flip operation. The Pauli matrices can be expressed as follows \cite{4_sx}:
\begin{align}
X=
\begin{bmatrix}
0 & 1\\1&0
\end{bmatrix}, \quad
Y=
\begin{bmatrix}
0 & -i\\i&0
\end{bmatrix}, \quad
Z=
\begin{bmatrix}
1 & 0\\0&-1
\end{bmatrix}.
\end{align}

Quantum gates such as the Pauli operators, which are represented by matrix forms, act linearly on quantum states that are described by state vectors. By stringing together sequences of these quantum gates, one can construct more complex operations and quantum circuits from simple gate building blocks. The output state from one gate serves as the input state for the next, creating a flow of quantum information through the circuit.

Quantum algorithms are designed as sequences of quantum circuits that systematically transform input states through repeated gate applications into output states. These output states encode the results of the computation, which can be extracted upon measurement.

\begin{figure}[H]
	\centering
	\[\Qcircuit @!C @!R {
        &\lstick{|x\rangle} &\ctrl{1}  & \rstick{|x\rangle}\qw && \lstick{|x\rangle} &\targ  & \rstick{|x \oplus y\rangle}\qw\\
        &\lstick{|y\rangle} &\targ  & \rstick{|y \oplus x\rangle}\qw&&\lstick{|y\rangle} &\ctrl{-1}  & \rstick{|y\rangle}\qw   }\]
	\caption{\label{6_1} CNOT gate circuit.}
\end{figure}
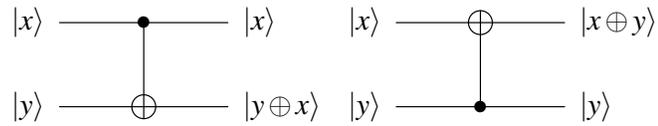

Among these gates, the CNOT, or Controlled NOT gate, stands out due to its ability to entangle two qubits. The circuit diagram for the CNOT gate is depicted in Figure \ref{6_1}.

CNOT with first qubit as the control:
\begin{equation}
|x\rangle \otimes |y\rangle \to |x\rangle \otimes (|y\rangle  \oplus |x\rangle).
\end{equation}
Similarly, a CNOT gate with the second qubit as the control qubit takes
\begin{equation}
|x\rangle \otimes |y\rangle \to (|x\rangle \oplus |y\rangle) \otimes |y\rangle.
\end{equation}

The CNOT gate is a two-qubit gate that performs a NOT operation on the second qubit if the first qubit is in the state $|1\rangle$. It is represented by the following matrix:
$$ \begin{pmatrix} 1 & 0 & 0 & 0 \\ 0 & 1 & 0 & 0 \\ 0 & 0 & 0 & 1 \\ 0 & 0 & 1 & 0 \\ 
\end{pmatrix}.$$

\begin{example}
The initial state is $|10\rangle$, where $|x\rangle$ is $|1\rangle$ and $|y\rangle$ is $|0\rangle$. After applying the CNOT gate, the final state becomes $|11\rangle$. The corresponding MATLAB \cite{2_19,2_20} code is as follows:

\begin{lstlisting}
clear,clc;

% Define the CNOT gate matrix representation
CNOT = [1 0 0 0;
0 1 0 0;
0 0 0 1;
0 0 1 0];

% Initialize the control qubit 'x' and target qubit 'y' states
state_x = [0; 1]; % Column vector for |x>
state_y = [1; 0]; % Column vector for |y>


% Using the Kronecker product (tensor product) to combine qubit states
initial_state = kron(state_x, state_y); % Combined state |xy>

% Apply the CNOT gate to the initial state
% Matrix multiplication to get the final state after the gate operation
final_state = CNOT * initial_state;

% Display the final state to the command window
disp('Final state:');
disp(final_state);
\end{lstlisting}
\end{example}

The CZ gate consists of 2 input qubits and 2 output qubits;
The coefficients of the output qubits are inverted only when both input qubits are 1:
\begin{equation}
CZ|00\rangle=|00\rangle, CZ|01\rangle=|01\rangle, CZ|10\rangle=|10\rangle, CZ|11\rangle=-|11\rangle,
\end{equation}
which can be represented by the following matrix \cite{2_1}:
$$ \begin{pmatrix} 1 & 0 & 0 & 0 \\ 0 & 1 & 0 & 0 \\ 0 & 0 & 1 & 0 \\ 0 & 0 & 0 & -1 \\ 
\end{pmatrix}.$$

\begin{figure}[H]
	\centering
	\[\Qcircuit @!C @R=1.5em {
         &\lstick{|x\rangle}&\ctrl{2} & \rstick{|x\rangle}\qw  & & &\lstick{|x\rangle} &\ctrl{2} &\rstick{|x\rangle}\qw\\
         &  & & & & =& & & & \\
         &\lstick{|y\rangle}&\control\qw& \rstick{(-1)^{xy}|y\rangle}\qw  & & &\lstick{|y\rangle} &\gate{Z} &\rstick{(-1)^{xy}|y\rangle}\qw
        }\]
	\caption{\label{6_5} Circuit diagram of CZ gate both represent the CZ gate.}
\end{figure}
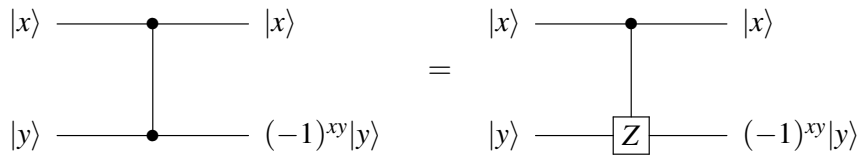

The CZ gate is a symmetric two-qubit gate, which is typically represented by a circuit diagram. As depicted in Figure \ref{6_5}.

Then we demonstrate how the controlled-Z (CZ) two-qubit quantum gate can be constructed using the controlled-NOT (CNOT) gate combined with single-qubit Hadamard gates. The Hadamard gate H transformation:
\begin{equation}
H|\pm\rangle = \frac{1}{\sqrt{2}}(|0\rangle \pm |1\rangle).
\end{equation}
In the basis of ${|0\rangle,|1\rangle}$ the matrix:
\begin{align}
H=\frac{1}{\sqrt{2}}
\begin{bmatrix}
1 & 1\\1 & -1
\end{bmatrix}.
\end{align}

\begin{figure}[H]
	\centering
	\[\Qcircuit @!C @!R {
        &\gate{H} &\gate{Z} &\gate{H}&\qw &=&&\gate{X}&\qw\\
        &\gate{H} &\gate{X} &\gate{H}&\qw &=&&\gate{Z}2&\qw  }\]
	\caption{\label{HZH} Mutual transformation of Gate X and Gate Z.}
\end{figure}
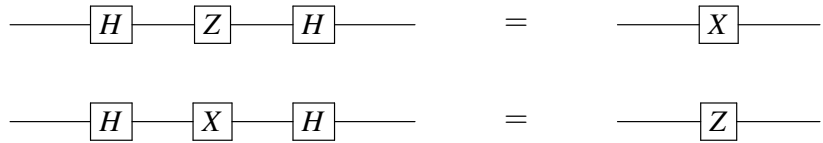

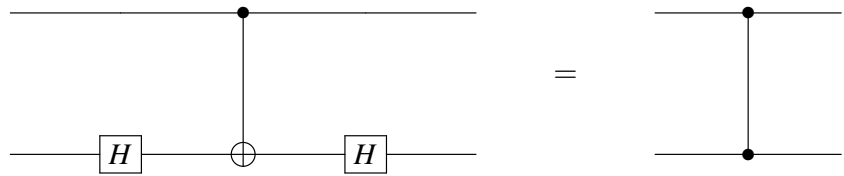
\begin{figure}[H]
	\centering
	\[\Qcircuit @C=3em @R=2em {
        &\lstick{}& \qw &\ctrl{2} &\qw &\qw & & &\ctrl{2} &\qw\\
        & &  &  &  &  & =& &  & \\
        &\lstick{}& \gate{H} & \targ &\gate{H}  
        &\qw & & &\control\qw &\qw
        }\]
	\caption{\label{6_12} CZ gate can be combined from H gate and CNOT gate.}
\end{figure}

An identity that is utilized states that enclosing a Pauli Z gate between two Hadamard gates results in a Pauli X gate: $HZH = X$. Similarly, $HXH = Z$, as illustrated in Figure \ref{HZH}.
 By using this property, a controlled-Z (CZ) operation can be constructed by placing Hadamard (H) gates on either side of a controlled-NOT (CNOT) gate on the target qubit: $H(CNOT)H = CZ$, as shown in Figure \ref{6_12}.

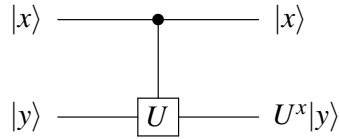
\begin{figure}[H]
		\centering
		\[\Qcircuit @!C @!R {
        &\lstick{|x\rangle}  &\ctrl{1}&\rstick{|x\rangle}\qw\\
        &\lstick{|y\rangle}  &\gate{U} &\rstick{U^{x}|y\rangle}\qw
        }\]
	\caption{\label{CU} The circuit diagram for controlled-U gate.}
\end{figure}

We then introduce the general concept of a controlled-U gate, where U is an arbitrary single-qubit unitary operation. In a controlled-U gate, the first qubit acts as a control for whether the U operation is applied to the second qubit. Mathematically, a controlled-U transforms $|x\rangle|y\rangle$ into $|x\rangle U^{x}|y\rangle$, where $U^{x}$ indicates that U is applied when $x=1$ and the identity operation is applied when $x=0$, as illustrated in Figure \ref{CU}..

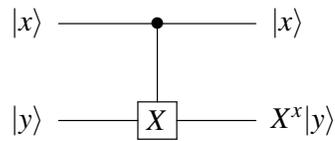
\begin{figure}[H]
		\centering
		\[\Qcircuit @!C @!R {
        &\lstick{|x\rangle}  &\ctrl{1}&\rstick{|x\rangle}\qw\\
        &\lstick{|y\rangle}  &\gate{X} &\rstick{X^{x}|y\rangle}\qw
        }\]
	\caption{\label{CX} The circuit diagram for controlled-X gate.}
\end{figure}

Some examples covered are controlled-NOT, which is actually a controlled-X gate, and controlled-Z. In controlled-NOT (CNOT), the X (NOT) operation flips the second qubit when the first qubit is $|1\rangle$, shown in Figure \ref{CX}.

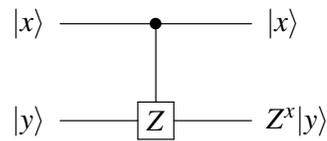
\begin{figure}[H]
		\centering
		\[\Qcircuit @!C @!R {
        &\lstick{|x\rangle}  &\ctrl{1}&\rstick{|x\rangle}\qw\\
        &\lstick{|y\rangle}  &\gate{Z} &\rstick{Z^{x}|y\rangle}\qw
        }\]
	\caption{\label{CZ_gate} The circuit diagram for controlled-Z gate.}
\end{figure}

In controlled-Z (CZ), the second qubit acquires a $-1$ phase factor when the first qubit is $|1\rangle$, shown in Figure \ref{CZ_gate}.

The construction of the controlled-U gate \cite{2_2,4_sx} embodies the fundamental principle of conditional quantum operations: it allows one quantum system to be manipulated based on the state of another. This quantum control is pivotal for establishing entanglement and facilitating information transfer between qubits. The fundamental operation of all such controlled gates is the conditional application of a unitary operation U. By selecting various unitary operations U, a diverse array of controlled multi-qubit gates can be designed.

This article will also present the operator and matrix representations of controlled-U gates. In the operator representation, the action of the controlled-U is described as the tensor product of two operators:
\begin{equation}
|0\rangle \langle0| \otimes I + |1\rangle \langle 1| \otimes U.
\end{equation}
The $2\times 2$ identity I if the control qubit is in state $|0\rangle$, and the single-qubit unitary U if the control is $|1\rangle$. This compactly captures the conditional application of U based on the control.
The matrix representation is:
$$ \begin{bmatrix} 1 & 0 & 0 & 0 \\ 0 & 1 & 0 & 0 \\ 0 & 0 & u_{00} & u_{01} \\ 0 & 0 & u_{10} & u_{11} \\ 
\end{bmatrix},$$
where U is expressed in its matrix form with entries $u_{ij}$,
\begin{align}
U=
\begin{bmatrix}
u_{00} & u_{01}\\u_{10} & u_{11}
\end{bmatrix}.
\end{align}
Here it is evident how U gets applied selectively only to states where the control qubit is $|1\rangle$.

This section demonstrates how an arbitrary controlled-U gate can be constructed solely using controlled-NOT (CNOT) and single-qubit gates. This further confirms that CNOT combined with single-qubit unitaries constitutes a universal gate set.

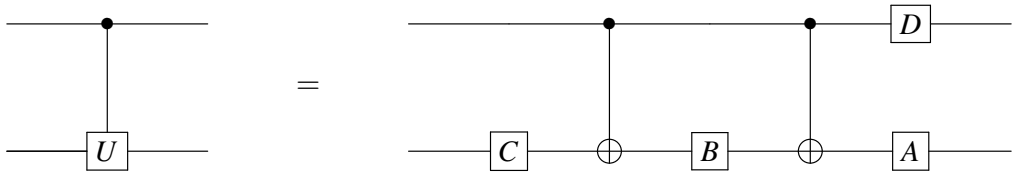
\begin{figure}[H]
	\centering
	\[\Qcircuit @!C @R=1.5em {
         &\ctrl{2} & \qw & & &\qw &\ctrl{2}&\qw&\ctrl{2}&\gate{D}&\qw\\
         &  &  & =& & & \\
         &\gate{U}\qw&\qw & & &\gate{C} &\targ&\gate{B}&\targ&\gate{A}&\qw
        }\]
	\caption{\label{CU1} Controlled-\textbf{U} from single-qubit unitaries and controlled-NOT.}
\end{figure}

The construction process involves intermediary work qubits and utilizes four single-qubit gates labeled A, B, C, D, in addition to two CNOT gates. These components are arranged as depicted in Figure \ref{CU1}, where 
\begin{align}
D=
\begin{bmatrix}
1 & 0\\ 0 & e^{i\alpha}
\end{bmatrix},
\end{align}
and \textbf{U}, $\alpha$, \textbf{A}, \textbf{B}, and \textbf{C} satisfy:
\begin{align}
\textbf{U} &= e^{i\alpha}\boldsymbol{AXBXC},\\ \nonumber
\boldsymbol{I} &= \boldsymbol{ABC}.
\end{align}
By choosing \textbf{U}, $\alpha$, \textbf{A}, \textbf{B}, and \textbf{C} appropriately, this circuit can implement any desired controlled-\textbf{U} operation on the bottom qubit with the top qubit as control.

\begin{example}
Here, we use CNOT gates and single-qubit gates to construct the $R_y(\theta)$ gate. The corresponding MATLAB code is as follows:

\begin{lstlisting}
clear,clc;

% Define CNOT gate
CNOT = [1 0 0 0; 0 1 0 0; 0 0 0 1; 0 0 1 0];

% Define the X gate
X = [0 1; 1 0]; % Define the X gate (Pauli X)

% Define the angle theta
theta = pi/3; % Set theta to 60 degrees, as an example

% Define RY gate
RY = [cos(theta/2) -sin(theta/2); sin(theta/2) cos(theta/2)];

% Define controlled-RY gate
CRY = [1 0 0 0; 0 1 0 0; 0 0 cos(theta/2) -sin(theta/2); 0 0 sin(theta/2) cos(theta/2)];

% Define the alpha
alpha = 0;

% Define single-qubit gates A, B, C, D
A = eye(2); % Define gate A 
B = [cos(-theta/4) -sin(-theta/4); sin(-theta/4) cos(-theta/4)]; % Define gate B 
C = [cos(theta/4) -sin(theta/4); sin(theta/4) cos(theta/4)]; % Define gate C
D = [1 0 ; 0 expm(1i*alpha)]; % Define gate D


% Compute exp(i*alpha)AXBXC which is the sequence for the decomposition of RY
eAXBXC = expm(1i*alpha) * (A * (X * (B * (X * C)))); 
decomposed_RY = eAXBXC; % Decompose RY using the sequence of alpha, A, X, B, X, and C

% Verify whether exp(i*alpha)AXBXC is equal to RY
assert(isequal(decomposed_RY, RY), 'decomposed RY is not equal to RY.'); % Assert that decomposed RY is equal to RY

% Create the decomposition of the controlled-RY gate
decomposed_CRY= kron(D, A) * CNOT * kron(eye(2), B) * CNOT * kron(eye(2), C); % Compute the decomposition using tensor products

% Verify if the decomposition equals the controlled-RY gate
assert(isequal(decomposed_CRY, CRY), 'Decomposed CRY is not equal to CRY.'); % Assert that the decomposed controlled-RY gate equals the defined controlled-RY gate matrix

disp('The decomposition has been verified!'); % Display message that the decomposition has been verified
\end{lstlisting}
\end{example}

\section{Universality}
Here we discuss how to implement the Toffoli gate using basic quantum gates. The Toffoli gate is a three-qubit gate that flips the state of the third qubit only when the first two qubits are in the $|1\rangle$ state. The matrix representation of Toffoli gate is \cite{2_2,4_sx}:
\begin{align}
Toffoli =
\begin{bmatrix}
1& 0 &0 &0 &0 &0 &0 &0 \\
0& 1 &0 &0 &0 &0 &0 &0 \\
0& 0 &1 &0 &0 &0 &0 &0 \\
0& 0 &0 &1 &0 &0 &0 &0 \\
0& 0 &0 &0 &1 &0 &0 &0 \\
0& 0 &0 &0 &0 &1 &0 &0 \\
0& 0 &0 &0 &0 &0 &0 &1 \\
0& 0 &0 &0 &0 &0 &1 &0 
\end{bmatrix}.
\end{align}

The first implementation shows how to construct the Toffoli gate using two-qubit controlled gates, such as the CNOT gate. By utilizing three qubits and a sequence of controlled gates, the desired Toffoli functionality can be achieved, as illustrated in Figure \ref{Toffoli1}. 
\begin{figure}[H]
	\centering
	\[\Qcircuit @C=1.5em @R=1em {
        &\lstick{}&\ctrl{2} &\qw &  & & \qw  &\ctrl{1} &\qw &\ctrl{1} &\ctrl{2} &\qw\\
        &\lstick{}&\ctrl{1} &\qw & =& & \ctrl{1} &\targ &\ctrl{1} &\targ &\qw &\qw\\
        &\lstick{}&\targ    &\qw &  & & \gate{\sqrt{X}}&\qw &\gate{\sqrt{X^{\dagger}}} &\qw &\gate{\sqrt{X}} &\qw
        }\]
	\caption{\label{Toffoli1} The two circuits are equivalent and both represent the Toffoli gate.}
\end{figure}
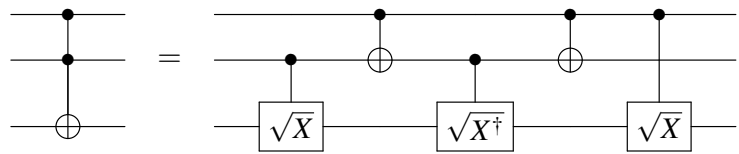
\begin{example}
 The MATLAB code for the circuit diagram shown in Figure \ref{Toffoli1} is:
 \begin{lstlisting}
clear,clc;

% Define the square root of Pauli-X gate (sqrt_X) and its conjugate transpose (sqrt_X_dagger)
sqrt_X = 0.5 * [1 + 1i, 1 - 1i; 1 - 1i, 1 + 1i];
sqrt_X_dagger = 0.5 * [1 - 1i, 1 + 1i; 1 + 1i, 1 - 1i];

% Define the identity and CNOT gate
I = eye(2);
CNOT = [1 0 0 0; 0 1 0 0; 0 0 0 1; 0 0 1 0];

% Define projectors for control bit being 0 and 1
not_control = [1 0; 0 0];  % Projector for control qubit 0
control = [0 0; 0 1];      % Projector for control qubit 1

% Controlled sqrt_X on the first and third qubits, with the first qubit as the control
Csqrt_X_13 = kron(control, kron(I, sqrt_X)) + kron(not_control, kron(I, I));

% CNOT gates with the first qubit as the control and the second qubit as the target
CNOT_12 = kron(CNOT, I);

% Controlled sqrt_X_dagger on the second and third qubits, with the second qubit as the control
Csqrt_X_dagger_23 = kron(kron(I, control), sqrt_X_dagger) + kron(kron(I, not_control), I);

% Controlled sqrt_X on the second and third qubits, with the second qubit as the control
Csqrt_X_23 = kron(kron(I, control), sqrt_X) + kron(kron(I, not_control), I);

% Combine the gates to get the final 3-qubit circuit
circuit = Csqrt_X_13 * CNOT_12 * Csqrt_X_dagger_23 * CNOT_12 * Csqrt_X_23;

% Display the matrix representation of the circuit
disp('The Toffoli gate matrix:');
disp(circuit);
 \end{lstlisting}
\end{example}

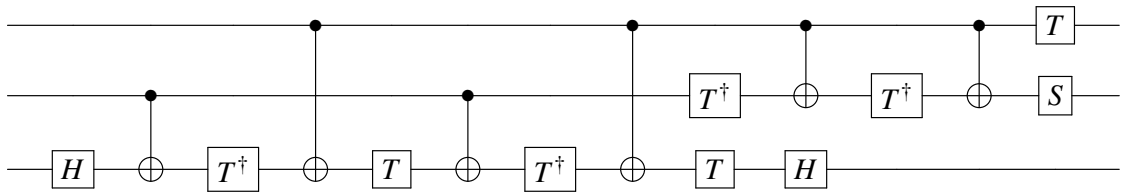
\begin{figure}[H]
	\centering
	\[\Qcircuit @C=1.5em @R=1em {
        &\lstick{}&\qw &\qw &\qw &\ctrl{2} &\qw &\qw&\qw &\ctrl{2} &\qw &\ctrl{1} &\qw& \ctrl{1} &\gate{T} &\qw\\
        &\lstick{}&\qw &\ctrl{1}&\qw &\qw &\qw &\ctrl{1}&\qw &\qw &\gate{T^{\dagger}} &\targ &\gate{T^{\dagger}} &\targ &\gate{S} &\qw\\
        &\lstick{}&\gate{H} &\targ&\gate{T^{\dagger}}&\targ &\gate{T}&\targ&\gate{T^{\dagger}} &\targ&\gate{T} &\gate{H} &\qw &\qw &\qw &\qw
        }\]
	\caption{\label{Toffoli2} The circuit represents the Toffoli gate.}
\end{figure}

The second implementation method employs even more fundamental single and two-qubit gates, such as Hadamard gates, phase gates, CNOT gates, and the $\pi/8$ gate \cite{4_sx}, as shown in Figure \ref{Toffoli2}. By carefully arranging these gates and combining them with the use of three qubits and reverse operation steps, the Toffoli gate operation can be simulated. This demonstrates that a universal set of simple gates is sufficient to construct complex quantum operations.

Both implementations provide insight into how complex quantum logic operations can be synthesized from simpler building blocks. This is important for the actual physical construction of quantum computers, where certain gate operations may be easier to implement technologically depending on the hardware platform. Understanding gate decompositions assists with the programming and control of quantum systems.
\begin{example}
 The MATLAB code for the circuit diagram shown in Figure \ref{Toffoli2} is:
 \begin{lstlisting}
clear,clc;

% Define the T gate and T dagger
T = [1 0; 0 expm(1i*pi/4)];
T_dagger  = [1 0; 0 expm(-1i*pi/4)];

% Define the S gate 
S = [1 0 ; 0 1i];

% Define the H gate
H = 1/sqrt(2) * [1 1; 1 -1];

% Define the identity, X and CNOT gate
I = eye(2);
CNOT = [1 0 0 0; 0 1 0 0; 0 0 0 1; 0 0 1 0];
X = [0 1 ; 1 0];

% Define projectors for control bit being 0 and 1
not_control = [1 0; 0 0];  % Projector for control qubit 0
control = [0 0; 0 1];      % Projector for control qubit 1

% CNOT gates with the first qubit as the control, the second qubit as
% the target and the third qubit as identity
CNOT_12 = kron(CNOT, I);

% CNOT gates with the second qubit as the control, the third qubit as
% the target and the first qubit as identity
CNOT_23 = kron(I, CNOT);

% CNOT gate on the first and third qubits, with the first qubit as the control
CNOT_13 = kron(control, kron(I, X)) + kron(not_control, kron(I, I));


% Combine the gates to get the final 3-qubit circuit
circuit = kron(kron(T, S), I) * CNOT_12 * kron(kron(I, T_dagger), I) * kron(CNOT, H) * kron(kron(I, T_dagger), T) * CNOT_13 * kron(kron(I, I), T_dagger) * CNOT_23 * kron(kron(I, I), T) * CNOT_13 * kron(kron(I, I), T_dagger) * CNOT_23 * kron(kron(I, I), H);

% Display the matrix representation of the circuit
disp('The Toffoli gate matrix:');
disp(circuit);
 \end{lstlisting}
\end{example}

This section discusses quantum measurements and their effect on quantum states. For a quantum state $|\psi\rangle = \sum\limits_x \alpha_x |x\rangle$, measuring in the basis set ${|x\rangle}$ will yield the state $|x\rangle$ with probability $p_x = |\alpha_x|^2$. The measurement projects the superposition state onto one of the basis states, as illustrated in Figure \ref{measure1},
\begin{figure}[H]
    \centering
        \[\Qcircuit @C=1.5em @R=1em {
        &\push{|\psi\rangle=\sum{\alpha_{x}\mid x\rangle}}& &{/}\qw &\meter&\cw&\rstick{|x\rangle}\cw
        }\]
    \caption{\label{measure1} Measurements $|\psi\rangle = \sum\alpha_{x}\mid x\rangle$.}
\end{figure}

\begin{figure}[H]
    \centering
        \[\Qcircuit @C=1.5em @R=1em {
        & & &{/}\qw &\meter&\cw&\rstick{|x\rangle}\cw\\
        &\push{|\psi\rangle=\sum{\alpha_{x}\mid x\rangle}}& & & & & \\
        & & &{/}\qw &\qw&\qw&\rstick{|\phi_x\rangle}\qw
        }\]
    \caption{\label{measure2} Measurements $|\psi\rangle = \sum\alpha_{x}\mid x\rangle|\phi_x\rangle$.}
\end{figure}
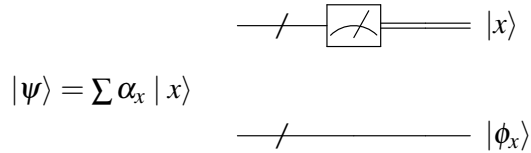

Similarly, for a two-register quantum state $|\psi\rangle = \sum\limits_x \alpha_x |x\rangle|\phi_x\rangle$, measuring the first register in the basis ${|x\rangle}$ will return the state $|x\rangle|\phi_x\rangle$ with probability $p_x = |\alpha_x|^2$. The second register will remain in the corresponding state $|\phi_x\rangle$ associated with the measurement result $x$, as illustrated in Figure \ref{measure2}. This demonstrates the concept of quantum manipulation—by measuring part of an entangled system, the state of the other part is projected or "steered" into a corresponding state.

Understanding quantum measurements and their probabilistic nature in projecting superposition states into definite outcomes is central to analyzing quantum algorithms and protocols. The statistical data obtained from measurement processes reveal relative amplitudes and phases within a quantum state, which are the origins of quantum interference effects. These effects are the backbone of quantum information processing. Thus, accurate modeling of the quantum measurement process is crucial, both in theoretical research and experimental practice.

In the realm of quantum computing research, a pivotal discovery has been made: any complex multi-qubit unitary operation can be decomposed and implemented using only single-qubit unitary gates and controlled-NOT (CNOT) gates. Specifically, by combining arbitrary single-qubit rotations with CNOT gates—which can entangle qubit pairs—any desired unitary transformation for any number of qubits can be constructed \cite{4_sx}.

This indicates that with the ability to perform arbitrary single-qubit operations and to use a two-qubit entangling gate like the CNOT, we possess a set of operations that is computationally ``universal" for quantum computers. No other types of interactions or gates are necessary. This greatly simplifies the physical requirements for constructing a quantum computer—control over individual qubits and the coupling between qubit pairs is sufficient.

Following this, we discuss the DiVincenzo criteria, which outline five fundamental requirements that a physical system must meet to serve as a quantum computer. These criteria provide a checklist for the essential capabilities and technical specifications needed to realize a fully functional, scalable quantum information processor.

The DiVincenzo criteria \cite{2_23} define five fundamental requirements for a quantum computer, forming important standards for evaluating the suitability of a physical system for quantum computation.

Firstly, the system must possess a scalable array of well-characterized qubits, the basic units of quantum memory. To achieve large-scale quantum computation, the number of qubits must be expandable, and their properties and interactions must be precisely calibrated and understood.

The second requirement is the ability to initialize qubits to a known fiducial state, typically the $|0\rangle$ state. This step is essential for clearing any previous states, ensuring a clean slate for quantum computation.

Thirdly, qubits must have decoherence times significantly longer than the duration of gate operations, allowing effective manipulation, interaction, and readout of the qubits before information loss and noise interference occur.

The fourth criterion is universality, meaning the system should possess a set of single-qubit and two-qubit gates that are universal for quantum computing, as discussed in the previous content.

Finally, the fifth criterion involves the need for qubit-specific measurements to read out computational results, ensuring accurate retrieval of computational information.

These five criteria not only outline the key characteristics of a quantum computer but also provide engineering guidance for implementing quantum algorithms. Current research focuses on overcoming the challenges of meeting these criteria simultaneously in practical physical platforms.

\section{The quantum Fourier transform}

The quantum Fourier transform $U_{FT}$ in terms of basis vectors \cite{2_24,2_25}:
\begin{equation}
U_{FT}|x\rangle = \frac{1}{2^{n/2}}\sum_{y=0}^{2^{n}-1}e^{2\pi i xy/N}|y\rangle,
\end{equation}
this leads to
\begin{equation}
U_{FT}(\sum_{x=0}^{2^{n}-1}\gamma(x)|x\rangle = \sum_{x=0}^{2^{n}-1} \tilde{\gamma}(x)|x\rangle,
\end{equation}
where
\begin{equation}
\tilde{\gamma}(x) = \frac{1}{2^{n/2}}\sum_{y=0}^{2^{n}-1}e^{2\pi i xy/N}\gamma(y).
\end{equation}

Classically, computing the discrete Fourier transform for a sequence of length $2^n$ can be done in $O(n2^n)$ time using what is known as the Fast Fourier transform \cite{4_sx}. However, that will not be fast enough for our purposes since it is still exponential in $n$. Instead, we will construct aquantum circuit that can construct a quantum state corresponding to the discrete Fourier transform in $O(n^2)$ time.

A collection of $N=2^n$ qubits has computational basis states,
\begin{equation}
|00 \cdots 00\rangle, |00 \cdots 01\rangle, |00 \cdots 10\rangle, \cdots, |11 \cdots 11\rangle.\nonumber
\end{equation}
So as to make our notation easier, we will treat these $t$ qubits as a register that contains a binary number from $0$ to $N-1=2^n-1$. In other words, we will treat the basis states as binary numbers, and we will abbreviate them by writing the corresponding decimal number instead. Hence we will write 
\begin{equation}
|0\rangle=|00 \cdots 00\rangle, |1\rangle=|00 \cdots 01\rangle, |2\rangle=|00 \cdots 10\rangle, \cdots, |2^n-1\rangle=|11 \cdots 11\rangle.\nonumber
\end{equation}
We now define the quantum Fourier transform(QFT) as follows.
The QFT acts on computational basis states by 
\begin{equation}
|j\rangle \rightarrow \frac{1}{\sqrt{N}} \sum_{k=0}^{N-1} e^{2\pi i jk/N}|k\rangle.
\end{equation}
This sends 
\begin{equation}
\sum_{j=0}^{N-1}x_j|j\rangle  \rightarrow \sum_{k=0}^{N-1}y_k|k\rangle.
\end{equation}

\begin{example}
Let $n=1$, $e^{2\pi i xy/N} = e^{\pi i xy}$. Then the QFT maps:
\begin{align}
\nonumber
&U_{FT}|0\rangle \rightarrow \frac{1}{\sqrt{2}}(|0\rangle+|1\rangle)\\\nonumber
&U_{FT}|1\rangle \rightarrow \frac{1}{\sqrt{2}}(|0\rangle-|1\rangle)\\
\end{align}
Let $n=2$, $e^{2\pi i xy/N} = e^{\pi i xy/2}$. Then the QFT maps:
\begin{align}
\nonumber
&U_{FT}|0\rangle \rightarrow \frac{1}{2}(|0\rangle+|1\rangle+|2\rangle+|3\rangle)\\\nonumber
&U_{FT}|1\rangle \rightarrow \frac{1}{2}(|0\rangle+i|1\rangle-|2\rangle-i|3\rangle)\\\nonumber
&U_{FT}|2\rangle \rightarrow \frac{1}{2}(|0\rangle-|1\rangle+|2\rangle-|3\rangle)\\
&U_{FT}|3\rangle \rightarrow \frac{1}{2}(|0\rangle-i|1\rangle-|2\rangle+i|3\rangle).
\end{align}
\end{example}

\begin{figure}[H]
		\centering
		\[\Qcircuit @!C @!R {
        &\lstick{} &\gate{H} &\gate{R_2}&\qw&\qswap \qw &\qw\\
        &\lstick{} &\qw &\ctrl{-1} &\gate{H}&\qswap \qwx&\qw
        }\]
	\caption{\label{QFT} The circuit diagram for a two-qubit quantum Fourier transform.}
\end{figure}
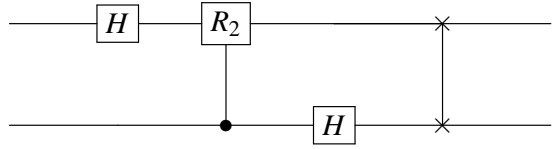

In the execution of a 2-qubit Quantum Fourier Transform (QFT), the steps are as follows: First, a Hadamard (H) gate is applied to the first qubit to create a superposition state. Next, a controlled-$R_2$ gate is used to apply a phase shift to the first qubit, with the second qubit acting as the control. Then, a Hadamard (H) gate is applied to the second qubit. Finally, a swap gate is used to exchange the states of the two qubits. This process is illustrated in Figure \ref{QFT}. The key gate used is the single qubit gate \cite{4_sx}:
\begin{align}
R_2 = S = \begin{bmatrix} 1 & 0 \\ 0 & i \end{bmatrix}.
\end{align}

To see this in action, let’s compute the output when $|01\rangle$ is fed into this circuit.\\
First we apply $H$ to the fist qubit, this sends, 
\begin{equation}
|0\rangle \rightarrow \frac{1}{\sqrt{2}}(|0\rangle+|1\rangle).
\end{equation}
Next, the second qubit is $|1\rangle$, so we apply the controlled $R_2$ gate to the fist qubit, this sends,
\begin{equation}
\frac{1}{\sqrt{2}}(|0\rangle+|1\rangle) \rightarrow \frac{1}{\sqrt{2}}(|0\rangle+i|1\rangle).
\end{equation}
Then we apply $H$ to the second qubit, which gives $\frac{1}{\sqrt{2}}(|0\rangle-|1\rangle)$. The two qubits now have the combined state,
\begin{equation}
\frac{1}{{2}}(|0\rangle+i|1\rangle) \otimes (|0\rangle-|1\rangle).
\end{equation}
Switching the order of the qubits then gives
\begin{equation}
\frac{1}{{2}}(|0\rangle-|1\rangle) \otimes (|0\rangle+i|1\rangle)=\frac{1}{{2}}(|00\rangle +i|01\rangle -|10\rangle -i|11\rangle).
\end{equation}
For n = 2, using binary representation:
\begin{align}
\nonumber
&U_{FT}|0\rangle \rightarrow \frac{1}{2}(|0\rangle+|1\rangle+|2\rangle+|3\rangle)=\frac{1}{{2}}(|0\rangle+|1\rangle) \otimes (|0\rangle+|1\rangle)\\\nonumber
&U_{FT}|1\rangle \rightarrow \frac{1}{2}(|0\rangle+i|1\rangle-|2\rangle-i|3\rangle)=\frac{1}{{2}}(|0\rangle-|1\rangle) \otimes (|0\rangle+i|1\rangle)\\\nonumber
&U_{FT}|2\rangle \rightarrow \frac{1}{2}(|0\rangle-|1\rangle+|2\rangle-|3\rangle)=\frac{1}{{2}}(|0\rangle+i|1\rangle) \otimes (|0\rangle-|1\rangle)\\
&U_{FT}|3\rangle \rightarrow \frac{1}{2}(|0\rangle-i|1\rangle-|2\rangle+i|3\rangle)=\frac{1}{{2}}(|0\rangle-|1\rangle) \otimes (|0\rangle-i|1\rangle).
\end{align}

For n = 3, the circuit for QFT shown in Figure \ref{QFT3}, the steps are as follows: First, a Hadamard (H) gate is applied to the first qubit to create a superposition state. Next, a controlled-$R_2$ gate is used to apply a phase shift to the first qubit, with the second qubit acting as the control, a controlled-$R_3$ gate is used to apply a phase shift to the first qubit, with the third qubit acting as the control. Then, a Hadamard (H) gate is applied to the second qubit, a controlled-$R_2$ gate is used to apply a phase shift to the second qubit, with the third qubit acting as the control. Then, a Hadamard (H) gate is applied to the third qubit. Finally, a swap gate is used to exchange the states of the first and third qubits. The key gate used is the single qubit gate:
\begin{align}
R_2 = S = \begin{bmatrix} 1 & 0 \\ 0 & i \end{bmatrix},\quad
R_3 = T = \begin{bmatrix} 1 & 0 \\ 0 & \text{exp}(\frac{\pi i}{4}) \end{bmatrix}.
\end{align}
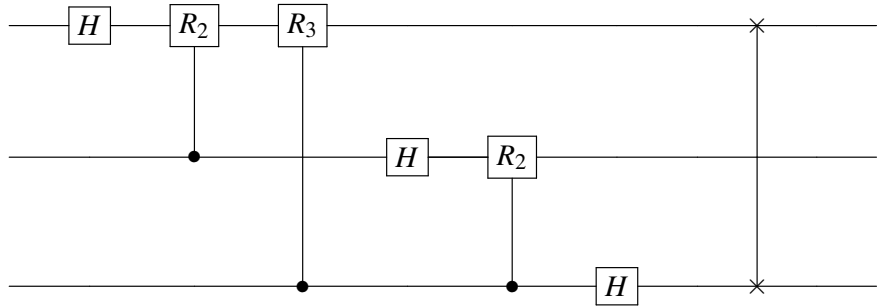
\begin{figure}[H]
	\centering
	\[\Qcircuit @C=2em @R=3em {
        &\lstick{} &\gate{H}&\gate{R_2} &\gate{R_{3}}&\qw&\qw&\qw&\qw&\qswap &\qw&\qw\\
        &\lstick{} &\qw&\ctrl{-1}&\qw&\gate{H} &\gate{R_{2}}\qw&\qw &\qw&\qw \qwx&\qw&\qw \\
        &\lstick{} &\qw&\qw & \ctrl{-2} &\qw&\ctrl{-1}&\gate{H}&\qw&\qswap \qwx&\qw&\qw
        }\]
	\caption{The circuit for QFT when $n=3$.}
 \label{QFT3}
\end{figure}
For n = 3, if the initial state is $|000\rangle$, the state after Quantum Fourier Transform (QFT) can be written as:
\begin{align}
&U_{FT}|000\rangle \rightarrow \frac{1}{2\sqrt{2}}\left(|000\rangle + |001\rangle + |010\rangle + |011\rangle + |100\rangle + |101\rangle + |110\rangle + |111\rangle\right).
\end{align}

\begin{figure}[H]
    \centering
    \includegraphics[scale=0.5]{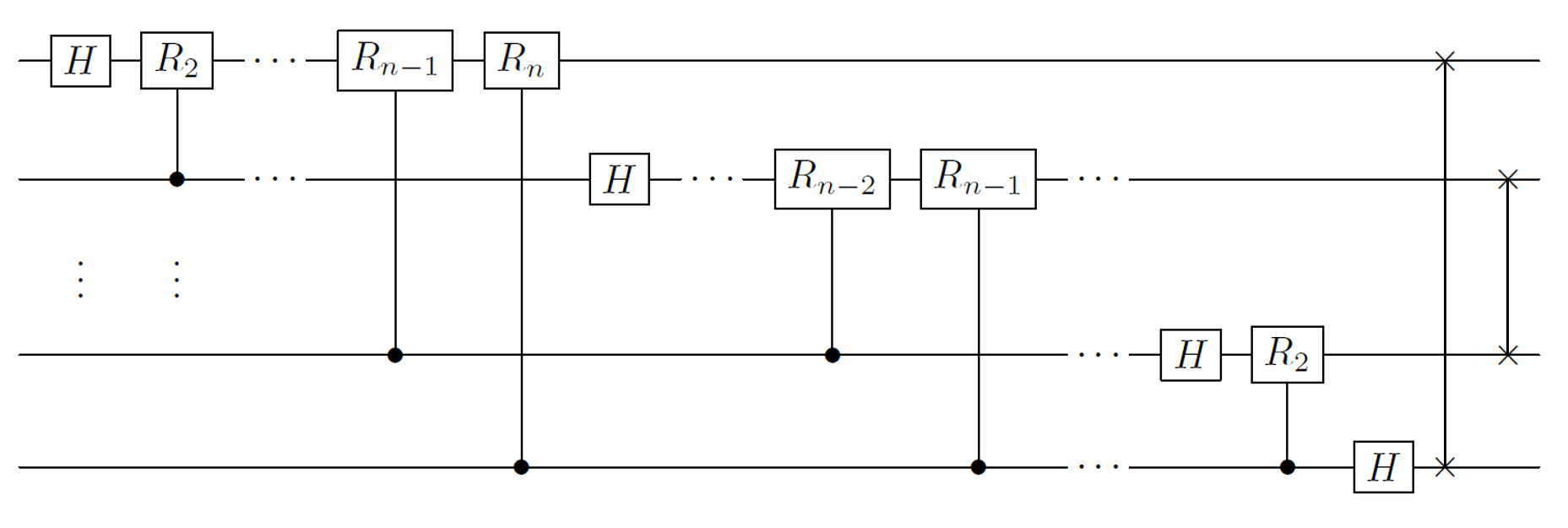}
    \caption{The circuit diagram for a $n$-qubit quantum Fourier transform.}
    \label{QFTn}
\end{figure}

For $n$ qubits, the circuit for QFT shown in Figure \ref{QFTn}. The quantum circuit for QFT can be constructed in a recursive manner. For each qubit $|q_j\rangle$, where $j$ ranges from $0$ to $n-1$ (assuming the first qubit is $|q_0\rangle$), the circuit can be described by the following steps \cite{2_24}:
\begin{enumerate}
    \item Apply the Hadamard gate to $|q_j\rangle$.
    \item For each qubit $|q_k\rangle$ that is lower than $j$, where $k$ ranges from $j+1$ to $n-1$, use the controlled rotation gate $CR_{k+1}$, which rotates $|q_j\rangle$ depending on the state of $|q_k\rangle$.
    \item Repeat the above steps until all qubits have been processed.
    \item Finally, a qubit reversal operation can be applied to obtain the correct order. 
\end{enumerate}
The key gate used is the single qubit gate:
\begin{align}
R_k = \begin{bmatrix} 1 & 0 \\ 0 & e^{2\pi i/2^{k}} \end{bmatrix}.
\end{align}
The efficiency of QFT lies in the fact that it only uses $O(n^2)$ basic quantum gates, which is a significant improvement compared to the classical FFT's $O(n2^n)$ \cite{4_sx}.

\section{Experiment of the quantum Fourier transform}

\begin{figure}[H]
    \centering
    \includegraphics[scale=0.5]{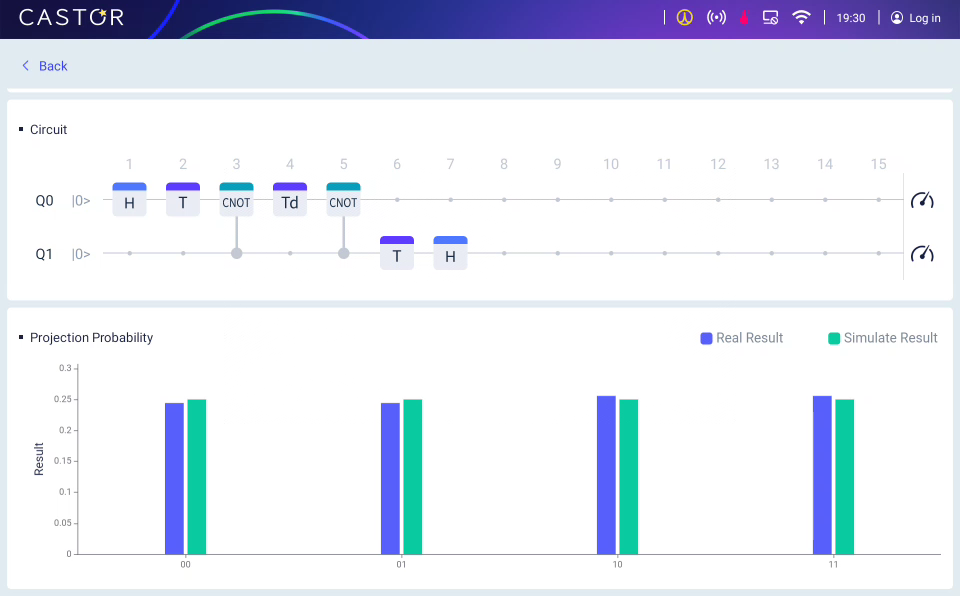}
    \caption{The experimental circuit and result diagram of a 2-qubit Quantum Fourier Transform on SpinQ Gemini mini.}
    \label{QFT_circuit}
\end{figure}

In this experiment, our objective was to implement and verify the effects of a 2-qubit quantum Fourier transform (QFT). QFT is one of the key algorithms in quantum computing \cite{4_sx}, efficiently transforming the representation basis of quantum states. It is a key component in the implementation of quantum algorithms such as Shor's algorithm \cite{2_26}. For a system of 2 qubits, QFT is defined as \( QFT|x\rangle = \frac{1}{\sqrt{4}} \sum_{k=0}^{3} e^{\frac{2\pi i x k}{4}} |k\rangle \), where \( |x\rangle \) represents the input state, and \( |k\rangle \) represents the output state. We began by initializing two qubits to the state \( |00\rangle \). By applying Hadamard gates and controlled phase rotation gates, we placed the qubits into a superposition state and created quantum entanglement between them. Following the experimental circuit diagram shown in Figure \ref{QFT_circuit}, where $R_2 = e^{\pi i/4}IXT_{d}XT$, we sequentially applied Hadamard gates and controlled rotation gates to each qubit in the quantum circuit. 
After the experiment was completed, we measured the state of all two qubits. The experimental result graph showed the probability distribution of the measurement outcomes shown in Figure \ref{QFT_circuit}, with each possible output state from \( |00\rangle \) to \( |11\rangle \) corresponding to a probability value, we confirmed the implementation of the 2-qubit QFT. 

\begin{figure}[H]
    \centering
    \includegraphics[scale=0.3]{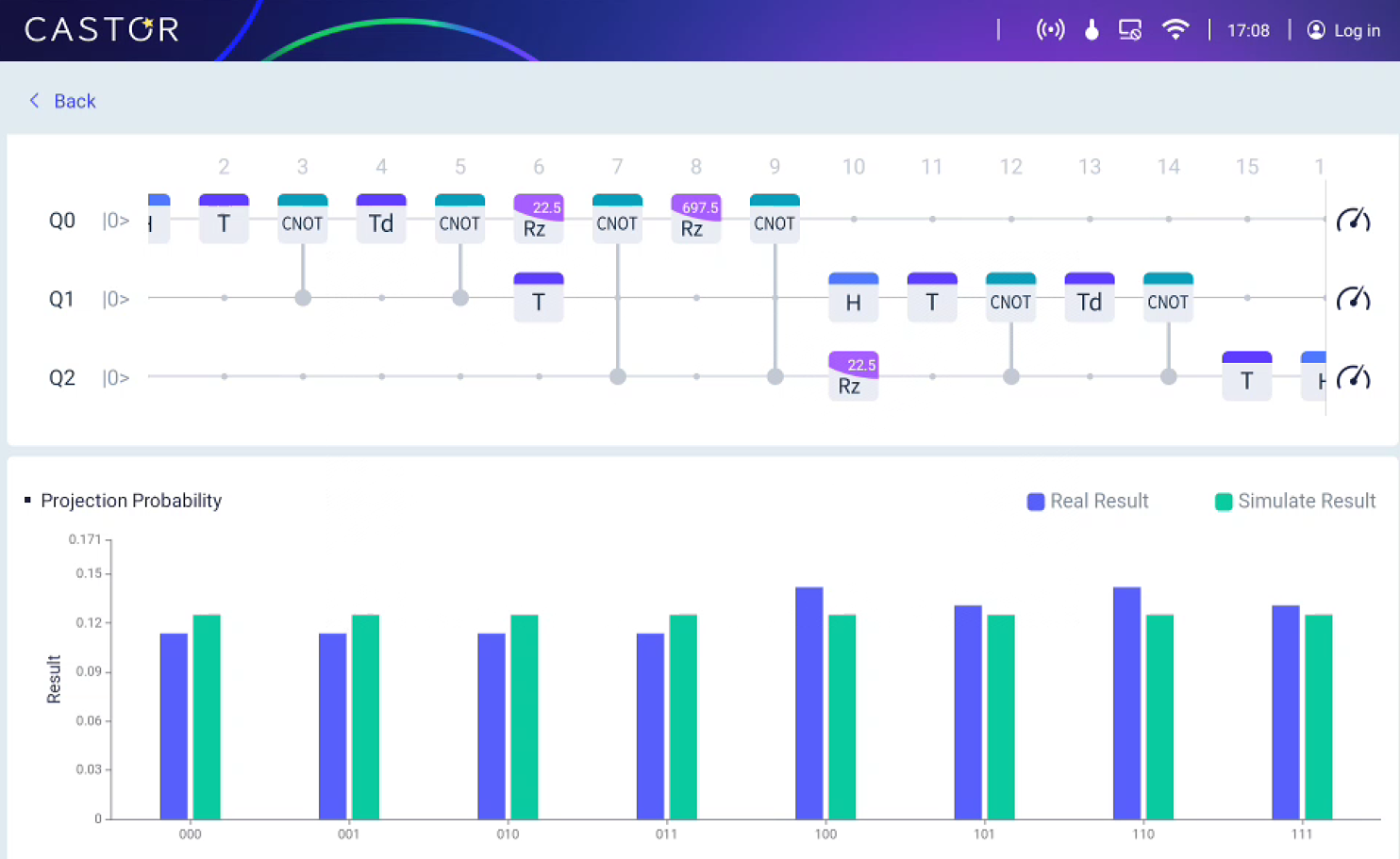}
    \caption{The experimental circuit and result diagram of a 3-qubit Quantum Fourier Transform on SpinQ Triangulum mini.}
    \label{QFT3_circuit}
\end{figure}

\begin{figure}[H]
	\centering
	\includegraphics[scale=0.4]{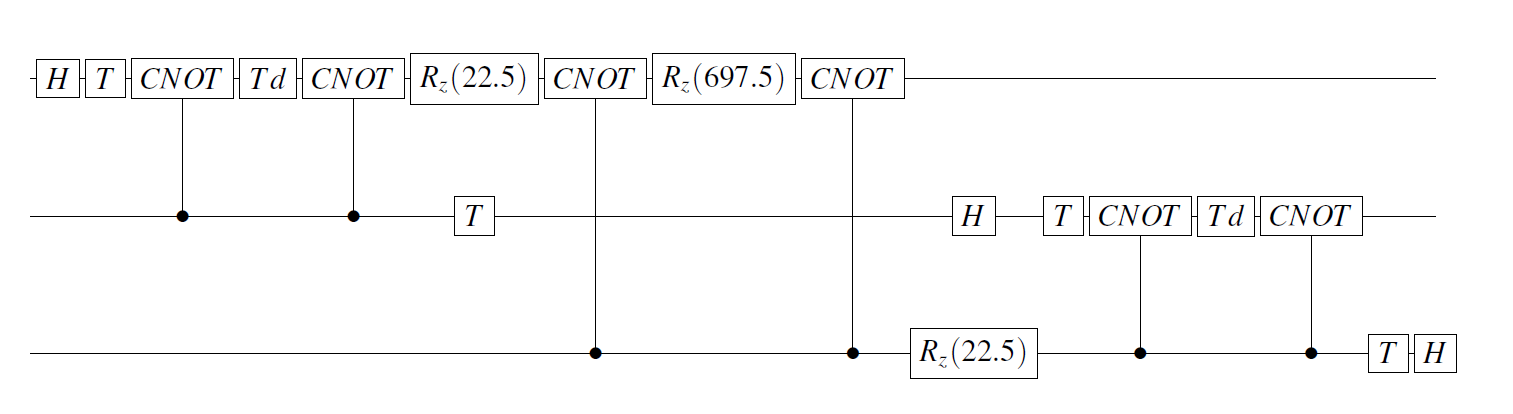}
	\caption{The circuit for QFT when $n=3$.}
 \label{QFT3_diagram}
\end{figure}

Similar to the 2 qubits QFT, for a system of 3 qubits, QFT is defined as \( QFT|x\rangle = \frac{1}{\sqrt{8}} \sum_{k=0}^{7} e^{\frac{2\pi i x k}{8}} |k\rangle \), where \( |x\rangle \) represents the input state, and \( |k\rangle \) represents the output state. We began by initializing three qubits to the state \( |000\rangle \). By applying Hadamard gates and controlled phase rotation gates, we placed the qubits into a superposition state and created quantum entanglement \cite{4_sx} between them. Following the experimental circuit diagram shown in Figure \ref{QFT3_circuit}, where $R_2 = e^{\pi i/4}IXT_{d}XT$,  $R_3 = e^{\pi i/8}IXR_z(-\pi/8)XR_z(\pi/8)$, we sequentially applied Hadamard gates and controlled rotation gates to each qubit in the quantum circuit. Due to the excessive depth of the experimental circuit, it is not fully displayed on the quantum computer shown in Figure \ref{QFT3_circuit}. The complete experimental circuit diagram is shown in Figure \ref{QFT3_diagram}.

After the experiment was completed, we measured the state of all three qubits. The experimental result graph showed the probability distribution of the measurement outcomes shown in Figure \ref{QFT3_circuit}, with each possible output state from \( |000\rangle \) to \( |111\rangle \) corresponding to a probability value, we confirmed the implementation of the 3-qubit QFT. 

However, it is worth noting that due to the presence of noise \cite{4_sx}, experiments conducted on actual quantum computers often yield results that differ from those obtained on simulators.

\section{Reading material: Shor's algorithm}
\begin{center}
\textbf{RSA Algorithm}
\end{center}

The RSA encryption algorithm is a widely-used asymmetric encryption algorithm, employing two different keys for encryption and decryption processes — hence the term ``asymmetry" \cite{2_27}. The algorithm works by first generating a pair of keys: a private key $(D, N)$, which is kept secret by the user, and a public key $(E, N)$, which can be openly distributed.

For security purposes, the RSA key length should be substantial. While a minimum of 500 bits is required, a length of 1024 bits is commonly recommended to ensure robust encryption \cite{2_29}.

The encryption process involves converting plaintext $m$ into ciphertext $c$ using the public key's exponent $E$, with the operation $c = m^E \mod N$. Conversely, the decryption process transforms ciphertext $c$ back into plaintext $m$ utilizing the private key's exponent $D$, calculated as $m = c^D \mod N$.

A critical component of the RSA algorithm is $N$, the product of two large prime numbers $P$ and $Q$ (i.e., $N=P \times Q$). The public key is comprised of $N$ and $E$, while the private key includes $D$, which can be easily derived if one knows $P$ and $Q$ \cite{2_27}. The security of RSA hinges on the difficulty of deducing $P$ and $Q$ from $N$ by classical computers; for a sufficiently large $N$, factoring it could take an impractical amount of time. For example, even with advanced computational resources like a supercomputer, breaking down a 2048-bit $N$ could require billions of years.

Therefore, the selection of a large integer for $N$ is crucial for RSA's security. The creation of an RSA key is a straightforward process: select two large prime numbers as $P$ and $Q$, compute their product $N$, and the algorithm is ready for use.

The RSA algorithm is essential for secure communication in computer systems and is notably used in encrypting SSH protocol data \cite{2_28}. The challenge in breaking RSA encryption lies in factoring the large integer $N$ from the public key back into its prime components $P$ and $Q$, upholding the relationship $N = P \times Q$.

\begin{center}
 \textbf{Classical Large Number Factorization Algorithm}   
\end{center}

Let us pose the following problem: Given an integer $N$, the task is to find two prime numbers $P$ and $Q$ such that $N = P \times Q$. Examples include:

\begin{enumerate}
    \item For $N = 15$, the solution is $P = 3$ and $Q = 5$.
    \item For $N = 221$, the primes are $P = 13$ and $Q = 17$.
    \item For $N = 12709189$, the factors are $P = 3559$ and $Q = 3571$.
\end{enumerate}

The straightforward approach to this problem involves iterating through numbers from 1 to $\sqrt{N}$ in a loop, incrementing by 1 each time, and checking whether the number divides $N$. However, this method has a computational complexity of $O(\sqrt{N}) = O(2^{n/2})$, where $n$ is the number of bits representing $N$.

We outline the steps of the algorithm as follows:

\begin{enumerate}
    \item Check if $N$ is even. If it is, then 2 is a factor of $N$.
    \item Test whether $N = p \times q$ where $p \geq 1$ and $q \geq 2$. If this holds true, then return $p$ as a factor. This step runs in polynomial time.
    \item Randomly select $a$ such that $1 < a \leq N-1$, and test whether $\text{gcd}(a, N) > 1$. If it is, then return $\text{gcd}(a, N)$ as a factor.
    \item Find the period $r$ of $x$ in the equation $a^x \equiv 1 \pmod{N}$.
    \item If $r$ is odd, or if $r$ is even but $a^{r/2} \equiv -1 \pmod{N}$, go back to step 3. Otherwise, compute both $\text{gcd}(a^{r/2}-1, N)$ and $\text{gcd}(a^{r/2}+1, N)$. One of these will be a factor of $N$.
\end{enumerate}

The function $\text{gcd}(M, N)$ computes the greatest common divisor of $M$ and $N$, with its time complexity being polynomial.

Below are some illustrative examples:

\begin{example} 
	Given $N=15$, we follow the algorithm steps:

\begin{description}
	\item[Step 1] As $N$ is not even, we proceed to the next step.
	
	\item[Step 2] We test if $N = p \times q$ for any $p$ and $q$ that satisfy the condition, and find that there are no such $p$ and $q$.
	
	\item[Step 3] We randomly select a number $a$ from 2 to 14. Let's say we choose $a=4$. We calculate $\text{GCD}(15, 4)$, which is 1, so we move on to the next step.
	
	\item[Step 4] We select $x$ starting from 0, calculate $a^x \mod N$, and find the period $r$. For $a=4$, the period $r=2$ is illustrated in Figure \ref{fig:N15_a4}.
	
	\begin{figure}[H]
		\centering
		\includegraphics[scale=0.4]{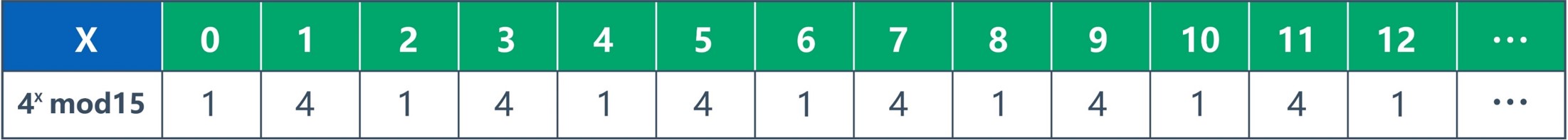}
		\caption{Calculation period table of $a^x \mod 15$, when $a=4$.}
		\label{fig:N15_a4}
	\end{figure}

	\item[Step 5] Since $a^{r/2} \mod N = 4^{1} \mod 15 = 4 \neq -1$, we compute the following:
	\begin{align*}
		\text{gcd}(a^{r/2}-1, N) &= \text{gcd}(4-1, 15) = \text{gcd}(3, 15) = 3, \\
		\text{gcd}(a^{r/2}+1, N) &= \text{gcd}(4+1, 15) = \text{gcd}(5, 15) = 5.
	\end{align*}
	It is clear that both 3 and 5 are factors of 15.
\end{description}

Alternatively, if we select $a=13$ in Step 3, the process would be as follows:

\begin{description}
	\item[Step 3] For $a=13$, we calculate $\text{GCD}(15, 13)=1$ and proceed to the next step.
	
	\item[Step 4] We find the period $r$ for $a=13$. This is shown in Figure \ref{fig:N15_a13}, where we determine that $r=4$.
	
	\begin{figure}[H]
		\centering
		\includegraphics[scale=0.4]{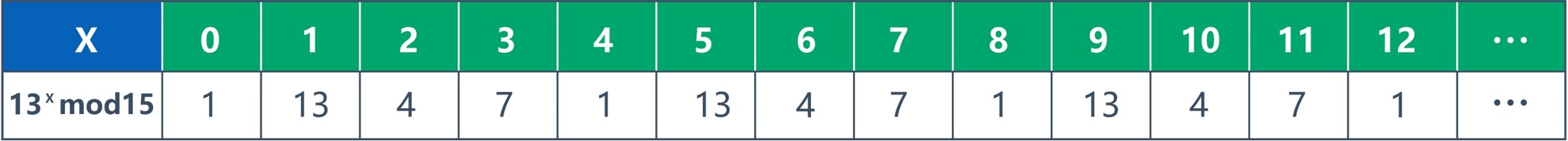}
		\caption{Calculation period table of $a^x \mod 15$, when $a=13$.}
		\label{fig:N15_a13}
	\end{figure}

	\item[Step 5] We calculate $a^{r/2} \mod N = 13^{2} \mod 15 = 4 \neq -1$, and then:
	\begin{align*}
		\text{gcd}(a^{r/2}-1, N) &= \text{gcd}(13^{2}-1, 15) = \text{gcd}(168, 15) = 3, \\
		\text{gcd}(a^{r/2}+1, N) &= \text{gcd}(13^{2}+1, 15) = \text{gcd}(170, 15) = 5.
	\end{align*}
\end{description}

Hence, we confirm again that 3 and 5 are factors of 15.
\end{example}

\begin{example} 
Consider the number $N=55$. We apply the following algorithm steps to find its factors:

\begin{enumerate}
	\item Confirm that $N$ is not even, which is true for $N=55$.
	
	\item Test whether $N$ can be expressed as $N=p^q$ for any integers $p$ and $q$. We find no such $p$ and $q$ that satisfy this condition for $N=55$.
	
	\item Randomly select a number $a$ from the range 2 to 54. For example, choose $a=12$ and calculate the greatest common divisor (GCD) of $N$ and $a$, which is $\text{GCD}(55, 12) = 1$. Since the GCD is 1, we proceed to the next step.
	
	\item Select $x$ starting from 0, calculate $a^x \mod N$, and find the period $r$. For $a=12$, the period is found to be $r=4$, as shown in Figure \ref{fig:N55_a12}.
	
	\begin{figure}[H]
		\centering
		\includegraphics[scale=0.4]{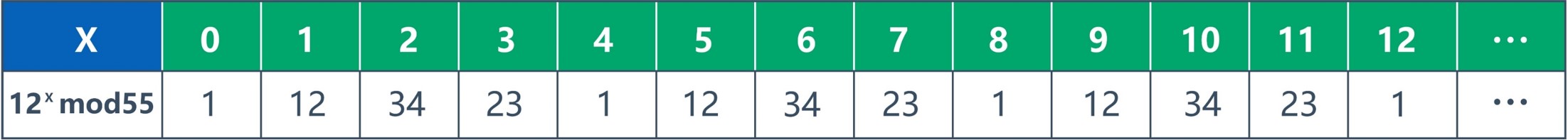}
		\caption{Calculation period table of $a^x \mod 55$, when $a=12$.}
		\label{fig:N55_a12}
	\end{figure}
	
	\item Calculate $a^{r/2} \mod N = 12^{4/2} \mod 55 = 34 \neq -1$. Next, compute the GCDs:
	\begin{align*}
		\text{gcd}(a^{r/2}-1, N) &= \text{gcd}(12^{2}-1, 55) = \text{gcd}(143, 55) = 1, \\
		\text{gcd}(a^{r/2}+1, N) &= \text{gcd}(12^{2}+1, 55) = \text{gcd}(145, 55) = 5.
	\end{align*}
	We find that 5 is a factor of $N$. The other prime factor can then be determined as $\frac{N}{5} = 11$.
\end{enumerate}

Therefore, the prime factors of $N=55$ are 5 and 11.
\end{example}

The complexity of classical algorithms for factoring large numbers can be analyzed by representing the number $N$ as an $n$-bit integer. The algorithmic complexity is expressed in terms of $n$ as follows \cite{4_sx}:

\begin{enumerate}
	\item The complexity of Step 1 is constant, denoted as $O(1)$.
	
	\item In Step 2, we face a polynomial complexity in terms of $n$, which is denoted as $O(p_1(n))$.
	
	\item Step 3 also has polynomial complexity, denoted as $O(p_2(n))$.
	
	\item The complexity of Step 4 is exponential in terms of $n$, specifically expressed as $O(n \cdot 2^{n/3})$.
	
	\item Finally, Step 5 has polynomial complexity, denoted as $O(p_3(n))$.
\end{enumerate}

Consequently, the total complexity of the algorithm is given by the sum $O(p_1(n)) + O(p_2(n)) + O(n \cdot 2^{n/3}) + O(p_3(n))$. However, due to the exponential term $O(n \cdot 2^{n/3})$, it significantly surpasses the complexities of the other polynomial terms. Therefore, in practice, the complexity is often approximated by the exponential term alone, $O(n \cdot 2^{n/3})$.

In theoretical computer science, problems that exhibit polynomial complexity are generally considered easier to solve than those with exponential complexity. When dealing with the factorization of large integers on classical computers, the exponential complexity step renders the problem challenging. As $n$ grows, the resources required for computation increase exponentially, reaching beyond the capabilities of current computers for sufficiently large integers. This exponential complexity is a cornerstone of RSA encryption's security.

The security of RSA encryption depends on the fact that no classical algorithm currently reduces the exponential complexity step to polynomial complexity. Should such an algorithm be discovered, the factorization of large numbers would become tractable in polynomial time, threatening the integrity of RSA encryption.

Quantum computing introduces Shor's algorithm, which theoretically reduces the factoring problem to polynomial time complexity. If a sufficiently advanced quantum computer were to run Shor's algorithm, it could potentially compromise the security of RSA encryption by enabling efficient factorization of large numbers.

\begin{center}
    \textbf{Shor's Algorithm}
\end{center}

Peter Shor, an eminent American mathematician and professor at the Massachusetts Institute of Technology (MIT), made a groundbreaking contribution to quantum computing in 1994 with the introduction of Shor's algorithm \cite{2_26}. This algorithm established that quantum computers have an exponential speed advantage over classical computers in the task of factoring large integers, which has significant implications for cryptography. Specifically, it showed the potential to decrypt encryptions, like RSA \cite{2_27}, that rely on the difficulty of factoring large semiprime numbers.

Shor's algorithm was the first to demonstrate the superiority of quantum algorithms over classical algorithms for certain problems. It showcased the power of quantum computing in number theory factorization, leveraging quantum parallel processing and quantum Fourier transforms \cite{4_sx}. Furthermore, Shor has been pivotal in the development of quantum error correction theory, advancing techniques critical for maintaining quantum information integrity and constructing scalable quantum computers.

Shor's contributions extend beyond theoretical work; he has engaged in collaborations with experimental physicists to transition quantum algorithms from theory to practice. His involvement in testing quantum computing prototypes and advocating for the scaling up of quantum systems has been instrumental in the field's progress.

As a visionary in quantum information science, Shor's profound insights and exceptional talent have sparked widespread interest and investment in the domain. Decades after his seminal work, Shor's algorithm continues to underpin the field of quantum computing and guide the ongoing quest to harness the full potential of quantum technology.

Shor's algorithm, which is essential for factoring integers in quantum computing, relies on determining the multiplicative order $r$ of an integer $a$ modulo another integer $N$. The goal is to find the smallest positive integer $x$ such that:

\begin{equation}
    a^x \equiv 1 \pmod{N}
\end{equation}

This integer $x$ is the period $r$ that the algorithm seeks. Quantum computing facilitates this process by employing a quantum circuit. The design of such a circuit, which is instrumental in finding the period, is illustrated in Figure \ref{17_shor}, can be decomposed into the following three main steps \cite{2_30}:
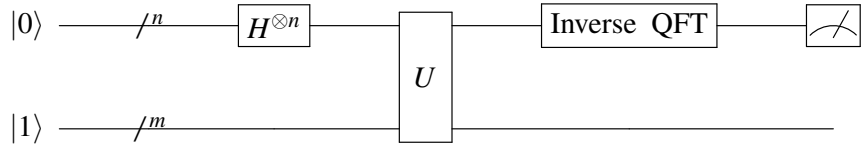
\begin{figure}[H]
	\centering
	\[\Qcircuit @C=3em @!R {
        &\lstick{|0\rangle}&{/^{n}}\qw &\gate{H^{\otimes n}} &\multigate{1}{U} &\gate{\text{Inverse \ QFT}}  &\meter\\
        &\lstick{|1\rangle}&{/^{m}}\qw &\qw &\ghost{U} &\qw &\qw
        }\]
	\caption{\label{17_shor} The circuit of shor algorithm.}
\end{figure}

\begin{enumerate}
    \item Initialize an $n$-qubit system to a superposition state, represented as $\sum_{x=0}^{2^{n}-1}|x\rangle$.
    \item For each quantum state $|x\rangle$, compute $a^x \mod N$.
    \item Apply the quantum fast Fourier transform to find the period $r$ of the function $a^x \mod N$. Afterwards, a classical computer uses the period $r$ to compute the greatest common divisor of either $a^{r/2}-1$ or $a^{r/2}+1$ with $N$, which yields the prime factors of $N$.
\end{enumerate}

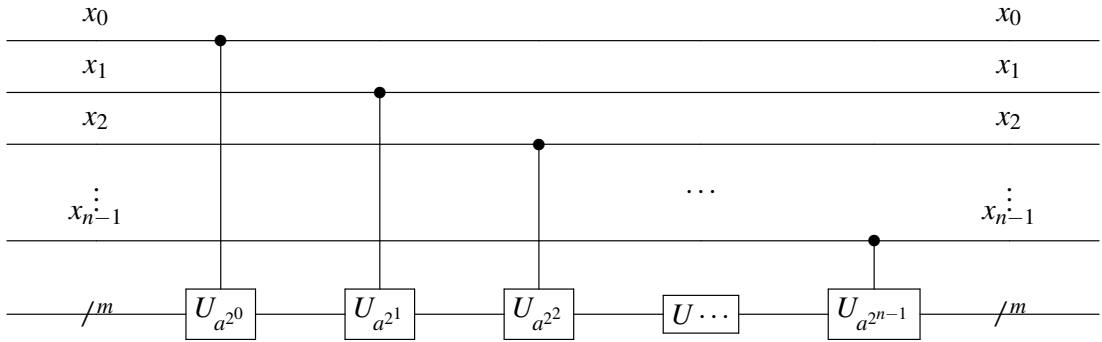
\begin{figure}[H]
	\centering
	\[\Qcircuit @C=3em @R=1.5em {
        &\ustick{x_{0}}\qw&\ctrl{5} &\qw &\qw &\qw&\qw&\ustick{x_{0}}\qw&\qw\\
        &\ustick{x_{1}}\qw&\qw &\ctrl{4} &\qw&\qw&\qw&\ustick{x_{1}}\qw&\qw\\
        &\ustick{x_{2}}\qw &\qw &\qw &\ctrl{3}&\qw&\qw&\ustick{x_{2}}\qw&\qw\\
        &\vdots &  &  & &\cdots & &\vdots\\
        &\ustick{x_{n-1}}\qw&\qw &\qw &\qw&\qw &\ctrl{1}&\ustick{x_{n-1}}\qw&\qw\\
        &{/^{m}}\qw&\gate{U_{a^{2^{0}}}} &  \gate{U_{a^{2^{1}}}} &\gate{U_{a^{2^{2}}}}& \gate{U\cdots}&\gate{U_{a^{2^{n-1}}}}&{/^{m}}\qw&\qw
        }\]
	\caption{\label{17_U} The circuit to achieve $a^x$ mod $N$.}
\end{figure}

In order to compute $a^x \mod N$, which is shown in Figure \ref{17_U}, the following steps are employed:

\begin{enumerate}
    \item Express the integer $x$ in binary form as $x_{n-1} x_{n-2} \cdots x_2 x_1 x_0$, where each $x_i$ is either 0 or 1.
    \item The binary representation allows us to write $x$ as the sum $x = x_{n-1} 2^{n-1} + x_{n-2} 2^{n-2} + \cdots + x_2 2^2 + x_1 2 + x_0$.
    \item Compute $a^x \mod N$ using the repeated squaring method: $a^x \mod N = (a^{2^{n-1}})^{x_{n-1}} \times (a^{2^{n-2}})^{x_{n-2}} \times \cdots \times (a^{2^2})^{x_2} \times (a^2)^{x_1} \times a^{x_0} \mod N$.
    \item This can be further simplified to $a^x \mod N = ((a^{2^{n-1}} \mod N)^{x_{n-1}} \times ((a^{2^{n-2}} \mod N)^{x_{n-2}} \times \cdots \times ((a^{2^2} \mod N)^{x_2} \times ((a^2 \mod N)^{x_1} \times (a \mod N)^{x_0} \mod N$.
\end{enumerate}

The computation of $a^x \mod N$ can be efficiently performed through a classical algorithm in $O(n)$ time complexity, where $n=\log N$. The specific steps are as follows:

\begin{enumerate}
    \item Calculate the initial value $y_0 = a^{2^0} \mod N$.
    \item Sequentially compute each $y_i$ for $i=1$ to $n-1$ by squaring the previous value and taking the modulus $N$, i.e., $y_i = (y_{i-1} \times y_{i-1}) \mod N$.
\end{enumerate}

\noindent Each of these steps can be completed in constant time, leading to a total time complexity of $O(n)$. However, a classical algorithm would require iterating over all values of $x$ from 0 to $N$, resulting in a time complexity of $O(n \times N) = O(n \times 2^n)$, which is exponential.

In contrast, quantum algorithms can prepare all states $|x\rangle$ for $x$ ranging from 0 to $2^n-1$ simultaneously, enabling polynomial-time execution of this step. This represents a significant computational speed-up offered by quantum algorithms. The following two points are crucial:

\begin{enumerate}
    \item Efficient implementation of the modular exponentiation function $U_{a^{2^i}}$ on a quantum computer. Although this function requires exponential time classically, Shor's algorithm can execute it in polynomial time by utilizing the parallel computational capabilities of quantum computers.
    \item Determination of the period $r$ of $a^x \mod N$. This challenge is addressed using the quantum fast Fourier transform to calculate the period in polynomial time.
\end{enumerate}

\begin{example}
Below, we demonstrate an example of how to factorize the number 15 using Shor's algorithm, a process that utilizes the Quantum Fourier Transform (QFT) \cite{4_sx}.

Firstly, we select a random number that is coprime with 15, such as 7. Next, we need to determine the order $r$ of this number modulo 15. This can be accomplished using the quantum order-finding algorithm, which starts from the state $\ket{0}\ket{0}$ and by applying Hadamard transforms to the first register, creates a superposition state that contains all possible values of $k$ from 0 to $2^{t}-1$. The value of $t$ is chosen to be 11 to ensure that the probability of error is no more than 1/4.

Subsequently, we calculate the function $f(k) = 7^k \mod 15$, with the results stored in the second register. This results in a new superposition state encompassing all possible values of $k$ and their corresponding $f(k)$ values.

In this process, the inverse Quantum Fourier Transform (QFT$^{\dagger}$) is applied to the first register, which is then measured. We can analyze the potential measurement outcomes by computing the reduced density matrix for the first register, applying QFT$^{\dagger}$ to it, and computing the measurement statistics. However, since no further operation is performed on the second register, we can employ the principle of implicit measurement, assuming that the second register has been measured to yield a random result of 1, 7, 4, or 13. Let's assume we obtain 4.

After applying the inverse QFT, we achieve a probability distribution associated with the state $\ket{2} + \ket{6} + \ket{10} + \ket{14} + \cdots$. Upon final measurement, with a probability close to an exact 1/4, we might get one of the values 0, 512, 1024, or 1536 shown in Figure \ref{shor_N_15}. Suppose the measurement result is 1536, we find the order $r$ by computing the continued fraction expansion. The ratio 1536/2048 can be represented as $1/(1 + (1/3))$, indicating that 3/4 is a convergent in the expansion, thus determining the order $r$ to be 4.
\begin{figure}[H]
\centering
\includegraphics[scale=0.5]{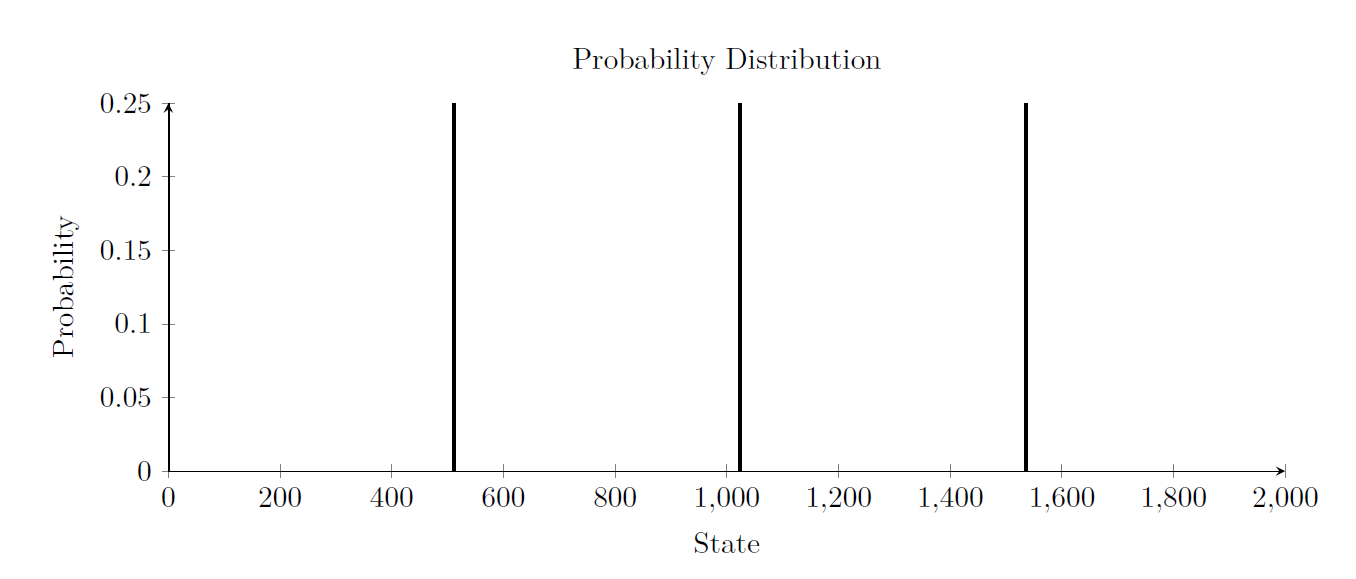}
 \caption{\label{shor_N_15} The measurement outcomes of the quantum state.}
\end{figure}

Since $r$ is even, and $7^{r/2} \mod 15 = 7^2 \mod 15 = 4$ is not equal to $-1 \mod 15$, the algorithm is effective. Lastly, we compute the greatest common divisors $\gcd(7^2 - 1, 15) = 3$ and $\gcd(7^2 + 1, 15) = 5$, revealing that 15 can be factored into the product of 3 and 5.

This procedure illustrates a practical application of the Quantum Fourier Transform in quantum algorithms. By leveraging the properties of quantum superposition and entanglement, QFT can efficiently process information, thereby enhancing the efficiency of certain computational tasks, such as factorizing large numbers in this example.
\end{example}

\newpage
\chapter{Week 7: A Brief Introduction to Quantum Mechanics}
\section{Quantum state, probability}

Quantum mechanics \cite{sun22dirac1926theory} offers a radically different approach to describing physical systems compared to classical physics. At the core of this theory are a series of postulates that reveal the non-classical and probabilistic nature of the quantum realm.

The first crucial postulate asserts that the complete description of a quantum system is encapsulated by its wavefunction. Unlike classical states, which are characterized by definite values, the wavefunction is an abstract mathematical representation that encompasses all possible information about the quantum system.

Other postulates focus on the outcomes of measurements. When a measurement is taken, the wavefunction is said to collapse to a specific basis state according to probability.

Dirac notation is a powerful mathematical tool for representing quantum states. It assigns each basis state $|x_i\rangle$ to a standard basis vector, which has a value of 1 at the $i^{\mathrm{th}}$ position and 0 at all others. For instance, the basis state $|x_0\rangle$ corresponds to the vector $[1,0,\ldots,0]^T$, the basis state $|x_1\rangle$ corresponds to the vector $[0,1,\ldots,0]^T$, and so on.

\begin{figure}[H]
	\centering
	\includegraphics[scale=0.5]{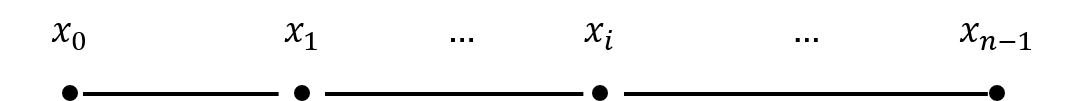}
	\caption{\label{18_dirac_notation} Dirac notation.}
\end{figure}

A general quantum state $|\psi\rangle$ can be expressed as a superposition of these basis states \cite{sun23wineland2013nobel}, with complex coefficients $c_i$ indicating the amplitude on each basis state. Mathematically, $|\psi\rangle = \sum_i c_i |x_i\rangle$. This quantum state $|\psi\rangle$ maps to a column vector containing the coefficients $[c_0,c_1,\ldots,c_{n-1}]^T$.

A key feature of quantum mechanics is the probabilistic nature of measurement outcomes. Consider a general quantum state $|\psi\rangle$ expressed as a superposition in some basis:

\begin{equation}
    |\psi\rangle = c_0|x_0\rangle + c_1|x_1\rangle + \cdots + c_{n-1}|x_{n-1}\rangle,
\end{equation}
where the $c_i$ are complex amplitudes. If a measurement is performed on this state, the probability of observing outcome $|x_i\rangle$ is given by:

\begin{equation}
    p(x_i) = \frac{|c_i|^2}{||\psi\rangle|^2} = \frac{|c_i|^2}{\sum_j |c_j|^2},
\end{equation}
we find $|\psi\rangle$ in one of the basis states, $|\psi\rangle$ becomes
\begin{equation}
|\psi\rangle \longrightarrow |x_i\rangle,
\end{equation}

Thus, the amplitudes $c_i$ determine the probability distribution over outcomes, a fundamentally quantum effect.

\begin{exercise}
Assume the particle can only be at the four points ${x_0,x_1,x_2,x_3}$.
Assume the state vector is
\begin{align}
|\psi\rangle=
\begin{bmatrix}
2+3i \\
-1+i \\
4 \\
-2-2i
\end{bmatrix},
\end{align}
Calculate the probability that the particle can be found at position $x_3$.
\end{exercise}

Quantum states can be added together to form new quantum states. Consider two quantum states $|\psi\rangle$ and $|\psi'\rangle$ expressed as superpositions in the same basis:

\begin{align}
	|\psi\rangle &= c_0|x_0\rangle + c_1|x_1\rangle + \cdots + c_{n-1}|x_{n-1}\rangle 
	= [c_0, c_1, \ldots, c_{n-1}]^T \nonumber\\
	|\psi'\rangle &= c'_0|x_0\rangle + c'_1|x_1\rangle + \cdots + c'_{n-1}|x_{n-1}\rangle 
	= [c'_0, c'_1, \ldots, c'_{n-1}]^T.
\end{align}

The sum of these two states is obtained by adding the respective coefficients on each basis state:
\begin{align}
	|\psi\rangle + |\psi'\rangle &= (c_0 + c'_0)|x_0\rangle + (c_1 + c'_1)|x_1\rangle + \cdots+ (c_{n-1} + c'_{n-1})|x_{n-1}\rangle \nonumber\\
	&= [c_0 + c'_0, c_1 + c'_1, \ldots, c_{n-1} + c'_{n-1}]^T.
\end{align}

Consider a quantum state $|\psi\rangle$ and a scalar $c$:
\begin{equation}
\begin{split}
	c|\psi\rangle &= cc_0|x_0\rangle + cc_1|x_1\rangle + \cdots + cc_{n-1}|x_{n-1}\rangle\\
	&= [cc_0, cc_1, \ldots, cc_{n-1}]^T.
 \end{split}
\end{equation}

Importantly, the measurement probabilities are unchanged up to a factor. 

Defining
\begin{equation}
    S = \sum_j |cc_j|^2 = |c|^2 \sum_j |c_j|^2,
\end{equation}
we have:
\begin{equation}
    p(x_j) = \frac{|cc_j|^2}{S} = \frac{|c_j|^2}{\sum_j |c_j|^2}.
\end{equation}
So $c|\psi\rangle$ describes the same physical system as $|\psi\rangle$.

Since scalar multiples of a quantum state describe the same physical system, the overall length (norm) of the state vector is irrelevant. We can thus work with normalized quantum states without loss of generality.
A normalized state $|\psi\rangle$ satisfies:
\begin{equation}
    \frac{|\psi\rangle}{||\psi\rangle|}.
\end{equation}

\begin{exercise}
	Normalize the vector 
 \begin{align*}
 |\psi\rangle = 
\begin{bmatrix}
1 - 2i \\
3 + i \\
-2 + 2i \\
0.5 - 1.5i
\end{bmatrix}.
 \end{align*}
\end{exercise}

With a normalized state, the measurement probabilities take the simplified form:

\begin{equation}
    p(x_i) = |c_i|^2.
\end{equation}

\begin{figure}[H]
	\centering
	\includegraphics[scale=0.5]{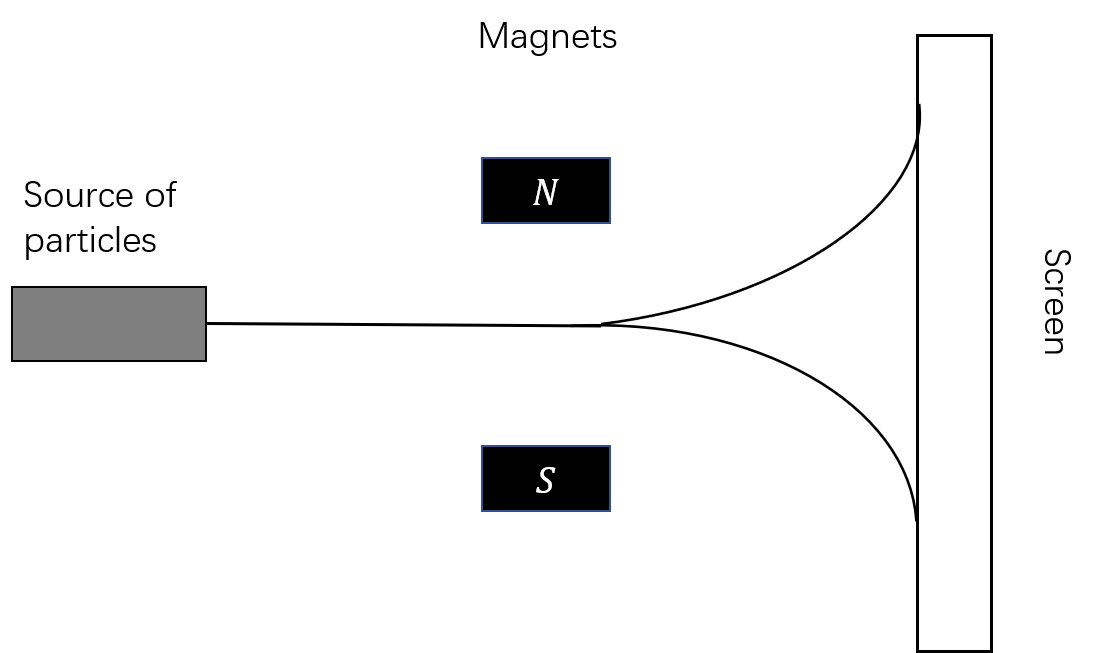}
	\caption{\label{18_stern_gerlach} The Stern-Gerlach experiment.}
\end{figure}

The Stern-Gerlach experiment \cite{sun24zhu2011improving} is a fundamental experiment in quantum mechanics that demonstrates the quantization of angular momentum and the discrete nature of spin angular momentum shown in Figure \ref{18_stern_gerlach}.

\begin{figure}[H]
	\centering
	\includegraphics[scale=0.5]{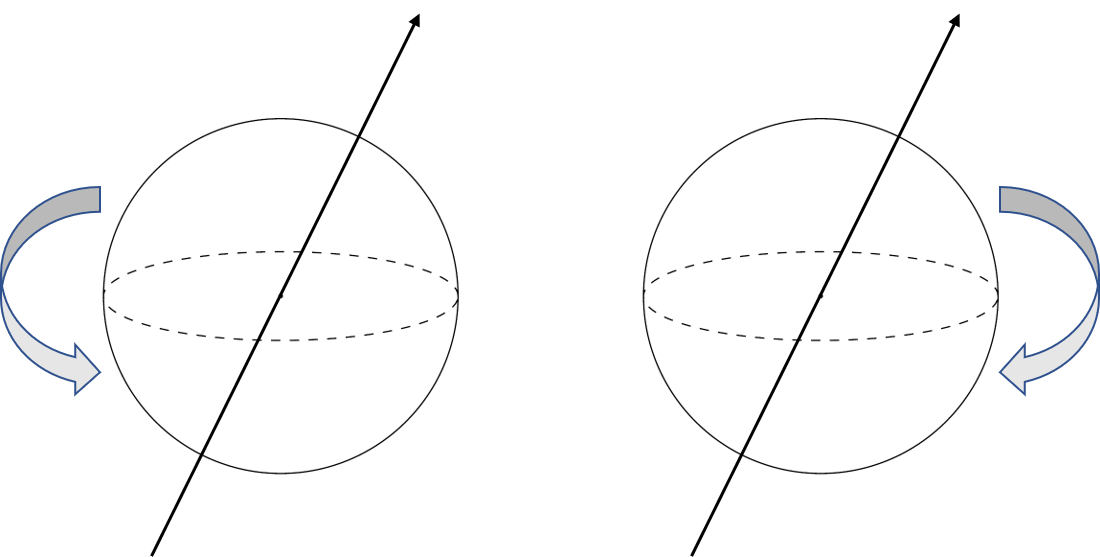}
	\caption{\label{18_spin} Particles with spin.}
\end{figure}

The spin states are denoted $|\uparrow\rangle$ for spin up and $|\downarrow\rangle$ for spin down shown in Figure \ref{18_spin}. These two basis states can be used to represent a quantum bit (qubit).
A general qubit state $|\psi_0\rangle$ is a superposition of the two spin states:

\begin{equation}
 |\psi_0\rangle = c_0|\uparrow\rangle + c_1|\downarrow\rangle, 
\end{equation}
where $c_0$ and $c_1$ are complex amplitudes.

\begin{exercise}
	Consider the vector 
 \begin{equation}
     |\psi\rangle = (2 + 6i)|\uparrow\rangle + (5 - 7i)|\downarrow\rangle.
 \end{equation}
 Calculate the probability of spin-up and spin-down.
\end{exercise}

In Dirac notation, quantum states are represented by ``ket" vectors and ``bra" covectors.

A ket vector $|\psi\rangle$ corresponds to a column vector containing the amplitudes $c_i$ on each basis state:

\begin{equation}
    |\psi\rangle = \begin{bmatrix} c_0 \\ c_1 \\ \vdots \\ c_{n-1} \end{bmatrix},
\end{equation}

The bra covector $\langle \psi|$ is defined as the Hermitian conjugate (adjoint) of the ket, giving a row vector:
\begin{equation}
    \langle \psi| = [c_0^\dagger, c_1^\dagger, \ldots, c_{n-1}^\dagger],
\end{equation}
where $\dagger$ denotes complex conjugation.
And
\begin{equation}
    \langle \psi| = |\psi\rangle^\dagger.
\end{equation}

\section{Dynamics}

The time evolution of a quantum systems is governed by unitary transformations. This is formalized in the fifth postulate of quantum mechanics:

If $|\psi(t)\rangle$ represents the state of a quantum system at time $t$, then:

\begin{equation}
    |\psi(t+1)\rangle = U|\psi(t)\rangle,
\end{equation}
where $U$ is a unitary operator that encapsulates the system dynamics.

The time evolution of a quantum system is governed by the Schrödinger equation. For a small $\Delta t$, this equation can be written as:

\begin{equation}
    |\psi(t + \Delta t)\rangle - |\psi(t)\rangle = -i\textbf{H}\Delta t|\psi(t)\rangle,
\end{equation}
Where \textbf{H} is the Hamiltonian operator representing the system's total energy. This differential equation can be integrated to give:

\begin{align}
|\psi(t + \Delta t)\rangle = (I - i\textbf{H}\Delta t)|\psi(t)\rangle &\approx (\cos(\textbf{H}\Delta t) -i\sin(\textbf{H}\Delta t))|\psi(t)\rangle \nonumber\\
\Rightarrow|\psi(t + \Delta t)\rangle &= e^{-i\textbf{H}\Delta t}|\psi(t)\rangle,
\end{align}
for the state evolution between times $t$ and $t + \Delta t$. Comparing with the general form:
\begin{equation}
    |\psi(t + \Delta t)\rangle = U|\psi(t)\rangle.
\end{equation}
We see that the unitary operator $U$ associated with the evolution is given by:
\begin{equation}
    U = e^{-i\textbf{H}\Delta t}.
\end{equation}

\begin{exercise}
Verify that
\begin{equation}
    U_1 = \begin{bmatrix} 0 & 1\\ 1 & 0 \end{bmatrix}, \quad U_2 = \begin{bmatrix} 1 & 0\\ 0 & -1 \end{bmatrix},
\end{equation}
are unitary matrices. Multiply them and verify that their product is
also unitary.
\end{exercise}

\begin{example}
The matrices
\begin{equation}
    X = \begin{bmatrix} 0 & 1\\ 1 & 0 \end{bmatrix}, \quad
Y = \begin{bmatrix} 0 & -i\\ i & 0 \end{bmatrix}, \quad
Z = \begin{bmatrix} 1 & 0\\ 0 & -1 \end{bmatrix},
\end{equation}
are unitary matrices. The content later will mention that they are also Hermitian matrices.
\end{example}

The time evolution of a closed quantum system can be modeled as a sequence of unitary transformations.

Consider a system at times $t_0, t_1, \ldots, t_{n-1}$ with associated unitary matrices $U_0, U_1, \ldots, U_{n-1}$ representing the evolution between successive time steps.

If $|\psi\rangle$ is the initial state of the system at time $t_0$, the state at the final time $t_{n-1}$ is given by:
\begin{equation}
    |\psi(t_{n-1})\rangle = U_{n-1}U_{n-2}\ldots U_1U_0|\psi\rangle.
\end{equation}

\begin{example}
For the initial state:
\begin{equation}
    |\psi\rangle = |1\rangle,
\end{equation}
and $U(t_0)=U_1, U(t_1)= U_2,$ where 
\begin{equation}
    U_1 = \frac{1}{\sqrt{2}}\begin{bmatrix} 1 & 1\\ 1 & -1 \end{bmatrix}, \quad U_2 = \begin{bmatrix} \cos(\pi/4) & -\sin(\pi/4)\\ \sin(\pi/4) & \cos(\pi/4) \end{bmatrix}.
\end{equation}
The evolution of $|\psi\rangle$ after $U_1, U_2$ is: 
\begin{equation}
    \frac{1}{\sqrt{2}}\begin{bmatrix} -\cos(\pi/4)-\sin(\pi/4) \\ \cos(\pi/4)-\sin(\pi/4)\end{bmatrix}.
\end{equation}
Here's the MATLAB code for this example:
\begin{lstlisting}
clear,clc;

% Define the initial state |1>
psi = [0; 1];

% Define the unitary operations U1 and U2
U1 = (1/sqrt(2)) * [1 -1; 1 1];
U2 = [cos(pi/4) -sin(pi/4); sin(pi/4) cos(pi/4)];

% Apply U1 to the initial state
psi_t0 = U1 * psi;

% Apply U2 to the state after U1
psi_t1 = U2 * psi_t0;

% Display the final state
disp('The state |psi> at time t1:');
disp(psi_t1);
\end{lstlisting}
\end{example}

\begin{exercise}
For the initial state:
\begin{equation}
    |\psi\rangle = |0\rangle,
\end{equation}
and $U(t_0)=U_1, U(t_1)= U_2,$ where 
\begin{equation}
    U_1 = \begin{bmatrix} 0 & 1\\ 1 & 0 \end{bmatrix}, \quad U_2 = \frac{1}{\sqrt{2}}\begin{bmatrix} 1 & 1\\ 1 & -1 \end{bmatrix}.
\end{equation}
How does $|\psi\rangle$ evolve in time?
\end{exercise}

The evolution of quantum systems over time follows precise mathematical laws, which are based on two fundamental principles:

Firstly, the evolution of quantum systems adheres to the principle of unitary transformations. Unitary transformations are reversible linear operations that maintain the inner product between quantum state vectors. This implies that the dynamics of a closed quantum system can be reversed, a property that is counterintuitive.

Secondly, the specific time evolution of a quantum system is governed by the Schrödinger equation. This equation is related to the system's Hamiltonian, which summarizes the total energy and intrinsic interactions. By solving the Schrödinger equation, one can predict how a quantum system will evolve from its initial state over time.

Combined, these principles offer a robust framework for the description of quantum phenomena. With knowledge of the Hamiltonian and the associated unitary transformations, the dynamic behavior of any closed quantum system can be accurately described. However, the act of measuring the system introduces probabilistic disturbances to this deterministic evolution, disturbances that are not characterized by unitary operations. Understanding quantum dynamics has profound implications for fields such as quantum computing, condensed matter physics, and beyond.

\section{Quantum entanglement}

Here we will introduce the mathematical rule for combining separate quantum systems into a joint quantum system. Specifically, if we have two independent systems $Q$ and $Q^{'}$ with state spaces $V$ and $V^{'}$, then the joint system has state space $V \otimes V^{'}$. 

\begin{figure}[H]
	\centering
	\includegraphics[scale=0.5]{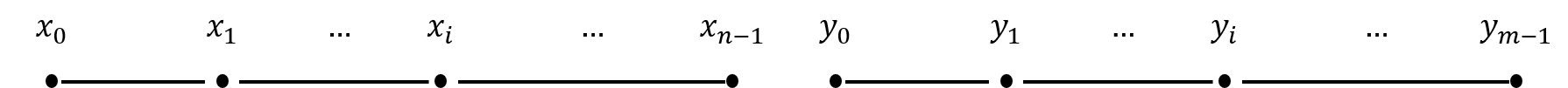}
	\caption{\label{vector2} Two quantum systems with orthonormal basis states.}
\end{figure}

This part introduces the tensor product, an important mathematical construct for describing composite quantum systems. 
Consider two quantum systems with orthonormal basis states, shown in Figure \ref{vector2}, ${|x_0\rangle, ..., |x_{n-1}\rangle}$ and ${|y_0\rangle, ..., |y_{m-1}\rangle}$ respectively. The tensor product expresses joint states across the two systems.

For example, the state $|x_i\rangle \otimes |y_j\rangle$ denotes the system being in state $|x_i\rangle$ for the first subsystem and $|y_j\rangle$ for the second. Taking linear combinations results in: 
\begin{equation}
|\psi\rangle = c_{0,0}|x_0\rangle\otimes|y_0\rangle+...+c_{i,j}|x_i\rangle\otimes|y_j\rangle+...+c_{n-1,m-1}|x_{n-1}\rangle\otimes|y_{m-1}\rangle.
\end{equation}

Then we discuss quantum entanglement, specifically the state:
\begin{equation}
|\psi\rangle = |x_0\rangle \otimes |y_0\rangle + |x_1\rangle \otimes |y_1\rangle,
\end{equation}
or 
\begin{equation}
|\psi\rangle = 1 \cdot |x_0\rangle \otimes |y_0\rangle + 0 \cdot |x_0\rangle \otimes |y_1\rangle+ 0 \cdot |x_1\rangle \otimes |y_0\rangle+ 1 \cdot |x_1\rangle \otimes |y_1\rangle,
\end{equation}
This state cannot be written in the separable form:
\begin{equation}
(c_0|x_0\rangle + c_1|x_1\rangle) \otimes (c'_0|y_0\rangle + c'_1|y_1\rangle),
\end{equation}
for any choice of coefficients $c_0, c_1, c'_0, c'_1$.
States that cannot be written in this separable form exhibit quantum entanglement. 

Here provides another example:
\begin{equation}
|\psi'\rangle = |x_0\rangle \otimes |y_0\rangle + |x_0\rangle \otimes |y_1\rangle + |x_1\rangle \otimes |y_0\rangle + |x_1\rangle \otimes |y_1\rangle.  
\end{equation}
Unlike the previous example, this state can be written in the separable form
\begin{equation}
|\psi'\rangle = (|x_0\rangle + |x_1\rangle) \otimes (|y_0\rangle + |y_1\rangle).
\end{equation}
States that can be expressed as a tensor product of two independent quantum states are called separable states \cite{sun25li2008classical}.

We will further clarify the distinction between separable and entangled quantum states~\cite{sun26doherty2002distinguishing}. A separable state can be decomposed as a tensor product of states from each constituent subsystem. For example:
\begin{equation}
|\psi'\rangle = (|x_0\rangle + |x_1\rangle) \otimes (|y_0\rangle + |y_1\rangle), \nonumber
\end{equation}
is a separable state. 
On the other hand, an entangled state cannot be broken down into tensor products of subsystem states. There is an inherent correlation between the subsystems that prevents such a decomposition. For example: 
\begin{equation}
|\psi\rangle = |x_0\rangle \otimes |y_0\rangle + |x_1\rangle \otimes |y_1\rangle, \nonumber
\end{equation}
is an entangled state.
This distinction between separability and entanglement is absolutely vital in quantum theory. 

\begin{figure}[H]
	\centering
	\includegraphics[scale=0.4]{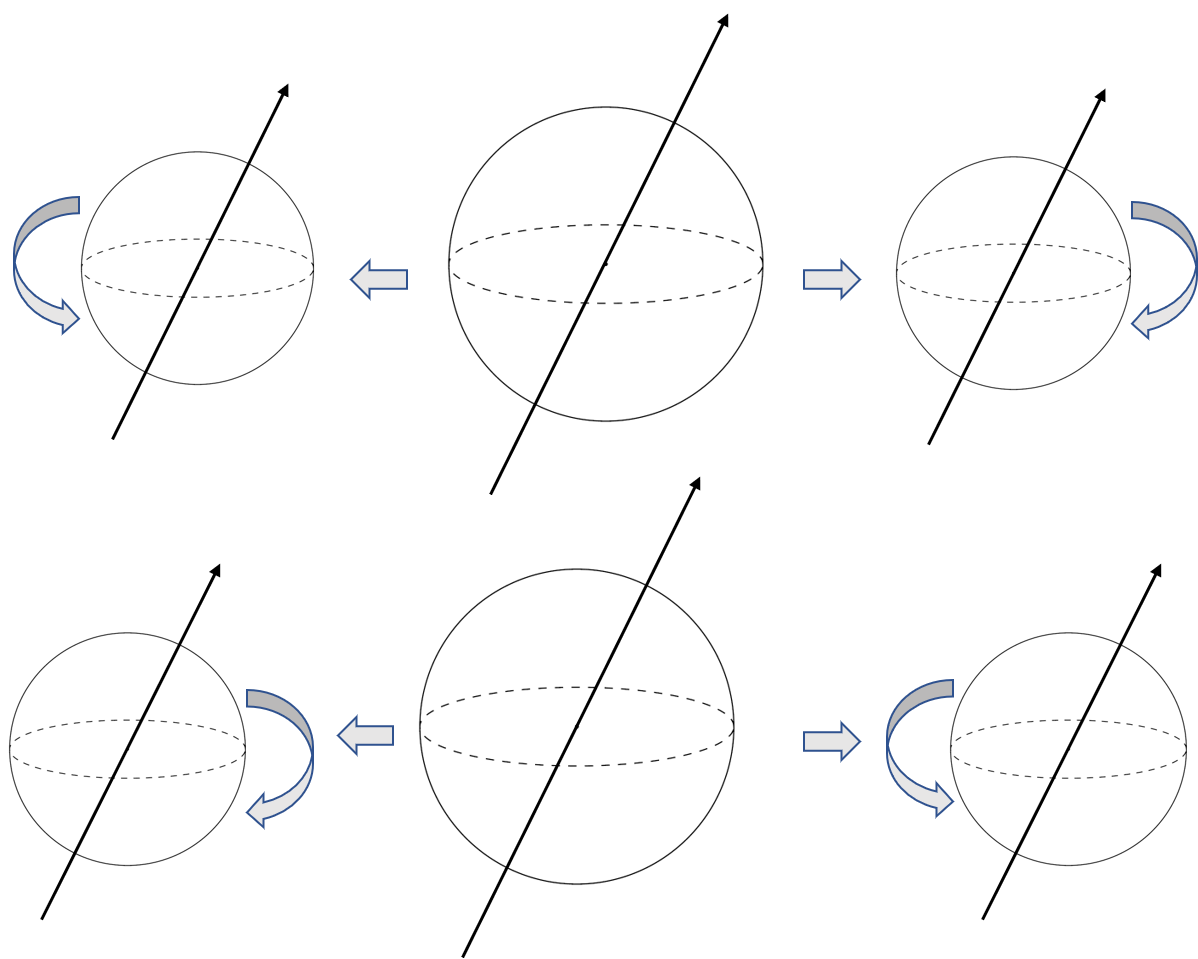}
	\caption{\label{EPR} A common example of an entangled quantum state.}
\end{figure}

And then consider another common example of an entangled quantum state~\cite{sun26doherty2002distinguishing} involving two spin systems \cite{sun27verstraete2004entanglement} labeled left (L) and right (R) shown in Figure \ref{EPR}:
\begin{equation}
\frac{1}{\sqrt{2}} (|\uparrow\rangle_L \otimes |\downarrow\rangle_R + |\downarrow\rangle_L \otimes |\uparrow\rangle_R).  
\end{equation}
This state is known as one of the Bell states or Einstein-Podolsky-Rosen (EPR) pairs\cite{sun28hagley1997generation}. 

The no-cloning theorem \cite{sun29wootters2009no}, first proposed by W.K. Wootters and W.H. Zurek in 1982, states that it is impossible to create an independent and identical copy of an arbitrary unknown quantum state. This theorem is fundamental to our understanding of quantum mechanics and has far-reaching implications.

The proof of the no-cloning theorem \cite{sun29wootters2009no} demonstrates the incompatibility between cloning an unknown quantum state and the linearity of quantum mechanics. It proceeds by contradiction as follows:

Suppose we could clone arbitrary unknown quantum states, so that $|\psi\rangle 
 \rightarrow |\psi\rangle|\psi\rangle$. This cloning transformation could be represented by a unitary operator $U$ acting on $|\psi\rangle|0\rangle$, such that 
 \begin{align}
 \nonumber
&U|0\rangle|0\rangle \rightarrow |0\rangle|0\rangle \\ 
&U|1\rangle|0\rangle \rightarrow |1\rangle|1\rangle.
 \end{align}
By the linearity of quantum mechanics, if $|\psi\rangle = \alpha|0\rangle+\beta|1\rangle$, then:
 \begin{align}
U(\alpha|0\rangle+\beta|1\rangle)|0\rangle \rightarrow \alpha|0\rangle|0\rangle+\beta|1\rangle|1\rangle.
 \end{align}
But we know that 
 \begin{align}
\alpha|0\rangle|0\rangle+\beta|1\rangle|1\rangle \neq (\alpha|0\rangle+\beta|1\rangle)(\alpha|0\rangle+\beta|1\rangle).
 \end{align}

This is a contradiction, since the linearity requirement is incompatible with the cloning requirement. Hence cloning of arbitrary unknown quantum states is impossible.
Therefore, the no-cloning theorem arises directly from the linear structure of quantum mechanics. This simple but profound result places strict limits on quantum information processing tasks.

The no-cloning theorem \cite{sun29wootters2009no} places fundamental limits on quantum operations and quantum information processing. For instance, it prohibits the possibility of perfectly copying an unknown quantum state. This has important implications in quantum cryptography, where security relies on the inability to exactly replicate quantum states. The no-cloning theorem is also closely related to the quantum no-deleting theorem and the no-hiding theorem. Together, these theorems encapsulate intrinsic features of quantum mechanics that distinguish it from classical mechanics.

In summary, several postulates in quantum mechanics can correspond to a few fundamental concepts in quantum computing as follows, shown in Table \ref{cor}.
\begin{table}[H]
	\centering
        \setlength{\tabcolsep}{5mm}
        \renewcommand\arraystretch{1.5}
	\begin{tabularx}{\textwidth}{|X|X|}
            \hline
		Postulates in quantum mechanics & Concepts in quantum computing \\\hline
The state of quantum mechanical system is completely specified by the wavefunction. &  The state of a qubit can be described using a two-dimensional complex vector, typically represented as $|\psi\rangle= \alpha|0\rangle + \beta |1\rangle$, where $|0\rangle$ and $|1\rangle$ are computational basis states, and $\alpha$ and $\beta$ are complex probability amplitudes. Thus, the wavefunction postulate corresponds to qubit in quantum computation.\\\hline
The evolution of a quantum system (that is not a measurement) is
given by a unitary operator or transformation. & In quantum computing, the evolution of qubits is controlled by quantum logic gates, which correspond to unitary transformations in the Hilbert space. Therefore, the evolution postulate corresponds to gates in quantum computing.\\\hline
When we measure a quantum system, the system collapses to one of its possible states, and the probability of each state occurring is given by the square of the modulus of the wavefunction. & In quantum computing, when we measure a qubit, we obtain a result of either $|0\rangle$ or $|1\rangle$. Thus, the measurement postulate corresponds to the readout in quantum computing.\\\hline
	\end{tabularx}
	\caption{\label{cor}Several postulates in quantum mechanics can correspond to a few fundamental concepts in quantum computing.}
\end{table}

\section{Experiments of Bell states and Bell's inequality}

The history of Bell's inequality \cite{sun30aspect1999bell} is deeply rooted in the fundamental differences between quantum mechanics and classical physics, and the exploration of realism in physics. In the early 20th century, the development of quantum mechanics challenged the worldview of classical physics. Despite the enormous success of quantum theory in predicting experimental outcomes, its indeterministic and probabilistic nature raised questions among some physicists. The most notable critique came from Einstein, who, along with his colleagues Boris Podolsky and Nathan Rosen, published a paper in 1935 presenting the famous Einstein-Podolsky-Rosen (EPR) paradox \cite{sun31reid2009colloquium}, questioning the completeness of quantum mechanics with respect to physical reality.

The EPR paper proposed a thought experiment involving two quantum particles that interact and then separate in such a way that a measurement on one particle seems to instantaneously affect the state of the other particle, even when they are far apart \cite{sun32fine2004einstein}. This is known as quantum entanglement. The EPR paper claimed that such instantaneous, long-distance effects, or ``spooky action at a distance" were incompatible with the theory of relativity, and suggested that quantum mechanics might be incomplete, with hidden variables potentially explaining the entanglement phenomenon.

It was not until 1964 that the Irish physicist John Bell proposed the so-called ``Bell's inequality", a set of inequalities that statistically distinguish the predictions of quantum mechanics from those of any local realistic theory with hidden variables. The essence of Bell's inequality  
\cite{sun30aspect1999bell} is that if local realism is correct, then the measurement results produced by quantum entanglement should obey certain statistical constraints, namely Bell's inequalities.

In the 1970s and 1980s, physicists such as Alain Aspect conducted experiments to test Bell's inequalities \cite{sun30aspect1999bell}. The experiments, which usually involved pairs of entangled particles, showed results that violated Bell's inequalities, supporting the predictions of quantum mechanics and challenging the concept of local realism. These experiments provided strong evidence for the reality of quantum entanglement and were a significant affirmation of the principles of quantum mechanics\cite{sun30aspect1999bell}.

The proposition and subsequent experimental verification of Bell's inequality not only deepened our understanding of the quantum world but also laid the groundwork for the development of quantum information science, including research in the fields of quantum computing, quantum communication, and quantum cryptography.

\begin{figure}[H]
	\includegraphics[scale=0.55]{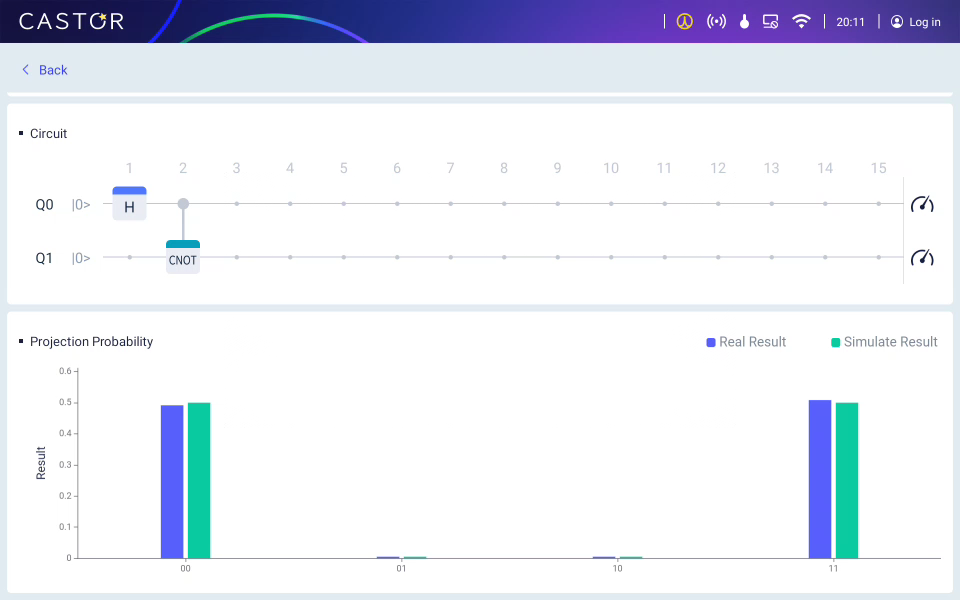}
	\centering
	\caption{\label{bell_state} The experimental circuit and result diagram of Bell states.}
\end{figure}

In this experiment, we endeavored to generate one of the Bell states, which are the epitome of quantum entanglement and foundational to quantum communication and computation. Our setup involved a quantum circuit that manipulates two qubits initially in the ground state $|00\rangle$. Through the application of a Hadamard gate on the first qubit, we created a superposition, followed by a CNOT gate to entangle the qubits, ideally resulting in the Bell state $|\phi^+\rangle = \frac{1}{\sqrt{2}}(|00\rangle + |11\rangle)$, shown in Figure \ref{bell_state}. Despite the presence of noise and imperfections within the quantum computer, the observed statistical distribution largely adhered to theoretical expectations.

\begin{figure}[H]
	\includegraphics[scale=1.2]{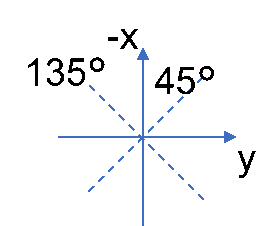}
	\centering
	\caption{\label{measurements} The four sets of measurements of Bell inequality verification experiment.}
\end{figure}

The verification of Bell's inequality involves measurements of quantum entangled states. This experiment requires four sets of measurements, each targeting specific directions for a pair of qubits (usually entangled particles). The measurement setups are as follows shown in Figure \ref{measurements}:
\begin{enumerate}
    \item First set of measurements: The first qubit is measured in the y-direction, and the second qubit is measured at a 45\textdegree angle.
    \item Second set of measurements: The first qubit is measured in the y-direction, and the second qubit is measured at a 135\textdegree angle.
    \item Third set of measurements: The first qubit is measured in the -x direction, and the second qubit is measured at a 45\textdegree angle.
    \item Fourth set of measurements: The first qubit is measured in the -x direction, and the second qubit is measured at a 135\textdegree angle.
\end{enumerate}

\begin{figure}[H]
	\centering
	\subfigure[The results for the first set of experiments.\label{Bell_inequality1}]{
		\includegraphics[scale=0.3]{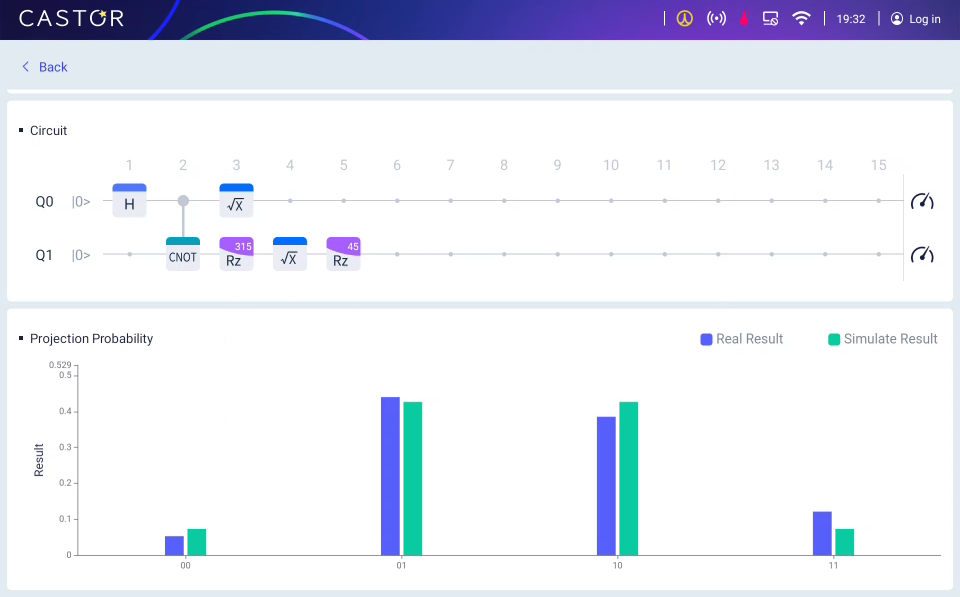}}
	\quad
	\subfigure[The results for the second set of experiments.\label{Bell_inequality2}]{
		\includegraphics[scale=0.3]{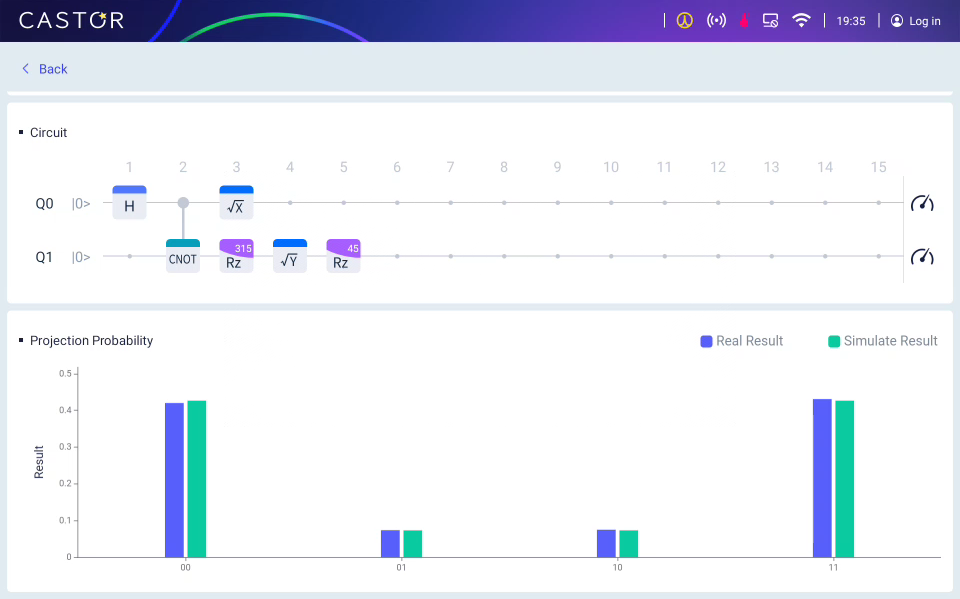}}
        \quad
	\subfigure[The results for the third set of experiments.\label{Bell_inequality3}]{
		\includegraphics[scale=0.3]{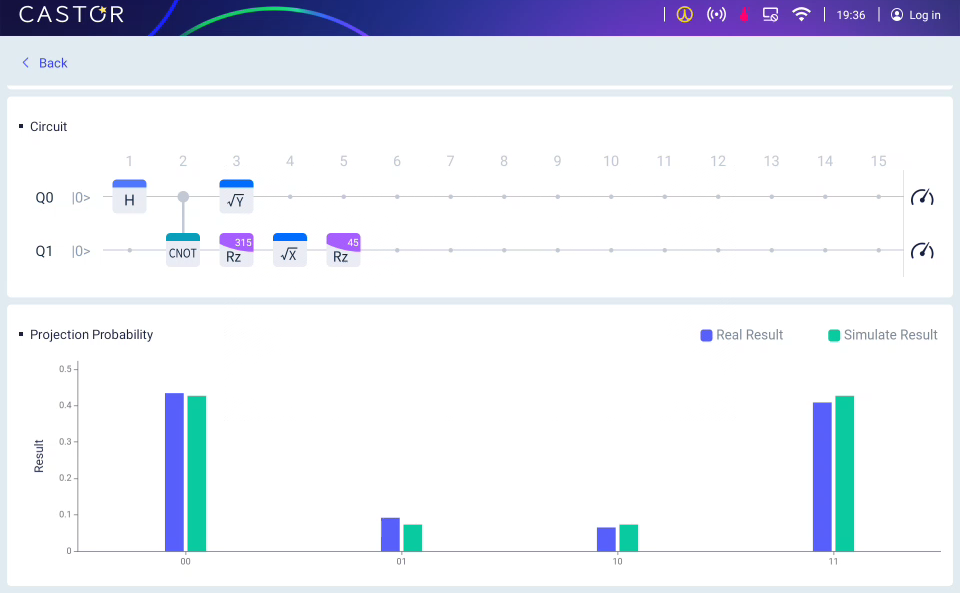}}
        \quad
	\subfigure[The results for the fourth set of experiments.\label{Bell_inequality4}]{
		\includegraphics[scale=0.3]{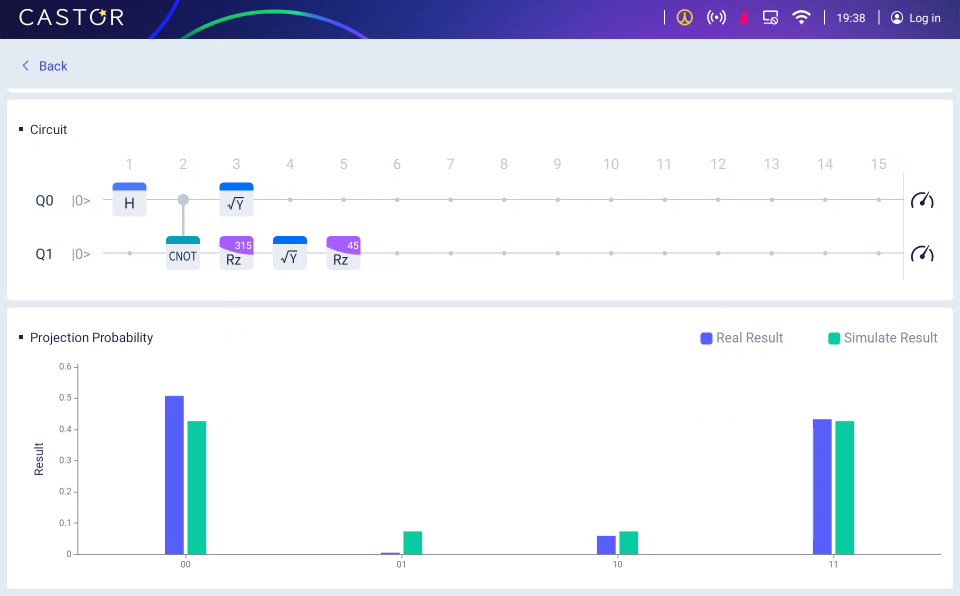}}
	\caption{The results for four sets of experiments.}\label{Bell_inequality}
\end{figure}

In each set of measurements, we record the probability distribution for the two qubits.
\begin{enumerate}
\item The results for the first set of experiments shown in Figure \ref{Bell_inequality1} are as follows:
\begin{equation*}
[ \rho_{1_{11}}, \rho_{1_{22}}, \rho_{1_{33}}, \rho_{1_{44}} ] = [0.053021, 0.440164, 0.385416, 0.121397],
\end{equation*}
\item The results for the second set of experiments shown in Figure \ref{Bell_inequality2} are as follows:
\begin{equation*}
[ \rho_{2_{11}}, \rho_{2_{22}}, \rho_{2_{33}}, \rho_{2_{44}} ] = [0.420338, 0.073635, 0.074921, 0.431103],
\end{equation*}
\item The results for the third set of experiments shown in Figure \ref{Bell_inequality3} are as follows:
\begin{equation*}
[ \rho_{3_{11}}, \rho_{3_{22}}, \rho_{3_{33}}, \rho_{3_{44}} ] = [0.4342, 0.091959, 0.065252, 0.408587],
\end{equation*}
\item The results for the fourth set of experiments shown in Figure \ref{Bell_inequality4} are as follows:
\begin{equation*}
[ \rho_{4_{11}}, \rho_{4_{22}}, \rho_{4_{33}}, \rho_{4_{44}} ] = [0.507454, 0.001001, 0.058879, 0.432663].
\end{equation*}
\end{enumerate}

Using these density matrix elements, we can calculate the Bell's inequality values:
\begin{align}
    \nonumber
    E_1 &= \rho_{1_{11}} - \rho_{1_{22}} - \rho_{1_{33}} + \rho_{1_{44}} = -0.651162, \\ \nonumber
    E_2 &= \rho_{2_{11}} - \rho_{2_{22}} - \rho_{2_{33}} + \rho_{2_{44}} = 0.702885, \\ \nonumber
    E_3 &= \rho_{3_{11}} - \rho_{3_{22}} - \rho_{3_{33}} + \rho_{3_{44}} = 0.685576, \\ 
    E_4 &= \rho_{4_{11}} - \rho_{4_{22}} - \rho_{4_{33}} + \rho_{4_{44}} = 0.880237. 
\end{align}
Bell's inequality predicts that for any theory based on local realism, the sum of the absolute values of these combined measurements should not exceed 2:
\begin{align}
\left| -E_1 + E_2 + E_3 + E_4 \right| \leq 2 .
\end{align}
However, our experimental results show:
\begin{align}
-E_1 + E_2 + E_3 + E_4 = 2.91986.
\end{align}
Even in the presence of noise in real quantum computers, this still exceeds the predictive range of Bell's inequality, thus supporting the non-locality of quantum mechanics and challenging the hypothesis of local hidden variable theories. This result is consistent with the nature of quantum entanglement and is in line with other similar experimental findings. Through these experiments, we further substantiate the predictions of quantum mechanics and deepen our understanding of the nature of physical reality.

\newpage

\chapter{Week 8: Building a Quantum Computer}
\section{Quantum gates and Hamiltonian}

The time evolution of a quantum system is governed by the Schrödinger equation. For a small $\Delta t$, this equation can be written as:
\begin{equation}
    |\psi(t + \Delta t)\rangle - |\psi(t)\rangle = -i\textbf{H}\Delta t|\psi(t)\rangle.
\end{equation}
Where \textbf{H} is the Hamiltonian operator representing the system's total energy. This differential equation can be integrated to give:
\begin{align}
	|\psi(t + \Delta t)\rangle = (I - i\textbf{H}\Delta t)|\psi(t)\rangle &\approx (\cos(\textbf{H}\Delta t) -i\sin(\textbf{H}\Delta t))|\psi(t)\rangle \nonumber\\
	\Rightarrow|\psi(t + \Delta t)\rangle &= e^{-i\textbf{H}\Delta t}|\psi(t)\rangle,
\end{align}
for the state evolution between times $t$ and $t + \Delta t$. Comparing with the general form:
\begin{equation}
    |\psi(t + \Delta t)\rangle = U|\psi(t)\rangle.
\end{equation}
We see that the unitary operator(quantum gate) $U$ associated with the evolution is given by:
\begin{equation}
    U = e^{-i\textbf{H}\Delta t}.
\end{equation}

The Hamiltonian (\textbf{H}) represents the total energy of a quantum system. It appears in the Schrödinger equation and determines the time evolution.\\
In classical physics, common Hamiltonian functions include:\\
(1) Kinetic energy: 
\begin{equation}
    \textbf{H} = \frac{p^2}{2m},
\end{equation}
(2) Gravitational potential energy:
\begin{equation}
    \textbf{H} = mgh,
\end{equation}
(3) Elastic potential energy:
\begin{equation}
    \textbf{H} = \frac{1}{2}kx^2.
\end{equation}

For a quantum particle with spin in a magnetic field $\vec{B}_0$, the Hamiltonian is:
\begin{equation}
    \textbf{H} = -\frac{1}{2}\hbar\gamma B_0 Z = -\frac{1}{2}\hbar\omega_0Z = 
\begin{bmatrix}
	-\frac{1}{2}\hbar \omega_0 & 0\\
	0 & \frac{1}{2}\hbar \omega_0
\end{bmatrix}.
\end{equation}

\section{Hermitian and unitary matrices}
An important class of matrices in quantum mechanics is symmetric matrix \cite{sun33taylor2001m}.
A matrix $A$ in $\mathbb{C}^{n\times n}$ is called symmetric if it satisfies:
\begin{equation}
    A^T = A.
\end{equation}
 Where $A^T$ represents the transpose of A.This means a symmetric matrix has the property:
\begin{equation}
    A[j,k] = A[k,j].
\end{equation}

Another important class of matrices in quantum mechanics is Hermitian matrix.
A matrix $A$ in $\mathbb{C}^{n\times n}$ is called Hermitian if it satisfies:
\begin{equation}
    A^\dagger = A.
\end{equation}
Where $A^\dagger$ represents the conjugate transpose of A. This means a Hermitian matrix has the property:
\begin{equation}
    A[j,k] = \overline{A[k,j]}.
\end{equation}

\begin{example}
The matrix
\begin{align}
\begin{bmatrix}
	9 & 3 - 8i & 4 - 3i\\
	3 + 8i & 5 & 18 + 2i\\
	4 + 3i & 18 - 2i & 3
\end{bmatrix}
\end{align}
is Hermitian.
\end{example}

\begin{example}
	Pauli Matrices
 \begin{equation}
    X = \begin{bmatrix} 0 & 1\\ 1 & 0 \end{bmatrix}, 
	Y = \begin{bmatrix} 0 & -i\\ i & 0 \end{bmatrix}, 
	Z = \begin{bmatrix} 1 & 0\\ 0 & -1 \end{bmatrix}
 \end{equation}
	are Hermitian.
\end{example}

Unitary matrices \cite{sun34li2013decomposition} are another fundamental class of matrices in quantum mechanics.
A matrix $U$ in $\mathbb{C}^{n\times n}$ is unitary if it satisfies:
\begin{equation}
    UU^{\dagger} = U^{\dagger}U = I.
\end{equation}
where $I$ is the identity matrix. 

\begin{example}
	Pauli Matrices
 \begin{equation}
    X = \begin{bmatrix} 0 & 1\\ 1 & 0 \end{bmatrix}, 
	Y = \begin{bmatrix} 0 & -i\\ i & 0 \end{bmatrix}, 
	Z = \begin{bmatrix} 1 & 0\\ 0 & -1 \end{bmatrix}
 \end{equation}
	are unitary and Hermitian.
\end{example}

An important relationship in quantum mechanics is that the matrix exponential of a Hermitian matrix yields a unitary matrix. Specifically, if $\textbf{H}$ is Hermitian, then:
    $e^{i\textbf{H}}$ is unitary, which can be verified by showing:
\begin{equation}
    (e^{i\textbf{H}})^\dagger e^{i\textbf{H}} = e^{-i\textbf{H}}e^{i\textbf{H}} = I,
\end{equation}
where $I$ is the identity matrix.

This relationship allows designing quantum logic gates of the form:
\begin{equation}
    U = e^{-i\textbf{H}\Delta t}.
\end{equation}
By engineering the Hermitian matrix \textbf{H} and evolution time $\Delta t$, we can produce the desired unitary transformation $U$. 

\section{The NMR system}

For NMR system, NMR largely satisfies the DiVincenzo criteria \cite{sun17perez2011quantum}. For example:
\begin{itemize}
\item Qubits: nuclear spin $\frac{1}{2}$ in $B_0$ field ($\uparrow$ and $\downarrow$ as 0 and 1).
\item Quantum gates: RF pulses and delay times.
\item Input: Boltzman distribution (room temperature).
\item Readout: detect spin states with RF coil.
\item Coherence times: easily several seconds.
\end{itemize}

\begin{figure}[H]
	\centering
	\subfigure[Single spin energy level difference.]{
		\includegraphics[scale=0.6]{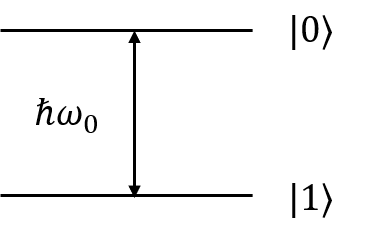}}
	\quad
	\subfigure[Single spin Bloch sphere.]{
		\includegraphics[scale=0.4]{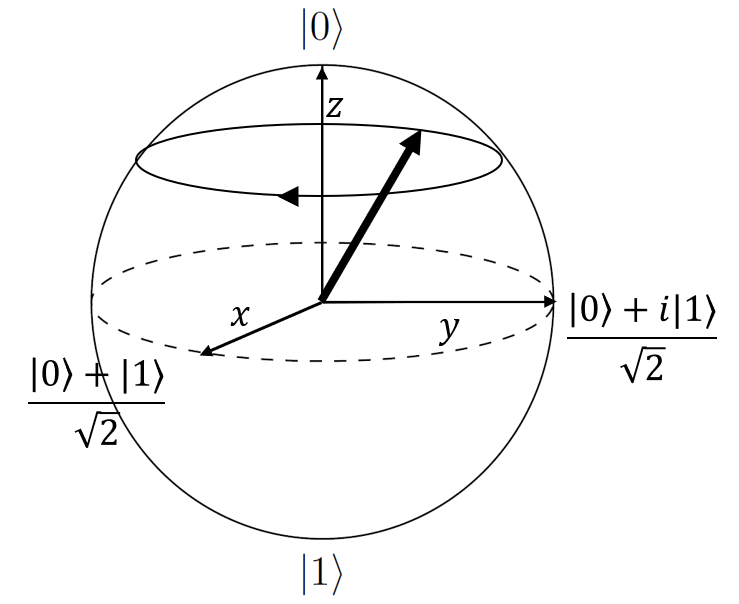}}
	\caption{The single spin system.}\label{18_single_spin}
\end{figure}

A fundamental quantum system is a single spin \cite{sun36jones2001nmr} shown in Figure \ref{18_single_spin} in an external magnetic field $\vec{B}_0$. The Hamiltonian for this models is the interaction energy between the spin and field:
\begin{equation}
    \textbf{H} = -\frac{1}{2}\hbar\gamma B_0 Z = -\frac{1}{2}\hbar\omega_0Z = 
\begin{bmatrix}
	-\frac{1}{2}\hbar \omega_0 & 0\\
	0 & \frac{1}{2}\hbar \omega_0
\end{bmatrix}.
\end{equation}

\begin{figure}[H]
	\centering
	\includegraphics[scale=0.5]{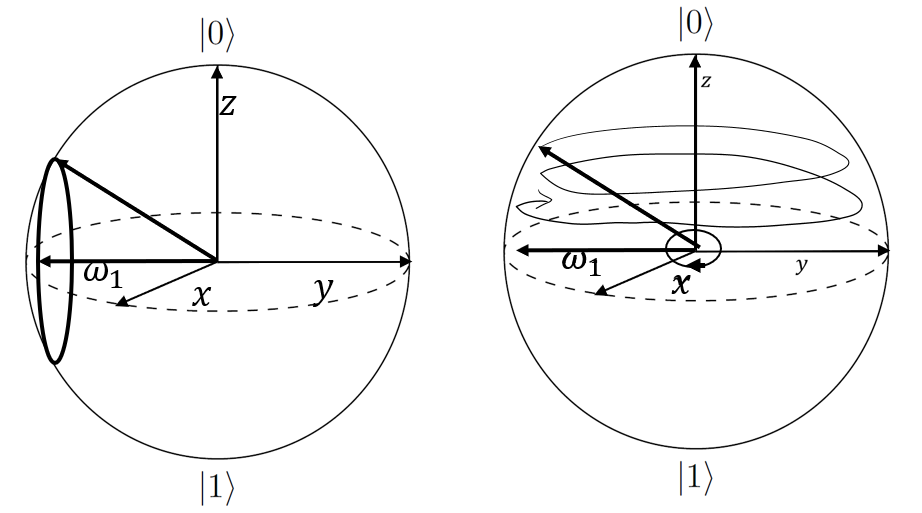}
	\caption{\label{18_control_H} The control Hamiltonian.}
\end{figure}

In quantum control, an additional controllable Hamiltonian \textbf{Hc} is applied to drive transitions and perform quantum gates \cite{sun37vandersypen2005nmr}. For a single spin system, this takes the form:
\begin{align}
\textbf{Hc} &= -\frac{1}{2}\hbar\omega_1(\cos(\phi)X + \sin(\phi)Y)\nonumber\\
&=-\frac{1}{2}\begin{bmatrix}
	0 & \cos(\phi)-i\sin(\phi)\\
	\cos(\phi)+i\sin(\phi) & 0
\end{bmatrix},
\end{align}
where $\omega_1$ is the control amplitude and $\phi$ is the control phase, both of which can be tuned experimentally shown in Figure \ref{18_control_H}. This applies a controllable rotation to the spin.

The time evolution operator generated by the control Hamiltonian for a duration $t_{pw}$ (the pulse width) is:
\begin{equation}
    \textbf{Hc}(t_{pw}) = e^{-i\textbf{Hc}t_{pw}} = e^{-i\frac{1}{2}\hbar\omega_1(\cos(\phi)X + \sin(\phi)Y)t_{pw}}.
\end{equation}

\begin{figure}[H]
	\centering
	\includegraphics[scale=0.3]{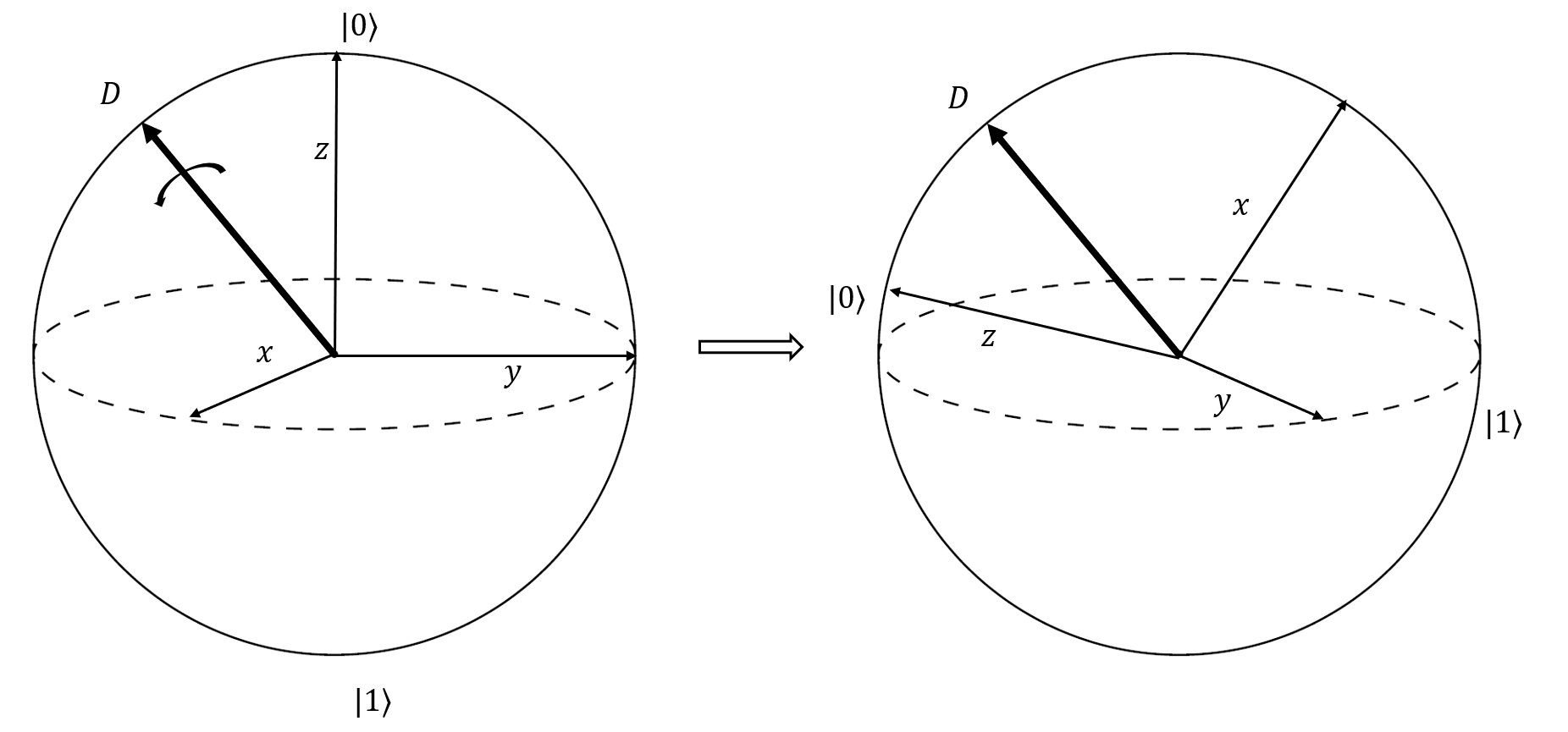}
	\caption{\label{17_rotation_D1} Rotation about the $D$ axis.}
\end{figure}

Common single-qubit quantum gates involve rotations generated by the Pauli matrices:
\begin{align}
	R_x(\theta) &= e^{-i\frac{\theta}{2}X} = \cos(\frac{\theta}{2})I - i\sin(\frac{\theta}{2})X,\nonumber\\
	R_y(\theta) &= e^{-i\frac{\theta}{2}Y} = \cos(\frac{\theta}{2})I - i\sin(\frac{\theta}{2})Y,\nonumber\\
	R_z(\theta) &= e^{-i\frac{\theta}{2}Z} = \cos(\frac{\theta}{2})I - i\sin(\frac{\theta}{2})Z,
\end{align}
which perform rotations about the x, y and z axes of the Bloch sphere. More generally, an arbitrary single-qubit rotation is given by:
\begin{equation}
    R_D(\theta) = \cos \frac{\theta}{2} I - i\sin \frac{\theta}{2} (D_x X+D_y Y+D_z Z) = \exp(-i\frac{\theta}{2}\vec{D}\cdot\vec{\sigma}),
\end{equation}
where $\vec{D} = (D_x, D_y, D_z)$ is the rotation axis shown in Figure \ref{17_rotation_D1}.

An important result is that any single-qubit unitary operator can be expressed in terms of rotations generated by the Pauli matrices. Specifically, for a $2\times2$ unitary $U$, there exist real parameters $\alpha, \beta, \gamma, \delta$ such that:
\begin{equation}
    U = e^{i\alpha} R_z(\beta) R_y(\gamma) R_z(\delta).
\end{equation}
where $R_i(\theta) = \exp(-i\frac{\theta}{2}\sigma_i)$ represents rotation about the $i$-axis. This gives:
\begin{equation}
    U=\begin{bmatrix} e^{i(\alpha-\frac{\beta}{2}-\frac{\delta}{2})}\cos(\frac{\gamma}{2}) & -e^{i(\alpha-\frac{\beta}{2}+\frac{\delta}{2})}\sin(\frac{\gamma}{2}) \\ e^{i(\alpha+\frac{\beta}{2}-\frac{\delta}{2})}\sin(\frac{\gamma}{2}) & e^{i(\alpha+\frac{\beta}{2}+\frac{\delta}{2})}\cos(\frac{\gamma}{2}) \end{bmatrix}.
\end{equation}
An alternative parameterization is:
\begin{equation}
    U = e^{i\alpha} R_x(\beta) R_y(\gamma) R_x(\delta).
\end{equation}
The interaction between two qubits can be described by the coupling Hamiltonian:
\begin{equation}
    {\textbf{H}_{J}} = \hbar\sum_{i<j}\frac{\pi}{2}J_{ij}Z_{i}Z_{j},
    \label{Hamiltonian}
\end{equation}
where $J_{ij}$ is the coupling strength between qubits $i$ and $j$.
For two qubits, the time evolution under the coupling Hamiltonian is given by:

\begin{equation}
	U_{J}(t) = e^{-i\hbar \frac{\pi}{2}JtZ\otimes Z},
\label{Uj}
\end{equation}
which can be explicitly written in matrix form as:
\begin{equation}
	U_{J}(t) = \begin{bmatrix}
		e^{-i\frac{\pi}{2}\hbar Jt} & 0 & 0 & 0 \\
		0 & e^{+i\frac{\pi}{2}\hbar Jt} & 0 & 0 \\
		0 & 0 & e^{+i\frac{\pi}{2}\hbar Jt} & 0 \\
		0 & 0 & 0 & e^{-i\frac{\pi}{2}\hbar Jt}
	\end{bmatrix}.
\end{equation}
At the special time $t = \frac{1}{2\hbar J}$, this becomes:
\begin{equation}
	U_{J}\left(\frac{1}{2\hbar J}\right) = e^{-i\frac{\pi}{4}Z\otimes Z} = \begin{bmatrix}
		e^{-i\frac{\pi}{4}} & 0 & 0 & 0 \\
		0 & e^{+i\frac{\pi}{4}} & 0 & 0 \\
		0 & 0 & e^{+i\frac{\pi}{4}} & 0 \\
		0 & 0 & 0 & e^{-i\frac{\pi}{4}}
	\end{bmatrix}.
\end{equation}

The controlled-Z gate can be implemented between two qubits by evolving under the coupling Hamiltonian for a specific time. First we define the single-qubit $S$ gate:
\begin{equation}
	S = \begin{bmatrix} 1 & 0 \\ 0 & i \end{bmatrix}.
\end{equation}
Then the controlled-Z(CZ) gate is given by:
\begin{equation}
	e^{i\frac{\pi}{4}}(S^{\dagger} \otimes S^{\dagger}) U_{J}\left(\frac{1}{2\hbar J}\right) =
	\begin{bmatrix}
		1 & 0 & 0 & 0 \\
		0 & 1 & 0 & 0\\
		0 & 0 & 1 & 0\\
		0 & 0 & 0 & -1
	\end{bmatrix}.
 \label{CZcompose}
\end{equation}\
\begin{example}
Here's the MATLAB code for implementing CZ gate:
\begin{lstlisting}
clear,clc;

% Define the matrix U_J at the special time t = 1/(2*hbar*J)
U_J = diag([exp(-1i*pi/4), exp(1i*pi/4), exp(1i*pi/4), exp(-1i*pi/4)]);

% Define the single-qubit S gate
S = [1 0; 0 1i];

% Calculate the adjoint (dagger) of S
S_dagger = [1 0; 0 -1i];

% Calculate the tensor product of S_dagger with itself
S_dagger_tensor_S_dagger = kron(S_dagger, S_dagger);

% Multiply by a global phase factor exp(i*pi/4)
global_phase_factor = exp(1i*pi/4);

% Implement the controlled-Z gate
CZ = global_phase_factor * S_dagger_tensor_S_dagger * U_J;

% Display the CZ gate
disp('The controlled-Z (CZ) gate:');
disp(CZ);
\end{lstlisting}
\end{example}

The controlled-NOT (CNOT) gate is another important entangling two-qubit gate. It can be constructed as:
\begin{equation}
	e^{i\frac{\pi}{4}}(I \otimes H)(S^{\dagger} \otimes S^{\dagger})U_{J}\left(\frac{1}{2\hbar J}\right)(I \otimes H) =
	\begin{bmatrix}
		1 & 0 & 0 & 0\\
		0 & 1 & 0 & 0\\
		0 & 0 & 0 & 1\\
		0 & 0 & 1 & 0
	\end{bmatrix},
 \label{CNOTcompose1}
\end{equation}
where $H$ is the Hadamard gate and $U_{J}(t)$ is the two-qubit coupling evolution. By appropriately choosing the interaction time and flanking the coupling evolution with single-qubit gates, we obtain the desired CNOT gate.
\begin{example}
Here's the MATLAB code for implementing CNOT gate:
\begin{lstlisting}
clear,clc;

% Define the matrix U_J at the special time t = 1/(2*hbar*J)
U_J = diag([exp(-1i*pi/4), exp(1i*pi/4), exp(1i*pi/4), exp(-1i*pi/4)]);

% Define the single-qubit S gate and its adjoint (dagger)
S = [1 0; 0 1i];
S_dagger = [1 0; 0 -1i];

% Define the Hadamard gate
H = (1/sqrt(2)) * [1 1; 1 -1];

% Calculate the tensor product of S_dagger with itself
S_dagger_tensor_S_dagger = kron(S_dagger, S_dagger);

% Calculate the tensor product of the identity I and Hadamard H
I_tensor_H = kron(eye(2), H);

% Apply global phase factor exp(i*pi/4)
global_phase_factor = exp(1i*pi/4);

% Implement the CNOT gate
CNOT = global_phase_factor * I_tensor_H * S_dagger_tensor_S_dagger * U_J * I_tensor_H;

% Display the CNOT gate
disp('The controlled-NOT (CNOT) gate:');
disp(CNOT);
\end{lstlisting}
\end{example}

\section{Experiment: realization of CNOT gate on a real quantum computer}

From the previous section, we  see that CNOT gate can be constructed. Here, we will see how the CNOT gate is realized on an NMR quantum computer.
Note that this experiment is realized on SpinQ Gemini \cite{sun38hou2021spinq}.

\begin{figure}[H]
	\centering
	\includegraphics[scale=0.5]{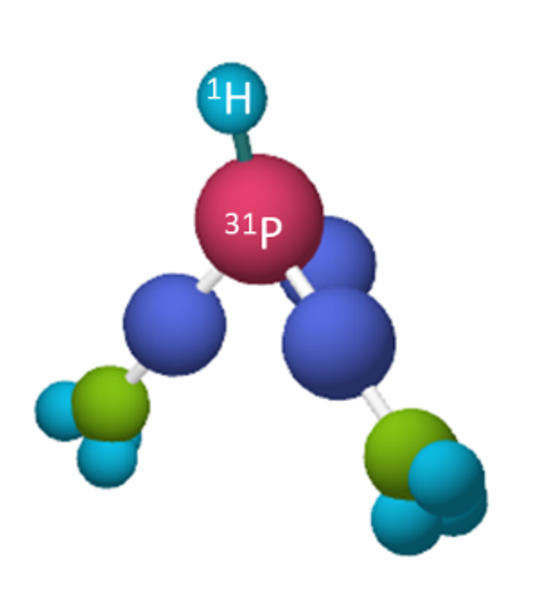}
	\caption{\label{molecule_qubit} In a quantum computer, the molecule used as qubits, where P is the second qubit and H is the first qubit.}
\end{figure}

In order to achieve a two-qubit CNOT gate where it is required to use the state of a qubit to control the state of the other qubit, these two qubits need to interact with each other. In the NMR system, qubits are nuclear spins \cite{sun36jones2001nmr}. There is the $J$ coupling interaction between two nuclear spins. Such coupling is carried by the shared electrons in the chemical bond between atoms, so the closer these two nuclear spins, the greater the $J$ coupling.
The molecule we use in the experiment is dimethyl phosphite shown in Figure \ref{molecule_qubit}, where P and H are used as two qubits. The coupling between these two nuclei is about 697.4 Hz.

For a two-qubit gate, it is the $J$ coupling interaction between nuclear spins that works \cite{4_sx}. In NMR, the implementation for the two-qubit gate is mentioned in Equation \ref{Hamiltonian} and Equation \ref{Uj}. In addition, CNOT gate can be achieved by combining $U_J$ and some RF pulses:
\begin{equation}
\text{CNOT} = e^{i\frac{\pi}{4}}R_{x}^{1}(90)R_{y}^{1}(90)R_{-x}^{1}(90)R_{x}^{2}(90)R_{-y}^{2}(90)U_{J}(\frac{1}{2J})R_{y}^{2}(90),
\label{cnot}
\end{equation}
where $U_{J}(\frac{1}{2J})$ refers to the free evolution under the effect of $J$ coupling for a duration of $1/2J$, which is about 717 \textmu s for our system. $R_y^2 (90)$ refers to the pulse applied on the second qubit (P) to rotate it around y axis by 90\textdegree. $R_{-x}^1 (90)$ refers to the pulse applied on the first qubit (H) to rotate it around x axis by 90\textdegree.

\begin{figure}[H]
	\centering
	\includegraphics[scale=0.2]{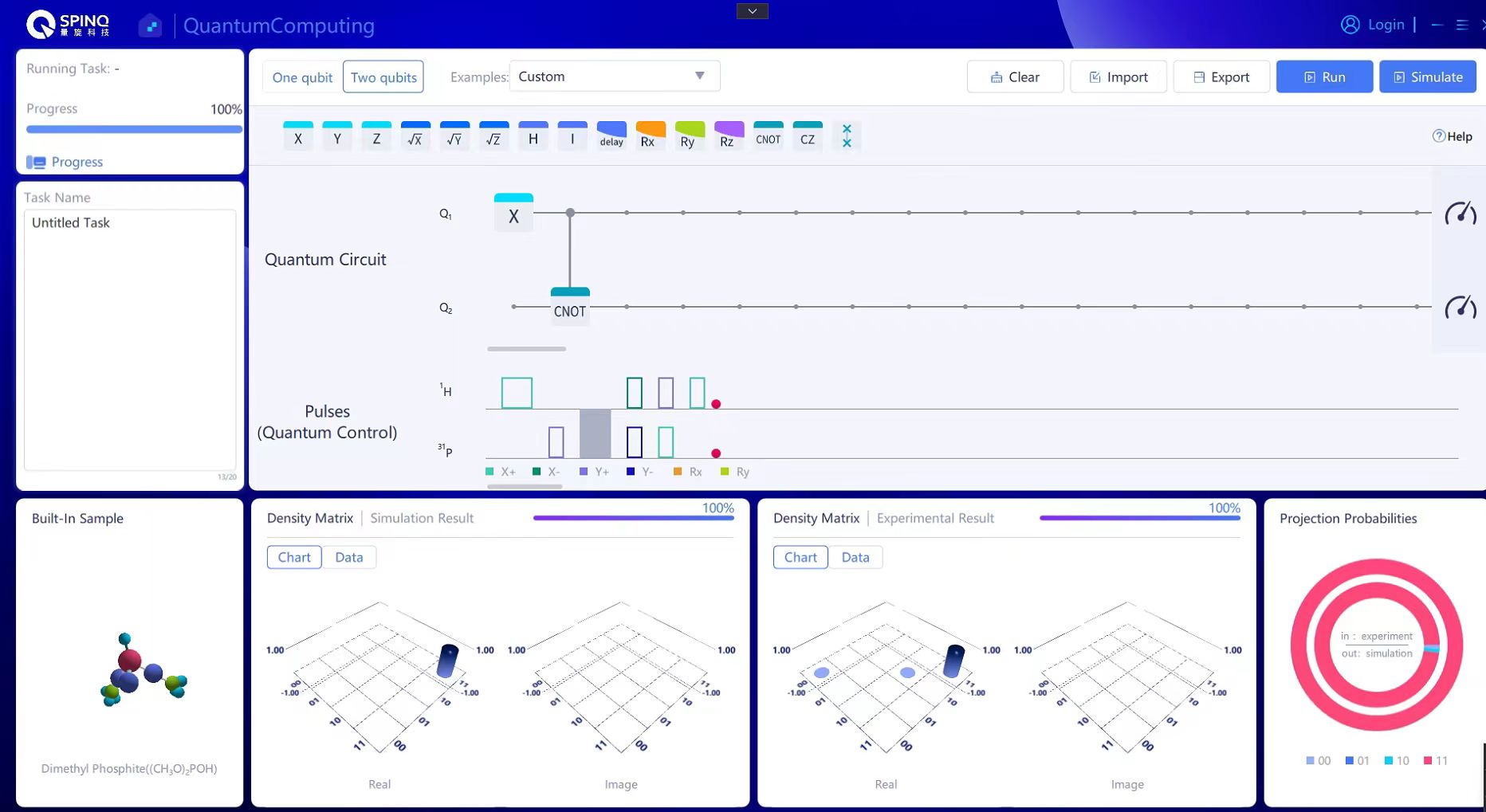}
	\caption{\label{cnot10} The experimental result of applying the CNOT gate pulse sequence to the state $|10\rangle$ on SpinQ Gemini.}
\end{figure}

\begin{figure}[H]
	\centering
	\includegraphics[scale=0.6]{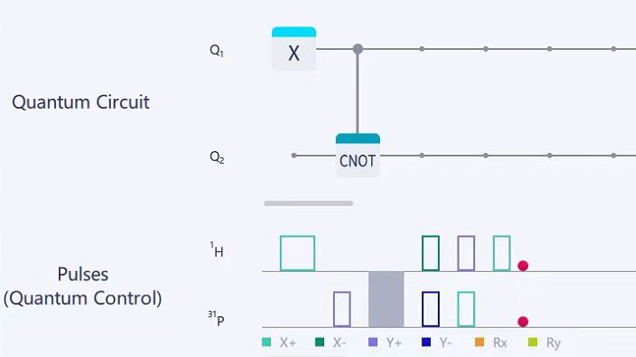}
	\caption{\label{pulse} The composite pulse sequence that constitutes the CNOT gate.}
\end{figure}

Then we implement the CNOT gate constructed from the pulse sequence of Equation \ref{cnot} on SpinQ Gemini. We choose to verify the result of the CNOT gate on the state $|10\rangle$, as shown in Figure \ref{cnot10}.
Figure \ref{pulse} displays a schematic of a quantum circuit along with the corresponding sequence of quantum control pulses. In the quantum circuit part, we can observe two qubits, designated as \( Q_1 \) and \( Q_2 \). At the beginning of the circuit, an X gate is applied to the \( Q_1 \) qubit, which is an operation that corresponds to a rotation of 180 degrees about the X-axis. This operation is represented by the first light green rectangle in the pulse sequence, and it changes the quantum state to \( |10\rangle \).
Subsequently, a Controlled-NOT gate (CNOT gate) is applied to the qubits \( Q_1 \) and \( Q_2 \). In the CNOT operation, the first qubit acts as the control qubit, while the second qubit is the target qubit. In the part illustrating the pulse sequence, there is a series of rectangles representing different pulse operations, with rectangles of different colors denoting rotations about different axes. As shown in Figure \ref{pulse}, initially there is a light blue rectangle on the second qubit, indicating a rotation of 90 degrees about the Y-axis in the positive direction; followed by a gray section, representing a specific interaction time corresponding to the coupling between the two quantum bits, here indicating a delay time of 717 $\mu$s; then there is a dark green rectangle on the first qubit, indicating a rotation of 90 degrees about the X-axis in the negative direction, and a dark blue rectangle on the second qubit, indicating a rotation of 90 degrees about the Y-axis in the negative direction; followed by a light blue rectangle on the first qubit, indicating a rotation of 90 degrees about the Y-axis in the positive direction, and a light green rectangle on the second qubit, indicating a rotation of 90 degrees about the X-axis in the positive direction; and finally, a light green rectangle on the first qubit, indicating a rotation of 90 degrees about the X-axis in the positive direction. These rectangles of different colors represent the various pulse operations that implement the CNOT gate.

As shown in the results in Figure \ref{cnot10}, the state $|10\rangle$ indeed changes to the state $|11\rangle$ after these pulses. In other words, this pulse sequence indeed constitutes a CNOT gate. 
Students can also perform similar experiments on SpinQ Gemini, 
applying the CNOT gate pulse sequence to the states $|00\rangle$, $|01\rangle$, $|10\rangle$, $|11\rangle$ and then 
compare the experimental results with the simulation results after running the experiments.

\bibliographystyle{plain}
\bibliography{sample.bib}